\documentclass[notitlepage,aps,pra,showpacs,amsmath,amssymb,amsfonts,lengthcheck,onecolumn, superscriptaddress]{revtex4-1}

\usepackage{dsfont}
\usepackage{graphicx} 
\usepackage[justification=raggedright,singlelinecheck=false]{caption}
\usepackage{graphics}
\usepackage{mathtools}
\usepackage{dcolumn}
\usepackage{bm}
\usepackage{graphicx}
\usepackage{amsmath}
\usepackage{latexsym}
\usepackage{amsfonts}
\usepackage{amssymb}
\usepackage{array}
\usepackage{epsfig}
\usepackage{txfonts}
\usepackage{xcolor}
\usepackage[colorlinks=true,linkcolor=blue,urlcolor=blue,citecolor=blue,pdfusetitle]{hyperref}
\usepackage{hyperref}
\usepackage[italicdiff]{physics}
\usepackage{mathtools}
\usepackage{siunitx}
\usepackage{multirow}

\newcommand{\subfigref}[2]{\ref{#1}\hyperref[#1]{(#2)}}

\DeclareMathOperator{\mvar}{Var}

\begin{document}
\title{Roadmap on Quantum Thermodynamics}


\author{Steve Campbell}
\address{School of Physics, University College Dublin, Belfield, Dublin 4, Ireland}
\address{Centre for Quantum Engineering, Science, and Technology, University College Dublin, Dublin 4, Ireland}

\author{Irene D'Amico}
\address{School of Physics, Engineering and Technology, University of York, York, YO10 5DD, United Kingdom}
\address{York Centre for Quantum Technologies, University of York, York, YO10 5DD, United Kingdom}

\author{Mario A. Ciampini}
\address{University of Vienna, Faculty of Physics, Vienna Center for Quantum Science and Technology (VCQ), Boltzmanngasse 5, A-1090 Vienna, Austria}

\author{Janet Anders}
\address{Institute of Physics and Astronomy, University of Potsdam, 14476 Potsdam, Germany}
\address{Department of Physics and Astronomy, University of Exeter, Exeter EX4 4QL, United Kingdom}

\author{Natalia Ares}
\address{Department of Engineering Science, University of Oxford, Parks Road, Oxford OX1 3PJ, United Kingdom}

\author{Simone Artini}
\address{Universit\`a degli Studi di Palermo, Dipartimento di Fisica e Chimica - Emilio Segr\`e, via Archirafi 36, I-90123 Palermo, Italy}

\author{Alexia Auff\`{e}ves}
\address{MajuLab, CNRS-UCA-SU-NUS-NTU International Joint Research Laboratory}
\address{Centre for Quantum Technologies, National University of Singapore, 117543 Singapore, Singapore}

\author{Lindsay Bassman Oftelie}
\address{NEST, Istituto Nanoscienze-CNR and Scuola Normale Superiore, 56127 Pisa, Italy}

\author{Laetitia P. Bettmann}
\address{School of Physics, Trinity College Dublin, College Green, Dublin 2, D02K8N4, Ireland}

\author{Marcus V. S. Bonan\c{c}a}
\address{Gleb Wataghin Physics Institute, The University of Campinas, 13083-859, Campinas, S\~{a}o Paulo, Brazil}

\author{Thomas Busch}
\address{Quantum Systems Unit, OIST Graduate University, Onna, Okinawa 904-0495, Japan}

\author{Michele Campisi}
\address{NEST, Istituto Nanoscienze-CNR and Scuola Normale Superiore, 56127 Pisa, Italy}

\author{Moallison F. Cavalcante}
\address{Department of Physics, University of Maryland, Baltimore County, Baltimore, MD 21250, USA}
\address{Quantum Science Institute, University of Maryland, Baltimore County, Baltimore, MD 21250, USA}
\address{Gleb Wataghin Physics Institute, The University of Campinas, 13083-859, Campinas, S\~{a}o Paulo, Brazil}

\author{Luis A. Correa}
\address{Instituto Universitario de Estudios Avanzados (IUdEA), Universidad de La Laguna, La Laguna 38203, Spain}
\address{Departamento de Física, Universidad de La Laguna, La Laguna 38203, Spain}

\author{Eloisa Cuestas}
\address{Quantum Systems Unit, OIST Graduate University, Onna, Okinawa 904-0495, Japan}
\address{Forschungszentrum J\"{u}lich, Institute of Quantum Control (PGI-8), D-52425 J\"{u}lich, Germany}

\author{Ceren B.~Dag}
\address{Department of Physics, Indiana University, Bloomington, Indiana 47405, USA}
\address{Department of Physics, Harvard University, 17 Oxford Street Cambridge, MA 02138, USA}

\author{Salamb\^{o} Dago}
\address{University of Vienna, Faculty of Physics, Vienna Center for Quantum Science and Technology (VCQ), Boltzmanngasse 5, A-1090 Vienna, Austria}

\author{Sebastian Deffner}
\address{Department of Physics, University of Maryland, Baltimore County, Baltimore, MD 21250, USA}
\address{Quantum Science Institute, University of Maryland, Baltimore County, Baltimore, MD 21250, USA}
\address{National Quantum Laboratory, College Park, MD 20740, USA}

\author{Adolfo Del Campo}
\address{Department of Physics and Materials Science, University of Luxembourg, L-1511 Luxembourg, Luxembourg}
\address{Donostia International Physics Center, E-20018 San Sebastian, Spain}

\author{Andreas Deutschmann-Olek}
\address{Automation and Control Institute, TU Wien, A-1040, Vienna, Austria}

\author{Sandro Donadi}
\address{Dipartimento di Ingegneria, Universit\`a degli Studi di Palermo, Viale delle Scienze, 90128 Palermo, Italy}

\author{Emery Doucet}
\address{Department of Physics, University of Maryland, Baltimore County, Baltimore, MD 21250, USA}
\address{Quantum Science Institute, University of Maryland, Baltimore County, Baltimore, MD 21250, USA}

\author{Cyril Elouard}
\address{Universit\'{e} de Lorraine, CNRS, LPCT, F-54000 Nancy, France}

\author{Klaus Ensslin}
\affiliation{Solid State Physics Laboratory and Quantum Center, ETH Zurich, 8093 Zurich, Switzerland}

\author{Paul Erker}
\address{Atominstitut, TU Wien, 1020 Vienna, Austria}
\address{IQOQI Vienna, Austrian Academy of Sciences, Boltzmanngasse 3, 1090 Vienna, Austria}

\author{Nicole Fabbri}
\address{Istituto Nazionale di Ottica del Consiglio Nazionale delle Ricerche (CNR-INO), and European Laboratory for Non-linear Spectroscopy (LENS), Università di Firenze, via Nello Carrara 1, I-50019, Sesto Fiorentino, Italy}

\author{Federico Fedele}
\address{Department of Engineering Science, University of Oxford, Parks Road, Oxford OX1 3PJ, United Kingdom}

\author{Guilherme Fiusa}
\address{Department of Physics and Astronomy, Center for Coherence and Quantum Science, University of Rochester, Rochester, New York 14627, USA}

\author{Thom\'{a}s Fogarty}
\address{Quantum Systems Unit, OIST Graduate University, Onna, Okinawa 904-0495, Japan}

\author{Joshua Folk}
\affiliation{Quantum Matter Institute, University of British Columbia, Vancouver, British Columbia, V6T 1Z4, Canada}
\affiliation{Department of Physics and Astronomy, University of British Columbia, Vancouver, British Columbia, V6T 1Z1, Canada}

\author{Giacomo Guarnieri}
\address{Department of Physics ``A. Volta", University of Pavia, Via Bassi 6, 27100, Pavia, Italy}
\address{INFN Sezione di Pavia, Via Agostino Bassi 6, I-27100, Pavia, Italy}

\author{Abhaya S. Hegde}
\address{Department of Physics and Astronomy, Center for Coherence and Quantum Science, University of Rochester, Rochester, New York 14627, USA}

\author{Santiago Hern\'{a}ndez-G\'{o}mez}
\address{Istituto Nazionale di Ottica del Consiglio Nazionale delle Ricerche (CNR-INO), and European Laboratory for Non-linear Spectroscopy (LENS), Università di Firenze, via Nello Carrara 1, I-50019, Sesto Fiorentino, Italy}

\author{Chang-Kang Hu}
\address{Shenzhen Institute for Quantum Science and Engineering, Southern University of Science and Technology, Shenzhen, Guangdong 518055, China}
\address{Guangdong Provincial Key Laboratory of Quantum Science and Engineering, Southern University of Science and Technology, Shenzhen, Guangdong 518055, China}
\address{International Quantum Academy, Futian District, Shenzhen, Guangdong 518048, China}

\author{Fernando Iemini}
\address{Instituto de F\'isica, Universidade Federal Fluminense, Av. Gal. Milton Tavares de Souza s/n, Gragoat\'a, 24210-346 Niter\'oi, Rio de Janeiro, Brazil}

\author{Bayan Karimi}
\address{Pritzker School of Molecular Engineering, University of Chicago, Chicago IL 60637, USA}
\address{PICO group, QTF Centre of Excellence, Department of Applied Physics, Aalto University, P.O. Box 15100, FI-00076 Aalto, Finland}

\author{Nikolai Kiesel}
\address{University of Vienna, Faculty of Physics, Vienna Center for Quantum Science and Technology (VCQ), Boltzmanngasse 5, A-1090 Vienna, Austria}

\author{Gabriel T. Landi}
\address{Department of Physics and Astronomy, Center for Coherence and Quantum Science, University of Rochester, Rochester, New York 14627, USA}

\author{Aleksander Lasek}
\address{Joint Center for Quantum Information and Computer Science, NIST and University of Maryland, College Park, Maryland 20742, USA}

\author{Sergei Lemziakov}
\address{PICO group, QTF Centre of Excellence, Department of Applied Physics, Aalto University, P.O. Box 15100, FI-00076 Aalto, Finland}

\author{Gabriele Lo Monaco}
\address{Universit\`a degli Studi di Palermo, Dipartimento di Fisica e Chimica - Emilio Segr\`e, via Archirafi 36, I-90123 Palermo, Italy}

\author{Eric Lutz}
\address{Institute for Theoretical Physics I, University of Stuttgart, D-70550 Stuttgart, Germany}

\author{Dmitrii Lvov}
\address{PICO group, QTF Centre of Excellence, Department of Applied Physics, Aalto University, P.O. Box 15100, FI-00076 Aalto, Finland}

\author{Olivier Maillet}
\address{Universit\'e Paris-Saclay, CEA, CNRS, SPEC, 91191 Gif-sur-Yvette, France}

\author{Mohammad Mehboudi}
\address{Atominstitut, TU Wien, 1020 Vienna, Austria}

\author{Taysa M. Mendon\c{c}a}
\address{Instituto de Física de São Carlos, Universidade de São Paulo, CP 369, 13560-970 São Carlos, São Paulo, Brazil}

\author{Harry J. D. Miller}
\address{Department of Physics and Astronomy, University of Manchester, Oxford Road, Manchester M13 9PL, United Kingdom}

\author{Andrew K. Mitchell}
\address{School of Physics, University College Dublin, Belfield, Dublin 4, Ireland}
\address{Centre for Quantum Engineering, Science, and Technology, University College Dublin, Dublin 4, Ireland}

\author{Mark T. Mitchison}
\address{School of Physics, Trinity College Dublin, College Green, Dublin 2, Ireland}
\address{Department of Physics, King’s College London, Strand, London, WC2R 2LS, United Kingdom}

\author{Victor Mukherjee}
\address{Department of Physical Sciences, IISER Berhampur, Berhampur 760003, India}

\author{Mauro Paternostro}
\address{Universit\`a degli Studi di Palermo, Dipartimento di Fisica e Chimica - Emilio Segr\`e, via Archirafi 36, I-90123 Palermo, Italy}
\address{Centre for Quantum Materials and Technologies, School of Mathematics and Physics,
Queen’s University Belfast, BT7 1NN, United Kingdom}

\author{Jukka Pekola}
\address{PICO group, QTF Centre of Excellence, Department of Applied Physics, Aalto University, P.O. Box 15100, FI-00076 Aalto, Finland}

\author{Mart\'{i} Perarnau-Llobet}
\address{F\'isica Te\`orica: Informaci\'o i Fen\`omens Qu\`antics, Department de F\'isica, Universitat Aut\`onoma de Barcelona, 08193 Bellaterra (Barcelona), Spain}

\author{Ulrich Poschinger}
\address{QUANTUM, Institut f\"{u}r Physik, Universit\"{a}t Mainz, D-55128 Mainz, Germany}

\author{Alberto Rolandi}
\address{Atominstitut, TU Wien, 1020 Vienna, Austria}

\author{Dario Rosa}
\address{ICTP South American Institute for Fundamental Research, Instituto de F\'{i}sica Te\'{o}rica, UNESP - Univ. Estadual Paulista, Rua Dr. Bento Teobaldo Ferraz 271, 01140-070, S\~{a}o Paulo, SP, Brazil}

\author{Rafael S\'anchez}
\address{Departamento de Física Te\'orica de la Materia Condensada, Condensed Matter Physics Center (IFIMAC), and Instituto Nicol\'as Cabrera (INC), Universidad Aut\'onoma de Madrid, 28049 Madrid, Spain\looseness=-1}

\author{Alan C. Santos}
\address{Instituto de Física Fundamental, Consejo Superior de Investigaciones Científicas, Calle Serrano 113b, 28006 Madrid, Spain}

\author{Roberto S. Sarthour}
\address{Centro Brasileiro de Pesquisas F\'isicas, 22290-180, Rio de Janeiro, RJ, Brazil}

\author{Eran Sela}
\affiliation{Raymond and Beverly Sackler School of Physics and Astronomy, Tel Aviv University, Tel Aviv 69978, Israel}

\author{Andrea Solfanelli}
\address{Max Planck Institute for the Physics of Complex Systems, Nöthnitzer Str. 38, 01187 Dresden, Germany}

\author{Alexandre M. Souza}
\address{Centro Brasileiro de Pesquisas F\'isicas, 22290-180, Rio de Janeiro, RJ, Brazil}

\author{Janine Splettstoesser}
\address{Department of Microtechnology and Nanoscience (MC2), Chalmers University of Technology, S-412 96 G\"oteborg, Sweden}

\author{Dian Tan}
\address{Shenzhen Institute for Quantum Science and Engineering, Southern University of Science and Technology, Shenzhen, Guangdong 518055, China}
\address{Guangdong Provincial Key Laboratory of Quantum Science and Engineering, Southern University of Science and Technology, Shenzhen, Guangdong 518055, China}
\address{International Quantum Academy, Futian District, Shenzhen, Guangdong 518048, China}
\address{Shenzhen Branch, Hefei National Laboratory, Shenzhen 518048, China}

\author{Ludovico Tesser}
\address{Department of Microtechnology and Nanoscience (MC2), Chalmers University of Technology, S-412 96 G\"oteborg, Sweden}

\author{Tan Van Vu}
\address{Center for Gravitational Physics and Quantum Information, Yukawa Institute for Theoretical Physics, Kyoto University, Kitashirakawa Oiwakecho, Sakyo-ku, Kyoto 606-8502, Japan}

\author{Artur Widera}
\address{Department of Physics and Research Center OPTIMAS, RPTU Kaiserslautern-Landau, Kaiserslautern, Germany}

\author{Nicole Yunger Halpern}
\address{Joint Center for Quantum Information and Computer Science, NIST and University of Maryland, College Park, Maryland 20742, USA}
\address{Institute for Physical Science and Technology, University of Maryland, College Park, MD 20742, USA}

\author{Krissia Zawadzki}
\address{Instituto de Física de São Carlos, Universidade de São Paulo, CP 369, 13560-970 São Carlos, São Paulo, Brazil}

\begin{abstract}
\vskip1cm
The last two decades have seen quantum thermodynamics become a well established field of research in its own right. In that time, it has demonstrated a remarkably broad applicability, ranging from providing foundational advances in the understanding of how thermodynamic principles apply at the nano-scale and in the presence of quantum coherence, to providing a guiding framework for the development of efficient quantum devices. Exquisite levels of control have allowed state-of-the-art experimental platforms to explore energetics and thermodynamics at the smallest scales which has in turn helped to drive theoretical advances. 
This Roadmap provides an overview of the recent developments across many of the field's sub-disciplines, assessing the key challenges and future prospects, providing a guide for its near term progress.
\end{abstract}

\maketitle

\clearpage

\tableofcontents

\section{Quantum thermodynamics in the 21$^{\text{st}}$ century}
\noindent
{\it Steve Campbell}

\noindent
{School of Physics, University College Dublin, Belfield, Dublin 4, Ireland, and}\\
{Centre for Quantum Engineering, Science, and Technology, University College Dublin, Dublin 4, Ireland}\\

\noindent
{\it Irene D'Amico}

\noindent
{School of Physics, Engineering and Technology, University of York, York, YO10 5DD, United Kingdom, and}\\
{York Centre for Quantum Technologies, University of York, York, YO10 5DD, United Kingdom}\\

\noindent
{\it Mario Ciampini}

\noindent
{University of Vienna, Faculty of Physics, Vienna Center for Quantum Science and Technology (VCQ), Boltzmanngasse 5, A-1090 Vienna, Austria}\\

\paragraph*{State-of-the-art.}
As a physical theory, thermodynamics is somewhat unique as its development was largely driven by practicality, and in particular, the need to optimise the efficiency of the first thermal machines which appeared in the late 19th and early 20th centuries. The theory is elegantly captured by the four basic laws of thermodynamics, whose range of applicability seems almost limitless. A seminal application (and one which arguably could be considered the start of quantum thermodynamics) came when Scovil and Schultz-DuBois examined the energetics of a three-level maser, establishing that its operation could be understood as a microscopic heat engine~\cite{Scovil1959}. While insightful, this paper remained hidden for decades, being academically interesting but of little practical value. The 21st century changed this. Efforts over the last few decades have established that the laws of thermodynamics hold at all scales, with remarkable demonstrations including using single atoms and ions as working substances in nanoscale heat engines~\cite{deffner2019qtddevices}. Technological progress has been marked by the rapid miniaturisation and digitisation of our world, bolstered by proposals for new devices to process, encode, and distribute information in fundamentally new ways. These devices are based on quantum systems operating out-of-equilibrium, with temperature affecting their functionalities. These transformative advances have led to a renewed interest in exploring the applicability of the core thermodynamic concepts at the nanoscale. This has ultimately led to the maturation of the vibrant field of quantum thermodynamics~\cite{Esposito2009, campisi2011colloquium, Vinjanampathy2016, Landi2021, Arrachea2023, DeChiaraPRXQ, StrasbergBook}. A comprehensive overview of the current state of the field was collated by the community in Ref.~\cite{binder2019bookqtd} and demonstrates the wide ranging impact of quantum thermodynamics.

The field has rapidly developed in the last 20 years. In a somewhat fitting reversal of roles, a sizable body of work started by considering how the now well-understood classical heat engines and thermodynamic cycles must be revisited when these devices are brought down to the nanoscale~\cite{deffner2019qtddevices, Mitchison2019, Mukherjee2021, Cangemi2024}. Concepts in classical thermodynamics and statistical mechanics have since been extensively reassessed under a quantum mechanical lens and although foundational issues persist (as will be explored in several of the perspectives in this Roadmap), we are now at a point where a robust framework for understanding the energetics and thermodynamics of quantum systems far from the traditional thermodynamic limit is established and therefore can be of pragmatic use in assessing and characterising new technologies. In many ways, thermodynamics has come full circle, once again being put to task in order to address practical considerations in determining the optimal performance of new devices. \\

\paragraph*{Current and future challenges.}
Independently, quantum theory and thermodynamics have demonstrated remarkable predictive power. Indeed, as recently noted by Alicki and Kosloff, “whenever the two theories have addressed the same problem, new insight has emerged”~\cite{binder2019bookqtd}. Nevertheless, challenges still remain and this Roadmap aims to provide a concise overview of the current status of many, but certainly not all, areas of focus in the community. Foundational aspects are explored in various guises related to fundamental concepts and approaches in Secs.~\ref{sec:foundations}, \ref{sec:thermogeometry}, \ref{sec:TURs}, \ref{sec:nonabelian}, \ref{sec:time}, \ref{sec:infopropagration}, \ref{sec:collapse}, thermodynamics in many-body systems in Secs.~\ref{sec:thermalisation}, \ref{sec:manybodyinteractions}, and tools for examining the thermodynamics of open quantum systems in Secs.~\ref{sec:trajectories}, \ref{sec:transport}, and \ref{sec:strongcoupling}. The exquisite levels of control achievable with quantum systems means there have been remarkable developments in a variety of experimental platforms testing and elucidating thermodynamic concepts. These are discussed in Secs~\ref{sec:superconductors}, \ref{sec:ultracold}, \ref{sec:quantumdots}, \ref{sec:NMR}, \ref{sec:trappedions}, \ref{sec:NVcentres}, and \ref{sec:optomechanics}. Such practical advances have allowed us to move beyond proof-of-principle settings and thermodynamics is now guiding the assessment and development of quantum technologies as discussed in Secs.~\ref{sec:control}, \ref{sec:batteries}, \ref{sec:thermometry}, and \ref{sec:computing}. The sections in this Roadmap are largely self-contained. The particular order of the contributions was chosen merely with the flow of the presentation in mind. \\

\paragraph*{Broader perspective and relevance to other fields.}
Thermodynamics has always been remarkable for its ability to impact diverse, even seemingly disconnected, fields and this remains true for quantum thermodynamics, as detailed in the various perspectives which follow in this Roadmap. One area that is worth highlighting from the outset, however, is the symbiosis between information theory and quantum thermodynamics~\cite{Goold2016, *livro_DeffnerSteve}. The insight provided by Landauer's principle, which first formalised the deep connection between two previously disparate theories, has continued to drive advances in quantum thermodynamics~\cite{Junior2025}. In this light, it is not surprising that since the turn of this century, many significant results in quantum thermodynamics have arisen thanks to the application of information-theoretic approaches to thermodynamic conundrums and problems which has also led to the development of resource theories~\cite{Lostaglio2019resource}. In hindsight, though, perhaps it was inevitable that information theory and quantum thermodynamics should be such suitable companions; from our very first introduction to quantum mechanics, we are taught the curious role that observation and measurements, i.e. the acquisition of information, have on the properties of a system. Information is patently physical in quantum systems, and quantum thermodynamics therefore seems particularly suited to reconciling some of the long-standing interpretational issues in quantum theory. Likewise, it also provides a framework for addressing persistent conceptual issues in the foundations of quantum mechanics, from the information flow in black holes to the collective behaviour of many body systems, to describing the quantum-to-classical boundary. However, as with its precursors, quantum thermodynamics continues to be a practical tool: born and embedded in the second quantum revolution~\cite{DeutschPRXQ}, it provides ways to assess and characterise the efficiency of emerging quantum technology devices and algorithms, as well as starting to deliver machines and protocols, such as energy transfer and cooling, able to help with the technologies themselves~\cite{Auf23}.  \\

\paragraph*{Concluding Remarks.}
This Roadmap collects together a range of perspectives on key developments in quantum thermodynamics. To balance accessibility and utility, each contribution follows a unified structure, with length and bibliographic constraints, such that each perspective should be viewed simply as a primer to start the interested reader in their exploration of the field. While the topics covered represent a cross-section of the activities in this exciting field, there are naturally areas omitted, e.g. entanglement engines~\cite{entanglementEngine}, thermodynamics of quantum correlations~\cite{huber2015thermodynamic}, and quantum thermodynamic resource theories~\cite{Lostaglio2019resource}. Our aim is that this Roadmap will serve as a starting point for the community and help catalyse and direct research on the fertile domain of Quantum Thermodynamics in the coming years. \\

\paragraph*{Acknowledgements.}
SC acknowledges support from the John Templeton Foundation Grant ID 62422 and 63626. SC and ID’A are grateful to the Royal Society International Exchanges Scheme IES$\backslash$R2$\backslash$242072. 

\section{Quantum energetics, foundations, applications}
\label{sec:foundations}

\noindent
{\it Cyril Elouard}

\noindent
{Universit\'{e} de Lorraine, CNRS, LPCT, F-54000 Nancy, France}\\

\noindent
{\it Alexia Auff\`{e}ves}

\noindent
{MajuLab, CNRS-UCA-SU-NUS-NTU International Joint Research Laboratory, and}\\
{Centre for Quantum Technologies, National University of Singapore, 117543 Singapore, Singapore}\\

\paragraph*{State-of-the-art.}
Quantum energetics is the youngest daughter of two sciences of randomness: quantum physics and stochastic thermodynamics. From the latter, it takes that engines can turn thermal noise into a resource, that work must be paid to control systems in the presence of noise, and that irreversibility captures the lack of control against noise, which prevents agents from reversing  any evolution at will. From the former, it recalls that fluctuations and noise do not necessarily come from thermal environments: it is enough to measure a quantum system, or to entangle it with another one, to increase its entropy. In that sense, quantum energetics builds on different premises than quantum thermodynamics, where thermal resources often play a central role. 

Quantum energetics started with the acknowledgement that measurement back-action has an energetic footprint. Measuring non-conserved quantities on quantum systems changes their energy, and increases their entropy if the measurement outcomes are not read~\cite{J09,EHC+17}. Hence, measurement channels play similar roles as hot sources, allowing to fuel engines with no classical equivalent. Such machines have been proposed and realized on various platforms~\cite{DYK+18,BCS+21,LDG+22}. This first body of results probed the fertility of the approach and contributed to formalizing the scope and the questions of the field.

Some research problems of quantum energetics are directly inspired by thermodynamics: Can quantum noise be turned into an energetic resource? What is the cost of control against quantum noise? What is irreversibility in quantum processes, and how does it relate to energy waste?  Can we gain new insights in complex quantum dynamics from the identification of energetic  resources unlocking otherwise forbidden evolutions? Owing to its quantum roots, quantum energetics also aspires to build energetic witnesses of quantumness like coherence, entanglement or quantum statistics. Singling out regimes of quantum energetic advantage, where quantum machines are proven to execute the same task as classical ones with less energy, is another motivation~\cite{FAC+23}. All questions are potentially impactful to optimize the energy cost of quantum technologies~\cite{Auf23}.

To address these questions, quantum energetics must build a new framework to analyze the nature of energy and entropy flows between quantum systems, and their relations. This goes beyond pioneer paradigms of quantum thermodynamics identifying work (heat) as the energy flow between a quantum system and a classical drive (a reservoir described via a master equation for the system).  In contrast, the systems of interest should not have a predetermined role (bath, battery or working substance), and sometimes the same system can play multiple roles (both a driving field and a noise source, for instance). This new framework should (i) be operational, i.e. propose concepts that are measurable in experiments, (ii) capture fundamental relations between energy and entropy flows, (iii) be useful for quantum technologies, just like thermodynamics has been a game changer for industrial revolution. These three requirements set the major challenges of the field we now elaborate on. \\

\paragraph*{Current and future challenges.} 
{\bf Building operational concepts.} Measuring energies in the quantum realm is a considerable challenge, because the measuring apparatus participates in the energy balance.  Thus, energy and entropy flows cannot be accessed in the system and must rather be tracked at their sources, inside the baths and batteries – and more generally, inside any other coupled quantum system. In this spirit, it was for instance suggested to measure work extraction directly inside 
microwave fields in superconducting circuits~\cite{CB19}. The viewpoint is thus shifting from open quantum systems to autonomous ensembles, i.e. isolated quantum systems formed of coupled subsystems~\cite{EL23,PMC+24}. 


In order to analyze energy flows within autonomous ensembles, a first natural step is to consider the case of two coupled, otherwise isolated quantum systems~\cite{PMC+24}.
These systems can exchange energy in two ways, through effective unitaries or through correlations, the former (the latter) being reminiscent of a work (a heat) flow. In the case of a qubit interacting with a bosonic field,  measuring the work-like and heat-like flows received by the field is remarkably simple, as it simply corresponds to the change of its coherent energy (the energy stored in the mean field amplitude) and of the incoherent energy (stored in the field fluctuations), respectively –- giving rise to recent experiments with quantum dots~\cite{MTM+23}. This very generic situation can serve as a basis to estimate the cost of fundamental quantum processes, such as quantum gates~\cite{Gea02,SSM+22} and pre-measurements~\cite{BCS+21}.

{\bf Assessing fundamental costs.} In thermodynamics, fundamental costs appear with irreversibility. For instance, running an isothermal process forward and backward leads to irreversible heat dissipation in the bath, $Q_{diss}=T \Delta_i S$, where $T$ is the bath temperature and $\Delta_i S$ the entropy produced along the process. However, relating entropy and energy flows becomes challenging when none of the systems in interaction have a well-defined temperature, which is exactly the type of situations quantum energetics aims at addressing. Finding new relations is a major challenge of the field. The framework introduced in~\cite{EL23} starts addressing this question for the case of a set of interacting quantum systems. Energy provided by one system to the others is systematically split into a contribution proportional to the system's entropy change, interpreted as heat, and an iso-entropic contribution, identified as work. These contributions verify a bound analogous to the Second law, constraining the direction of heat exchanges and the efficiency of heat-to-work conversions, while work expenditure is a resource allowing to decrease entropy. 

An even more formidable challenge is to account for irreversibility occurring at the quantum-to-classical border. This issue is blatant when it comes to measurements. Their irreversible nature has been known since Eddington, and quantified as the increase of the von Neumann entropy of the measured system~\cite{EHC+17}. Conversely, only unitary pre-measurements can be modeled with standard quantum formalism and experimentally studied. Being in principle reversible operation, their costs are not fundamental. In contrast, true quantum irreversibility happens at Heisenberg's cut, such that capturing the fundamental energy cost of the measurement channel may require to close the quantum formalism in the first place, by providing a physical model of the world encompassing both the quantum and the classical level.  Partial results in this direction have been obtained, e.g. from unitary equilibration arguments~\cite{ERW+24}, physical models of the measurement apparatus including reservoirs and dephasing sources~\cite{LE25} or algebraic properties of large system Hilbert spaces~\cite{VG25}. 

In the same way, a fair assessment of the cost of any quantum process should take into account the cost of isolating quantum systems such that they keep their quantum properties, while controlling them from the external, classical world. This is a non-equilibrium situation, equivalent to keeping a Schr\"{o}dinger cat dead and alive while controlling it. Hence, assessing the fundamental energy cost of quantum processes is equivalent to assessing the cost of the box trapping the cat - calling for the closure of quantum theory. \\

\paragraph*{Broader perspective and relevance to other fields.}
Quantum energetics is naturally connected to many research fields already identified in the community of quantum thermodynamics. In particular, its first frameworks \cite{EL23,PMC+24} provide natural bases for the research developed in the field of quantum batteries~(see Sec.~\ref{sec:batteries}) and could shed new light on strong coupling thermodynamics (see Sec.~\ref{sec:strongcoupling}). 

Quantum energetics aims at playing the same role for quantum technologies as classical thermodynamics played for the first industrial revolution~\cite{Auf23}, for instance, by bringing out material to build standards of energy-efficiency for quantum technologies, and by providing the methodologies to optimize them. A promising strategy in this direction could be to exploit quantum optimal control algorithms \cite{BSS21}, using metrics from quantum energetics as cost functions to be optimized, or as optimization constraints~\cite{ADA+24}.  In the case of quantum computing, there are preliminary evidences~\cite{FAC+23} that energy savings at quantum scales (e.g., at the level of quantum processors) impact energy savings at the macroscopic ones (for the full stack of a quantum computer) - an important justification of the interest of the fundamental quantum energetics for technological purposes. The search for quantum advantages of energetic nature, whether at the level of quantum processes~\cite{AG24} or at the level of the full stack~\cite{FAC+23} are also highly relevant for all quantum technologies, with the potential to drastically steer roadmaps. Noticeably, full-stack energetic analyses of quantum communication protocols have recently been delivered~\cite{YPP+25}, showing the quick progress of this field of research.

Finally, by showing the deep relation between the capacity to optimize a full-stack computer and the resolution of open foundational questions like the closure of the quantum formalism, quantum energetics sheds a new light on quantum foundations, making them of highly topical and practical relevance.  \\

\paragraph*{Concluding Remarks.}
Quantum energetics is a new research field at the crossroad between quantum foundations and quantum technologies. This proximity holds the promise to repeat the miracles of classical thermodynamics: in the same way optimizing heat engines brought out the thermodynamic time arrow, optimizing the efficiency of quantum tasks may lead to solving still open problems of quantum theory. \\

\paragraph*{Acknowledgements.}
C.E. acknowledges funding by the European Union under ERC grant 101163469. Views and opinions expressed are however those of the author(s) only and do not necessarily reflect those of the European Union or the European Research Council Executive Agency. Neither the European Union nor the granting authority can be held responsible for them. A.A. acknowledges the National Research Foundation, Singapore through the National Quantum Office, hosted in A*STAR, under its Centre for Quantum Technologies Funding Initiative (S24Q2d0009), the Plan France 2030 through the projects NISQ2LSQ (Grant ANR-22-PETQ-0006), OQuLus (Grant ANR-22-PETQ-0013), and BACQ (Grant ANR-22-QMET-0002), the ANR Research Collaborative Project  “Qu-DICE” (Grant ANR-PRC-CES47), and the ANR Research Collaborative Project ``QuRes" (Grant ANR-PRC-CES47-0019).

\clearpage
\section{Superconducting circuits as a platform for quantum thermodynamics}
\label{sec:superconductors}
\noindent
{\it Bayan Karimi}

\noindent
{Pritzker School of Molecular Engineering, University of Chicago, Chicago IL 60637, USA, and}\\
{PICO Group, QTF Centre of Excellence, Department of Applied Physics, Aalto University, P.O. Box 15100, FI-00076 Aalto, Finland}\\

\noindent
{\it Jukka Pekola}

\noindent
{PICO Group, QTF Centre of Excellence, Department of Applied Physics, Aalto University, P.O. Box 15100, FI-00076 Aalto, Finland}\\

\begin{figure}[b]
  \centering
  \begin{minipage}{0.5\textwidth}
    \includegraphics[width=\linewidth]{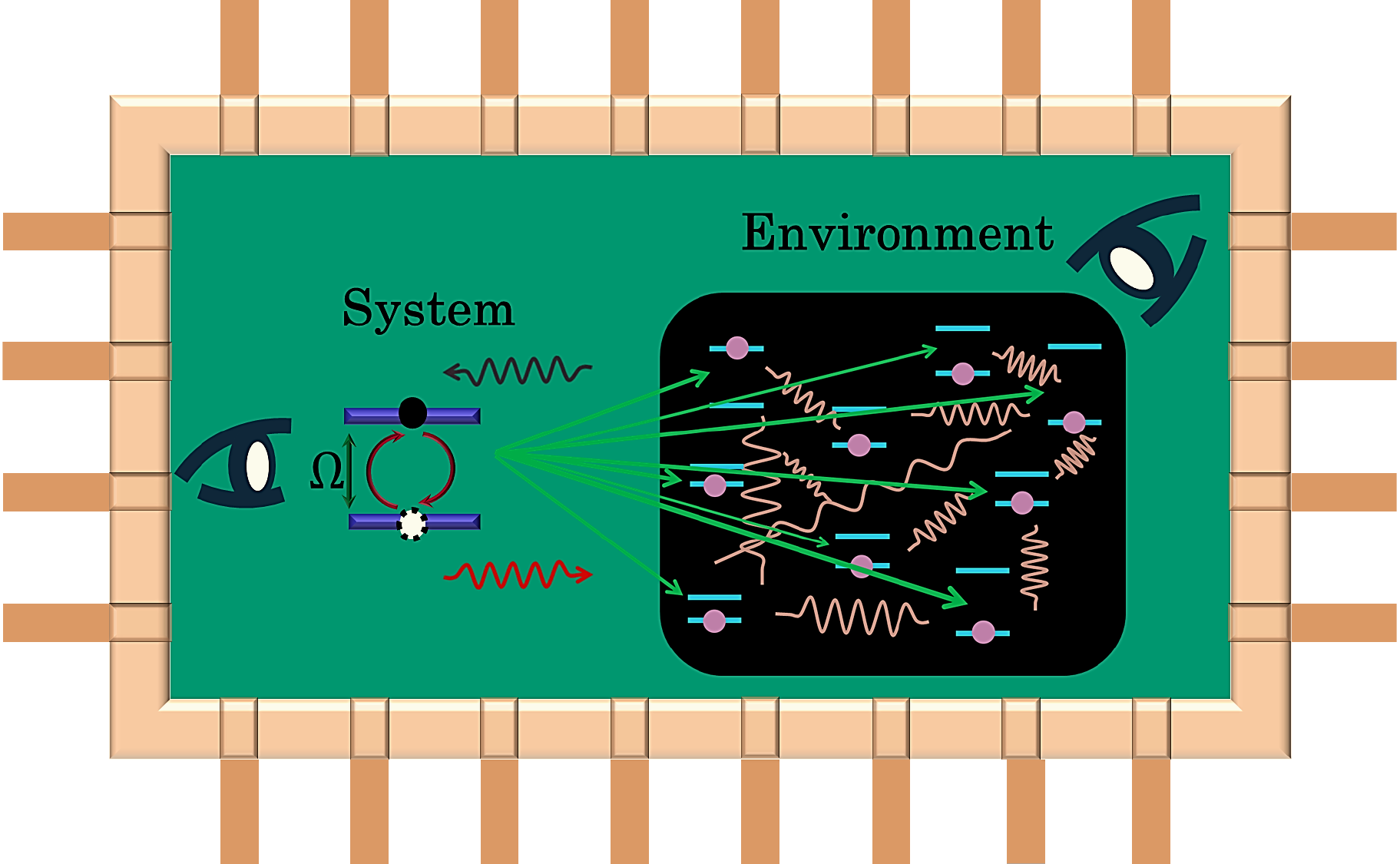}
  \end{minipage}%
  \hfill
  \begin{minipage}{0.35\textwidth}
\captionof{figure}{An open quantum system, here visualized as a device on a chip, is composed of the system itself (here symbolically a qubit) interacting with its environment often formed of a resistor in a circuit. This environment, which can typically be described using a weak-coupling master equation, is often treated as classical in nature, acting as a heat bath or a collection of multiple baths. To understand the dynamics of such a device, one can measure either the quantum system itself or its surrounding environment. \label{SysEnv}}
  \end{minipage}
\end{figure}

\paragraph*{State-of-the-art.}
The two well-established fields, quantum mechanics and thermodynamics, are complementary frameworks for description of physical systems. Roughly speaking, thermodynamics governs our visible world and quantum mechanics defines the microscopic realm. While the first one is the preferred framework for large, high-temperature systems, the later one is more applicable to small, low-temperature ones, such as superconducting setups. The interaction between these two descriptions of nature presents thermodynamics of quantum systems and processes. This domain will have a significant impact in the development of technologies as they are miniaturized to the quantum scale, where controlling heat dissipation becomes a key challenge for device functionality. Superconducting quantum circuits integrated on a chip (Fig.~\ref{SysEnv}) serve as nearly ideal building blocks for fundamental studies of this field, as they can be precisely engineered with predefined parameters, enabling direct measurement of their thermal and electrical properties. These systems can also be developed as ultrasensitive and nearly noninvasive detectors because of their small size and low operation temperature. Advancements in micro- and nanofabrication have created a significant opportunity for new classes of experiments to explore the largely uncharted field of ``quantum thermodynamics on a chip'', which we have coined as circuit quantum thermodynamics (cQTD)~\cite{PK-cQTD}. This field describes phenomena and devices that map the physics of open quantum systems into concrete quantum circuits, typically qubits and cavities, coupled to heat baths made of mesoscopic electron conductors and phonons on the chip.

The emergence of quantum thermodynamics experiments at low temperatures is often linked to stochastic thermodynamics studies on quantum dot and single-electron box circuits. While their quantum nature is feeble due to their low-frequency dynamics following classical rate equations, these setups allow precise tracking of individual charged particles using ultrasensitive electrometers. These experiments have provided highly accurate verification of fluctuation relations, demonstrated both non-autonomous and autonomous Maxwell’s demons~\cite{KoskiPekola2014}, and explored the link between information and energy~\cite{Pendry1983}, and the minimum energy cost of computation, i.e., the Landauer bound~\cite{Landauer1961}. The key challenging question is determining whether quantum information has a thermodynamic value beyond its classical Landauer erasure energy of $k_BT\ln(2)$, where $k_B$ is the Boltzmann constant and $T$ represents the temperature. Furthermore, although Maxwell’s demon presently produces only a minimal power output, in the future experiments it may potentially minimize local dissipation using feedback mechanisms. \\

\paragraph*{Current and future challenges.}
We will next list the key unanswered questions and the fundamental challenges in modern quantum thermodynamics. The selection is naturally the authors' subjective view and does not necessarily present all the important quests. Does a quantum system thermalize on its own? In other words, does a system reach a Gibbs-like distribution within timescales short enough to be considered isolated from the rest of the world? Under what circumstances does this occur? What makes a reservoir a thermal bath, i.e. a “swamp” of energy, thus exhibiting an almost infinite Poincaré time, with no revivals within any realistic timescale~\cite{PK-heatbath,Mori2018}? 

Quantum heat transport, heat engines, and refrigerators are among the active domains in cQTD devices. Despite extensive theoretical research, experimental realizations are still limited, and fully functional quantum heat engines and refrigerators remain purely conceptual at this stage. A crucial question is whether quantum systems can outperform classical devices in power and efficiency, which can in the first place be tested against theoretical modeling only. However, since generating quantum coherence requires energy, it is still unclear whether coherent quantum dynamics would provide a real advantage. Another question is whether avoiding coherences (``quantum friction'') is necessary or if tailored designs and driving protocols can be used to eliminate unwanted coherences. It is also essential to keep in mind that in driven open quantum systems, the phase or phases of the quantum state can cause interference effects, with Landau-Zener-Stuckelberg being a prime example~\cite{SHEVCHENKO20101}. In open systems, phase also influences thermal transport properties, see e.g.,~\cite{Giazotto2012}, and it is expected to affect the power and efficiency of thermal machines. Can quantum interference be harnessed to achieve quantum supremacy in refrigeration?

Beyond the basic thermal microwave photon mediated heat transport mechanism~\cite{PK-cQTD}, there are important subtle processes that have not yet been observed experimentally. The first one of them is a next order effect that can be described as "co-tunneling" or "Kondo effect"~\cite{Saito2013,Luca2025}, meaning virtual excitation with simultaneous relaxation carrying heat equal to the energy level separation of the mediating element, which can be a qubit or resonator. Although a next order effect, it is expected to become the dominant one at low temperatures, not vanishing exponentially unlike the basic transport processes. Another important, still unexplored question is the influence of Dicke superradiance~\cite{Dicke1954} on heat transport via a system of $N$ parallel qubits coupled to the same reservoirs. It is likely that the thermal transport would be enhanced in this situation beyond that of $N$ uncoupled qubits in parallel. 

Thermal machines generally rely on external power sources for operation. Low efficiencies of refrigerators, i.e., conversion ratio of work done to extracted heat from the object to be cooled, results in a significant amount of wasted energy. This raises the natural question whether it is possible to harness this wasted energy in form of thermal fluctuations and use them as a driving field to enable efficient cooling on a chip~\cite{Pekola-Hekking}? Is it possible to use the latest advancements in heat transport and quantum thermodynamics to develop efficient cQTD devices powered by thermal energy? Doubtlessly a timely objective in cQTD devices is to build ultrasensitive and noninvasive detectors of energy exchanged between the quantum system and its environment in a continuous manner with low self-heating and only minute increase of relaxation rates of the quantum system. This approach allows us to address intriguing questions, especially in the largely unexplored experimental domains of quantum heat transport and noise e.g., experiments on the noise produced by a quantum system mediating the heat current. An even greater challenge lies in detecting energy exchange at the level of individual quanta, allowing us to map the quantum dynamics in time domain. This could pave the way towards a new research area, potentially even an entire field in circuit-based studies, namely stochastic quantum thermodynamics, sQTD. This then leads us to face another key challenge in modern quantum thermodynamics, the role of fluctuations~\cite{Pekola-Averin2010}. While their impact is well established in classical stochastic thermodynamics, it remains unclear how these fundamental fluctuation relations are realized in cQTD systems, where the measurement and its apparatus affect the system rather than simply observing it.

\paragraph*{Broader perspective and relevance to other fields.}
Ultimately quantum thermodynamics is a field where similar scientific questions can be addressed on various different platforms described in this roadmap collection. The question then remains, which one of these realizations is most suitable for a given test or task. The great advantage of superconducting circuits is that they can be controlled accurately, and they can be integrated into large ensembles of qubits and other building blocks such as resonators (harmonic oscillators). As described in this article, one can also realize heat baths directly on the chip, and the temperatures can be controlled and measured locally and accurately. The low temperature of these systems allows direct measurements of heat by thermometry, which is unreachable in many systems, where only indirect measurements of thermal properties are possible. 

\paragraph*{Concluding Remarks.}
Precise and local thermometry, e.g., by using superconductor-normal metal tunnel junctions~\cite{PK-cQTD} is a central asset in experimental quantum thermodynamics on a superconducting platform. Thermometry can be tailored for both steady-state and time-domain experiments, in the latter case enabling thermal single-quantum detection in the future. Similarly, normal conductors on a chip can act as ideal sources of thermal noise, both in classical and quantum regimes. There are recent very promising, but still indirect quantum thermodynamics experiments~\cite{Mikko2025,Gasparinetti2025} realized using synthetic noise and qubit state measurements. While indirect methods offer valuable insights, we expect experiments to utilize natural thermal noise and direct temperature measurements in the near future quantum thermodynamics studies. Finally, the era of superconducting quantum circuits started by the experimental demonstration of coherent oscillations in a qubit in 1999~\cite{Nakamura1999}. Since then, the circuits' performance has improved immensely: the relevant lifetimes (coherence times) have increased from nanoseconds up to a millisecond in the best realizations, i.e. by a factor of one million. This tremendous progress, thanks to the massive investigation and investments in developing quantum processors based on superconducting circuits, has facilitated, or one could say even enabled, the emergence of research on quantum thermodynamics on superconducting platforms. Yet experimental realization of some of the ideas presented in this article are still limited by the fact that superconducting circuits are, after all, open quantum systems rather than isolated ones. 

\paragraph*{Acknowledgements.} This work has received funding from the European Union’s Research and Innovation Programme, Horizon Europe, under the Marie Sk\l{}odowska-Curie Grant Agreement No. 101150440 (TcQTD). We acknowledge the QuantERA II Programme that has received funding from the EU's H2020 research and innovation programme under the GA No 101017733, and the Research Council of Finland Centre of Excellence programme grant 336810 and grant 349601 (THEPOW).

\clearpage

\section{Quantum thermodynamics with ultracold atoms}
\label{sec:ultracold}
\noindent
{\it Thom\'as Fogarty and Thomas Busch}

\noindent
{Quantum Systems Unit, OIST Graduate University, Onna, Okinawa 904-0495, Japan}\\

\noindent
{\it Eloisa Cuestas}

\noindent
{Quantum Systems Unit, OIST Graduate University, Onna, Okinawa 904-0495, Japan, and}\\
{Forschungszentrum J\"{u}lich, Institute of Quantum Control (PGI-8), D-52425 J\"{u}lich, Germany}\\

\noindent
{\it Artur Widera}

\noindent
{Department of Physics and Research Center OPTIMAS, RPTU Kaiserslautern-Landau, Kaiserslautern, Germany}\\

\paragraph*{State-of-the-art.}
Over the past two decades, advances in experimental techniques for cold atomic gases have established them as ideal testbeds for simulating complex quantum systems and exploring fundamental physics. Modern experimental setups offer remarkable precision in confining and controlling their degrees of freedom, including the control over interactions using Feshbach resonances. Species-dependent traps enable the realization of atomic mixtures with individual control over each component and the number of atoms or impurities in a system can be freely adjusted, providing a unique platform to explore the crossover between few- and many-body physics.

In addition to their high degree of controllability, cold atom systems can be probed with a variety of measurement techniques. Probability densities can be extracted using time-of-flight absorption imaging, while atomic gas microscopes enable single-atom detection in optical lattices and beyond. These measurements provide direct access to thermodynamic quantities such as pressure, isothermal compressibility, and internal energy density and have been successfully demonstrated in interacting Bose and Fermi gases \cite{Nascimbene_2010}. Techniques such as momentum resolved microwave spectroscopy \cite{Li2024} allow for direct measurement of the spectral function, offering insights into elementary excitations in complex quantum systems. Furthermore, Ramsey interferometry has been used to implement a two-point measurement scheme to determine the work probability distribution of a driven non-equilibrium state \cite{Cetina2016}, granting direct access to both its excitation spectrum and thermodynamic properties.

However, precise thermometry of ultracold gases has long been a challenge, as traditional methods rely on destructive time-of-flight measurements to fit the high-momentum tails to infer temperature. Recently, impurity probes embedded within ultracold gases have emerged as a promising non-destructive alternative where enhanced sensitivity can be achieved by the creation of both impurity-gas correlations and gas mediated impurity-impurity correlations. Several approaches have been suggested. In one of them the impurities are allowed to thermalize with the ultracold gas and the temperature is extracted by measuring the position and momentum of the impurities, potentially achieving sub-nano-kelvin precision \cite{Khan2022}. Another approach relies on using Ramsey interferometry to monitor the decoherence of impurities following an interaction quench with the surrounding gas \cite{Mitchison2020}, with the temperature inferred from this decay rate. This method has been successfully demonstrated in experiments using Cs impurities in a Rb gas \cite{Adam2022}. Moreover, mapping of thermal information on the quantum spin-levels via spin-exchange interactions between impurity and bath has paved the way to obtain information beyond the equilibrium paradigm of standard thermometry \cite{PhysRevX.10.011018}.

\begin{figure}[t]
    \centering
    \includegraphics[width=\linewidth]{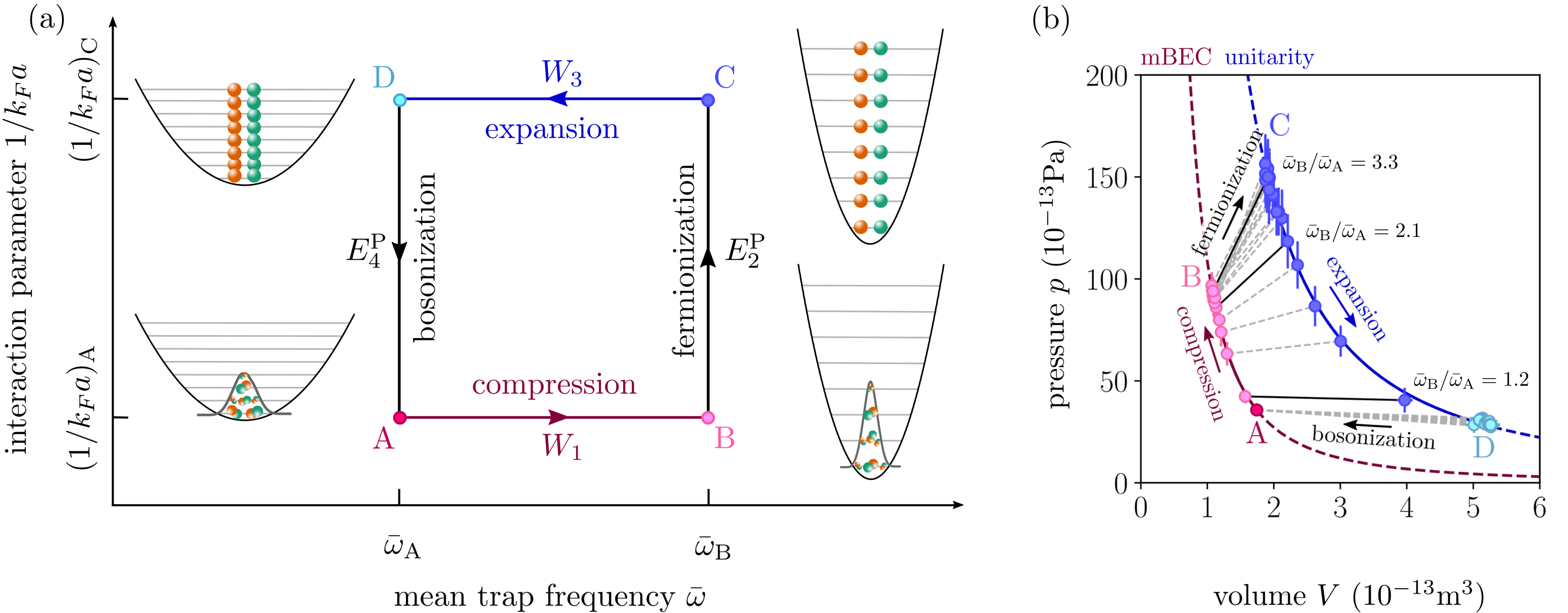}
    \caption{(a) Schematic of the Pauli engine cycle which consist of two work strokes and two statistical strokes \cite{Koch2023}. \textbf{A}$\rightarrow$\textbf{B}: the trap compression in the molecular BEC phase does work $W_1$. \textbf{B}$\rightarrow$\textbf{C}: through an interaction ramp the molecular bosons are broken up and driven into the unitary Fermi gas limit. This results in a change in energy due to the change in particle statistics $E_2^P$, which is called Pauli energy. \textbf{C}$\rightarrow$\textbf{D}: the unitary Fermi gas is expanded by reducing the trap frequency and extracting work $W_3$. \textbf{D}$\rightarrow$\textbf{A}: the interaction strength is ramped back to its initial value for the gas to return to a molecular BEC state at the cost of Pauli energy $E^P_4$. (b) Pressure-volume diagram of the Pauli cycle for different compression ratios $\bar{\omega}_B/\bar{\omega}_A$. The fermionization process increases the pressure of the gas and allows for work to be extracted from the engine.}
    \label{fig:PauliEngine}
\end{figure}

One of the advantages of ultracold atoms when it comes to driving thermodynamic engine cycles is that they can be used to realize both single-atom and manybody machines with novel and creative mechanisms. The former allows tracing energy transfer atom-by-atom and quantum-by-quantum with full-counting statistics, while the latter allows exploiting the entire portfolio of collective quantum properties developed by cold gas experiments.
A quantum heat engine based on Cs impurities coupled to a Rb bath was realized in Ref.~\cite{Bouton2021}, where inelastic spin-exchange collisions facilitated heat transfer between the working medium and the bath, allowing to implement a full quantum Otto cycle with high efficiency and power. Beyond these conventional engine cycles, cold atom systems have also enabled deeper exploration of how quantum statistics influence thermodynamics \cite{Koch2023}. By tuning interactions, a unitary Fermi gas can be transformed into molecular bosons in the BEC-BCS crossover, fundamentally altering the single-particle distribution function and leading to distinct physical behaviors in both regimes. Incorporating this process into an Otto-like engine cycle leads to the realization of a novel quantum machine driven purely by changes in quantum statistics instead of requiring traditional heat baths, see Fig.~\ref{fig:PauliEngine}(a). The stark difference in Fermi and Bose statistics results in an enhanced work output, which can be attributed to the Fermi pressure that originates from the Pauli exclusion principle, see Fig.~\ref{fig:PauliEngine}(b). Crucially, this effect is inherently quantum and vanishes at high temperatures where both fermionic and bosonic systems attain classical Boltzmann statistics. A similar engine using changes in the chemical potential of the atomic gas has also been experimentally realised \cite{Simmons2023}.  \\

\paragraph*{Current and future challenges.}
While current experimental techniques allow for the measurement and inference of many thermodynamic quantities by having access to averaged local densities, they generally provide limited insight into quantum correlations and entanglement. The latter play an important role in quantum thermodynamics and their measurement requires full access to the density matrix of the many-body state. Recently, this has become experimentally feasible in few-body systems, where the full quantum state of three atoms in an optical tweezer setup was reconstructed via in-situ, spin-resolved position and density measurements \cite{Becher_2020}. However, trying to measure correlations in few-body systems can only be an intermediate  step, as for a true many-body system this is a Herculean task as the information grows exponentially with system size. Thus, it would be important to base predictions of quantum thermodynamics on few variables that stem from quantum correlation and quantum statistical arguments. It is therefore important to test if the standard thermodynamic variables still hold or if new or additional ones emerge. This is particularly relevant to the understanding of the third law of thermodynamics which can limit the ability of current cold atom experiments to further cool systems to sub-nano-Kelvin temperatures due to vanishing cooling rates \cite{Masanes2017}.

Another key challenge for cold atomic systems is the direct extraction of work produced by quantum engines or stored in quantum batteries. In recent cold atom experiments, work strokes of engine cycles are implemented by modulating the power of the trapping laser, compressing and expanding the trapped gas. However, the work output in these cases is inferred from the energy difference between initial and final states rather than being directly measured. To obtain a more direct quantification of usable work, coupling the system to an external work load is essential. While several methods have been proposed, their feasibility depends on the specific system and the efficiency of energy transfer. One method involves coupling an atomic heat engine to an optical cavity with an oscillating mirror, enabling mechanical work extraction \cite{Carollo2020}. Another promising method is the use of quantum flywheels, where work is transferred to a different degree of freedom within the system, as demonstrated in a recent trapped ion experiment \cite{flywheel2019}. Furthermore, for cyclical energy exchange processes, such as in quantum batteries, while the unitary operator that maximizes work extraction can be calculated, it is typically challenging to implement it experimentally. Alternative approaches must then be used, for example employing variational techniques to approximate the optimal transformation with physically realizable unitaries which could enable practical work extraction \cite{Brown_2016}.\\

\paragraph*{Broader perspective and relevance to other fields.}
The versatility of cold atomic systems together with the ability to carry out high fidelity operations have already led to their utilization in traditional quantum technologies. The next step is therefore to exploit their quantum thermodynamical properties for studying fundamental physics, but also for adding another pillar to future applications of quantum technologies. In particular their many-body aspect connects their thermodynamic behaviour to long-standing questions in condensed matter and statistical physics, ranging from non-equilibrium dynamics to quantum phase transitions and crossover physics. Examples for crossover into other fields are the question of thermalization in isolated systems or the appearance of many-body localization, quantum many-body scars and Hilbert space fragmentation.

The precise dynamical control that is paramount for the efficient operation of quantum thermal machines connects quantum thermodynamics with cold atoms to the broader area of optimal quantum control. Here a wealth of literature exists describing shortcuts to adiabaticity, which allow to mimic adiabatic dynamics in short timescales \cite{GueryOdelin19}. These techniques have been devised for both discrete and continuous quantum systems, and have been successfully applied to a wide range of cold atomic systems, allowing for fast compression or expansion of the trap frequency or changes in the interaction strength. Extending the existing knowledge and developing STAs protocols for strongly interacting many- or few-body systems is of particular interest in order to control and optimize the performance of new machines such as the recent Pauli engine. While exact STAs are possible, usually they require the implementation of non-local operators that would be difficult to implement experimentally. However, new techniques that supplement approximate local STAs with time-dependent control fields, such as enhanced STAs \cite{Whitty_2020} and counterdiabatic optimized local driving \cite{COLD2023}, have been proposed and have the potential to attain high fidelities with current cold atom setups. Combining these adiabatic control techniques with recent advances in shortcuts to equilibration for open systems~\cite{Dann2019,Boubakour_2025}, could also further optimize quantum thermal machines and significantly improve their practicality.\\

\paragraph*{Concluding Remarks.}
Cold atomic systems are highly versatile and offer many opportunities to further explore thermodynamics in the quantum regime and the development of new quantum devices. In these systems quantum correlations can be controlled via highly controllable interaction effects, such as short range s-wave, long range dipolar, synthetic spin-orbit and environment mediated interactions. The experimental control of several degrees of freedom in a clean way makes possible not only to explore the effects of symmetry, interactions, dimensionality, etc. for quantum machines but also to address fundamental questions such as equilibration and thermalization in complex many-body systems, many-body localization in systems with disorder, to study chaotic dynamics, and to challenge the usual definitions of work and heat in quantum systems. The use of impurities in cold atoms systems has opened the path for improved thermometry protocols and also the possibility to experimentally realize a controlled thermal bath. In that context cold atoms have emerged not only as an unique and interesting platform to test many-body physics but are also at the heart of a rich and growing interplay between quantum thermodynamics, quantum control, and quantum simulations, requiring strong collaborative work connecting these different areas for the development of emergent quantum technologies.  \\

\paragraph*{Acknowledgements.}
We would like to thank Jennifer Koch for stimulating discussions and for sharing the design for figures. T.F. acknowledges support from JSPS KAKENHI Grant No. JP23K03290. T.F. and T.B. are also supported by the Okinawa Institute of Science and Technology and the JST Grant No. JPMJPF2221. E.C. was supported by JSPS KAKENHI grant number JP23K13035 and the Horizon Europe programme HORIZON-CL4-2022-QUANTUM-02-SGA via the project 101113690 PASQuanS 2.1. A.W. acknowledges funding from Deutsche Forschungsgemeinschaft (DFG, German science foundation) via Sonderforschungsbereich SFB/TRR 185, project no. 277625399.

\clearpage

\section{Quantum thermodynamics in quantum dot devices}
\label{sec:quantumdots}
\noindent
{\it Andrew K. Mitchell}

\noindent
{School of Physics, University College Dublin, Belfield, Dublin 4, Ireland, and}\\
{Centre for Quantum Engineering, Science, and Technology, University College Dublin, Dublin 4, Ireland}\\

\noindent
{\it  Eran Sela}

\noindent
{Raymond and Beverly Sackler School of Physics and Astronomy, Tel Aviv University, Tel Aviv 69978, Israel}\\

\noindent
{\it  Joshua Folk}

\noindent
{Quantum Matter Institute, University of British Columbia, Vancouver, British Columbia, V6T 1Z4, Canada, and}\\
{Department of Physics and Astronomy, University of British Columbia, Vancouver, British Columbia, V6T 1Z1, Canada}\\

\noindent
{\it Klaus Ensslin}

\noindent
{Solid State Physics Laboratory and Quantum Center, ETH Zurich, 8093 Zurich, Switzerland}\\

\begin{figure}[t]
\centering
\includegraphics[width=0.85\columnwidth]{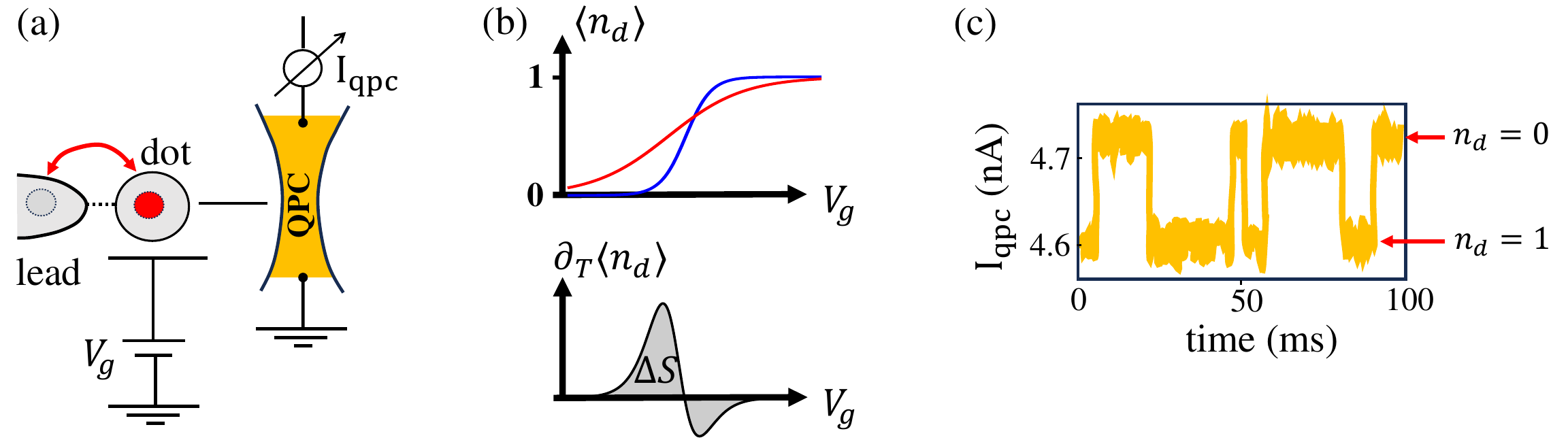}
\caption{(a) Quantum dot and charge-detector setup. The charging of a QD by a single electron is controlled by the gate voltage $V_g$ and can be probed via the current through a nearby quantum point contact (QPC). (b) From the charging curve and its temperature dependence, the entropy change is extracted using a Maxwell relation. (c) The time-resolved measurement of QPC current constitutes a weak continuous measurement of the QD charge and can be interpreted as a quantum trajectory. 
}\label{fig:QDs}
\label{fig:W3}
\end{figure}

\paragraph*{State-of-the-art.}
Quantum dot (QD) devices are tunable nanoelectronic circuits in which one or more confined regions are tunnel-coupled to metallic leads, see Fig.~\ref{fig:QDs}a. They can be lithographically defined in the two-dimensional electron gas of semiconductor heterostructures, or realized in 2D materials such as graphene. QD devices have proven to be a unique platform for studying the physics of complex open quantum systems, and have been used to realize exotic many-body states of quantum matter. QD degrees of freedom may be regarded as the `system' whereas the continuum baths of electrons in the leads constitute an `environment', which can be strongly coupled and produce highly non-Markovian dynamics, with strong system-environment entanglement building up at low temperatures. Device properties can be controlled \textit{in situ} by application of gate voltages and magnetic fields, providing single-electron transistor functionality. The devices are typically operated at low temperatures, $T\sim 10$mK--$1$K.

Although quantum transport through the QD is a more common experimental observable, recently the detection of electronic \textit{charge} on the QD~\cite{field1993measurements,gustavsson2006counting,piquard2023observing} has opened the door to detailed studies of quantum thermodynamics in mesoscopic systems. The QD charge is typically detected via changes in the current through a nearby, capacitively-coupled circuit (Fig.~\ref{fig:QDs}a), which in most cases does not disturb the quantum dynamics of the system being probed. We focus on such charge-based thermodynamic measurements in this perspective.

In the equilibrium setting, the time-averaged QD charge $\langle \hat{n}_d\rangle$ is related to the total thermodynamic entropy $S$ of the system through a local Maxwell relation~\cite{hartman2018direct}. Assuming that the gate voltage $V_g$ controls the QD charge through a coupling term in the Hamiltonian $\hat{H}=\hat{H}_0+V_g\hat{n}_d$ then the thermodynamic identity $\partial_T \langle \hat{n}_d\rangle = -\partial_{V_g} S$ holds. The change in entropy between two configurations is therefore related to the temperature-derivative of the measured charge, see Fig.~\ref{fig:QDs}b.
Recently, this approach was used to measure the occupation-dependent entropy of a QD in the strongly-coupled regime~\cite{child2022entropy}, demonstrating both the $k_B\ln 2$ spin entropy for an electron fully localized in the QD, as well as the reduction in entropy relative to the classical result at the QD charge transition, due to the quantum coherent hybridization between QD and lead electrons. In another example, charge detection enabled the ground state degeneracy of a two-electron quantum dot in bilayer graphene to be determined~\cite{adam2024entropy}. 

Thermodynamic characteristics can also be determined from time-resolved charge detection, which in principle yields richer information~\cite{bayer2025real}. A weak continuous measurement, in which electrons tunneling on and off the QD appear as a series of charge jumps in the time-trace (Fig.~\ref{fig:QDs}c), can be viewed as a \textit{quantum trajectory} in the context of stochastic quantum thermodynamics. In non-equilibrium conditions, where the QD level potential is driven, real-time charge detection has been used to verify fundamental fluctuation theorems~\cite{saira2012test,kung2012irreversibility,koski2013distribution}
and to extract the work distribution function~\cite{hofmann2016equilibrium,*hofmann2017heat}. Optimized protocols for Landauer information erasure have  been developed~\cite{scandi2022minimally}; and by conditioning gate control on measurement outcomes, Maxwell's demon and information-to-work conversion have also been demonstrated~\cite{annby2020maxwell,barker2022experimental}.\\

\paragraph*{Current and future challenges.}
Advances in quantum control and manipulation at the single electron level in QD devices, together with thermodynamic characterizations via charge detection, have provided a unique window on interacting open quantum systems in a non-Markovian, strong-coupling setting. However, challenges for both experiment and theory remain.

A major future objective is to obtain insights into exotic quantum states from thermodynamic measurements, beyond what can be deduced from transport \cite{kleeorin2019measure}. For example, the thermodynamic characterization of states with a fractional entropy would provide a discriminating perspective.
In particular, quantum criticality in QD devices arising due to frustrated Kondo interactions is predicted from theory to give a fractional residual entropy contribution, corresponding to an emergent anyonic quasiparticle localized on the QD: $k_B\ln\sqrt{2}$ for two-channel-Kondo hosting a Majorana fermion; $k_B\ln{\phi}$ with $\phi$ the golden ratio for three-channel-Kondo hosting a Fibonacci anyon; or $k_B\ln\sqrt{3}$ for the double-charge-Kondo system hosting a $Z_3$ parafermion. Direct observation of these smoking-gun entropy signatures has so far remained elusive, despite remarkable agreement between experimental transport measurements and quantum critical scaling predictions from theory in all of these systems~\cite{iftikhar2018tunable,karki2023z}.  The need to go to very low temperatures to realize exotic coherent states, and also to estimate temperature \textit{derivatives} of charge measurements, makes such experiments very challenging. A possible route to improve charge detection efficiency and sensitivity is to increase the coupling between the QD and charge detector, with the precaution that the many-body coherence underpinning the Kondo effect might then be deteriorated by a measurement backaction. Understanding and mitigating noise sources is also crucial. Another difficulty is that disorder and defects may contribute as unwanted sources of entropy.  Isolating the entropy contribution from the QD itself to better than a  $0.1 k_B$ level is, in general, challenging.

Going beyond the measurement of thermodynamic entropy for an equilibrium thermal state, the characterization of entropy \textit{production} in strongly non-equilibrium conditions presents considerable further challenges -- especially in the coherent regime. One would like to be able to study non-equilibrium quantum thermodynamics in coherent QD devices subject to an applied bias voltage, real-time driving and quantum control, or evolving under quantum measurements.
Although real-time charge detection during QD driving has been used to study the work distribution~\cite{hofmann2016equilibrium, *hofmann2017heat}, the quantum coherent regime has not been accessible. 
Indeed, there may be quantum limitations on testing non-equilibrium fluctuation theorems in such setups~\cite{han2024quantum}. 
Detailed theoretical predictions have been made for the work statistics in linear response for coherent QD systems, including those near criticality~\cite{ma2025quantum}, but these have yet to be verified experimentally. Aside from reducing backaction effects, one issue is the need for faster charge detection and increased resolution, to match the timescales for coherent dynamics in QD setups. One possibility is to utilize resonator circuits or cavities to boost signal strength and time resolution. On the theory side, a major challenge is to extend the existing predictions for quantum thermodynamics and work statistics in QD systems into the strongly non-equilibrium regime.

Finally, we remark that it is currently difficult to make \textit{projective} measurements in coherent QD systems. 
Overcoming this problem would enable the study of stochastic quantum thermodynamics, and could allow entanglement to be exploited as a resource in QD setups.\\


\paragraph*{Broader perspective and relevance to other fields.}
In the field of quantum thermodynamics, theory has in many cases outpaced experiment, as the topics covered in this \textit{Roadmap} article attest. Theoretical predictions have often proved difficult to verify experimentally due to the idealizations made in toy-model studies, and the unavoidable practical complexities of physical realizations. Yet remarkable demonstrations have still been achieved in various platforms, including superconducting circuits, ultracold atoms, NMR systems, trapped ions, and NV centers, as well as in QD devices (Secs.~\ref{sec:superconductors}--\ref{sec:NVcentres}).

QD devices offer some specific advantages and novelties, but also have their limitations. In contrast to the other common platforms, QD devices offer the opportunity to study fermionic problems in a regime of strong system-bath coupling. Whereas many theoretical studies have employed a weak-coupling/Markovian approximation, the low-temperature quantum dynamics of QD devices is typically beyond that describable by Master equations or simple collision models. QD devices are also characterized by strong electron-electron interactions, which produce non-perturbative many-body physics. As such QD systems provide a playground for the study of thermodynamics in non-trivial settings. The high degree of synergy between experiment and theory, together with the sophisticated numerical and analytical techniques developed to treat the generalized `quantum impurity models' describing QD devices, has led to remarkably precise tests of predictions for these systems.

QD nanoelectronic devices are versatile and tunable, meaning they can be used as \textit{quantum simulators} to realize fundamental models; for example systems hosting non-Abelian anyons~\cite{iftikhar2018tunable,karki2023z} which are the building-blocks for topological quantum computation. QDs can also be realized in existing CMOS technology, providing a possible route to scalability and integration.

Limitations include the difficulty of implementing projective measurements or non-unitary dynamics and the inability to prepare arbitrary non-equilibrium states. Indeed, quantum control is generally limited to manipulation of electrical and magnetic fields; more complex interactions would be  needed e.g.~for optimal control protocols.

QD devices are of central importance for studies of quantum transport and thermal machines (see Sec.~\ref{sec:transport}) and for testing thermodynamic/kinetic uncertainty relations (Sec.~\ref{sec:TURs}). In the context of metrology (Sec.~\ref{sec:thermometry}) QDs have been proposed as high-precision quantum sensors, and they are also natural candidates for practical quantum batteries (Sec.~\ref{sec:batteries}). Finally, we note that quantum trajectories (Sec.~\ref{sec:trajectories}) may be obtained via real-time charge detection in QD setups.\\


\paragraph*{Concluding Remarks.}
QD devices provide a platform that is uniquely suited to exploring the quantum thermodynamics of strongly-coupled, non-Markovian open systems in a strongly-interacting, many-body setting. Nontrivial states of matter can be engineered, including fractionalized quantum critical states. Thermodynamic observables provide a novel and revealing viewpoint for such systems, complementing more standard electronic transport measurements.  
The use of charge detection is currently being explored as a tool for thermodynamic characterization.

Arguably the key questions to be addressed with future studies relate to the crossover between classical and quantum thermodynamics in QD systems. What are the uniquely \textit{quantum} properties arising due to coherences? In this regard, the observation of the fractional entropy of intrinsically quantum anyonic degrees of freedom would be a significant breakthrough. Likewise, elucidating the role of coherence in the non-equilibrium thermodynamics of QD systems, and demonstrating universal scaling as a QD device is driven across a quantum critical point, would constitute major advances.\\


\paragraph*{Acknowledgements.}
We gratefully acknowledge useful discussions and input from Frederic Pierre, Thomas Ihn and Yigal Meir. AKM acknowledges financial support from Science Foundation Ireland through Grant 21/RP-2TF/10019. ES, JF and KE acknowledge financial support from the European Research Council (ERC) under the European Union Horizon 2020 research and innovation programme under grant agreement No. 951541.

\clearpage

\section{Quantum thermodynamics in NMR systems}
\label{sec:NMR}
\noindent
{\it Taysa M. Mendon\c{c}a}

\noindent
{Instituto de F\'{i}sica de S\~{a}o Carlos, Universidade de S\~{a}o Paulo, CP 369, 13560-970, S\~{a}o Carlos, SP, Brazil}\\

\noindent
{\it Roberto S. Sarthour and Alexandre M. Souza}

\noindent
{Centro Brasileiro de Pesquisas F\'isicas, 22290-180, Rio de Janeiro, RJ, Brazil}\\

\paragraph*{State-of-the-art.} 
The field of quantum thermodynamics has deepened our understanding of thermodynamic laws in the quantum realm, where quantum coherence, non-classical correlations, and inherent fluctuations play a crucial role. On the experimental side, Nuclear Magnetic Resonance (NMR) has been a key platform for testing quantum thermodynamic concepts, due to the long coherence time of nuclear spins and their precise manipulation using radiofrequency (RF) fields \cite{IvanLivroNMR}. These features have enabled various proof-of-principle experiments \cite{Batalhao2014,Peterson2019,Assis2019,Serra2019,Mendonca2020}.

\begin{figure}[b]
  \centering
  \begin{minipage}{0.55\textwidth}
    \includegraphics[width=\linewidth]{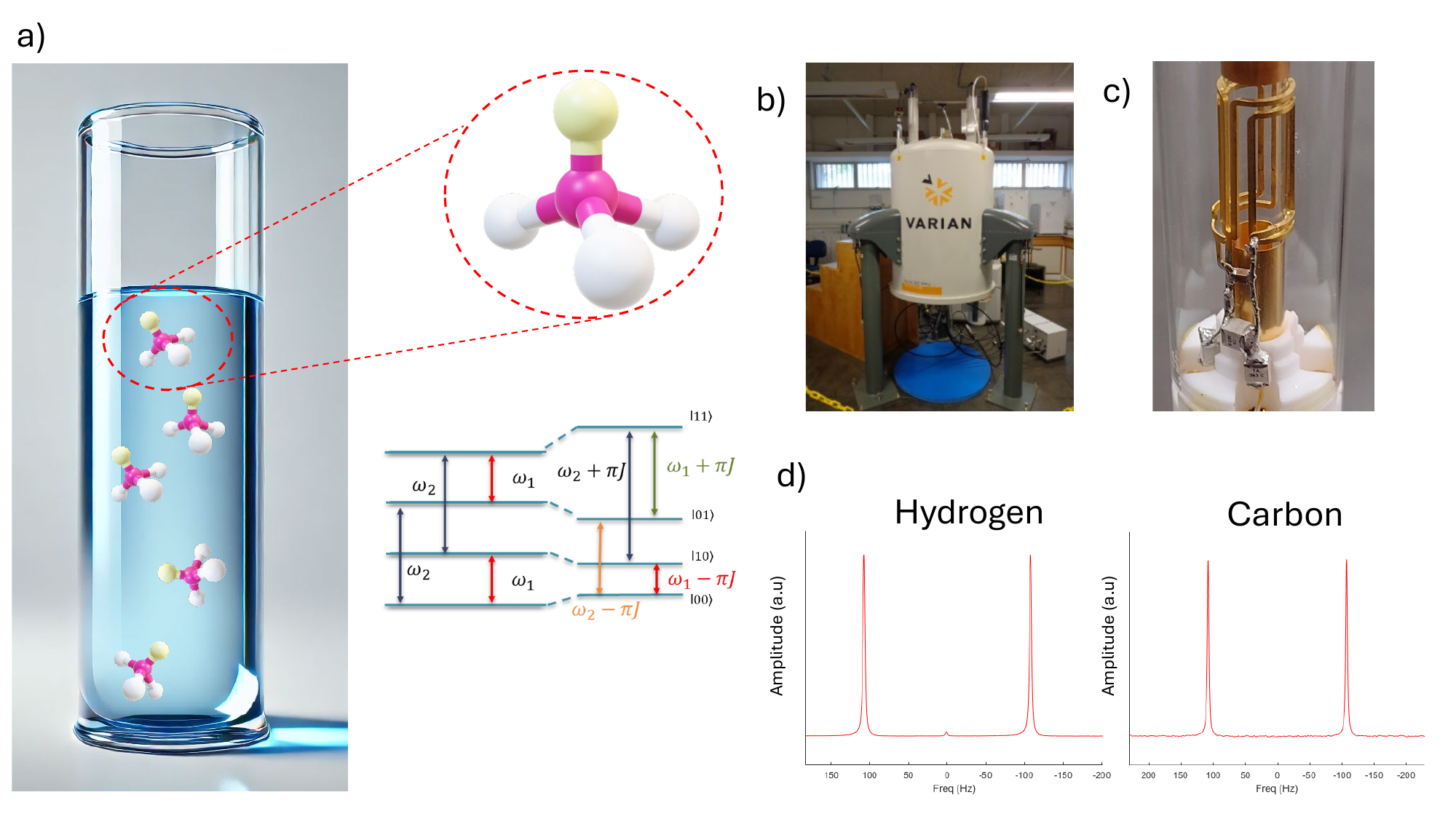}
  \end{minipage}%
  \hskip0.1cm
  \begin{minipage}{0.4\textwidth}
\captionof{figure}{Experimental setup: a) A sample containing a large number of molecules, represented here by the chloroform molecule, along with an illustration of its energy levels under a magnetic field. b) A typical cryostat that contains the superconducting magnet generating the static magnetic field. c) A typical RF coil used to apply radio-frequency pulses and detect the NMR signal. d) Observed NMR spectra of hydrogen and carbon for the chloroform molecules, where each line reveals partial information about the system density matrix. \label{fig:Magneto}}
  \end{minipage}
\end{figure}

The NMR technique is a spectroscopic method that investigates the response of atomic nuclei with non-zero angular momentum when subjected to magnetic fields. In NMR experiments, a static magnetic field is complemented by the application of oscillating RF fields that induce transitions of nuclear spin levels \cite{SlichterLivro1990}, Figure \ref{fig:Magneto} illustrates the configuration of the system. 
Most NMR experiments in quantum thermodynamics are performed on liquid-state samples at room temperature. In this regime, rapid molecular motion averages out dipolar interactions, while spins within each molecule remain coupled through scalar interactions. Under these conditions, the thermal energy far exceeds the magnetic energy associated with the nuclear spin levels. When a strong external magnetic field is applied, the nuclear spins experience Zeeman splitting of their energy levels. Although the population difference between these levels is extremely small, it is sufficient to generate a measurable net magnetization.
This small polarization deviation enables the preparation and manipulation of effectively pure subspaces by means of RF pulses.  Furthermore,  it is also possible to improve the signal-to-noise ratio using hyperpolarization techniques \cite{Verhulst2001} and algorithm cooling \cite{Boykin2002,Lin2024,Molpeceres2025}.
Importantly, all observable NMR signals arise from this pure component of the density matrix. Consequently, despite the high-temperature nature of the ensemble, coherent quantum processes can be implemented and controlled with high precision.

Hamiltonian simulation, a method in which one quantum system emulates another, is particularly valuable in NMR-based quantum thermodynamics. For example, thermal contact between two spins can be simulated by combining RF pulses with the free evolution of spin systems to generate an effective Hamiltonian $H_{eff}$ \cite{Knill1998}. This capability makes liquid-state NMR a powerful platform for exploring fundamental aspects of quantum information processing and quantum thermodynamics under well-controlled experimental conditions, allowing the simulation of spin ensembles over a wide range of effective temperatures. Although $H_{eff}$ governs unitary evolution, tracing out other qubits results in a non-unitary map, mimicking the thermalization of thermal reservoirs \cite{Batalhao2014,Peterson2019,Serra2019,Assis2019} or using the qubits as a true thermal environment \cite{Mendonca2020}.
Solid-state NMR can also be a valuable tool for quantum thermodynamics; these systems enable the study of complex many-body systems, as dipolar coupling becomes the dominant interaction among spins in solids \cite{SlichterLivro1990}. Although individual spins cannot be directly manipulated, solid-state NMR techniques developed over the years for Hamiltonian simulation and spin decoupling allow useful manipulation of the collective spin dynamics to study thermalization processes and non-equilibrium dynamics~\cite{Suter2010,Barret2018,Suter2021,Beatrez2023}.

\paragraph*{Current and future challenges.}
The usefulness of NMR for testing quantum thermodynamics has been demonstrated in both liquid and solid. In liquids, individual spins can be precisely manipulated, though the number of controllable spins is limited. Although most experiments involve up to three spins, advances in NMR quantum computing suggest that full control of around ten spins is feasible \cite{Laflamme2008}. A key advantage of NMR over superconducting circuits or trapped ions is its ability to operate at room temperature with more accessible equipment.

In dipolar-coupled solid-state NMR, individual spin control is lost, but dipolar interactions naturally create a complex 3D many-body system of thousands of spins. Standard pulse techniques allow the quantification of interacting spins, the modification of the interaction strength, and the implementation of various Hamiltonian dynamics.
This enables the study of the collective spin dynamics.

A striking example observed in solid-state NMR is many-body localization, where a system fails to thermalize despite interactions, defying conventional equilibrium statistical mechanics \cite{Suter2010}. Using the theory of effective Hamiltonian, it is also possible to simulate the thermalization of an ensemble of single spins in contact with an environment at negative temperatures \cite{Mendonca2020}. Additionally, periodic pulse applications in dipolar-coupled systems have revealed signatures of time crystals \cite{Barret2018} and Floquet prethermalization \cite{Beatrez2023}. Here, it is important to note that even before recent interest in many-body physics, NMR had already observed signs of collective spins dynamics such as the Floquet prethermalization, previously termed quasi-equilibrium in the NMR literature \cite{Suter2021}. Given the rich set of physical phenomena and potential applications in quantum information devices, solid-state NMR is expected to be further explored in the future.

In the scenarios discussed above, we have considered spin ensembles at room temperature. However, single-spin magnetic resonance can also be implemented using Nitrogen Vacancy (NV) centers in diamonds. NV centers are point defects in the diamond lattice, where a nitrogen atom replaces a carbon adjacent to a vacancy. In its negatively charged state, the ground state of the six-electron system (two from Nitrogen, three from Carbon, and one additional captured electron) has spin $S=1$, which can be manipulated using microwave irradiation. In crystals with sufficiently low defect concentrations, individual defects can be observed via a technique known as Optically Detected Magnetic Resonance (ODMR). This setup can be further used for quantum thermodynamics in the future, as it allows experimental tests to be performed on a single spin \cite{Hernandez2022}.   \\

\paragraph*{Broader perspective and relevance to other fields.}
The ability to manipulate spin systems with high precision makes NMR an ideal platform for studying fundamental thermodynamic processes at the quantum scale. The control techniques developed over years of research have allowed using NMR to design and experimentally test concepts of quantum thermodynamics, such as designing quantum engines, building thermal reservoirs, studying fluctuation theorem, work production and entropy of thermodynamic cycles. 

Apart from its relevance for fundamental physics, NMR based quantum thermodynamics has also the potential to drive advancements in quantum technology in different fields. For example, analyzing the thermodynamic cost of quantum information processing can help to optimize the operation of quantum computers. A deeper understanding of energy dissipation and entropy generation in quantum systems could lead to better designs for quantum hardware \cite{livro_DeffnerSteve}. Quantum thermodynamics may also contribute to the development of novel devices in the realm of quantum technologies. Quantum batteries and quantum engines are examples of proposed devices whose performance can be enhanced by concepts developed within quantum thermodynamics. The lessons learned from NMR could advance such technologies in a similar way to that NMR quantum information processing has done for quantum computing \cite{Jones2011}.

Quantum many-body physics is another field in which NMR serves as a valuable experimental platform. The dynamics of interacting quantum particles are central to understanding non-equilibrium quantum matter, quantum information propagation, and fundamental aspects of condensed matter physics, such as the mechanisms of thermalization.

A particularly intriguing phenomenon in this context is the study of time crystals, which hold potential applications in quantum sensing. Solid-state NMR techniques applied to dipolar-coupled nuclear spins enable analog quantum simulations of many-body physics, as demonstrated in recent experiments \cite{Suter2010,Barret2018,Suter2021,Beatrez2023}. As interest in quantum many-body physics continues to grow within the condensed matter and quantum information communities, NMR is expected to play an increasingly valuable role as an experimental technique in this field of research.

\paragraph*{Concluding Remarks.}
Magnetic resonance is one of the most successful experimental techniques used in both academia and industry. In particular, nuclear magnetic resonance has an exceptionally wide range of applications. To cite just a few examples, NMR plays a crucial role in chemical analysis, medical imaging, drug discovery, and oil exploration.

After decades of development, NMR has evolved into a robust and precise technique for manipulating nuclear spins. Beyond its traditional applications, NMR has also been established as a valuable tool for advancing quantum control techniques. Notably, it was the first experimental technique to demonstrate the implementation of a quantum algorithm, marking a significant milestone in the field of quantum computing.

More recently, NMR has emerged as an excellent testbed for the rapidly growing field of quantum thermodynamics, providing a controllable and accessible platform for investigating fundamental thermodynamic principles at the quantum level. Additionally, it has proven to be a powerful tool for studying quantum many-body systems, enabling experimental exploration of complex quantum interactions and non-equilibrium dynamics. As interest in these areas continues to grow, along with potential applications in quantum information technology, we expect NMR to remain a versatile and indispensable platform for experiments in quantum physics. 

\paragraph*{Acknowledgements.}
The authors acknowledge support from  CNPq and National Institute of Science and Technology of Quantum Information (Grant No. 465 469/2 014-0). TMM thanks the S\~ao Paulo Research Foundation (FAPESP) (Processes No. 2021/01277-2 and No. 2022/09219-4) and CNPq (Grant No. 446576/2023-9). AMS thanks FAPERJ (Pocess. No. E-26/203.946/2024). 

\clearpage

\section{Quantum thermodynamics with trapped ions}
\label{sec:trappedions}
\noindent
{\it Giacomo Guarnieri}

\noindent
{Department of Physics ``A. Volta", University of Pavia, Via Bassi 6, 27100, Pavia, Italy, and}\\
{INFN Sezione di Pavia, Via Agostino Bassi 6, I-27100, Pavia, Italy}\\

\noindent
{\it Ulrich Poschinger}

\noindent
{QUANTUM, Institut f\"{u}r Physik, Universit\"{a}t Mainz, D-55128 Mainz, Germany}\\

\paragraph*{State-of-the-art.}
The technological progress of the past century has culminated to the construction of the first quantum devices, whose promise is to exploit genuinely quantum features to outperform their classical counterparts. These devices, ranging from high-precision sensors to quantum computers, are realized through a variety of physical platforms, each presenting their own distinct advantages and challenges.
In this rich panorama, quantum thermodynamics stands out as a powerful compass: through the formulation of fundamental bounds and universal equalities, its aim is to characterize their performance through the physically meaningful lens of the 'thermodynamic cost' associated with their operation ~\cite{Landi2021}.
Recent developments of quantum thermodynamics are in fact devoted to precisely quantify the impact of quantum properties such as coherence, entanglement, non-Markovian memory effects and even quantum measurements on thermodynamic quantities such as the extractable output power or the dissipated heat, as well as their fluctuations. This framework ultimately provides guiding principles to optimize the operational regimes and energetic footprint of next-generation quantum technologies.

Among the existing quantum platforms, trapped ions have gained the spotlight as one of the most suitable systems for bottom-up realizations of protocols from quantum thermodynamics, with the ultimate goal of probing it beyond the boundaries of current theoretical predictions. Radiofrequency or Penning traps are employed to store individual atomic ions, typically group II or rare-earth atoms, each encoding a two-level system (qubit) on pairs of long-lived states, either Zeeman or hyperfine sub-levels of the electronic ground state and/or an optically excited metastable electronic state. Crucially, each ion is tightly confined in well-defined potential wells, often arranged in linear chains, and accurately manipulated using radiofrequency-, microwave- and optical fields. Measurements are carried out by detecting state-dependent laser-induced fluorescence. The availability of optical transitions enables the convenient engineering of dissipative channels, which is particularly neat in the context of quantum thermodynamics. The high degree of control enables the realization of engineered (and also time-dependent) Hamiltonians and baths that are useful for simulating thermodynamic processes in the quantum regime. Additionally, laser cooling techniques can bring ions to the motional ground state, which gives access to coherent manipulation and probing of harmonic oscillator degrees of freedom. This enables the realization of a wealth of protocols within quantum thermodynamics, for instance a continuous degree of freedom can act as a work repository~\cite{flywheel2019} (or \emph{quantum battery}). 

Trapped ion platforms exhibit long coherence times and yet unparalleled precision in quantum state control and high-fidelity operations. This is precisely what rendered them ideal testbeds for conducting groundbreaking experimental investigations in quantum thermodynamics over the past decade. These include the realization of single-spin quantum heat engines~\cite{Bouton2021} and heat-leak detectors~\cite{Pijn2022}, the investigation of Landauer's principle~\cite{Yan2018} and of work fluctuation relations and entropy production's quantum distribution~\cite{Onishchenko2024}. Finally, trapped ions offer capabilities for high-fidelity in-sequence readout and coherent real-time feedback operations, which is of particular importance for realizing variations of Maxwell's demon in the quantum realm \cite{stahl2024demonstrationenergyextractiongain}. \\

\paragraph*{Current and future challenges.}
One of the first challenges posed by quantum thermodynamics is the proper definition and precise measurement of work and heat in the quantum realm. A common approach is to define them (conceptually and operatively) through the so-called two-point measurement (TPM) scheme, according to which the quantities entering the energy book-keeping equation known as the First Law require a consecutive projective measurement of energy at different times. This in turn requires the possibility to perform high-fidelity quantum non-demolition measurements, which are available on trapped ion platforms~\cite{erhard2021entangling}. In trapped ion systems, the action of external fields, e.g. arbitrary waveform generators, lasers or microwave sources, effectively result in time-dependent
Hamiltonian driving and thus can be properly interpreted in terms of mechanical work which modify the energy levels of the system.

Most importantly, trapped ions are able to preserve quantum coherence and entanglement for comparatively long times and with high accuracy. This makes them the perfect candidate to study the impact of genuine quantum properties (such as quantum friction) on thermodynamic quantities such as work and entropy production, not just at the level of averages but also at the full stochastic level. Furthermore, the typically short duration of the laser pulses, used to emulate Hamiltonian quenches and unitary gates, allows to implement many operations on the system before having to reset the qubit, thus granting the possibility to investigate  thermodynamics even in the slow-driving regime.

By using e.g. laser-driven interactions, researchers can furthermore create entangled states that serve as resources in quantum thermodynamic protocols, such as quantum-enhanced work extraction or refrigeration ~\cite{maslennikov2019quantum}. All these experiments help to elucidate the role of quantum coherence in thermodynamic transformations, paving the way to elucidating how classical thermodynamics emerges from quantum mechanics.

Despite the high degree of isolation of trapped ions from the surrounding environment, these systems have been used to detect heat leaks~\cite{Pijn2022} and even to explore fundamental questions such as the emergence of the second law of thermodynamics in small quantum systems, the efficiency of quantum heat engines and fluctuation relations ~\cite{Hu2020}. In order to achieve this, additional laser fields were successfully employed in order to mimic the effects of external thermal baths onto the dynamics of the ions: for example, amplitude-damping or depolarizing channels can readily be realized by means of incomplete optical pumping. Dephasing channels can be realized by generating entanglement with ancilla qubits, which are subsequently traced out.
Importantly, the concept of temperature, one of the most fundamental ones in thermodynamics, is likewise mimicked owing to the fact that trapped ions encode qubits whose populations can always be made to match those given by a thermal Boltzmann-Gibbs ensemble.

One of the main technological challenges that trapped-ion platforms have faced insofar is scalability.
Although recent progress has been made in this respect~\cite{chen2024benchmarking}, there are technical limits to the maximum number of qubits that can be stored and controlled in a single Paul or Penning trap. Different approaches to extend scalability have been explored and implemented, such as the so-called Quantum Charge Coupled Device (QCCD) architecture, which aims at proving increased scalability via multiple storage potentials and ion shuttling.
Fully overcoming this technological limitation would open the door to study quantum thermodynamics with trapped ions in the many-body regime, where emergent thermodynamic phenomena such as phase transitions are expected to onset. \\

\paragraph*{Broader perspective and relevance to other fields.}
The exquisite level of control and versatility offered by trapped-ion systems has not only made them instrumental in order to advance quantum technologies in general as well as to investigate a plethora of other physical scenarios.

Trapped ion platforms are in fact one of the leading contenders for realizing universal quantum computers, both in  academia and industry. Owing to the pioneering work by Landauer and Bennett, quantum computation and quantum thermodynamics are considered to be profoundly intertwined, and studies of Landauer's erasure at the quantum stochastic level or quantum thermodynamic uncertainty relations are finding increasing applications to quantum circuits and to quantify the thermodynamic cost of mid-circuit measurements. Moreover, there is a fundamental link between quantum error correcting codes and quantum heat engines~\cite{danageozian2022thermodynamic}.
The synergetic development of both quantum thermodynamics, especially of its stochastic version, and of trapped ion platforms, has the potential to foster an improved understanding of the notoriously intricate question for which problems and under which conditions a genuine quantum advantage can be expected.  

However, the applicability of both trapped ion systems and stochastic thermodynamics branches out beyond quantum computing. Laser-cooled ions have been used to emulate nano-contacts and to provide insights into friction processes~\cite{Kiethe2017}, ranging from earthquakes and wear/crack propagation, to fibrous composite materials, DNA strands sliding and protein propagation. All these scenarios are in turn notoriously framed and quantitatively described by stochastic thermodynamics results, such as Jarzynski's and Crook's fluctuation relations.

Further applications of trapped ions include simulations of spin-lattices relevant to condensed matter systems and to strongly-coupled quantum field theories, with applications to high-energy particle and nuclear physics~\cite{Bauer2024}.
Last but not least, trapped ions also find applications in high-precision sensing ~\cite{gilmore2021quantum} and atomic clocks~\cite{Ludlow2015}. Serendipitously, one of the main developments of quantum thermodynamics nowadays follows its applications to quantum metrology and sensing, as well as to quantum clocks~\cite{Pearson2021}. \\

\paragraph*{Concluding Remarks.}
Trapped ion systems have established themselves as a cornerstone in the advancement of quantum technologies, offering unparalleled precision in the control and measurement of individual quantum systems. Their exceptional capabilities have facilitated significant progress across various domains, including quantum computing, precision metrology, and quantum simulation.

Owing to significant developments over recent years, these platforms have also established themselves as a powerful and reliable testbed for quantum thermodynamics, enabling the investigation of fundamental thermodynamic principles at the quantum scale. Experiments with trapped ions have provided critical insights into quantum heat engines, work fluctuations, entropy production, and the thermodynamic role of coherence and entanglement.

As the field of quantum thermodynamics continues to expand, alongside its growing relevance to quantum information science and quantum technologies, trapped ion platforms are poised to remain at the forefront of experimental quantum thermodynamics. \\

\paragraph*{Acknowledgements.}

G.G. acknowledges support from the Italian Ministry of Research (MUR) under the grant 'Rita Levi-Montalcini'.

\clearpage

\section{NV centers in diamond for quantum thermodynamics}
\label{sec:NVcentres}
\noindent
{\it Santiago Hern\'{a}ndez-G\'{o}mez and Nicole Fabbri}

\noindent
{Istituto Nazionale di Ottica del Consiglio Nazionale delle Ricerche (CNR-INO), and European Laboratory for Non-linear Spectroscopy (LENS), Università di Firenze, via Nello Carrara 1, I-50019, Sesto Fiorentino, Italy}\\

\paragraph*{State-of-the-art.}
While quantum thermodynamics has made enormous strides in theoretical research, experimental progress is challenged by the need of precise control of thermodynamic processes in quantum systems at the nanoscale. In recent years, advances in  quantum technologies have provided new test beds for quantum thermodynamics, ranging from superconducting circuits (Sec.~\ref{sec:superconductors}) to  ultracold atoms (Sec.~\ref{sec:ultracold}), to quantum dots (Sec.\ref{sec:quantumdots}), NMR systems (Sec.~\ref{sec:NMR}), trapped ions (Sec.~\ref{sec:trappedions}), and, more recently, color centers in diamond, as discussed in this Section. Color centers are fluorescent quantum defects in solid state materials, with local electronic wavefunctions that mimic the behavior of single trapped atoms. Among these centers, nitrogen-vacancy (NV) centers in diamond – consisting of a nitrogen atom adjacent to a vacancy in the diamond lattice – stand out due to their remarkable properties.  Their success in quantum technology applications spanning the three areas of quantum sensing~\cite{Barry2020}, computing~\cite{Neumann2008, Abobeih2022}, and communication~\cite{Sipahigil2012}, is largely due to their exceptionally long coherence times for both electronic and nuclear spins. Remarkably, coherent spin control, as well as initialization and readout at the single qubit level can be achieved under ambient conditions.
The spin energy levels of the NV electronic triplet ground state can be used to implement both spin qubits and qutrits. Optical initialization into a pure state allows for preparation in both mixed and pure states. The interaction with a rich environment enables the realization of two- or multi- qubit dynamics, and the implementation of unitary, unital, and dissipative maps. Moreover, the solid-state nature of diamond color centers enables integration into chip-scale architectures, and compatibility with nanofabrication techniques, crucial aspects for device scalability.

\begin{itemize}
\item Nuclear spin bath. The NV qubit interacts with a complex environment, composed of randomly distributed $^{13}$C nuclear spins
and other paramagnetic impurities. The large ensemble of unresolved $^{13}$C can be treated as a collective bath, where the ratio between the environment internal energy and its coupling to the NV center can be tuned by varying the strength of an applied external magnetic field, allowing exploration of different  regimes~\cite{Reinhard2012, Hernandez2018}: 
\begin{itemize}
\item[(i)]	Thermal bath. In the weak coupling regime, the unpolarized nuclear spin ensemble can be described as a time-varying mean field with stochastic amplitude and phase, and the interaction of the NV electronic spin and this bath takes the form of an effective dephasing Hamiltonian. 
\item[(ii)]	Quantum bath. In the strong coupling regime, the dynamics of the nuclear spin bath is influenced by the controlled NV dynamics, due to the back action of the NV spin onto the bath itself. The electron-nuclear dipole-dipole interaction results in entangled states, which contribute to decoherence by destroying the off-diagonal elements of the NV density matrix.
\end{itemize}
\item Coupling to single nearby nuclear spins. The controlled manipulation of entangled states in strongly coupled electron-nuclear spin systems opens up opportunities for creating hybrid spin registers demonstrated nowadays with up to 50 qubits~\cite{vandeStolpe2024}, developing advanced readout schemes~\cite{Neumann2010}, and implementing interferometric protocols that make use of ancillary qubits~\cite{Hernandez2025}.
\item Engineered laser-induced dissipation and projective measurements. The application of short laser pulses can be used to combine quantum projective measurements and tunable optical pumping, providing new avenues for exploring quantum dissipative processes in the presence of feedback mechanisms~\cite{Hernandez2022}.
\end{itemize}

The combination of the precise  control on electronic and nuclear spins at the single qubit level and the versatility in implementing a large variety of open quantum system scenarios makes NV centers in diamond an emerging ideal platform for investigating thermodynamic processes at the nanoscale. Recent groundbreaking results include the experimental demonstration of power advantage in a quantum heat engine~\cite{Klatzow2019}, the realization of a novel type of autonomous Maxwell’s demon acting on a dissipative channel~\cite{Hernandez2022},  the observation of quasi-Floquet prethermalization~\cite{He2023a}, the observation of coherence signatures in entropy production~\cite{Hernandez2023} and anomalous work extraction~\cite{Hernandez2024, Hernandez2025}. \\

\paragraph*{Current and future challenges.}
Single NV centers have been used to investigate  energy transfer mechanisms in open quantum systems, particularly within the framework of Jarzynski-like quantum fluctuation relations using the two-point measurement (TPM) protocol~\cite{Hernandez2022, Hernandez2020}. However, experimentally verifying quantum fluctuation relations in driven open quantum system remains a significant challenge. Progress has been made with the demonstration of the quantum fluctuation relation in its integral form, in the special cases of an effective infinite-temperature reservoir, and when the total work vanishes at stroboscopic times despite non-zero delivered power~\cite{Hernandez2021}. Nevertheless, an experimental demonstration in more general scenarios is still lacking, primarily due to the conceptual and operative difficulty of distinguishing work and heat. An interferometric method has been theoretically proposed to obtain the work statistics in open driven systems, even in the strongly dissipative regime~\cite{Campisi2013} but the experimental implementation is still missing.

Recent approaches were also aimed at clarifying the contribution of quantum coherence and multi-time correlations, which are not captured by standard TPM protocols. The end-point measurement (EPM) approach~\cite{Gherardini2021} was used to characterize the entropy production associated to quantum coherence in the initial  state of a driven open quantum system~\cite{Hernandez2023}. Additionally, the measurement of Kirkwood-Dirac quasiprobability (KDQ) distributions of work in closed systems was achieved via a weak-TPM~\cite{Hernandez2024} and an ancilla-assisted interferometric scheme~\cite{Hernandez2025}, enabling the measurement of two-time correlations between incompatible observables, the observation of anomalous work extraction and the demonstration of the Robertson-Schr\"{o}dinger uncertainty relation. However, the measurement of KDQ distributions in open quantum systems remains an open issue.

NV centers can also be used to explore new regimes in quantum heat engines, where not only efficiency but also power output can be optimized. NV ensembles have been used to implement quantum heat engines, demonstrating a power advantage due to coherence~\cite{Klatzow2019}. Exploring the effects of out-of-equilibrium dynamics at stroke transitions would enable exploration of thermalizing and non-thermalizing heat strokes, which are expected to result in higher work extraction~\cite{Alicki2015}.
 
More broadly, current thermodynamics experiments on diamond platforms have been currently limited to single qubits or qutrits, or two-qubit interactions. While single NV centers provide a rich platform for quantum thermodynamics, many practical applications require larger, multi-qubit systems, where collective effects and entanglement can play a more pronounced role. 
Dipole-dipole interaction in hybrid electronic-nuclear spin systems or electron-electron spin systems (NV-NV or NV-P1) could be used to study non-trivial many body phenomena, such as chaotic dynamics and realizations of the Eigenstate Thermalization Hypothesis
(Sec.~\ref{sec:thermalisation}).
This requires overcoming technical challenges related to the control and the protection of coherence of multi-qubit systems, addressable in low temperature experiments, and exploiting advanced coherent control tool such as dynamical decoupling~\cite{Bar-Gill2013}. \\

\paragraph*{Broader perspective and relevance to other fields.}
The high degree of control and the versatility of the NV platform enables the investigation of a variety of open quantum system dynamics and explore energy exchange mechanisms at the nanoscales. Remarkably, NV centers have recently been used to demonstrate KDQ as a promising distribution for understanding work statistics. Extending previous experimental findings to many-body dynamics opens the door to exploring regimes where correlations between subsystems become crucial. Measuring multi-time correlation functions not only provides insights into the statistics of individual observables but also reveals their trajectory-like statistics. This approach addresses a central challenge in quantum thermodynamics: defining non-state variables like Work, Heat, and Entropy. It paves the way for studying new efficiency and power regimes in heat engines, investigating irreversibility and the arrow of time in quantum systems, and testing fundamental limits such as the Leggett–Garg and Bell inequalities for observables measured at different times. 

We also anticipate that a solid, experimentally-grounded understanding of quantum thermodynamic processes at the nanoscale will drive advancements in energy harvesting and the management of dissipation and thermalization in quantum devices. This could lead to the energetic optimization of future quantum technologies through more informed design, a topic of growing importance in light of industrial adoption and sustainability concerns. The discovery of novel behaviors may spark technological breakthroughs, such as optimized quantum gates, innovative sensing schemes, on-chip engines and refrigerators, or new energy storage methods. Furthermore, the research in this field could provide valuable insights into the experimental validation of foundational concepts like work in the quantum realm, quantum fluctuations, and autonomous systems, with far-reaching applications in quantum sensing and computing. \\

\paragraph*{Concluding Remarks.}
The long coherence times of diamond spins and the ability to address them individually, along with the interactions of the NV center with the surrounding nuclear spin environment, and the possibility to engineer unitary, unital, or dissipative maps by applying laser and mw pulses, provide a rich testbed for simulating complex quantum systems and investigating novel quantum thermodynamic effects. In many quantum technology applications in the areas of sensing, computing and communication, decoherence and dissipation are significant obstacles, as they degrade the fidelity of quantum gates and diminish the sensitivity of quantum sensors. 
Understanding the role of thermodynamic aspects — such as work and heat fluctuations, entropy production, and thermalization — in the operation of quantum devices, including their impact on sensing precision, is of fundamental interest. This could reveal how thermodynamic processes set ultimate limits on device performances and inform the design of energetically optimized quantum devices.
\\

\paragraph*{Acknowledgements.}
The authors acknowledge financial support from the European Union’s Next Generation EU Programme through the IR0000016 I-PHOQS Infrastructure, through PE0000023 NQSTI, and through PRIN 2022 QUASAR. The work was also supported by the European Union’s Research and Innovation Programme Horizon Europe G.A. no. 101070546 MUQUABIS, and by ASI through the INO-ASI Joint Laboratory for Quantum Technologies.

\clearpage

\section{Thermodynamics for enhanced mechanical quantum control}
\label{sec:optomechanics}
\noindent
{\it Nikolai Kiesel and Salamb\^{o} Dago}

\noindent
{University of Vienna, Faculty of Physics, Vienna Center for Quantum Science and Technology (VCQ), Boltzmanngasse 5, A-1090 Vienna, Austria}\\

\noindent
{\it Eric Lutz}

\noindent
{Institute for Theoretical Physics I, University of Stuttgart, D-70550 Stuttgart, Germany}\\

\noindent
{\it Andreas Deutschmann-Olek}

\noindent
{Automation and Control Institute, TU Wien, A-1040, Vienna, Austria}\\

\paragraph*{State-of-the-art.}
Advanced real-time feedback control schemes are essential in modern technology and are nearly ubiquitous in diverse applications, ranging from autonomous driving to stabilizing power grids. The integration (or co-design) of systems and control algorithms is known to be key for high-performance applications, as exemplified by the field of mechatronics \cite{munnig_schmidt_design_2014}. 

Over the last decades, methods from automatic control have also been applied to quantum systems \cite{wiseman_quantum_2009} with striking success, e.g., to generate non-classical states in Rydberg atom arrays \cite{omran_generation_2019} by pre-calculated control signals or to stabilize photon-number states in a cavity \cite{sayrin_real-time_2011} through measurement-based feedback schemes. 

However, the integration of control algorithms into the design of quantum systems still leaves much to explore and constitutes a highly promising avenue to genuinely novel and innovative quantum platforms. Section~\ref{sec:control} of this roadmap describes open-loop control (like shortcuts to adiabaticity). In contrast, our perspective focusses on measurement based closed-loop control and on optomechanical systems \cite{RevModPhys}. Here, real-time optimal feedback has so far been successfully applied to effectively counteract thermal and quantum back-action noise to prepare nearly pure Gaussian quantum states via ground-state cooling, e.g. \cite{RossiSchliesser, magrini_real-time_2021, tebbenjohanns_quantum_2021}. Looking ahead, especially experiments based on levitation are poised to extend operations with dynamically shaped nonlinear potentials into the quantum non-Gaussian regime \cite{Review_Levitation}.  With experimental capabilities improving rapidly, also the demand for more complex control tasks in the quantum regime increases and the role of measurement and quantum fluctuations becomes central. Accordingly, a natural question emerges: Can quantum thermodynamics inform the design of experiments and real-time control strategies? To answer this question, one has to bridge the gap between theory in stochastic and quantum thermodynamics and control engineering.

From a thermodynamic perspective, measurement-based feedback schemes essentially implement a classical Maxwell demon,  embodied by detectors and a computer, that can monitor the mechanical motion and act accordingly on the system. Such a scenario has been implemented and analyzed on a variety of experimental platforms \cite{Parrondo2015Feb, gieseler_millen_levitated_2018}, for example, to create virtual potential landscapes or for the comparably simple task of motional feedback cooling. Thermodynamics provides limits to the performance of measurement-based feedback control \cite{tourchette, bechhoefer} in the form of fluctuation theorems that incorporate information
gain. Such extensions of the second law \cite{SagawaUeda} restrict the efficiency of information-to-work conversion and the amount of work that can be extracted from the measured system. They also set limits on the minimal amount of dissipation required to implement the feedback control \cite{tourchette, bechhoefer,SagawaUeda}. The influence of imperfect measurements has been analyzed in \cite{ferri-cortes_conditional_2025}. In the quantum setting, unavoidable measurement back action, that perturbs the state of the system and is hence associated with a thermodynamic cost \cite{jordan_measurement_2020}, strongly impacts the action of the feedback protocol, either enhancing or suppressing it \cite{jordan_measurement_2020}.

\paragraph*{Current and future challenges.}
Yet, a mismatch exists between the description of Maxwell's demon in information thermodynamics and optimal feedback control even in the simple classical linear case: Thermodynamic bounds that include information terms typically assume control based on instantaneous measurements. This is well approximated in many experimental scenarios that have been investigated. However, control strategies in engineering usually rely on optimal state estimators (i.e., filters) that combine all available information from the history of measurements, making the resulting control law inherently non-Markovian. Integrating such feedback into standard stochastic master equations (SMEs) is particularly challenging, as incorporating non-Markovian effect into SMEs is nontrivial. This challenge extends to thermodynamics, where non-Markovianity has a critical impact on the second law and the analysis of fluctuation theorems even for linear feedback with delayed measurement \cite{Debiossac}. The ultimate thermodynamic bounds derived for a system under measurement-based feedback should be based on the optimal use of all available information, as is done in optimal control.

In turn, thermodynamic bounds tailored to available experimental resources could inform experiments that utilize feedback schemes about feasible operations and assess control performance as they would constitute fundamental benchmarks. This is particularly intriguing in the quantum regime, where the act of measurement fundamentally alters the state of the system through localization and backaction. For a linear feedback scenario, the contribution of quantum measurements to entropy production has been observed in a recent experiment~\cite{Kumasaki2025}. Such linear feedback strategies do not necessitate a dedicated quantum model as well-established stochastically optimal control strategies apply and quantum backaction only appears as correlations of measurement and process noise \cite{Belavkin}. However, experiments in the linear regime are intrinsically limited to Gaussian states. Today, new schemes enable access to nonlinear dynamics where genuine quantum features such as Wigner negativity appear, posing new opportunities and challenges both for quantum thermodynamics and measurement-based feedback control schemes. For example, the choice of measurement becomes crucial as it determines which quantum features, like Wigner negativity, can be preserved, and thus which states remain accessible. Specific strategies, such as restricting measurement to certain quadratures, can help preserve non-Gaussian features while mitigating excessive entropy production. At the same time, such selective measurements inherently limit the information available to control algorithms. 

In addition to these general constraints, the systematic design of new experiments also requires a mathematical formulation of the control objective. Specifically, this objective is implemented by a cost function that the control algorithm attempts to minimize based on the available system model, measurement data, and control parameters. For instance, in feedback cooling, the cost function is often chosen as the energy of a harmonic oscillator. In more advanced applications, the choice of cost function is less obvious. Consider, for instance, the preparation of entanglement between two mechanical oscillators in a common situation where only weak position measurements are available. Obviously, neither minimizing energy nor maximizing state purity will necessarily maximize the entanglement, but could also lead to independently cooled oscillators. Recently, it has been shown that EPR correlations provide a suitable cost function to enhance achievable entanglement \cite{engtanglement}. However, it may not be optimal given the available resources and constraints in any given experimental setting. The situation is even less clear for the generation of non-Gaussian states like cat states. Quantum thermodynamics may provide the right point of view to guide and evaluate the systematic utilization of measurement and model information in actively controlled open quantum systems. \\

\paragraph*{Broader perspective and relevance to other fields.}
The fundamental limits of control imposed by thermodynamics have received surprisingly little attention in the control community so far. However, they may become crucial when operating in the quantum regime. For experiments, feedback control is a powerful tool, enabling the manipulation of a system’s dynamics beyond what is naturally accessible. This results in a nonequilibrium steady state or process, where entropy production, decoherence, and entropy pumping by the feedback must be carefully managed. A framework that embraces both the control engineering and the thermodynamics perspective would significantly contribute to methods on the control of continuously monitored systems including applications in quantum optomechanics.

Measurement-based feedback control offers distinct benefits for mechanical sensing. First, keeping the system within a well-characterized operating range enhances sensor fidelity. Second, stabilizing naturally unstable dynamics improves the signal-to-noise ratio by amplifying the sensor’s response to external forces relative to detector noise. Also, time-dependent driving in combination with measurement-based control can prepare specific quantum states, which allow to push the sensing performance beyond the Heisenberg limit. This advantage is already evident in the stabilization of squeezed states (e.g., via optical cavities) and could be even more pronounced with non-Gaussian states.

Some fundamental tests of quantum physics require significantly extending the coherence length of increasingly massive quantum systems. As mass increases, the free expansion rate of initially pure quantum states decreases, such that decoherence requirements become difficult to manage. To speed up the expansion, the use of repulsive potentials has been proposed. Combining this approach with measurement-based feedback may offer a way to control the nonequilibrium process of preparing the desired quantum state in such unstable regimes, balancing the competing effects of entropy production, decoherence, and entropy pumping. While these constraints limit what can be achieved, they also define a structured landscape of possibilities, guiding the design of optimal control strategies to reach otherwise inaccessible quantum states.\\

\paragraph*{Concluding Remarks.}
The design of novel quantum experiments is driven by the creativity of researchers who work with an ever-expanding toolbox of methods. Classical control engineering is not only becoming an increasingly important part of this toolbox but also introduces a perspective that emphasizes the goal-driven co-design of experimental setups and control algorithms. 
While this brief perspective cannot do full justice to the wealth of groundbreaking work in quantum control and quantum thermodynamics, its aim is to highlight key opportunities and challenges in adopting such an interdisciplinary approach, using optomechanical systems as an example. We have argued that once the frameworks of quantum thermodynamics and control engineering are well-aligned, quantum thermodynamics can play a direct role in experiment design by benchmarking and informing control strategies. Beyond this, exploring the thermodynamics of quantum control in nonlinear systems presents a wealth of open questions that can be tackled through a joint effort at the intersection of both fields. \\

\paragraph*{Acknowledgements.}
A.D. acknowledges support by the Austrian Science Fund (FWF) [10.55776/COE1, PAT9140723] and the European Union – NextGenerationEU. S.D.  acknowledges support from the European Union under the Horizon Europe Marie Sk\l{}odowska-Curie Actions Postdoctoral Fellowship (Grant Agreement no. 101106514, FLIP). E.L. is supported by the German Research Foundation (DFG) [FOR 2724]. NK acknowledges support from the Austrian Science Fund (FWF, START project) [10.55776/Y952]

\clearpage

\section{Thermodynamics of quantum trajectories}
\label{sec:trajectories}
\noindent
{\it Guilherme Fiusa, Abhaya S. Hegde, and Gabriel T. Landi}

\noindent
{Department of Physics and Astronomy, Center for Coherence and Quantum Science, University of Rochester, Rochester, New York 14627, USA}\\

\begin{figure}[b]
    \centering
    \includegraphics[width=0.5\linewidth]{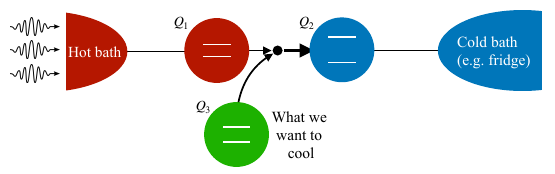}
    \caption{Schematics of a quantum absorption refrigerator with 3 qubits. Heat naturally flows from the hot bath to the cold bath,  through qubits $Q_1$ and $Q_2$. However, the qubits' frequencies $\omega_{1/2}$ are designed to be far off-resonance blocking the natural flow. 
    A third qubit $Q_3$ is introduced with a gap $\omega_3$, designed so that $\omega_1+\omega_3 = \omega_2$. 
    This resonance condition allows for two excitations, one in $Q_1$ and one in $Q_3$, to be converted into a single excitation in $Q_2$. 
    Any excitation in $Q_3$ will therefore be sucked off by the hot/cold temperature gradient. 
    The device therefore operates like an autonomous refrigerator helping qubit $Q_3$ cool.
    }
    \label{fig:absorption_refrigerator}
\end{figure}

\paragraph*{State-of-the-art.}
Fig.~\ref{fig:absorption_refrigerator} depicts a 3-qubit quantum absorption refrigerator, a device that has been central to the development of quantum thermodynamics (QT) in the last two decades~\cite{Mitchison2019}. Recently, this device was experimentally implemented using superconducting circuits~\cite{Gasparinetti2025}, which represents a paradigm shift in the practical uses of QT for modern quantum experiments. The basic idea of Ref.~\cite{Gasparinetti2025} was that qubit $Q_3$ is part of a quantum computer, and will therefore be involved in a variety of tasks. Whenever $Q_3$ needs to be reset, a microwave drive mimicking a hot bath is used to populate $Q_1$. Any excitation in $Q_3$ will then be combined through a resonance condition to create an  excitation in $Q_2$, which is then dumped to the cold bath. The authors have shown that this method is competitive to state-of-the-art qubit reset techniques. 

This experiment is the first concrete practical application of quantum thermodynamics. And it highlights new paradigms that will shape the community in the coming years. For example, in thermodynamics heat baths are traditionally assumed to be cheap resources. In quantum systems cold baths occur naturally, but hot baths do not. In fact, in Ref.~\cite{Gasparinetti2025} the authors used a microwave drive as a substitute. This suggests that quantum thermodynamicists might need to reevaluate the use of hot baths as cheap resources. 

The other paradigm shift introduced is the idea of \emph{cooling on demand}~\cite{Gasparinetti2025}. Most QT studies have focused on nonequilibrium steady states (NESSs), where steady currents of heat/work flow from one part of the system to another. For the scenario described in Fig.~\ref{fig:absorption_refrigerator} the steady state is irrelevant. Instead, the correct question is \emph{``how much time will it take until $Q_3$ is cooled?''} or \emph{``if I perform a cooling protocol with a fixed duration $\Delta t$, what is the probability that $Q_3$ cools?''} Addressing these questions requires going beyond steady states. This is a single-shot problem, which should therefore be studied from the perspective of quantum trajectories~\cite{Landi2024}. Even though quantum trajectories are, by now, fairly well understood, there are still several open questions concerning their thermodynamic properties. In this article, we discuss these challenges and the new and interesting research directions they might lead to. \\

\paragraph*{Current and future challenges.}
Unlike classical stochastic thermodynamics, quantum systems are prone to measurement-induced back-action and fluctuations, so analyzing the thermodynamics of quantum trajectories requires a new paradigm. This challenges the fundamental nature of entropy, work, and heat at the single-trajectory level~\cite{Hekking2013,Alonso2016}. Over the last decade, various experiments have studied thermodynamics at the level of single quantum trajectories~\cite{Rossi2020, Murch2013}. However, in these papers, the thermodynamic properties were \emph{assessed} from the trajectories and not directly measured since such measurements are extremely challenging even at the ensemble level~\cite{Ro_nagel2016, Pekola2013}. 

Many theoretical questions remain open, particularly regarding the role of quantum coherence in continuously monitored systems~\cite{Manzano2022}. In a coherent quantum trajectory, the system evolves in superpositions of energy eigenstates, raising concerns about the definitions of heat and work in this context. This reflects the Bayesian nature of quantum trajectories, where the conditional state is our best estimate based on available information. In contrast, classical energy is a tangible resource consumed to perform work. This disparity prompts a crucial question: how does energy in a quantum trajectory transition to a usable resource?

The challenge becomes even more pronounced when considering entropy production, a measure of irreversibility. Significant efforts have explored its quantum nature through Fluctuation Theorems (FTs)~\cite{Landi2021}.
The prevailing approach, which remains widely used, is that these should be based on a two-point measurement scheme, where a quantum system is measured at the beginning and the end of a protocol. However, quantum trajectories are inferred from continuous weak measurements which never let the system fully collapse, necessitating more sophisticated formulations of FTs.

The description of thermodynamic currents at the trajectory level must go beyond the steady state paradigm. Originally developed for long-time statistics, full counting statistics (FCS) provides a powerful framework for this analysis~\cite{Esposito2009}. Several groups, including ours, have recently sought to extend FCS to describe thermodynamics at the trajectory level~\cite{Hegde2024, Monsel2025}. For instance, consider the cooling protocol in Fig.~\ref{fig:absorption_refrigerator} as a single stochastic event. Thermal and quantum fluctuations lead to three possible outcomes: successful cooling $S$, a failure $F$ without cooling, or a disaster $D$ where $Q_3$ is heated instead.
Their probabilities sum to unity: $P_{S} + P_F + P_{D}=1$. Ideally, one would desire $P_{S}=1$, but this is unattainable in practice. Eliminating disastrous events ($P_D = 0$) may come at the cost of increasing failures $P_F$, raising a key question: Should a machine prioritize avoiding disasters at the cost of frequent failures, or allow a small probability of disaster to improve success? These trade-offs are central to thermodynamics in stochastic systems. 
While experiments ultimately reveal the answers, a robust theoretical framework is needed to assess these probabilities realistically. In addition, the \emph{time} associated with these events is also crucial. The duration of these stochastic events follows a waiting-time distribution that can be different for $S$, $F$, and $D$. Notably, if failures occur on short timescales, their impact may be mitigated by the ability to quickly attempt the cooling process again. Understanding these temporal aspects is key to practical cooling strategies. \\

\paragraph*{Broader perspective and relevance to other fields.} 
The ability to manipulate quantum trajectories is a powerful tool for understanding entropy production, fluctuations, and energetics of quantum systems leading to various new applications. Cooling --- as in Fig.~\ref{fig:absorption_refrigerator} --- is just one example. Another is quantum control and metrology discussed in Secs.~\ref{sec:optomechanics} and \ref{sec:control}. Understanding thermodynamic properties at the trajectory level can inform optimal strategies for manipulating quantum states with e.g. minimal energy dissipation. This has implications in quantum sensing, where thermodynamic considerations provide a way to improve sensitivity and precision,   see Secs.~\ref{sec:TURs} and \ref{sec:time}. Monitoring quantum trajectories and their statistics also provides a unique way to enhance feedback mechanisms that are crucial to e.g. stabilize quantum states in a quantum computer.
 
The characterization of quantum trajectories is also intimately related to technologies such as quantum heat engines and quantum batteries. In general, characterizing the steady state of such devices is not enough, since there are trade-offs between heating capacity and heating precision. In the same spirit, understanding the thermodynamics at the trajectory level can provide better charging protocols, see Sec.~\ref{sec:batteries}. If there is any ambition that quantum thermodynamic devices may be useful in the future, then it surely depends on our understanding of quantum trajectories.

Finally, experiments on small nonequilibrium systems very often lack spatial or temporal resolution to keep track of all relevant degrees of freedom, which renders the measurement of thermodynamic quantities very difficult. This has been broadly studied in the context of coarse-grained dynamics~\cite{Esposito2012}. Accounting for coarse-graining in the thermodynamic variables has long been an active topic of study in the stochastic thermodynamics community.  The presence of quantum coherence introduces new challenges. 

Quantum thermodynamics has long addressed fundamental questions, which become even more significant in the context of quantum trajectories. Classically, energy is a readily available resource whereas in quantum mechanics it is manifested as the quantity appearing in the exponential factor $e^{- i E t}$, essentially representing just a frequency. Bridging the gap between this abstract notion and the concept of energy as a usable resource is a non-trivial task that involves an amplification mechanism. Making this connection meaningful would be much easier if guided by experiments (see Secs.~\ref{sec:quantumdots} and \ref{sec:transport}). \\

\paragraph*{Concluding Remarks.}
We provided a brief overview of the thermodynamics of quantum trajectories. This field has seen exciting developments recently both in theory and experiment. It is also driven by the potential use of quantum thermodynamics as a genuinely useful tool for modern quantum-coherent applications. We have highlighted key challenges that work in unison with theory and experiment. Namely, (a) how to define and monitor trajectory-level thermodynamic quantities in a model-agnostic way; (b) how to extend fluctuation theorems for the quantum realm; (c) how to characterize the role of coherence and superposition in assessing the energetics quantum trajectories; and (d) how to use trajectories to properly and fully characterize out of equilibrium thermodynamic quantities. Overall, the insights on thermodynamical aspects of quantum trajectories are likely to contribute significantly in shaping the next generation of quantum technologies. \\

\paragraph*{Acknowledgements.}
This material is based upon work supported by the U.S. Department of Energy, Office of Science, Office of Basic Energy Sciences under Award Number DE-SC0025516.

\clearpage

\section{Thermodynamics of quantum transport: Energy conversion and transport spectroscopy}
\label{sec:transport}
\noindent
{\it Olivier Maillet}

\noindent
{Universit\'e Paris-Saclay, CEA, CNRS, SPEC, 91191 Gif-sur-Yvette, France}\\

\noindent
{\it Rafael S\'anchez}

\noindent
{Departamento de Física Te\'orica de la Materia Condensada, Condensed Matter Physics Center (IFIMAC), and Instituto Nicol\'as Cabrera (INC), Universidad Aut\'onoma de Madrid, 28049 Madrid, Spain\looseness=-1}\\

\noindent
{\it Janine Splettstoesser and Ludovico Tesser}

\noindent
{Department of Microtechnology and Nanoscience (MC2), Chalmers University of Technology, S-412 96 G\"oteborg, Sweden}\\

\paragraph*{State-of-the-art.}
Energetic properties of electron transport at the nanoscale are of importance in a wide range of fields reaching from the development of practical applications to fundamental questions in quantum transport and thermodynamics. Nanostructuring provides opportunities for energy filtering that can significantly improve thermoelectrics~\cite{Heremans2013Jul} or photovoltaics. 
\begin{figure}[b]
  \includegraphics[width=.8\linewidth]{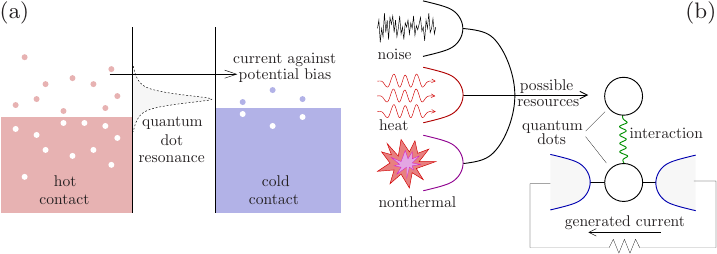}
  \caption{(a) A nanostructure acting as an energy filter generates a thermoelectric current. (b) A current can be generated in an isothermal and electrically isolated conductor (bottom) by interaction-mediated conversion of a resource (up) that can be heat but also noise or a nonthermal distribution. }
  \label{fig:scheme}
\end{figure}
Recently, similar energy-filtering mechanisms have  been used in a very different context, namely for \textit{energy conversion} in quantum-transport realizations of heat engines, which convert a heat flow into tiny amounts of electrical power but at high efficiencies, on the level of single-electron processes~\cite{Josefsson2018Oct,Thierschmann2015Oct}, or for cooling with mesoscopic devices~\cite{Giazotto2006}. In steady-state quantum-dot heat engines, for example, as sketched in Fig.~\ref{fig:scheme}, energy transfer occurs via energy-selective electron tunneling~\cite{Josefsson2018Oct}, Fig.~\subfigref{fig:scheme}{a}, or via Coulomb interaction between capacitively coupled quantum dots~\cite{Thierschmann2015Oct}, Fig.~\subfigref{fig:scheme}{b}. These experiments require careful design at the nanoscale to control the coupling to the environment. Crucially, the low energy scales of quantum devices ($\lesssim 1$K), restricts experiments to very low temperatures. This imposes additional hurdles, as couplings of electrons to thermal baths (phonons and photons) decay as power laws in temperature, which makes things even more challenging for small  sample volumes $\sim(10-100~\mathrm{nm})^3$~\cite{Giazotto2006}. Hence, controlling and fine-tuning electron temperatures remains a difficult task.

But low-temperature nanoelectronic setups also provide novel strategies for energy conversion that are not available in standard macroscopic engines~\cite{Benenti2017Jun}. 
Energy-filtering due to quantum interference~\cite{Benenti2017Jun} or even strong correlations~\cite{Chiaracane2020Jan} are at the basis of quantum thermoelectric effects, tunable via electromagnetic fields impacting the phase of coherent electron states. Strong magnetic fields also lead to time-reversal symmetry breaking in quantum Hall setups, where, similar to other topological systems, chiral electron transport increases control over heat and charge transport~\cite{Sivre2018} and even allows for switching between engine operation principles~\cite{Benenti2017Jun}. Furthermore, correlations induced between single-electron processes via capacitively coupled device elements provide a platform for implementing information-driven engines~\cite{Parrondo2015Feb}. 

\begin{figure}[b]
  \includegraphics[width=.8\linewidth]{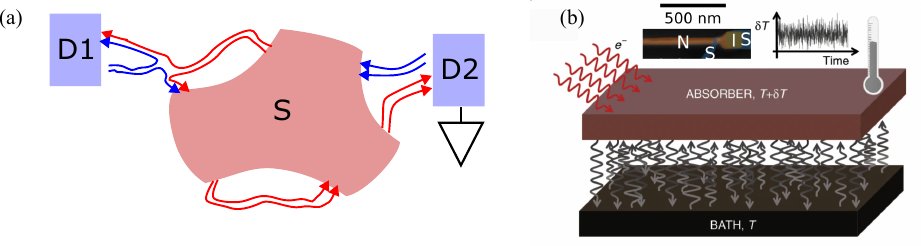}
  \caption{(a) Heat Coulomb Blockade experiment (representing the device used in~\cite{Sivre2018}): the macroscopic charge degree of freedom of a metallic island at temperature $T_\Omega$ is frozen due to strong Coulomb interactions. This leads to exactly one suppressed quantum of thermal conductance towards a drain reservoir D via chiral ballistic channels. (b) Temperature fluctuations monitored in a nano-absorber N (adapted from \cite{Karimi2020NatComms}) through proximity superconductor thermometry (S and I parts). At low mean calorimeter temperature, the lowest bound in temperature sensitivity at equilibrium is reached.}
  \label{fig:exp}
\end{figure}

From a different perspective, quantum thermodynamics provides a novel tool for quantum \textit{transport spectroscopy}, giving insights into quantum many-body effects, which are not accessible from pure charge transport analyses. Measurements of the Seebeck coefficient give access to Kondo physics~\cite{Dutta2019Jan} or energy current measurements are proposed to access strong electron correlations dominated by Coulomb interactions. For ballistic systems, the signature of these interactions are deviations from the otherwise quantized heat flow, like in heat Coulomb blockade~\cite{Sivre2018} where one quantum of thermal conductance is exactly suppressed [Fig.~\subfigref{fig:exp}{a}]. Such transport spectroscopy is made possible through  noise thermometry, exploiting the fluctuation-dissipation relation (FDR) between the voltage noise spectrum $S_\mathrm{V}$ of a mesoscopic quantum system at frequencies much lower than $k_BT/\hbar$, and its electron temperature $T$, $S_V=4k_\text{B}TR$. \\

\paragraph*{Current and future challenges.}
The intriguing features of the heat Coulomb blockade reveal hindered thermalization at time-scales relevant for heat transport~\cite{Sivre2018}. This absence of thermalization is relevant also for \textit{energy-conversion} processes in nanoscale devices~\cite{Sanchez2019Nov}. Indeed, one of the key differences between macroscopic and nanoscale engines lies in the available diverse resources, going beyond the standard resources (heat or work). The quantum-dot experiment~\cite{Thierschmann2015Oct} sketched in Fig.~\subfigref{fig:scheme}{b} basically rectifies environmental fluctuations as resource, which are thermal in~\cite{Thierschmann2015Oct}, but could in general be of different nature, e.g., when generated by nearby working devices on a chip. Nonthermal distributions~\cite{Sanchez2019Nov}, for example due to hindered equilibration of the electronic distributions in the presence of competing environments, quantum correlations in the bath, or correlations induced by the bath coupling~\cite{Carrega2024Jun} can be exploited to perform a useful task without the heat flow from a resource that a standard heat engine would require. A pressing question at the interface between quantum thermodynamics and quantum transport is hence how one can best make use of such ``nonthermal" resources~\cite{Monsel2025Jan,Yamazaki2025Sep}.  An important requirement to identify ``how useful" a given energy conversion process would be is the understanding of how to best quantify the resource of such non-standard engines (and hence the resulting efficiencies) and how to connect them to relevant quantum-transport observables. 

One of the benefits of nonthermal resources for \textit{energy conversion} could be their impact on the precision of the desired output, in the style of squeezing that improves sensing. Indeed, noise in quantum transport is not only a ``tool" for transport spectroscopy or a resource that can be rectified into useful power or used for cooling---it importantly also limits the precision of a desired output~\cite{Pietzonka2018May,Tesser2024May}. While this is typically not of relevance in macroscopic engines, energy and charge fluctuations at the nanoscale are easily of the same order of magnitude as average currents \cite{MailletPRL2019}. 

Therefore also for detection of small ($<$ mK) temperature increases (due to, e.g., single particle absorption or emission), as required for \textit{transport spectroscopy}, precision is a bottleneck. Nanoscale systems typically possess a small specific heat which makes them suited for calorimetry, but they also undergo significant temperature fluctuations at equilibrium, that limit their resolution. The lower bound on such fluctuations, which is dictated by the FDR for heat currents, was reached recently \cite{Karimi2020NatComms} [see Fig.~\subfigref{fig:exp}{b}]. An associated fundamental question is when a violation of the FDR at finite frequency $\hbar\omega\gtrsim k_BT$ occurs. Indeed, at high-frequency (GHz or higher) temperature may be ill-defined~\cite{Giazotto2006}. Combined with the technical difficulties of high-frequency noise measurements, this makes detecting such a violation difficult. \\

\paragraph*{Broader perspective and relevance to other fields.}

The amount of fluctuations that are to be expected or that need to be accepted in an \textit{energy-conversion} process realized in a quantum-transport setting can be shown to fulfill constraints in the spirit of thermodynamic or kinetic uncertainty relations, see Sec. \ref{sec:TURs}. The constraints on the precision of currents and output power, which result from quantum-transport calculations~\cite{Tesser2024May}, hence relate to research in (quantum) stochastic thermodynamics, which typically relies on an analysis of stochastic processes~\cite{Pietzonka2018May} in terms of stochastic trajectories, see also Sec.~\ref{sec:trajectories}. The challenge in combining these two areas lies in the way in which strong coupling and quantum effects, which play an important role in energy-conversion in quantum transport, are treated. An important task for the future will hence be to make meaningful connections between the quantum-transport and the stochastic-trajectory descriptions. From the perspective of applications and experiments, it will become important to optimize energy-conversion with respect to precision as well as to actually measure the precision of relevant transport quantities and how they relate to predicted bounds. 

Beyond the measurement of noise even the full counting statistics of electron transfer can be accessed in quantum-transport measurements via charge-counting experiments \cite{MailletPRL2019}. This was recently shown to give insights into the entropy of quantum systems, thereby opening a new direction in quantum transport spectroscopy combined with thermodynamics, see Sec.~\ref{sec:quantumdots}. Indeed, strongly interacting systems possess an entropy that is not obtained through simple counting of ground state microstates, making its measurement an appealing probe of correlations.   This  entropy measurement, using Maxwell relations on the charge response to temperature changes, is an achievement that is only possible through reliable control of local temperature and heating.

While we have in this section highlighted some recent experiments in quantum-dot and semiconductor devices, both for energy conversion and for thermodynamics in quantum transport spectroscopy, the described achievements and challenges are highly relevant also in other types of platforms. We would in particular like to mention superconducting devices, see Sec.~\ref{sec:superconductors}, where additional phase-coherent control of heat currents can be achieved, or hybrid devices, where electron quantum transport is impacted or induced by coupling to photon or phonon heat resources. Exotic properties of new materials may also be exploited, for instance the quasiparticle nature determines the transport properties in strongly correlated materials, see for example Sec.~\ref{sec:batteries}, and finite spectral Berry curvature opens transverse channels for the thermoelectric effect in systems with nontrivial topology. \\

\paragraph*{Concluding Remarks.}
We focused here on two aspects where quantum transport and thermodynamics meet: Energy conversion and transport spectroscopy. The former allows to use transport settings as heat engines at the nanoscale, where additional resources, such as quantum coherence or nonthermal distributions, become available. The latter exploits quantum thermodynamics to reveal additional insights into quantum systems, such as many-body effects. Both aspects come with new opportunities and challenges. On the one hand, one might envision energy converting devices providing power or refrigeration to other quantum technologies, or being used to manage excess heat. However, these devices are still limited to low temperatures ($\lesssim 1$ K), making it necessary to extend their operation to higher temperatures for broader application. On the other hand, transport spectroscopy would benefit from having higher control on temperatures and access to their fluctuations. This would allow to develop novel transport spectroscopy tools, where inferring heat statistics and entropy production and testing for violations of the heat FDR at high frequencies are some examples. \\

\paragraph*{Acknowledgements.}
O.M. acknowledges funding from the ANR (ANR-23-CE47-0002 CRAQUANT). R.S. acknowledges funding from the Spanish Ministerio de Ciencia e Innovaci\'on via grant No. PID2022-142911NB-I00, and through the ``Mar\'{i}a de Maeztu'' Programme for Units of Excellence in R{\&}D CEX2023-001316-M. J.S. and L.T. acknowledge funding from the European Research Council (ERC) under the European Union’s Horizon Europe research and innovation program (101088169/NanoRecycle) and from the Knut and Alice Wallenberg foundation via the fellowship program.
\clearpage

\section{Thermalization in quantum many-body systems and its intersection with quantum thermodynamics}
\label{sec:thermalisation}
\noindent
{\it Ceren B.~Dag}

\noindent
{Department of Physics, Indiana University, Bloomington, Indiana 47405, USA, and}\\
{Department of Physics, Harvard University, 17 Oxford Street Cambridge, MA 02138, USA}\\

\paragraph*{State-of-the-art.}
The central questions in the field of thermalization of quantum many-body systems are (i) whether an isolated quantum many-body system thermalizes, and (ii) if it does, what is the mechanism behind thermalization? In this sense, the field tries to answer the fundamental questions of whether temperature could emerge from the microscopical details of a quantum many-body system and its dynamics, and how a quantum bath becomes a bath in the first place. In this perspective, our aim is to explore the intersection of many-body quantum thermalization with quantum thermodynamics, and determine the current and future challenges at this uncharted overlap. In this section, we start with a brief review of quantum thermalization. 

Although the dynamics of isolated quantum many-body systems is reversible and unitary, they still exhibit a form of equilibration which is defined as `the dynamical process where a time-dependent
observable evolves to some equilibrium value and remains close to this value for most times during the time evolution' \cite{Gogolin_2016}. Given an arbitrary initial state $\ket{\psi(0)}$ and an evolution Hamiltonian $\mathcal{H}$ with an eigenbasis $\lbrace E_m,\ket{\phi_m} \rbrace$, the initial state in the basis of the evolution Hamiltonian is $\ket{\psi(0)} = \sum_m c_m \ket{\phi_m}$. Let the system have a local operator $\mathcal{\hat O}$, where $\mathcal{O}_{nm} = \langle \phi_n \vert \mathcal{\hat O} \vert \phi_m \rangle $ is the eigenstate expectation values. The observables at later times are determined by $\langle \mathcal{\hat O} (t)\rangle  = \sum_{mn}c_{m}^* c_{n} \exp\left[-i(E_m-E_n)t\right] \mathcal{O}_{nm}$. Then mathematically the equilibration is defined as
\begin{eqnarray}
    \lim_{T\rightarrow \infty}\frac{1}{T} \int_0^{T} \langle \mathcal{\hat O} (t)\rangle dt = \langle \mathcal{\hat O} (t \rightarrow \infty)\rangle = \sum_{m} |c_{m}|^2 \mathcal{O}_{m m}, \label{Eq.1.1}
\end{eqnarray}
where the second equality holds for a non-degenerate spectrum, and is called the prediction of diagonal ensemble \cite{rigol2008thermalization}. 
Hence, we observe that the equilibration in a unitary and reversible many-body quantum system is simply phase decoherence over time. For thermalization to happen, the equilibration stated above must be captured by a statistical ensemble. This leads to the celebrated \textit{eigenstate thermalization hypothesis} (ETH) \cite{Srednicki_1996,rigol2008thermalization,Gogolin_2016,d2016quantum},
\begin{eqnarray}
    \mathcal{O}_{mn} = \mathcal{O}(\bar E) \delta_{mn} + \text{max} |\mathcal{O}_{mn}| f_{\mathcal{O}} (\bar{E},E_m-E_n) R_{mn}, \label{Eq.1.2}
\end{eqnarray}
where $\bar E = (E_m+E_n)/2$ is the center of an energy window $E_k \in \Delta E$ of size $N_{\rm int}$; $\overline{|R_{mn}|^2}=1$ are real or complex random variables depending on the symmetries of $\mathcal{H}$; $O(\bar E)$ and $ f_{O} (\bar{E},E_m-E_n)$ are smooth functions of their arguments. Let us unpack Eq.~\eqref{Eq.1.2}. Due to the smoothness assumption of $\mathcal{O}(\bar E)$, we can Taylor expand the diagonal elements $\mathcal{O}_{mm}$ around $\bar E$ and utilizing the prediction of diagonal ensemble defined above, one can derive a bound on how large the energy window must be such that $\mathcal{O}_{mm}=\mathcal{O}(\bar E)$ holds \cite{Srednicki_1996}. This importantly shows that the equilibrium properties of the system is independent of the properties of the initial state, except its energy $\bar{E} = \sum_m \vert c_m \vert^2 E_m$. Hence, the first term is also where we can invoke statistical mechanics. Since the system is isolated, we expect $\mathcal{O}(\bar E)$ is to be predicted by the microcanonical ensemble, i.e.,~$\mathcal{O}_{mm}=\mathcal{O}(\bar E)=\text{Tr} \lbrace \rho_{\rm mc} \mathcal{\hat O} \rbrace $ where $\rho_{\rm mc} = \frac{1}{N_{\text{int}}} \sum_{\phi_n \in \Delta E} \vert \phi_{k}\rangle \langle \phi_{k}\vert$.

The second term states that the off-diagonal elements of $\mathcal{O}_{mn}$ are negligibly small compared to the diagonal elements $\mathcal{O}_{mm}$, and follow the distributions of random matrix theory (RMT) \cite{Srednicki_1996,d2016quantum}. This term is also the connection of ETH to quantum chaos, which is described by RMT and Berry's random-wave conjecture for eigenstates \cite{berry1977regular}. The essence of this hypothesis is that one eigenstate alone encodes the equilibrium (long-time) properties of $\mathcal{H}$, and therefore $\ket{\phi_k}$ are thermal eigenstates. 

In its strong form of ETH, the many-body system will thermalize starting from any initial state, which also implies an arbitrary size for the energy window over the spectrum \cite{PhysRevE.90.052105} and can be understood within the bound mentioned above for the window. However, ETH can still hold, albeit in its weak form, if a vanishingly small fraction of the spectrum is nonthermal, leading to the fact that not all initial states can result in thermalization. This observation led to a construction where a single nonthermal many-body state can be embedded in a thermal spectrum \cite{PhysRevLett.119.030601} breaking the strong form of ETH, which is the first example of weak ergodicity breaking, often called quantum many-body scarring \cite{Serbyn_2021}. While in this perspective we focus on ETH as a mechanism for quantum thermalization, let us remark that quantum thermalization can also be understood within the so-called canonical typicality \cite{popescu2006entanglement, *GoldsteinPRL2006, Gogolin_2016}. In particular, Refs.~\cite{popescu2006entanglement, *GoldsteinPRL2006} show that for almost all pure states of an isolated system, any sufficiently smaller subsystem must attain a canonical thermal state — implying that thermalization can emerge solely from the entanglement between the system and its environment, without the requirement of a statistical ensemble of states as in ETH.\\

\paragraph*{Current and future challenges.}
In classical and quantum thermodynamics, the presence of a bath with infinitely many degrees of freedom is typically assumed in thermodynamic processes and cycles (see Secs.~\ref{sec:superconductors}, \ref{sec:ultracold}). Even when a quantum bath is considered, e.g.,~in open quantum systems and quantum thermodynamics, such a bath is chosen to be non-interacting many-body system, and assumed to be in a Gibbs state with a well-defined temperature \cite{scully2003extracting}. However, non-interacting isolated many-body systems are known to not thermalize or satisfy ETH in its strong form \cite{d2016quantum}. Then the most immediate question that arises at the intersection of many-body quantum thermalization and thermodynamics is whether ETH could lead to quantum baths to be utilized in thermodynamic processes. 

Although so far no ETH bath has been used in a thermodynamic process, such as a heat engine or refrigeration cycle, significant steps have recently been taken toward utilizing the theory of ETH in quantum thermodynamics. Specifically, Ref.~\cite{PhysRevE.92.022104} derived a quantum master equation for a single qubit that is coupled to an ETH bath. In this analytical treatment, the ETH bath is assumed to be in the state $\rho_{\rm mc}$, which is consistent with $\mathcal{O}_{mm}=\text{Tr} \lbrace \rho_{\rm mc} \mathcal{\hat O} \rbrace$. The notion has been formally extended to a large class of pure initial states in Ref.~\cite{odonovan2024quantummasterequationeigenstate}, which was numerically shown in \cite{PhysRevE.92.022104} to hold for a model of cold bosonic atoms loaded in a two-band double-well potential. Therefore, these works along with more recent references in \cite{odonovan2024quantummasterequationeigenstate} reveal the wisdom that the diagonal part of the ETH leads to a shift in the energy levels of the system qubit, while the RMT, apparent in the non-diagonal part of the ETH, is responsible for both the loss of coherence and the thermalization of the system, i.e.,~emergence of Markovianity, so long as the system and the bath can also exchange energy \cite{PhysRevE.92.022104}. 

Despite the system being set to be a single qubit so far, the ETH bath exhibits energy and temperature fluctuations in time, due to the fact that it is a finite-size bath \cite{PhysRevE.92.022104,odonovan2024quantummasterequationeigenstate}. What do the temperature fluctuations depend on in addition to the system size? For example, if the system is considered to be two qubits with correlations, the information encoded in the set of two qubits will be lost only to the local probes due to information scrambling \cite{PRXQuantum.5.010201}, still existing distributed over the bath degrees of freedom. Furthermore, although common intuition demands a much smaller system size than the size of the bath to ensure Markovianity, this intuition might not always hold. If a long chain is significantly disordered to exhibit many-body localization, even a small RMT bath of three qubits can drive the larger system to delocalize \cite{PhysRevLett.119.150602}. It is important to understand the interplay of system and ETH bath in depth, not only for fundamental reasons but also for practical reasons. For example, once we utilize a many-body ETH bath in a thermodynamic cycle, the temperature fluctuations in the bath will likely alter its efficiency \cite{PhysRevLett.131.220405}. \\

\paragraph*{Broader perspective and relevance to other fields.}
Typical physical systems have mixed spectra having nonthermal regions in their Hilbert spaces, hence effectively forming only a weak ETH bath. In this sense, it will also be helpful to understand how the rare regular regions embedded in the thermal spectra alter the Markovianity of the bath and to examine the memory effects induced by the weak ETH in the system. 

Thermodynamic cycles can be implemented on quantum simulation platforms with Floquet engineering \cite{PhysRevApplied.21.044050}. It is also long known that periodically driving a quantum many-body system at certain frequency regimes leads to the emergence of quantum chaos and thermalization of local observables \cite{PhysRevX.4.041048,PhysRevE.90.012110}. Given that the emergence of Markovianity mainly relies on the random matrix theory, as discussed in the previous section, could Floquet-driven many-body systems also act as faithful quantum baths? Designing drives to simultaneously implement quantum baths and thermodynamic cycles in many-body systems, which are spatially separated into different zones of baths and the system, is an exciting direction. In a similar spirit, recently the spatially deformed Hamiltonians have been proposed to cool down a half of a quantum many-body system to its ground state by treating the other half as a `bath', and hence effectively creating a temperature gradient between the two sides for the entropy to flow \cite{PhysRevLett.126.103401}. Given that their many-body system is a non-integrable model, which likely thermalizes via the ETH, their setup can instead be considered as two quantum baths exchanging energy to lower the temperature of one. Such a design is passive, whereas Ref.~\cite{mi2024stable} periodically resets the bath degrees of freedom, which are a set of non-interacting qubits. The latter setup is reminiscent of single-bath quantum thermodynamic cycle \cite{scully2003extracting} with an important difference that bath degrees of freedom deterministically couple to the cooling system, hence possibly lacking Markovianity. Nevertheless,~\cite{mi2024stable} successfully shows that the technology is ripe to implement genuine quantum heat baths and construct thermodynamic cycles with them. \\

\paragraph*{Concluding Remarks.}
In this perspective, we discussed the current and future prospects of finite-size quantum heat baths, --a topic at the intersection of many-body quantum thermalization and quantum thermodynamics. More specifically, we explained how finite-size quantum baths can be understood within the framework of eigenstate thermalization hypothesis, and why they exhibit the characteristics of genuine Markovian baths. The limitations of these novel baths that harness the power of random matrix theory and how their ability to thermalize and its efficiency depend on the quantum correlations, random disorder and size of the system to thermalize remain to be explored. In an equally intriguing final note, they await being utilized in thermodynamic devices. \\

\paragraph*{Acknowledgements.}~C.B.D. acknowledges support from Faculty 100 Initiative at Indiana University. 

\clearpage

\section{Thermodynamics of many-body interaction effects}
\label{sec:manybodyinteractions}
\noindent
{\it Andrea Solfanelli}

\noindent
{Max Planck Institute for the Physics of Complex Systems, Nöthnitzer Str. 38, 01187 Dresden, Germany}\\

\noindent
{\it Fernando Iemini}

\noindent
{Instituto de F\'isica, Universidade Federal Fluminense, Av. Gal. Milton Tavares de Souza s/n, Gragoat\'a, 24210-346 Niter\'oi, Rio de Janeiro, Brazil}\\

\noindent
{\it Victor Mukherjee}

\noindent
{Department of Physical Sciences, IISER Berhampur, Berhampur 760003, India}\\

\noindent
{\it Krissia Zawadzki}

\noindent
{Instituto de Física de São Carlos, Universidade de São Paulo, CP 369, 13560-970 São Carlos, São Paulo, Brazil}\\

\paragraph*{State-of-the-art.}
Quantum thermodynamics has gained considerable attention during the recent years, due to its significant fundamental as well as practical importance. On one hand, the field helps us to understand if the well-established laws of classical thermodynamics are valid in the quantum regime, and whether additional universal features, such as in the statistics of quantum machines output \cite{goold18the}, exist for quantum systems. On the other hand, advancements in laws governing the thermodynamics of quantum systems enable us to design machines which harness quantum physics for their operation. 

The development of quantum machines for real-world purposes would necessitate scaling up of quantum technologies to many-body systems \cite{mukherjee2024promises}. For example, already several works have reported quantum simulators based on hundreds of atoms \cite{altman21quantum}, regarded as driven dissipative many-body systems. The effectiveness of such simulators crucially depends on their ability to sustain non-equilibrium states while evading thermalization and heating, which inevitably lead to the loss of locally stored quantum information. Understanding the thermodynamics of such systems is therefore crucial to mitigate the effects of noise and dissipation.

 Simulators showing quantum advantage can bring about a disruptive change in the existing technologies \cite{altman21quantum}. Similarly, many-body engines and batteries, especially those showing quantum advantage \cite{Rossini:20} can be highly relevant for the development of energy-efficient machines. Several many-body effects have been shown to be beneficial in this respect. For example, phase transitions may allow us to develop engines which operate with non-zero efficiency even close to the Canot limit \cite{campisi16the}, while superabsorption may enable us to design high-performing many-body quantum batteries \cite{quach22superabsorption}. Statistics of quantum particles have been used to enhance the performance of engines \cite{watanabe20quantum}, and experimentally realize a novel quantum engine based on BEC-BCS crossover \cite{Koch2023}. Long-range interactions have proven useful to boost quantum thermal devices by reducing non-adiabatic losses and enhancing the power-to-efficiency ratio \cite{solfanelli2025universal}.  Collective effects, where multiple spins are collectively coupled to dissipative baths, have been shown to enhance the output work of engines \cite{niedenzu18cooperative}, while boundary time crystals, a relatively newly discovered phase of matter, can be used to design high-precision quantum sensors \cite{gribben2024boundarytimecrystalsac} and time-keeping devices~\cite{Ludmila2025}. In addition, several works have shown the crucial role that many-body effects can play in the thermodynamics of quantum systems, and in the development of quantum machines \cite{binder2019bookqtd, mukherjee2024promises}. Below, we delve more into this topic, addressing the challenges and connections to other areas in quantum science and technologies. \\

\paragraph*{Current and future challenges.}
While many-body effects hold promise to harness quantum thermodynamic advantage, exploring them in real-world devices remains challenging \cite{mukherjee2024promises, deffner2019qtddevices}. In the following, we highlight a few critical theoretical and experimental issues and discuss potential strategies to overcome them. 

{Theoretical and numerical methods: } 
more accurate methods are needed to model out-of-equilibrium many-body systems in the presence of dissipation. Currently, tensor networks are among the most powerful numerical techniques for studying strongly correlated phenomena with controlled precision, especially in one dimension. Recent improvements have enabled finite-temperature and time-dependent calculations, as well as extensions to open and higher-dimensional systems. However, the study of work distributions and entropy production has been restricted to simplified models  - exactly solvable due to some integrability and/or symmetries - or limited to unsatisfactorily small systems. These difficulties have inspired a few approximation schemes tailored to specific regimes. For example, mean-field and density functional theories~\cite{Zawadzki2022} have proven to be useful in weakly interacting systems, or within linear response and perturbation
theory valid in sudden quenches and slow-driven regimes\cite{deffner2019qtddevices}.

{ Experimental scalability: } the success of quantum thermodynamics has extended beyond its theoretical foundation, thanks to experimental efforts, from which it was possible to verify laws and fluctuation theorems, and realize heat engines, in atomic-scale setups. Only recently, batteries \cite{quach22superabsorption} and cycles \cite{Koch2023} fueled by many-body effects started to be explored in cold atom platforms. A remarkable experiment implemented an Otto cycle with the working stages crossing the BEC-BCS transition \cite{Koch2023}. The efficiency, much lower than Carnot's limit, could be improved with shortcuts-to-adiabaticity \cite{beau16scaling}, a strategy hard to implement in strongly correlated regimes, especially at criticality. Different platforms to implement many-body quantum thermodynamics are nevertheless still lacking. Time crystals are emerging as promising candidates due to their intrinsic collective effects, and their potential as quantum heat engines \cite{Carollo_2020_2}. In addition, these phases have been already realised on a variety of platforms ~\cite{mukherjee2024promises}. 

Two-point measurements and tomography, two techniques to access quantities such as work fluctuations and entropy production, are difficult to adapt for many-body setups; here, it would require an exponentially large number of projections in highly entangled states, which are difficult to prepare and measure. Weak measurements and spectroscopies in the linear-response regime would offer an interesting alternative for probing the characteristic function of work. Shadow tomography \cite{Huang2020shadows} could be explored to obtain quantities defined in terms of distances between quantum states, including entropy production. \\

\paragraph*{Broader perspective and relevance to other fields.}
The fields of quantum thermodynamics and quantum technologies are inherently connected to other fields, such as quantum control, open systems, and many-body physics. The motivation to build high-performing finite-time quantum engines can lead to the introduction of novel quantum control protocols \cite{beau16scaling}. Scaling up of quantum engines to multi-particle systems necessitates studies on the dynamics of many-body open quantum systems \cite{niedenzu18cooperative}. Further pursuing these topics would result in more efficient protocols to manipulate and sustain many-body correlations, opening new possibilities for achieving different phases of matter in- and out-of-equilibrium.  

Development of technologies in the quantum regime may necessitate precise measurement of different parameters, such as temperatures and magnetic fields. This has boosted the advancement of the field of many-body effects enhanced sensitivity \cite{binder2019bookqtd, gribben2024boundarytimecrystalsac}, especially near criticality and in the presence of collective dynamics \cite{mukherjee2024promises}. 

Energy storage in the quantum regime can have a significant impact on the field of many-body quantum batteries. For example, finding practical ways to store significant amounts of ergotropy or work capacity, has led to studies on quantum batteries which are reliable, operate with high charging power, and are robust against dissipation \cite{binder2019bookqtd, Rossini:20, quach22superabsorption}. 

Several experimental platforms relevant to quantum technologies naturally host long-range interactions, where two-body potentials decay as a function of the interparticle distance $r$ following a power law $V(r) \propto r^{-\alpha}$. Notable examples include dipolar interactions in Rydberg atom arrays ($\alpha = 3$ or $6$), quantum gases coupled to optical cavities ($\alpha = 0$) and trapped-ions ($\alpha\approx 0-3$). Such systems have proven useful in quantum metrology, search algorithms, and quantum batteries. Recent work indicates that long-range interactions can reduce non-adiabatic effects in finite-time processes \cite{solfanelli2025universal}. A comprehensive framework for the quantum thermodynamics of these systems would therefore be of broad relevance.

Finally, the idea that quantum thermodynamic advantages can be obtained through many-body effects can inspire novel developments in quantum computing. At the hardware level, new strategies for cooling and controlling qubits are crucial for resilient quantum computation. A recent proposal of an autonomously driven refrigerator based on three-body interactions has demonstrated high efficiency in resetting superconducting qubits \cite{Gasparinetti2025}. At the software level, the possibility to simulate thermodynamics experiments with many qubits can lead not only to better quantum algorithms \cite{altman21quantum}, but also provide valuable insights into how quantum correlations can serve as a thermodynamic resource. Quantum variational algorithms to maximise ergotropy through entangling operations could improve thermodynamic protocols yielding maximum work.  \\

\paragraph*{Concluding Remarks.}
In conclusion, thermodynamics of many-body systems is a vibrant field of study, which on one hand contributes to fundamental research in quantum thermodynamics, and on the other hand, can be crucial for the development of high-performing practical quantum technologies \cite{mukherjee2024promises}. In addition, research on many-body quantum technologies, which operate following the laws of thermodynamics in the quantum regime, can play an important role in varied fields of study, including quantum control and many-body physics. The study of energy transfer and thermodynamics in many-particle models can also help unveil the quantum-to-classical crossover, and benefit other fields, ranging from chemistry to biology. Recent experiments on many-body quantum engines \cite{Koch2023} and quantum batteries \cite{quach22superabsorption} seem very promising. However, in light of the challenges discussed above, more intensive research is needed for understanding the thermodynamics of many-body systems, and for developing many-body effects assisted quantum technologies. \\

\paragraph*{Acknowledgements.}
VM acknowledges support from the Science and Engineering Research Board (SERB) through MATRICS (Project No. MTR/2021/000055). F.I. acknowledges financial support from the Brazilian funding agencies CAPES, CNPQ, and FAPERJ (No. 151064/2022-9, and No. E-26/201.365/ 2022)  and by the Serrapilheira Institute (Grant No. Serra 2211-42166).

\clearpage

\section{Strong coupling thermodynamics}
\label{sec:strongcoupling}
\noindent
{\it Janet Anders}

\noindent
{Institute of Physics and Astronomy, University of Potsdam, 14476 Potsdam, Germany, and}\\
{Department of Physics and Astronomy, University of Exeter, Exeter EX4 4QL, United Kingdom}\\

\paragraph*{State-of-the-art.}
Strong coupling thermodynamics extends standard thermodynamics to explicitly include a system's interaction with its bath in all considerations \cite{Seifert2016, Jarzynski2017}. While irrelevant for macroscopic systems where only the bath's temperature is felt by the system,  the details of the system-bath coupling can be an essential factor in determining the dynamics and steady state of microscopic and quantum systems. To illustrate how this  arises, one can consider the case where the total system and bath have come to equilibrium at inverse temperature $\beta$. Their canonical state then is $\tau_{tot} = {e^{-\beta H_{tot}} / \tr[e^{-\beta H_{tot}}]}$ with total Hamiltonian $H_{tot} = H_S + H_B + \lambda \, V_{int}$  where $H_S$ describes the bare system,  $H_B$ the bare bath, and $\lambda \, V_{int}$ the system-bath interaction. The properties of the system are then described by the reduced state $\tr_B[\tau_{tot}] =: \tau_{MF}(\lambda)$. When the system-bath coupling is ultraweak, i.e. $\lambda \to 0$,   it is easy to see that  the system state simplifies to the standard Gibbs state $\tau_{MF}(\lambda \to 0) = {e^{-\beta H_S} / \tr[e^{-\beta H_S}]}=:\tau_G$. However, for non-negligible $\lambda$, the system's reduced equilibrium state $\tau_{MF}(\lambda)$, often referred to as the {\it mean force (Gibbs) state} \cite{Trushechkin2022, *Purkayastha2024},  will generally differ from $\tau_G$. 

Thermodynamic arguments assume that the equilibrium state is $\tau_G$. The presence of non-negligible system-bath coupling, and the modification of the system equilibrium state, requires a fundamental rethink of thermodynamic accounting and arguments. Classical stochastic thermodynamic frameworks have been developed \cite{Seifert2016, Jarzynski2017}, which consistently include the system-bath coupling in the splitting of energies into work and heat, as well as the definition of entropy. Classical and quantum fluctuation relations have been extended \cite{Jarzynski2004, Campisi2009}, and non-Markovian dynamics arising from strong system-bath coupling has been shown to be embeddable into a Markovian description under certain assumptions of timescale separation \cite{Strasberg2017}. While the bath impacts on both, open classical and quantum systems, their resulting MF states are not the same. A quantum-classical correspondence of the equilibrium state was proven for an uncoupled spin (closed system) in the 1970s, and has now been extended to an open system for the first time \cite{Cerisola2024}. 

Moreover, the emergence of the above constructed MF state as the dynamical steady state of an open quantum system has been shown for a number of cases. This includes general analytical arguments for the so-called {\it weak coupling regime} \cite{Mori2008}, where expansion to order $\lambda^2$ is sufficient, and for the {\it ultrastrong coupling regime} \cite{Trushechkin2021,Latune2022}, where taking the $0$-th order in the limit $1/\lambda \to 0$ is sufficient. For a harmonic oscillator, which is linearly coupled to a bath of harmonic oscillators, it was proven that system observables, as well as multi-time correlation functions, relax to their corresponding values for the MF state \cite{Subasi2012}. A second widely used open system model is the spin-boson model, with Hamiltonian $H_{tot} = - {\hbar \omega_L \over 2} \sigma_z + {1\over 2} \int_0^{\infty} d \omega \, \left(P^2_\omega + \omega^2 X^2_\omega \right) + \lambda \, {\hbar \over 2} \sigma_\theta\, \int_0^{\infty} d \omega \, C_\omega \, X_\omega$ where  $\sigma_{x,y,z}$ are the Pauli matrices for a spin-1/2, and a $\theta$-angled direction is chosen for the interaction, i.e. $\sigma_\theta = \cos \theta \, \sigma_z - \sin \theta \,  \sigma_x$. Coupling is with coupling function $C_\omega$ to the bosonic bath, with operators $P_\omega$ and $X_\omega$, at frequencies $\omega$. Numerical solution of its dynamics has consistently shown a match between steady state and MF state  at all coupling strengths \cite{Cerisola2024}, while an analytical proof is missing. 
Furthermore, the connection between such open system thermalisation and thermalisation in closed many-body systems, see Sec.~\ref{sec:thermalisation}, is only beginning to be explored \cite{Trushechkin2022, *Purkayastha2024}.

\begin{figure}[t]
    \centering
    \includegraphics[width=0.6\columnwidth]{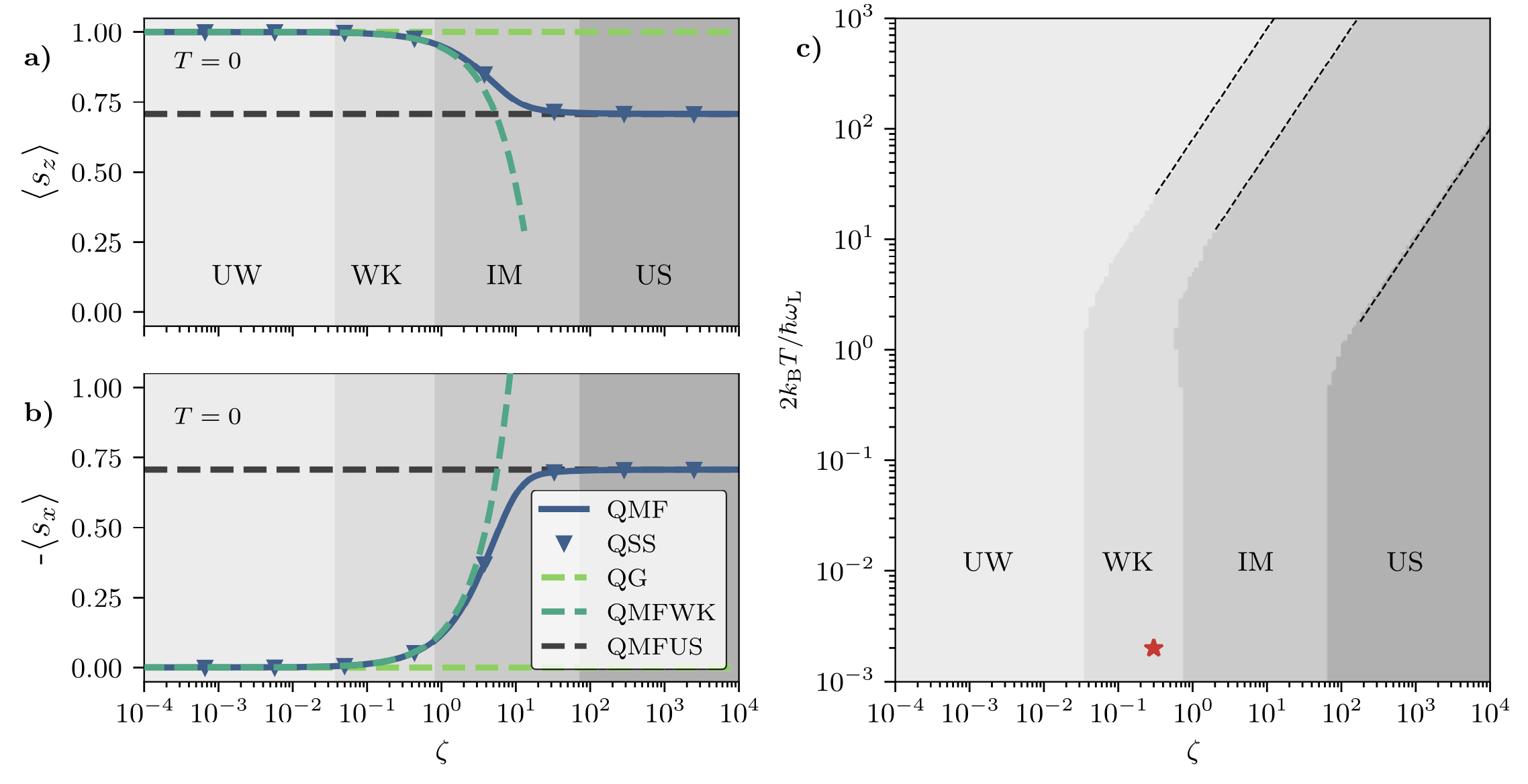}
    \caption{
    Left:~Expectation values of $\sigma_{z}$ (in panel a) and  $\sigma_{x}$ (in panel b) of the $\theta$-angled spin-boson model at $\theta = 45^\circ$ and $T=0$ for a Lorentzian function $C_\omega$, see \cite{Cerisola2024}, as a function of overall coupling strength $\zeta \equiv \lambda^2$. The match between the numerically exact quantum MF state (blue solid lines) with the quantum Gibbs state (light green dashed), and the analytically known weak coupling  MF state (dark green dashed) and  ultrastrong MF state (dark grey dashed) is used to identify four coupling regimes: ultraweak (UW), weak (WK), ultrastrong (US) and intermediate (IM). Note also the match between the numerically found dynamical steady state (blue triangles) and the MF state (blue solid).  
    Right:~Coupling regime boundaries as a function of $\zeta$ and temperature $T$, evidencing the tendency of returning to lesser coupling regimes at higher temperatures. 
    This figure is for the quantum model, while a similar figure exists for the classical model. 
    Figure adapted from \cite{Cerisola2024}. 
    }
    \label{fig:couplingregimes}
\end{figure}

\paragraph*{Current and future challenges.}
Despite considerable progress, an extensive range of questions are open \cite{Trushechkin2022}. Here we give a few pointers for the way ahead.

Whether the maximum efficiency/power of thermal machines is affected by  increased coupling to the bath is an ongoing issue of debate. For an Otto cycle the efficiency was argued to be increased \cite{Latune2023, Kaneyasu2023}, while for Carnot cycles it was found to be at most on par with the Carnot efficiency \cite{Liu2024}. Similarly, for thermometry, it has been shown that using a probe that is strongly coupled to the bath whose temperature is to be determined can be advantageous at low temperatures  \cite{Correa2017}. However, more research is needed to fully explore system-bath coupling benefits for thermodynamic cycles and practically relevant thermometry, see also {Secs.~\ref{sec:control} and \ref{sec:thermometry}}. 

``Strong coupling thermodynamics'' is the umbrella term  whenever the coupling $\lambda \, V_{int}$ plays a non-negligible thermodynamic role. Within, various {\it coupling regimes} can be identified which range from weak (the regime of validity of many master equations, including Lindblad and Redfield) to ultrastrong. For the spin-boson model, the boundaries of these regimes have recently been identified \cite{Cerisola2024}, see Fig.~\ref{fig:couplingregimes}. However, a more general understanding of  how to gauge whether a system is in a particular coupling regime is missing. (Note, that the interaction term $\lambda \, V_{int}$ acts on both, the system and bath, and a general operator norm comparison with the bare system Hamiltonian $H_S$ is futile.)

To explicitly express a quantum system's MF state $\tau_{MF}(\lambda)$ in terms of system operators alone, a key mathematical challenge is to analytically carry out the trace over the bath. For a generic system coupled to a single bosonic bath, this has been achieved \cite{Cresser2021} in the weak and ultrastrong limit. A first route to characterise the intermediate regime has been opened by the reaction-coordinate polaron-transform (RCPT) framework developed in~\cite{Anto-Sztrikacs2023, *Anto-Sztrikacs2023b}. Extensions of these ideas to different bath choices, including fermionic baths, are currently missing. Moreover, there is the possibility of several non-commuting operators of the system to couple to multiple baths, which can lead to significantly increased system-bath entanglement \cite{Hogg2024}. The exploration of non-commuting coupling operators is still in its infancy, and bridges to non-Abelian thermal states (NATS) discussed in {Sec.~\ref{sec:nonabelian}}.

Furthermore, beyond the weak/ultrastrong coupling limit, and beyond a handful of specific models, there is a need for general proofs that the  dynamical steady state of an open system (if it exists and is unique) is in fact the MF state. Related to this, it is important to realise that most master equation derivations make a range of approximations to fix the steady state to be the Gibbs state. However, if the system-bath dynamics should result in the system relaxing to the MF state, then approximate (master) equations should reproduce this too. Exactly that is achieved by the canonically consistent master equation (CCQME), which is constructed with a dissipator that has the MF state as the steady state  \cite{Becker2022}.  Interestingly, for a few test models, the CCQME has also produced early time dynamics that are closer to the exact dynamics than standard master equations (Lindblad, Redfield). This indicates a new potential to construct efficiently solvable master equations that can better describe the full dynamics. Master equations are widely used, from quantum optics and quantum technologies, to the modeling of chemical reaction rates. 

Finally, much anticipated are future experiments that show clear signatures of a system's coupling to its environment beyond the ultraweak limit, such as the observation of non-Markovian dynamics arising from bath-induced memory kernels, and the observation of signatures of bath-induced changes to the equilibrium state. Impurities in cold gases are one possible platform, while other experimental platforms where microscopic systems are strongly affected by their environment should be explored for quantitative tests.

\paragraph*{Broader perspective and relevance to other fields.}
``Strong coupling'' is a phrase frequently attached to the neighbouring area of light-matter interactions, where two quantum systems (coherently) interact with each other, while coupling less with a dissipative environment. The last two decades have seen huge advances here, in both theory and experiment, where regimes range from weak to ``deep-strong'' \cite{Yoshihara2017}. The difference in ``strong coupling thermodynamics'' is that the system (consisting of one or more parts) here couples significantly to a bath which supports a range of frequencies. Instead of coherent energetic exchange between two quantum systems, one here has the (coherent) energetic exchange between a system and its bath, which affects the system's {\it thermodynamic} properties, including its energy, heat exchange, heat capacity, equilibration behaviour, steady state, and so on. Ongoing experimental advances mean that both, system and bath, can be characterised in finer and finer detail. Soon it will  become possible to quantitatively verify/falsify  system-bath models that have been widely used for more than 40 years. This emerging capability to compare experiment and theory brings the exciting potential of discovery of new physics. 

\paragraph*{Concluding Remarks.}
All physics undergraduates learn about the Boltzmann distribution (a.k.a. the Gibbs state for the quantum case). The question of how realistic it is that a microscopic system interacts with a bath and comes to equilibrium with it, while not interacting with it so much that there are strong coupling effects, was hardly ever asked. The ongoing developments in this subfield have the power of discovering new physics and changing undergraduate textbooks on thermodynamics and statistical physics.

\paragraph*{Acknowledgements.}
J.A. gratefully acknowledges funding from the Deutsche Forschungsgemeinschaft (DFG, German Research Foundation) under Grants No. 384846402 and No. 510943930, and from the Engineering and Physical Sciences Research Council (EPSRC) (Grant No. EP/R045577/1), and thanks the Royal Society for support. 
 
\clearpage

\section{Quantum thermodynamic geometry}
\label{sec:thermogeometry}

\noindent
{\it Laetitia P. Bettmann}

\noindent
{School of Physics, Trinity College Dublin, College Green, Dublin 2, D02K8N4, Ireland}\\

\noindent
{\it Harry J. D. Miller}

\noindent
{Department of Physics and Astronomy, University of Manchester, Oxford Road, Manchester M13 9PL, United Kingdom}\\

\noindent
{\it Mart\'i Perarnau-Llobet}

\noindent
{F\'isica Te\`orica: Informaci\'o i Fen\`omens Qu\`antics, Department de F\'isica, Universitat Aut\`onoma de Barcelona, 08193 Bellaterra (Barcelona), Spain}\\

\noindent
{\it Alberto Rolandi}

\noindent
{Atominstitut, TU Wien, 1020 Vienna, Austria}\\

\begin{figure}[b]
  \centering
  \begin{minipage}{0.4\textwidth}
    \includegraphics[width=\linewidth]{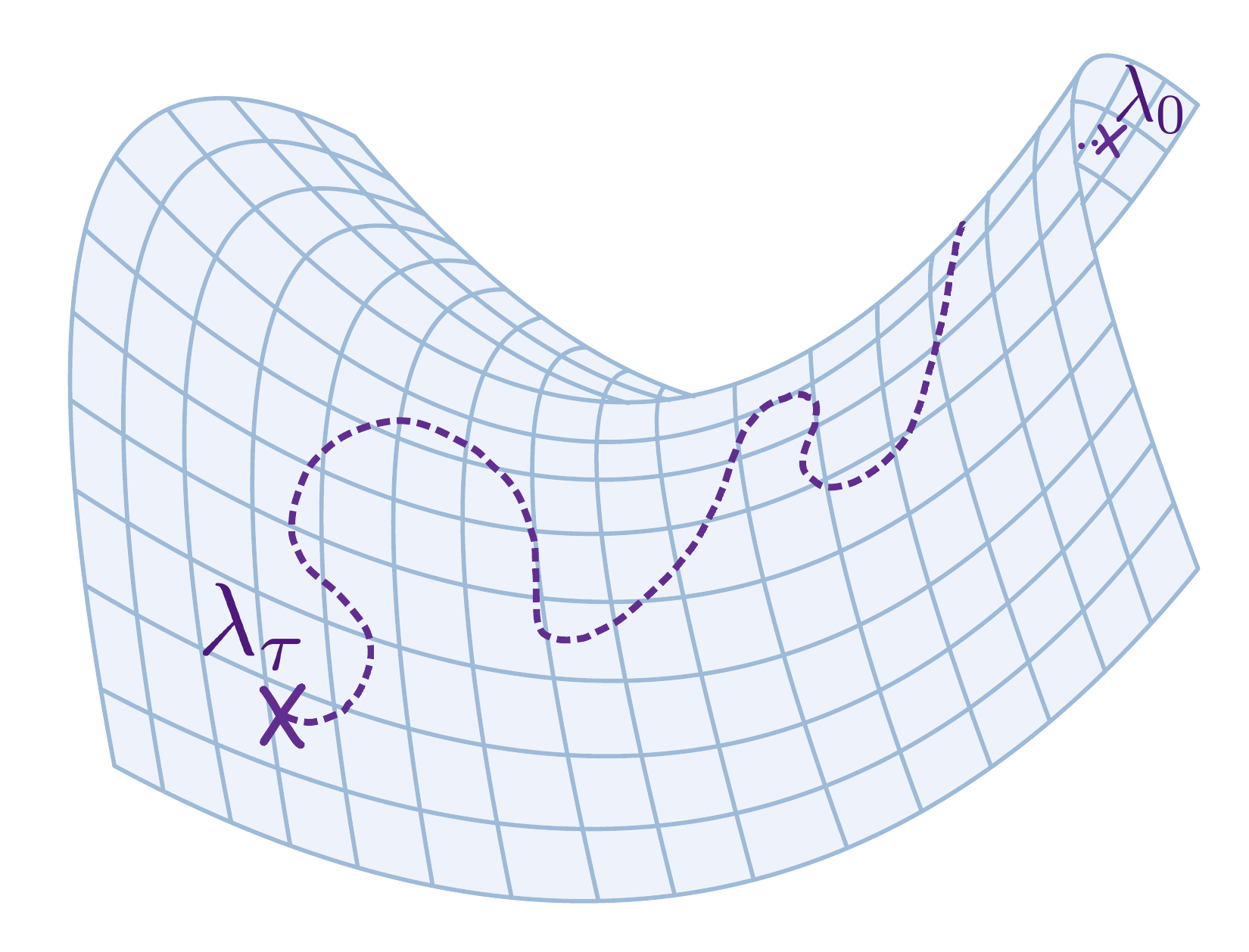}
  \end{minipage}%
  \hskip1cm
  \begin{minipage}{0.35\textwidth}
\captionof{figure}{Illustration of the parameter curved space of the control variable $\lambda$ with an example of a (non-optimal) protocol. \label{fig:param_space}}
  \end{minipage}
\end{figure}

\paragraph*{State-of-the-art.}
Idealised thermodynamic transformations proceed quasi-statically with negligible entropy production, allowing for simple characterisations of variables such as efficiency and extractable work. However, most realistic processes occur over finite timescales in which the assumption of instantaneous equilibrium is not justified, raising the question of how best to reduce the impact of entropy production. A formalism known as \textit{thermodynamic geometry} was first introduced in the 1980s to deal with finite-time corrections to classical, endoreversible thermodynamics~\cite{salamon1983thermodynamic}. The guiding principle is the following geometric lower bound on the entropy production $\Sigma$ produced in time $\tau$, 
\begin{align}
    \Sigma\geq \frac{\mathcal{L}^2(\vec{\lambda}_0,\vec{\lambda}_\tau)}{\tau}
    \label{eq:thermolength}
\end{align}
which is valid for slow but finite-time processes that stay close to equilibrium at all times. Here $\mathcal{L}(\vec{\lambda}_0,\vec{\lambda}_\tau)$ is the \textit{thermodynamic length} connecting some initial and final configuration of time-dependent control variables $\vec{\lambda}_t$ such as the temperature, chemical potential or Hamiltonian parameters, see Fig.~\ref{fig:param_space}. In general, the metric that determines this length encodes information about the equilibrium state space as well as the non-equilibrium relaxation dynamics stemming from interactions with an environment. Through the tools of differential geometry, minimization of the entropy production is achieved by ensuring the control variables follow a geodesic path with respect to this thermodynamic metric, i.e., the path of least action.       


This geometric approach to thermodynamics is particularly useful for controlling dissipation in microscopic systems out of equilibrium, and so in the past two decades it has been adapted for use in classical stochastic thermodynamics~\cite{sivak2012thermodynamic}. Applications here have included the optimal control of molecular motors and the reduction of energetic costs in classical bit erasure (see~\cite{blaber2023optimal} for a review). 

Recently, it was extended into the quantum regime to address the thermodynamic optimisation of slowly driven open quantum systems (see review \cite{abiuso2020geometric} and references therein). This has been used to solve the problem of maximising efficiency and power in low-dissipation quantum Carnot cycles \cite{abiuso2020geometric} and Stirling cycles \cite{brandner2020thermodynamic}. For quantum thermal machines, another important performance metric is the stochastic work fluctuations, and the geometric approach has also been used as a tool for the multi-objective optimisation of both dissipation and fluctuations simultaneously~\cite{miller2020geometry}. While simple low-dimensional systems were considered initially, recent developments have explored applications of thermodynamic length in many-body systems, such as the enhancement of a quantum heat engine undergoing Bose-Einstein condensation \cite{eglinton2023thermodynamic} and how to exploit collective effects to obtain sub-extensive entropy production~\cite{rolandi2023collective}. Another emblematic application is in information erasure, where Landauer's limit asserts that a minimum energy of $k_B T \ln 2 $ is required to erase one bit of information. In this case, the geometric approach has been used to derive a finite-time correction to such a fundamental bound  (see~\cite{rolandi2023finite} and references therein), showing how the minimal work cost scales with the quality of the erasure process. Most recently, thermodynamic geometry has been extended to describe slow transitions between quantum non-equilibrium steady states. In that case the thermodynamic length can be used to characterise the non-adiabatic entropy production \cite{lacerda2025information}. Parallel developments have established connections between information geometry and quantum stochastic thermodynamics, revealing that the quantum Fisher information with respect to time is closely linked to entropy dynamics and thus provides a geometric perspective on irreversibility, even far from equilibrium~\cite{Bettmann2025}.\\

\paragraph*{Current and future challenges.}
After extensive theoretical work on quantum thermodynamic geometry, experimental implementations of the formalism to optimal control are now beginning to take place. Experiments have applied minimally dissipative protocols to optimise the cooling of a dilute atomic gas \cite{mayer2020nonequilibrium} and the erasure of information in a quantum dot \cite{scandi2022minimally}. However, optimal thermodynamic control in the presence of quantum coherence remains an experimental challenge. As coherent control of platforms such as trapped ions~\ref{sec:trappedions}, cold atoms~\ref{sec:ultracold} and quantum dots~\ref{sec:quantumdots} 
are now rapidly developing, we anticipate such experiments will soon take place. There are a number of key challenges that remain in order to move from theory to experiment. Firstly, to optimise a given system one needs information about its thermodynamic geometry; this implicitly requires a  measurement of the thermodynamic metric tensor, which may be a difficult task in complex systems. Secondly, it is not known how robust optimal protocols are to classical noise in the control parameters, and an important development would be to quantify and account for these additional sources of irreversibility.  

Another challenge lies in the fact that the current approaches to geometric control in quantum thermodynamics are limited to slow, close-to-equilibrium transformations, and a more general approach that is applicable to arbitrarily fast non-equilibrium dynamics is still missing. Some significant progress has been made in this direction with the development of the so-called quantum Wasserstein distance as a generalised notion of thermodynamic length \cite{van2021geometrical,PRXVanVu2023}. This approach has been successful at deriving geometric finite-time bounds on entropy production akin to~\eqref{eq:thermolength}, though questions still remain as to how to saturate the bound with a particular control protocol. Further mathematical developments will be needed to understand how to compute the quantum Wasserstein distance and its associated geodesic paths. In this sense, it would also be beneficial to connect thermodynamic geometry to numerical approaches to thermodynamic control, such as reinforcement learning \cite{erdman2022identifying} (see also Sec.~\ref{sec:control}).

Thus far thermodynamic length and optimal control have primarily been applied to weakly-coupled open quantum systems that are adequately described by a Markovian master equation. This is another limitation as many microscopic systems are heavily influenced by interactions with the environment and their non-equilibrium dynamics may be non-Markovian in general, see Sec.~\ref{sec:strongcoupling}. An important future challenge will be to understand how to utilise  geometric control in these regimes, which inevitably requires the use of more sophisticated dynamical modeling. This would greatly extend the experimental applicability of thermodynamic length and allow one to minimise dissipation stemming from strong coupling to the environment. A first example of using thermodynamic length to reduce dissipation in a strongly coupled fermionic system was recently explored in~\cite{rolandi2023finite}.\\

\paragraph*{Broader perspective and relevance to other fields.}
The mathematical framework of quantum thermodynamic geometry has broad applications across quantum physics, from many-body systems to quantum information. One of its most impactful results is the derivation of speed limits, such as Eq.~\eqref{eq:thermolength}, which establish a fundamental trade-off between dissipation and the time required to complete a given operation. These speed limits originate from the deep ties between the thermodynamic metric and Fisher information~\cite{Scandi2025}, and are particularly relevant in thermodynamic cooling and computing. While we have mentioned their direct application to Landauer's limit, future work is expected to also use them for other computational tasks. In particular, this framework can find new applications in quantum computing and thermodynamic computing, where the relaxation dynamics of dissipative systems are  leveraged to perform logical operations, see e.g.~\cite{Aifer2024}.   

Deep links between thermodynamic geometry and many-body dynamics are also anticipated. While thermodynamic geometry has been formulated for open systems, there is no fundamental reason to expect that it cannot extend beyond this regime.  Indeed, insights from the behavior of isolated quantum many-body systems suggest that sufficiently complex many-body systems admit a thermal description for most times and observables, see Sec.~\ref{sec:thermalisation}. 
We therefore expect that thermodynamic geometry can also be applied to closed many-body systems. This would considerably increase its regime of applicability, while  providing an efficiently tool for controlling many-body systems in an energetically efficient manner. 

Beyond closed many-body systems, thermodynamic geometry can also find applications in non-equilibrium steady states of open-driven dissipative systems, where a novel class of phase transitions has been identified. While these phases have been characterized in the context of information geometry (see \cite{carollo_2020} and references therein), a thermodynamic geometry approach remains unexplored. This could, for instance, elucidate the finite-size and finite-time scaling of the minimum thermodynamic cost associated with crossing or approaching the phase transition.

Finally, thermodynamic geometry can play a relevant role in metrology, where a system's sensitivity to control parameters determines its effectiveness in parameter estimation. For example, thermodynamic geometry can be leveraged to find  optimal protocols for estimating free energy~\cite{blaber2023optimal}. Further applications in thermometry, as well as other sensing tasks involving open quantum systems, are also envisioned (see Sec.~\ref{sec:thermometry}). \\

\paragraph*{Concluding Remarks.}
Thermodynamic geometry has a long history, starting with seminal works in the 1980s where the concept of thermodynamic length was developed for macroscopic systems \cite{salamon1983thermodynamic}, along with its deep connection with dissipation and irreversibility. Since then, this framework has been extended to nanoscale systems described within stochastic thermodynamics~\cite{sivak2012thermodynamic} , and more recently to quantum systems \cite{abiuso2020geometric, lacerda2025information}. Nowadays, it represents a versatile tool to optimally control open quantum systems with minimal energy cost, while also providing new fundamental insights into the  nature of fluctuations~\cite{miller2020geometry}. As such, this framework has already found its use in the characterisation and optimisation of microscopic engines \cite{brandner2020thermodynamic,miller2020geometry,Bhandari2020}, cooling protocols~\cite{mayer2020nonequilibrium}, information erasure~\cite{rolandi2023finite,scandi2022minimally}, as well as free energy estimation~\cite{blaber2023optimal}. In the near future, we expect new applications in the optimal control of many-body systems and particularly systems close to criticality,  non-equilibrium sensing tasks, as well as in analog computing (see Sec.~\ref{sec:computing}). \\

\paragraph*{Acknowledgements.}
A.R. is supported by the Swiss National Science Foundation through a Postdoc.Mobility (Grant No. P500PT225461.
 M.P.-L.
acknowledges support from the Spanish Agencia Estatal de Investigación through the Grant ATR2024-154621
funded by MICIU/AEI/10.13039/501100011033.
L.P.B. acknowledges Research Ireland for support through the Frontiers for the Future project. H. J. D. M. acknowledges
funding from a Royal Society Research Fellowship (No. URF/R1/231394).
\clearpage

\section{Quantum thermodynamics of precision: Thermodynamic and kinetic uncertainty relations}
\label{sec:TURs}
\noindent
{\it Tan Van Vu}

\noindent
{Center for Gravitational Physics and Quantum Information, Yukawa Institute for Theoretical Physics, \\Kyoto University, Kitashirakawa Oiwakecho, Sakyo-ku, Kyoto 606-8502, Japan}\\

\paragraph*{State-of-the-art.}Achieving high precision demands high costs---an intuitive concept that has recently been formalized through a class of uncertainty relations for mesoscopic systems \cite{Horowitz.2020.NP}. Specifically, it has been rigorously demonstrated that the relative fluctuation of currents is constrained from below by thermodynamic and kinetic costs, explicitly expressed as
\begin{equation}
F_J\coloneqq\frac{\tau\mvar[J]}{\ev{J}^2}\ge\ell(\sigma,a).\label{eq:class.URs}
\end{equation}
Here, $\ev{J}$ and $\mvar[J]$ represent the mean and variance, respectively, of a stochastic current $J$ over a time interval $\tau$, while $\sigma$ denotes the entropy production rate and $a$ is the dynamical activity rate. Currents, such as heat flux, particle current, and the displacement of molecular motors, are time-integrated observables that are odd under time reversal. The lower bound $\ell$ can take various forms: $\ell_1(x,y)\coloneqq 2/x$ corresponds to the thermodynamic uncertainty relation (TUR) \cite{Gingrich.2016.PRL}, $\ell_2(x,y)\coloneqq 1/y$ represents the kinetic uncertainty relation (KUR) \cite{Terlizzi.2019.JPA}, and $\ell_3(x,y)\coloneqq (4y/x^2)\Phi(x/2y)^2$ provides a refined thermo-kinetic bound (TKUR) \cite{Vo.2022.JPA}, where $\Phi(x)$ denotes the inverse function of $x\tanh x$. Historically, these uncertainty relations were developed for classical Markov jump processes and overdamped Langevin systems. Despite their significance, they are not universally valid, as violations have been identified in underdamped cases and non-Markovian dynamics. Notably, a generalized TUR can be derived solely from fluctuation theorems for generic time-antisymmetric observables. Although this generalization is applicable to a broad class of dynamics, it comes at the cost of significantly relaxing the lower bound $\ell$, which becomes exponentially less stringent than the original \cite{Hasegawa.2019.PRL,Timpanaro.2019.PRL}. These relations not only deepen our understanding of the interplay between precision and costs but also enable diverse applications in nonequilibrium physics.

In quantum domains, classical uncertainty relations \eqref{eq:class.URs} have been numerically and experimentally observed to be violated \cite{Ptaszynski.2018.PRB,Saryal.2019.PRE,Brandner.2025.PRL}, prompting significant progress in their generalization to open quantum systems. While some results are available for relevant setups \cite{Guarnieri.2019.PRR,Hasegawa.2021.PRL,Miller.2021.PRL.TUR}, this perspective focuses specifically on quantum generalizations restricted to Markovian open quantum systems \cite{Hasegawa.2020.PRL,Vu.2022.PRL.TUR,Prech.2025.PRL,Vu.2025.PRXQ,Kwon.2024.arxiv}, drawing an analogy to classical cases. The relevant findings are summarized in Table \ref{table:res.sum}. In essence, the bounds for quantum systems generally include a quantum contribution, which characterizes the potential violation due to quantum effects. An intuitive explanation for this violation is that the creation of quantum coherence facilitates the consecutive occurrence of identical quantum jumps, thereby enhancing precision. In contrast, such a mechanism is absent in classical systems.
{\renewcommand{\arraystretch}{2}
\begin{table*}[h]
\begin{tabular}{|c|c|c|c|}
\hline
Generalizations & Formulation & Applicable observables & References\\[1.0ex]
\hline\hline
\multirow{2}{*}{\raisebox{-2.0ex}{Quantum KUR}} & $F_J\ge 1/(a+q)$ & counting observables & \cite{Hasegawa.2020.PRL,Vu.2022.PRL.TUR} \\[1.0ex]
\cline{2-4}
 & $F_J\ge(1+\psi_J)^2/a$ & counting observables & \cite{Prech.2025.PRL} \\[1.0ex]
\hline
Quantum TKUR & $F_J\ge(1+\delta_J)^2\ell_3(\sigma,a)$ & currents & \cite{Vu.2025.PRXQ} \\ [1.0ex]
\hline
\end{tabular}
\centering
\caption{Summary of quantum generalizations for Markovian dynamics with quantum jump unraveling. Here, $q$ denotes a quantum contribution from coherent dynamics, whereas $\psi_J$ and $\delta_J$ are quantum contributions to the observable average. All these contributions vanish in the classical limit. Note that $\ell_3\ge\max(\ell_1,\ell_2)$ since $\Phi(x)\ge\max(\sqrt{x},x)$. \label{table:res.sum}}
\end{table*}} \\

\paragraph*{Current and future challenges.}
Thus far, it has been established that the precision of observables is constrained not only by thermodynamic and kinetic costs but also by quantum coherence. Several quantum generalizations of classical bounds have been derived for Markovian dynamics, shedding light on the precise role of quantum coherence. However, numerous challenges remain to be addressed in future research.

The first challenge lies in extending these quantum bounds to non-Markovian and strong-coupling regimes, where memory effects and environmental correlations significantly influence system dynamics. Current frameworks are predominantly centered on Markovian dynamics, leaving a vast array of open questions for systems with complex interactions. A promising direction involves exploring a composite setup in which the target system and its environment evolve under a unitary transformation, with continuous measurements performed on the environment \cite{Hasegawa.2021.PRL}. Alternatively, a two-point measurement scheme---where observables are determined by the initial and final outcomes---may provide another viable approach \cite{Vu.2025.arxiv}. Identifying the relevant costs that constrain the precision of observables in such scenarios remains a critical challenge.

Elucidating the role of information flow in enhancing the precision of observables is another important task in the quantum regime. In multipartite systems, information flow continuously occurs between subsystems. In classical cases, it is well-established that, alongside energetic costs, information flow plays a vital role in improving the precision of currents. However, the interplay between information flow, energetic costs, and precision in quantum domains remains an open question. This challenge extends to measurement and feedback control processes, where entropy can be reduced without heat dissipation, as exemplified by the Maxwell demon.

A different yet closely related aspect of observables is their response to parameter perturbations. Recent studies have demonstrated that the response precision of observables is constrained by thermodynamic costs \cite{Ptaszynski.2024.PRL}. In quantum systems, while the response precision of observables has been shown to be limited by dynamical activity \cite{Vu.2025.PRXQ}, it remains an open question how the response precision of currents is constrained by entropy production. Investigating this direction could provide a unified understanding of quantum thermodynamics of precision and response.

Another notable challenge is investigating the precision of the first-passage time (FPT). In the classical domain, the relative fluctuation of the FPT for currents has been shown to be constrained by entropy production in the large-threshold limit \cite{Gingrich.2017.PRL}. While an analogous bound in terms of dynamical activity has been derived \cite{Vu.2022.PRL.TUR}, formulating a similar quantum bound in terms of entropy production remains a significant challenge.\\

\paragraph*{Broader perspective and relevance to other fields.}
Looking more broadly, the study of the TUR and KUR provides valuable insights and has relevance to other fields. One notable application is in quantum heat engines, where a key question is whether the Carnot efficiency can be achieved at finite power. For classical engines, a direct implication of the TUR is that achieving the Carnot efficiency at finite power is possible only if power fluctuations diverge. In the quantum regime, however, it has been shown that quantum coherence between degenerate energy eigenstates can be exploited to attain the Carnot efficiency at finite power. Moreover, the quantum TKUR \cite{Vu.2025.PRXQ}, which includes a quantum contribution absent in its classical counterpart, does not impose the same restriction. This suggests the intriguing possibility of high-performance quantum heat engines without divergent power fluctuations. Further exploration of this direction is crucial for advancing our understanding and design of efficient quantum heat engines.

Another closely related field is quantum metrology, which focuses on estimating parameters---such as phase shifts, frequencies, or coupling strengths---with the highest possible precision. The ultimate precision limit is set by the quantum Cram{\'e}r-Rao bound, which is determined by the quantum Fisher information (see also Sec.~\ref{sec:thermometry}). Both quantum metrology and uncertainty relations share a fundamental trade-off: achieving higher precision---whether in parameter estimation or current fluctuations---requires expending a resource, such as quantum coherence and quantum entanglement in metrology or energetic dissipation in the TUR. From a methodological perspective, these fields are also closely linked, as the quantum TUR and KUR can be derived from the quantum Cram{\'e}r-Rao bound. Exploring this connection is highly relevant, as it highlights how resource costs, such as dissipation or other thermodynamic quantities, fundamentally constrain the achievable precision in quantum sensing. Advancing this direction could provide insights into a key problem: What is the minimum energetic cost required to achieve a desired measurement precision in a quantum sensor?

Other promising areas for the development and application of uncertainty relations include quantum transport (see Sec.~\ref{sec:transport}), quantum measurement, quantum clocks (see Sec.~\ref{sec:time}), and quantum computation (see Sec.~\ref{sec:computing}), where precision serves as a key performance metric. It remains largely unexplored how these relations constrain the performance of quantum clocks and quantum computing, as well as how quantum coherence can be leveraged to enhance their accuracy.\\

\paragraph*{Concluding Remarks.}
In this perspective, we have briefly outlined the recently developed framework on the thermodynamics of precision, with a particular focus on the TUR and KUR in Markovian dynamics. While the findings summarized in Table \ref{table:res.sum} are based on quantum jump unraveling, similar results can be obtained for quantum diffusion unraveling. Beyond the Markovian regime, the notion of stochastic trajectories becomes less well-defined, leaving open the question of how thermodynamic costs constrain precision in such cases. The challenges and connections to other fields discussed above are by no means exhaustive, as the trade-off between precision and cost is a fundamental aspect of nature. Developing variations of these uncertainty relations for different setups and contexts is therefore a crucial direction for future research.

From a different viewpoint, the TUR can be interpreted as a refined version of the second law of thermodynamics. Similarly, a trade-off between time, cost, and precision also arises in the context of the third law. This raises an intriguing question: How can the framework of finite-time quantum thermodynamics be systematically characterized through trade-off relations in which precision plays a central role? Exploring this possibility could offer deeper insights into the fundamental limits of thermodynamic processes. \\

\paragraph*{Acknowledgements.}
T.V.V. was supported by JSPS KAKENHI Grant No. JP23K13032.

\clearpage

\section{Non-Abelian thermodynamics}
\label{sec:nonabelian}
\noindent
{\it Aleksander Lasek}

\noindent
{Joint Center for Quantum Information \& Computer Science, NIST and University of Maryland, College Park, Maryland, USA}\\

\noindent
{\it Nicole Yunger Halpern}

\noindent
{Joint Center for Quantum Information \& Computer Science, NIST and University of Maryland, College Park, Maryland, USA}\\
{Institute for Physical Science and Technology, University of Maryland, College Park, MD 20742, USA}\\

\paragraph*{State-of-the-art.}
Conserved quantities, called \textit{charges}, are  important in thermodynamics, as they restrict the available Hilbert space. Common examples of charges include energy and particles. A Hamiltonian conserves charges globally and can transport them between subsystems. For decades, thermodynamic charges were implicitly assumed to commute with each other, e.g., in derivations of the thermal state's form. Researchers discovered this implicit assumption approximately a decade ago, working at the intersection of quantum information theory and quantum thermodynamics \cite{Guryanova_2016_thermo,Lostaglio_2017,NYH_2016_NATS}. Yet the noncommutation of operators leads to quintessentially quantum phenomena such as the uncertainty principle. One must therefore ask, \textit{what happens to thermodynamic results if charges fail to commute with each other?} The subfield concerning the answer was dubbed the \textit{thermodynamics of noncommuting charges}, or \textit{non-Abelian thermodynamics}.

We illustrate noncommuting thermodynamic charges using a chain of qubits interacting via a Heisenberg Hamiltonian \cite{NYH_2020_BeverlandPRE, Majidy_2022_Hamiltonian, Kranzl_NATS_2023}: $H_{\rm Heis} = \sum_{ i, j } J_{ij}\ \vec{\sigma}^{(i)} \cdot \vec{\sigma}^{(j)}$. The $\vec{\sigma}^{(i)}$ denotes the conventional vector of Pauli operators at site $i$. $J_{ij}$ denotes the hopping frequency. The interaction conserves the  components $\alpha=x,y,z$ of the total spin, $\sum_i  \sigma_\alpha^{(i)}$. The charges do not commute: $[\sigma_{\alpha},\sigma_{\alpha'}] \neq 0 \; \forall \alpha \neq \alpha' \, $. Two qubits can form a system of interest, while the rest form an effective environment. One can study how the charges affect this setup's thermodynamics \cite{NYH_2020_BeverlandPRE,Kranzl_NATS_2023,Majidy_NATSReview_2023}.

Charges' noncommutation alters several thermodynamic results \cite{Majidy_NATSReview_2023}.  We now overview three examples, beginning with the \textit{eigenstate thermalization hypothesis} (ETH). The ETH explains how closed quantum many-body systems thermalize internally (Sec.~\ref{sec:thermalisation}). Noncommuting charges violate the ETH, Murthy \textit{et al.} found \cite{Murthy_NAETH_2023}. The authors therefore proposed a \textit{non-Abelian ETH} \cite{Murthy_NAETH_2023}. It implies that, under certain conditions, systems can locally thermalize to the same extent as under the conventional ETH (in the absence of noncommuting charges): let $N$ denote the total system size. According to the ordinary ETH, a generic local operator’s time-averaged expectation value equals the thermal expectation value to within $O(N^{-1})$ corrections. Under other conditions, though, the corrections may be polynomially larger, scaling as $O(N^{-1/2})$. Charges' noncommutation may prevent subsystems from thermalizing as much as usual in finite many-body systems.

Second, charges' noncommutation causes derivations of the thermal state's form to break down \cite{NYH_2016_NATS,Majidy_NATSReview_2023}. Suppose that a small system $S$ and an environment $E$ exchange only energy and particles. One can calculate $S$'s thermal state as follows: assume that $S\!E$ is in a \textit{microcanonical state}, with a well-defined number of particles and a fairly well-defined amount of energy. Trace out $E$, and assume that $S$ couples to $E$ only weakly. The small system's reduced state turns out to be the grand canonical state, the thermal state that arises under global energy conservation and particle-number conservation. Now, suppose that $S$ exchanges noncommuting charges with $E$. The charges cannot necessarily have well-defined values simultaneously, so microcanonical states may not exist. To rescue the derivation, researchers generalized the microcanonical state \cite{NYH_2016_NATS}.  Doing so yields the \textit{non-Abelian thermal state} (NATS), $\rho \propto  e^{-\beta \left(H - \sum_{\alpha} \mu_{\alpha} Q_{\alpha}\right)} $ \cite{Guryanova_2016_thermo,Lostaglio_2017,NYH_2016_NATS}. The $\beta$ denotes the inverse temperature, and $\mu_{\alpha}$ denotes the effective chemical potential of charge $Q_\alpha$.
Researchers observed signatures of the NATS in a trapped-ion system evolved under a long-range Heisenberg Hamiltonian \cite{Kranzl_NATS_2023}. (For more quantum thermodynamics of trapped ions, see Sec.~\ref{sec:trappedions}.) This experiment marked the first test of non-Abelian thermodynamics.

Third, charges' noncommutation can change entanglement entropy and thermodynamic-entropy production. As a quantum many-body system thermalizes internally, its constituent particles entangle. To pinpoint how charges' noncommutation affects entanglement, researchers built two models \cite{Majidy_Entropy_2023}. Each is a one-dimensional chain of two-qubit sites. The models parallel each other, such as by having the same number of charges, which have the same eigenvalues. However, one model's charges commute, and the other model's do not. The noncommuting-charge model achieved a greater average bipartite entanglement entropy than its counterpart. However, charges' noncommutation can decrease average thermodynamic entropy production, which quantifies  irreversibility ~\cite{Landi_2022_Entropy,NYH_2024_Entropy}. These two results raise the question of whether charges' noncommutation aids or disrupts thermalization, as detailed in the next section.\\

\paragraph*{Current and future challenges.} Non-Abelian thermodynamics includes many open problems \cite{Majidy_NATSReview_2023}, three of which we outline. First, thermalization and charges' noncommutation participate in a paradox. Charges' noncommutation hinders at least six features of thermalization but enhances at least four others \cite{Majidy_NATSReview_2023}. These results, following from different setups, do not contradict each other technically. Yet they disagree conceptually and so need reconciling. Possible tools include the parallel models that isolate effects of charges' noncommutation \cite{Majidy_Entropy_2023}. Another possible tool is a method for constructing Hamiltonians that conserve noncommuting charges globally while transporting them locally \cite{Majidy_2022_Hamiltonian}.

Second, the non-Abelian ETH's predictions need to be observed. As mentioned above, the non-Abelian ETH enables abnormally large corrections to thermal predictions \cite{Murthy_NAETH_2023}. Such \textit{anomalous thermalization} would prevent systems from thermalizing as much as usual. Thermalization draws all states toward a fixed point. Therefore, anomalous thermalization may enable systems to retain extra information about their initial conditions. Such systems may serve as quantum memories. Any such application must follow numerical and experimental observations of the prediction.

Finally, non-Abelian thermodynamics demands more experimental testing. Kranzl \textit{et al.} performed the first test, using trapped ions  \cite{Kranzl_NATS_2023}. Other feasible platforms include superconducting qubits, quantum dots, and ultracold atoms \cite{NYH_2020_BeverlandPRE}. Many theoretical results merit testing---for example, anomalous thermalization \cite{Murthy_NAETH_2023} and predictions about thermodynamic and entanglement entropies \cite{Landi_2022_Entropy,NYH_2024_Entropy,Majidy_Entropy_2023}. Also, we should check whether charges' noncommutation slows thermalization down in time \cite{Kranzl_NATS_2023}.  Decoherence threatens such experiments to an unusual degree: the system of interest may lose not only information, but also charges of multiple types, to its surroundings. However, dynamical decoupling overcame such decoherence in \cite{Kranzl_NATS_2023}.\\

\paragraph*{Broader perspective and relevance to other fields.} 
Noncommuting thermodynamic charges are relevant to several fields; we present four examples \cite{Majidy_NATSReview_2023}. Gauge theories offer one. They can model condensed matter and fundamental interactions between particles. For example, quantum chromodynamics, which models the strong force, has an SU(3) symmetry.  Gauge theories contain extraneous degrees of freedom, eliminated by gauge fixing.  Transformations between gauges can form non-Abelian Lie groups. Therefore, particle physics may exhibit non-Abelian thermodynamics \cite{Majidy_2022_Hamiltonian, Ott_2022_Gauge}. To find out, we must reconcile gauge symmetries' local nature with the purely global symmetry attributed to non-Abelian thermodynamics so far.

Second, noncommuting charges can disrupt \textit{many-body localization} (MBL) \cite{Vasseur_2016_MBL}. MBL can occur in quantum many-body systems subject to strong disorder. Information and particles take a long time to disperse across such systems. If one introduces a non-Abelian symmetry into an MBL Hamiltonian, any symmetry-breaking perturbation can destabilize the MBL, hastening thermalization.

Third, integrable systems can have noncommuting charges. An integrable system has extensively many charges and so does not thermalize. It relaxes to a \textit{generalized Gibbs ensemble}, which is a NATS if the charges fail to commute.
Recall the Heisenberg chain described above, and suppose that the nonzero couplings $J_{ij}$ are nearest-neighbor. This system is integrable and exhibits \textit{anomalous diffusion}: the diffusion constant scales as the system size's square-root \cite{Znidaric_2011_Integrable}. Other properties of the system, such as out-of-time-ordered correlators, exhibit anomalous behaviors that need explaining.

Finally, noncommuting charges arise in hydrodynamics \cite{Lucas_2021_Hydro}. Hydrodynamics describes locally equilibrated fluids' long-range properties in terms of charge flows. Applications include condensed matter and  heavy-ion collisions. In hydrodynamics, noncommuting charges can affect conductivity and entropy currents. One might hope that non-Abelian thermodynamics can explain why heavy ions thermalize unexpectedly quickly during collision processes.\\

\paragraph*{Concluding Remarks.} 
Researchers recognized only recently that charges' noncommutation can conflict with derivations, and alter results, in thermodynamics. Examples of such results include the ETH, derivations of thermal states, and MBL. The emerging field of non-Abelian thermodynamics is dedicated to analyzing  these effects and their implications for neighboring fields. Many experimental and theoretical opportunities in non-Abelian thermodynamics call for investigation. Experiments have already begun. Outside the subfield itself, noncommuting thermodynamic charges may influence gauge theories, many-body localization, hydrodynamics, and conventional integrability. Redolent of quantum uncertainty and measurement disturbance, noncommuting charges help put the \textit{quantum} in \textit{quantum thermodynamics}.\\

\paragraph*{Acknowledgements.}
This project was supported by the John Templeton Foundation (award no. 62422). The opinions expressed in this publication are those of the authors and do not necessarily reflect the views of the John Templeton Foundation.

\clearpage

\section{Thermodynamics of timekeeping in the quantum regime}
\label{sec:time}

\noindent
{\it Natalia Ares and Federico Fedele}

\noindent
{Department of Engineering Science, University of Oxford, Parks Road, Oxford OX1 3PJ, United Kingdom}\\

\noindent
{\it Paul Erker}

\noindent
{Atominstitut, Technische Universit\"at Wien, 1020 Vienna, Austria, and}\\
{IQOQI Vienna, Austrian Academy of Sciences, Boltzmanngasse 3, 1090 Vienna, Austria}\\

\noindent
{\it Mark T. Mitchison}

\noindent
{School of Physics, Trinity College Dublin, College Green, Dublin 2, Ireland, and}\\
{Department of Physics, King’s College London, Strand, London, WC2R 2LS, United Kingdom}\\

\paragraph*{State-of-the-art.}
The fires of the Industrial Revolution gave birth to more than just thermodynamics. The rise of industrial capitalism brought not only wealth but also a pressing need to synchronise labour, driving the widespread adoption of modern timepieces. While these clocks consumed almost no power compared to the factories and steam engines they regulated, the same cannot be said for quantum technologies. Indeed, the power consumed by the circuitry or optics needed for precisely timed control in any quantum experiment generally eclipses the energetics of the quantum system itself. Therefore, a proper accounting of thermodynamics in the quantum regime cannot neglect the cost of timekeeping.

To that end, much recent work has been devoted to understanding the physical limits on clocks. By thinking about clocks as machines, whose task is not to produce power but rather to generate a regular series of \textit{ticks}, both information-theoretic and thermodynamic constraints on timekeeping have been elucidated.  Interestingly, further exploration of these limits has revealed that certain features of quantum clocks unlock dramatic advantages over their classical counterparts. 

Clock performance can be quantified by the accuracy $\mathcal{N}$---the number of reliable ticks a clock can produce---and the resolution $\nu$---the average rate at which a clock ticks. The second law of thermodynamics limits accuracy~\cite{Erker2017, Milburn2020}, in the sense that $\mathcal{N}$ grows (at most) linearly with the entropy produced per tick of a clock undergoing incoherent dynamics. Remarkably, however, coherent quantum dynamics can be exploited to yield an accuracy that scales \textit{exponentially} with entropy production~\cite{Meier2024a}. Considering the information-theoretic dimension (i.e., the number of distinguishable states) as a resource, quantum clocks were found to achieve a quadratic accuracy advantage over classical discrete clocks with the same dimension~\cite{Woods2022}. A fundamental accuracy-resolution tradeoff has also been identified, stating that increased accuracy comes at the expense of resolution~\cite{Meier2023}. Here, again, quantum clocks can achieve a quadratic improvement $\mathcal{N}\sim \nu^{-2}$ compared to their discrete classical counterparts, which are limited to linear scaling $\mathcal{N}\sim \nu^{-1}$ by a recently discovered clock uncertainty relation~\cite{Prech2024,Macieszczak2024}.

Recent experiments have begun to probe these limits. The first experiment showing the link between entropy production and clock accuracy was performed using a electromechanical resonator driven by white noise, albeit in a classical regime~\cite{Pearson2021}. The effect of quantum measurement backaction on a quantum clock was explored in a superconducting circuit~\cite{He2023}. Finally, Ref.~\cite{Wadhia2025} measured the entropy dissipated when reading out the ticks of a clock realised in a semiconductor quantum-dot device, showing that this is the dominant thermodynamic cost for timekeeping on the quantum scale.\\

\paragraph*{Current and future challenges.}
While numerous promising results have been obtained, it remains an outstanding problem to experimentally demonstrate quantum-thermodynamic advantages in timekeeping. A key desideratum is autonomy, i.e.~a genuine clock should be self-contained and independent of any external, time-dependent control that would necessitate another clock to implement. A longer-term goal would be to incorporate nanoscale autonomous clocks into other quantum devices, where they could implement control operations without the need for energetically expensive classical control~\cite{Woods2023}. The potential of autonomous machines for quantum technologies has already been demonstrated, e.g.~for high-fidelity qubit preparation~\cite{Gasparinetti2025}. 

However, avoiding the macroscopic energy cost associated with measurement~\cite{Wadhia2025} would require direct coupling between the clock and the system to be controlled. This scenario has been thoroughly investigated from a theoretical perspective~\cite{Woods2019a}, but experimental implementation of these ideas is difficult, leaving ample opportunity for theoretical and technological progress towards a feasible architecture. 

Among the challenges worth investigating is the thermodynamics of clock synchronisation, a fundamental problem spanning diverse fields, from quantum information to biological systems. Recent experimental progress has been made towards quantifying the thermodynamic resources required to synchronise two autonomous clocks~\cite{Yang_2024}. This experiment, which employed two membranes inside an optical cavity, suggests that synchronisation does not monotonically improve with increased entropy production; instead, there appears to be an optimal point or sweet spot. However, understanding entropy flows in these autonomous multipartite systems remains a non-trivial challenge. The thermodynamics of clock synchronisation, therefore, merits further theoretical and experimental investigation.

Another key challenge is to understand the thermodynamic limits on timekeeping at the precision frontier. Atomic clocks provide the best measurements of time ever achieved~\cite{Aeppli2024}, and have become an increasingly important tool for probing new fundamental physics. Yet a complete thermodynamic description of an atomic clock, including key ingredients such as feedback stabilisation and frequency downconversion, is still missing. Presumably, atomic clock designs already unwittingly exploit quantum coherence to exponentially increase their energy efficiency relative to classical clocks~\cite{Meier2024a}, given that they far exceed thermodynamic precision bounds for classical stochastic systems~\cite{Pietzonka2024}. An interesting question for future work is the extent to which further efficiency gains would be achievable or desirable, e.g.~if more compact and portable clocks are needed either for terrestrial or space-based applications. \\

\paragraph*{Broader perspective and relevance to other fields.}
The thermodynamics of timekeeping and clock synchronization can offer useful insights into the energy dynamics of mechanisms in biological and bio-inspired systems. Understanding the energy costs of coupling-induced synchronization may reveal the trade-offs and fundamental limits of maintaining coherent timing and cooperative dynamics. These insights could also have applications in neuromorphic models as well as stochastic and thermodynamic computing. Moreover, many biological processes act effectively as clocks, e.g.~cyclic molecular motors or biochemical oscillators. Studying the physics of these processes has revealed thermodynamic~\cite{Pietzonka2024} and frenetic~\cite{Prech2024,Macieszczak2024} constraints on classical stochastic timekeeping, which are closely analogous to results obtained from a quantum thermodynamic perspective (see Sec.~\ref{sec:TURs}). Further research on the thermodynamics of nanoscale clocks can therefore serve as a bridge between the quantum thermodynamics and statistical biophysics communities. 

Understanding the physical limits on clocks is also relevant for numerous foundational questions in quantum information and quantum gravity.  Certain approaches to canonical quantum gravity involve the quantisation of space-time, motivating the search for a fully quantum description of time. One prominent framework is the so-called Page-Wootters mechanism, whereby dynamical time is recovered from a timeless (stationary) state via an explicit quantum clock degree of freedom~\cite{Altaie2022}. In addition, recent research has uncovered the possibility of quantum operations with indefinite causal order or time direction~\cite{Taranto2025}, with potential consequences for the unification of quantum mechanics and general relativity. Examining the physical resources needed to operate quantum clocks in these contexts may reveal constraints on future theories in which quantised time or gravity play an explicit role. Likewise, investigating the thermodynamics of timekeeping in the context of hybrid classical-quantum theories of gravity~\cite{Oppenheim2023postquantum} (see also Sec.~\ref{sec:collapse}) could yield further insights. \\

\paragraph*{Concluding Remarks.}
The study of timekeeping in the quantum regime has uncovered intriguing thermodynamic constraints and potential advantages that differ fundamentally from those in classical systems. The realisation of autonomous quantum clocks, the study of clock synchronisation, and the integration of clocks with coherent control mechanisms are promising directions for future research, with implications for both fundamental physics and practical applications in quantum technologies.

A key challenge moving forward is to bridge the gap between abstract theoretical models and experimentally feasible implementations. Achieving this would be essential for significantly deepening our understanding of the energetic costs associated with timekeeping and control in quantum systems, as well as in emerging computational approaches and bio-inspired systems. \\

\paragraph*{Acknowledgements.} M.T.M. is supported by a Royal Society University Research Fellowship. N.A. acknowledges support from the European Research Council (grant agreement 948932) and the Royal Society (URF-R1-191150). P.E. acknowledges funding from the Austrian Federal Ministry of Education, Science, and Research via the Austrian Research Promotion Agency (FFG) through Quantum Austria. This project is co-funded by the European Union (Quantum Flagship project ASPECTS, Grant Agreement No.\ 101080167) and UK Research and Innovation (UKRI). Views and opinions expressed are however those of the authors only and do not necessarily reflect those of the European Union, Research Executive Agency or UKRI. Neither the European Union nor UKRI can be held responsible for them.

\clearpage

\section{Thermodynamics of information propagation}
\label{sec:infopropagration}
\noindent
{\it Moallison F. Cavalcante}

\noindent
{Department of Physics, University of Maryland, Baltimore County, Baltimore, MD 21250, USA}\\
{Quantum Science Institute, University of Maryland, Baltimore County, Baltimore, MD 21250, USA}\\
{Gleb Wataghin Physics Institute, The University of Campinas, 13083-859, Campinas, S\~{a}o Paulo, Brazil}\\

\noindent
{\it Emery Doucet}

\noindent
{Department of Physics, University of Maryland, Baltimore County, Baltimore, MD 21250, USA}\\
{Quantum Science Institute, University of Maryland, Baltimore County, Baltimore, MD 21250, USA}\\

\noindent
{\it Marcus V. S. Bonan\c{c}a}

\noindent
{Gleb Wataghin Physics Institute, The University of Campinas, 13083-859, Campinas, S\~{a}o Paulo, Brazil}\\

\noindent
{\it Sebastian Deffner}

\noindent
{Department of Physics, University of Maryland, Baltimore County, Baltimore, MD 21250, USA}\\
{Quantum Science Institute, University of Maryland, Baltimore County, Baltimore, MD 21250, USA}\\
{National Quantum Laboratory, College Park, MD 20740, USA}\\

\paragraph*{State-of-the-art.}
Since the inception of Maxwell's demon \cite{leff1990maxwell}, the study of the physical ramifications of processing information has been an integral part of thermodynamics \cite{Parrondo2015Feb}. Motivated by the advent of digital computers, the emphasis has historically been on quantifying the minimal thermodynamic costs associated with the writing and erasing of information \cite{Boyd2016PRL}. Curiously, information theory itself was born from the study of communication \cite{cover1999elements}, which in its essence refers to the directed and controlled propagation of information through a physical substrate. In this context, see, e.g., Ref.~\cite{Gyongyosi2018IEEE} for a recent review on quantum information transmission.

While statements of the second law for writing and erasing information, such as Landauer's principle, have been largely noncontroversial, whether or not there is a minimal cost for communicating information has been hotly debated \cite{Pendry1983,Landauer1987APL}. This debate appears far from settled, as even today quantifying the thermodynamic cost of reliable communication is a topic of current research efforts \cite{Yadav2024arXiv}. 

The situation is even less clear for the uncontrolled propagation of information, which in complex quantum many body systems may lead to scrambling or even quantum chaos \cite{Touil2024EPL}. For both controlled as well as uncontrolled propagation of information, notions and concepts from (quantum) stochastic thermodynamics appear uniquely suited to shine light on whether information travels freely or whether propagation incurs a thermodynamic price \cite{Touil2024EPL}. \\

\begin{figure}
    \centering
    \includegraphics[width=0.4\linewidth]{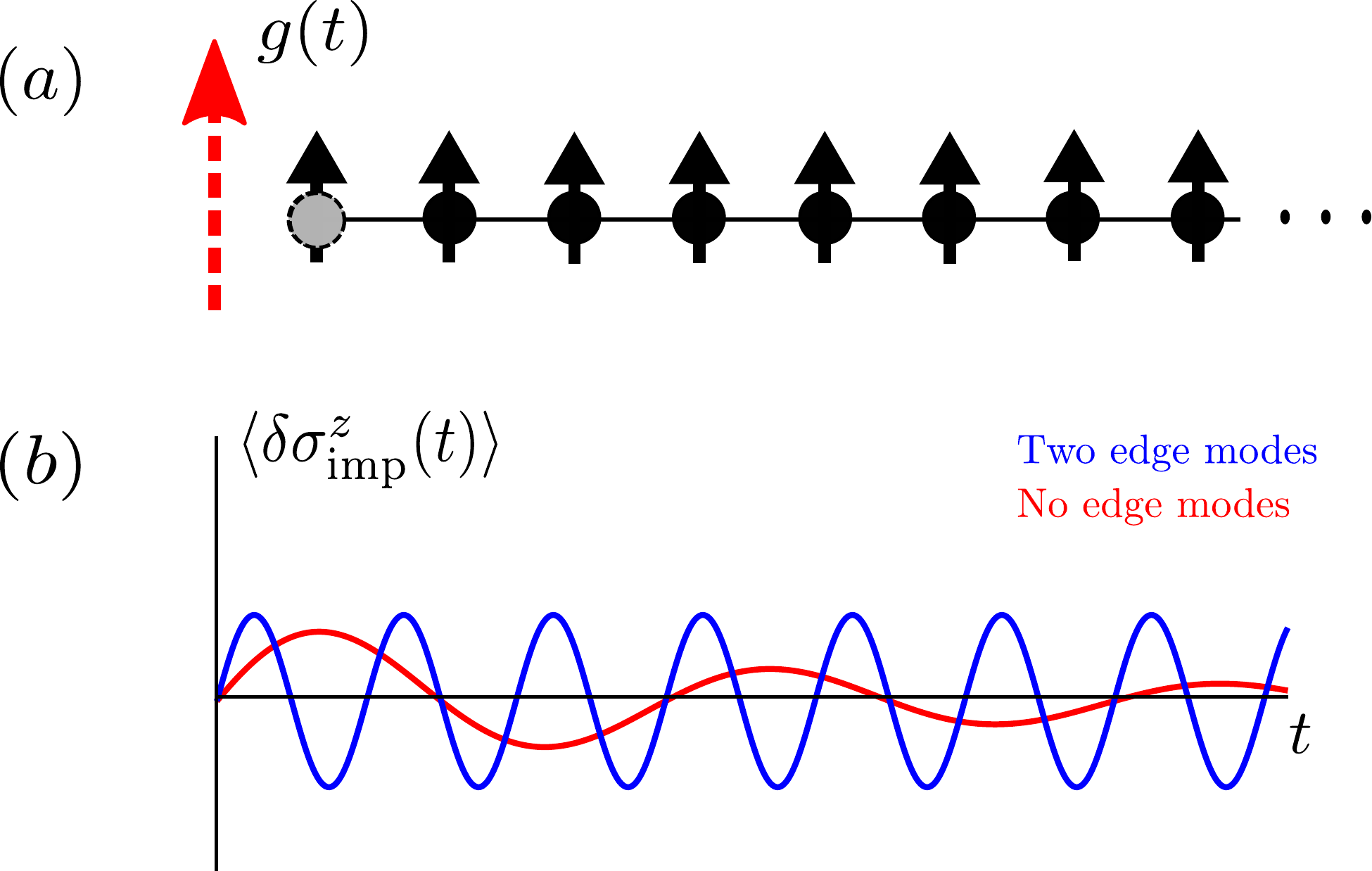}
    \caption{(a) The quantum Ising chain with an impurity (gray ball) at its edge.  The large red arrow represents the local control field $g(t)$ acting only at the impurity. (b) Time-dependent deviation of the impurity spin from its zero-temperature equilibrium value. The red line, where no edge modes are present, shows the perfect propagation of information into the bulk. In contrast, the blue line, where there are two edge modes, shows the partially trapped information around the impurity. }
\label{fig:model}
\end{figure}

\paragraph*{Current and future challenges.}
The complexity of the issue is best illustrated with a specific example. In a recent study, some of us analyzed how information is shared throughout a complex quantum many-body system with an impurity \cite{cavalcante2025}, see Fig.~\ref{fig:model} (a). For specificity, consider the system to be driven at the impurity, which then acts as a witness how information propagates throughout the system.  

Conventionally, such analyses rely on the Lieb-Robinson bound, which characterizes the effective speed of sound \cite{Lieb_1972} or the ``butterfly'' velocity \cite{Roberts2016PRL}. However, more genuinely thermodynamic information can be extracted from the response functions. To this end, imagine the complex system to be ``kicked'' at the impurity, and we observe how information about the perturbation propagates. This system shows a rich boundary phase diagram, with up to two edge modes localized around the impurity.  Remarkably, we found that within the phase with two edge modes both the time-ordered and the out-of-time ordered response functions show persistent long-time coherent oscillations after the bulk modes undergo relaxation, see Fig.~\ref{fig:model} (b). This behavior reflects the trapping of the excitations, hence information, into the localized edge modes. Thus, the impurity allows us to have partial control over the information injected into the system by the local field. 

In order to better understand this trapped information, we looked at the density operator of the system. We showed that, in the long-time limit, the bulk mode contribution to this operator decays rapidly, leaving behind an edge mode contribution that has a $X$-state form~\cite{Rau_2009}. We further characterized our $X$-state by calculating information quantities, such as purity, entanglement, and discord, as a function of the impurity strength. The results showed that the quantumness of the aforementioned $X$-state can be enhanced by increasing the intensity of the applied local control field.

This means that depending on the topological properties of the physical substrate, the properties of information propagation can be fundamentally different. Naturally, one has to conclude that then also the thermodynamic resources have to intimately depend on the physical properties of the information carriers. In particular in quantum systems, statements of the second law may have to be formulated to be specific for the properties of the complex systems used as communication channels, see also Pendry's early insight \cite{Pendry1983}. \\

\paragraph*{Broader perspective and relevance to other fields.}
Even more generally, understanding the costs, constraints, and properties of information dynamics transcends virtually all areas of modern research. For example, a fruitful line of inquiry into understanding and explaining the black hole information paradox comes from treating a black hole as a quantum channel and an information scrambler \cite{Hayden2007}.

In a more prosaic but no less important arena, quantum information propagation is critical to explaining how classical reality emerges in a quantum universe. Consider multiple observers, each of which performs some measurement of one common quantum system. The randomness inherent in quantum measurements means that it is not guaranteed that the inferences these observers draw from their results agree. Non-repeatability or incompatibility is an important feature of quantum measurements which is critical to many applications, but it is also clearly at odds with everyday classical experience. In the classical world, measurements are repeatable and many observers can all learn the same information about a system. That is, classical reality is objective.

Quantum Darwinism \cite{Zurek22} provides a framework to reconcile these statements, based on the key insight that all real measurements are indirect through an environment. The measured system  interacts with an environment, which observers then capture fragments of from which they infer the system state. Effectively, the environment is a channel connecting the system and observers. Whether measurement results are objective depends on the details of that communication channel, and in particular on how the system information is encoded into the environment. If the system information is redundantly encoded in the environment such that small fragments of the environment allow it to be inferred, many observers can reconstruct that information and hence it is objective. Interestingly, only classical information about the system in a certain ``pointer’’ basis can be rendered objective – quantum correlations can never be objective. Further, the only system-environment states that support objectivity are of a specific ``singly-branching’’ form, which only certain models can generate \cite{Doucet24}. 

The deep connections Quantum Darwinism reveals between emergent classicality and quantum information dynamics lead to a number of interesting avenues for future study. Considering Landauer’s principle and the potential costs of quantum communications on indirect measurements, one key questions arises: is there a thermodynamic price associated with the emergence of classicality? Such possibilities underscore the importance of understanding the thermodynamic properties of quantum information propagation, and show that advances in this direction can both raise and answer a range of broad, foundational questions. \\

\paragraph*{Concluding Remarks.}
Recent research has demonstrated that quantum many-body systems are excellent platforms to test the conjectures and develop notions of a ``thermodynamics of information propagation'', as we have illustrated this with the example of the quantum Ising chain with an impurity. Its different phases provide an easily accessible framework to study vastly different scenarios within the same systems. In other words, quantum many-body systems consist of excellent ``physical substrates" for the phenomena of information propagation, storage or scrambling. Additionally, we believe that our previous work has already revealed a thermodynamic aspect of this by showing the parallel between controlling information and excitations. The obvious next step will have to be the development of a comprehensive set of statements of the second law of thermodynamics for quantum information propagation.

\paragraph*{Acknowledgements.}

S.D. acknowledges support from the John Templeton Foundation under Grant No. 63626. This work was supported by the U.S. Department of Energy, Office of Basic Energy Sciences, Quantum Information Science program in Chemical Sciences, Geosciences, and Biosciences, under Award No. DE-SC0025997. This work was partially supported by Conselho Nacional de Desenvolvimento Cient\'ifico e Tecnol\'ogico (CNPq), Brazil, through grant No. 200267/2023-0. M.V.S.B. acknowledges the support of CNPq, under Grant No. 304120/2022-7, and the S\~ao Paulo Research Foundation (FAPESP), Brazil, Process Number 2022/15453-0.

\clearpage

\section{Thermodynamic consistency of quantum collapse models}
\label{sec:collapse}
\noindent
{\it Simone Artini and Gabriele Lo Monaco}

\noindent
{Universit\`a degli Studi di Palermo, Dipartimento di Fisica e Chimica - Emilio Segr\`e, via Archirafi 36, I-90123 Palermo, Italy}\\

\noindent
{\it Sandro Donadi}

\noindent
{Dipartimento di Ingegneria, Universit\`a degli Studi di Palermo, Viale delle Scienze, 90128 Palermo, Italy} \\

\noindent
{\it Mauro Paternostro}

\noindent
{Universit\`a degli Studi di Palermo, Dipartimento di Fisica e Chimica - Emilio Segr\`e, via Archirafi 36, I-90123 Palermo, Italy, and}
{Centre for Quantum Materials and Technologies, School of Mathematics and Physics,
Queen’s University Belfast, BT7 1NN, United Kingdom}\\

\paragraph*{State-of-the-art.}
Quantum mechanics accurately describes the behavior of microscopic systems. Yet, since its inception, the theory has been plagued by the measurement problem. This issue arises because the standard (Copenhagen) formulation of the theory postulates two dynamics for the state vector: the linear and deterministic Schrödinger evolution, and the non-linear and stochastic wave function collapse. These two dynamics are fundamentally different, and the measurement problem arises because the theory fails to unambiguously specify which postulate to apply for a given system, leading to paradoxes such as Schrödinger’s cat.

Collapse models solve the measurement problem by merging the two dynamics (Schrödinger and collapse) into a single evolution~\cite{ghirardi1986unified,bassi2003dynamical}. More precisely, they modify the Schrödinger equation by adding a non-linear interaction with classical noise, which is responsible for inducing collapse in space. The effects of the non-linear terms are negligible for microscopic systems, but, through an amplification mechanism, they become dominant for macroscopic systems. In contrast to other solutions to the measurement problem, such as Bohmian mechanics
 or the many-worlds theory, which make the same predictions of standard quantum mechanics, collapse models make different predictions and can thus be tested in experiments~\cite{donadi2021underground, carlesso2022present}.

Among all collapse models, the two most studied in the literature are the Continuous Spontaneous Localization (CSL) model~\cite{ghirardi1990markov} and the Di\'osi-Penrose (DP) model~\cite{diosi1989models,penrose1996gravity}. Both models predict steady spontaneous heating due to the interaction of systems with the noise responsible for the collapse. Such heating, for typical choices of the model's parameters, is very low. For example, in the CSL model, where predictions depend on two phenomenological parameters $\lambda$ and $r_C$, for the values originally suggested ($\lambda = 10^{-16}$ s and $r_C =10^{-7}$ m), the heating rate of a monoatomic gas is of the order of $10^{-15}$ K/year~\cite{bassi2003dynamical}. 

However, the prediction of a steady heating rate, irrespective of the energy of the system, is unphysical. It corresponds to the interaction with a bath at infinite temperature, while one would expect that the noise responsible for the collapse is associated with some fundamental field in nature, which, to be realistic, must have finite temperature. 
To mitigate this heating, dissipative extensions of the models have been introduced for both the CSL~\cite{Smirne2015dissipative} and the DP~\cite{bahrami2014role,dio2023linear} models. In these models, the system thermalizes at a given temperature. However, in none of these models the thermodynamic consistency of the out-of-equilibrium dynamics is considered in detail. In particular, a quantum-to-classical transition model must adhere to the Second law of Thermodynamics to be considered a proper physical model and the sole addition of a dissipative mechanism is not enough to guarantee positive entropy production. Below, we discuss the current literature addressing this issue. \\

\paragraph*{Current and future challenges.}
The first study of the entropy dynamics of a Collapse model was performed in Ref.~\cite{artiniS2023characterizing}, where the CSL model is considered. There, it is shown that the effect of the collapse can be seen in the phase-space representation of the dynamics as a diffusion on the momentum of the particle. This diffusion, which is not balanced by any dissipation, leads to an indefinite heating. If this is assumed to be the result of an interaction with an infinite temperature noise-field, the entropy production rate as defined via the Wigner entropy remains positive for all times and the dynamics can be therefore regarded as thermodynamically consistent despite the system remains out-of-equilibrium indefinitely. A similar approach, that however does not address the entropy production, is present in Ref.~\cite{Te2021master}. There, it is showed that a many-particle system subject to the collapse does not appraoch a state with homogeneous temperature, suggesting that this mechanism cannot be responsible for the emergence of thermodynamic equilibrium.

In Ref.~\cite{artini2025nonequilibriumthermodynamicsgravitationalobjectivecollapse} the same analysis of the entropy dynamics is carried out on the DP model which, as expected, shows the same qualitative behavior. Furthermore, the dissipative extension of the model proposed in Ref.~\cite{dio2023linear} is also considered. In the low friction regime, the dynamics is still gaussian and, in particular, has the same functional form of a Klein-Kramers equation of Brownian motion in terms of the Wigner function of the system in phase-space. Hence, in this regime, the system reaches a thermal steady state with finite temperature remaining consistent with the Second Law of thermodynamics. The complete dynamics, on the other hand, involves non-gaussian terms that give rise to higher-than-second order derivatives in the phase-space representation. To investigate this regime, a small-time linearization of the evolution has been used and this revealed negative entropy production rate for certain values of the parameters, thus violating the Second law. Even so, it is shown that such linearized approach could lead to non-physical evolutions even for simpler dynamics. Since a way to benchmark the validity of the approximation of the dynamics at hand is lacking, definitive conclusions cannot be drawn. 

The main challenge that emerges from this analysis is thus the characterization of the entropy dynamics for models leading to non-gaussian evolutions, which is the case for other dissipative generalizations of the collapse dynamics~\cite{Smirne2015dissipative}. In order to assess the thermodynamic consistency of such models it is paramount to develop more advanced techniques that are able to characterize the Wigner entropy. A possibility is to use perturbative techniques that rely on known Green's function methods for the solution of higher-order partial differential equations. These techniques are also necessary to study potentials more than quadratic in the position and/or momentum of the particle. Such potentials, in fact, lead to quantum corrections to the Poisson brackets that govern the system's evolution in the phase-space which include higher-than-second-order derivatives. \\

\paragraph*{Broader perspective and relevance to other fields.}
A critical assessment along the lines of what has been summarized here in regard to quantum collapse models would be crucial in other areas of fundamental physics as well. Of particular interest are the gravity-related models that have recently gained much attention as potential candidates for solving the measurement problem and/or describing, at least phenomenologically, the effect of gravity on quantum systems, typically predicting decoherence induced by gravity. Such approach was introduced for the first time by Károlyházy, who suggested that gravitational effects could induce quantum decoherence. His work was followed by the already mentioned DP model and, since then, many models predicting gravitational decoherence have been proposed, (see for example Ref.~\cite{Bassi2017gravitational} for a comprehensive review). 

A different approach is proposed in Ref.~\cite{Oppenheim2023postquantum} where space-time is treated as fundamentally classical and matter as fundamentally quantum. The corresponding theory is proposed as an embodiment for an effective theory resulting from taking the classical limit of a fully quantum theory of gravity. Coupling quantum systems to classical ones requires great care. In addition to ensuring standard properties such as the positivity of the dynamics, it is necessary to introduce fundamental stochasticity in the dynamics to prevent the possibility of faster-than-light signaling (e.g., in EPR-like setups). All of this is ensured by the dynamics introduced in Ref.~\cite{Oppenheim2023postquantum}. However, as such classical-quantum limit should be performed carefully -- and there is no unique way of performing it -- it would be crucial to investigate the thermodynamic consistency of such  an effective theory. \\

\paragraph*{Concluding Remarks.}
We have investigated the necessity of {\it putting thermodynamics into the collapse dynamics}, and more generally phenomenological models for the quantum-to-classical transition of open quantum systems, by investigating two of the most celebrated theories in this context. Our investigation, which can and should be furthered in various directions (only some of them having been mentioned here), would provide a thermodynamically motivated   -- and thus intrinsically fundamental -- framework that should accompany standard assessments of the tenability of phenomenological models based on complete positivity of the corresponding dynamical maps. \\

\paragraph*{Acknowledgements.}
We acknowledge support from the European Union’s Horizon Europe EIC-Pathfinder project QuCoM (101046973), the UK EPSRC (grants EP/T028424/1 and EP/X021505/1), the Department for the Economy of Northern Ireland under the US-Ireland R\&D Partnership Programme, the ``Italian National Quantum Science and Technology Institute (NQSTI)" (PE0000023) - SPOKE 2 through project ASpEQCt, the “National Centre for HPC, Big Data and Quantum Computing (HPC)” (CN00000013) – SPOKE 10 through project HyQELM, the Italian Ministry of University and Research under PNRR - M4C2-I1.3 Project PE-00000019 ``HEAL ITALIA" (CUP B73C22001250006). SD acknowledges support from Istituto Nazionale di Fisica Nucleare (INFN).

\clearpage
\section{Quantum control, thermodynamics, and machines}
\label{sec:control}
\noindent
{\it Adolfo Del Campo}

\noindent
{Department of Physics and Materials Science, University of Luxembourg, L-1511 Luxembourg, Luxembourg, and}\\
{Donostia International Physics Center, E-20018 San Sebastian, Spain}\\

\paragraph*{State-of-the-art.}
Early research on quantum machines focused on minimal models with a single-particle working medium \cite{Kosloff17,Mitchison2019}. This left out essential features 
such as many-body quantum correlations and collective phenomena. Developing many-particle quantum machines enables the harnessing of uniquely multipartite features that have no single-particle counterpart. Focusing on the working medium, these include critical phenomena, quantum indistinguishability and quantum statistics, the interplay of integrability and quantum chaos, and the tuning of interparticle interactions~\cite{Chen2019, watanabe20quantum, Koch2023}.

While scaling up quantum machines offers tantalizing opportunities, it comes with the additional challenge of reducing quantum friction in a many-body setting \cite{delCampo2018}. This is closely tied to the possibility of sustaining adiabaticity in a driven and possibly open many-body quantum system and provides an enticing frontier in quantum thermodynamics that has barely been explored. 

At the single-particle level, studies of quantum lubrication identified model-specific strokes that cancel nonadiabatic excitations, effectively reducing quantum friction to zero \cite{Kosloff17}. Such strokes, also known as accidental shortcuts, can be used to build a quantum cycle involving isentropic strokes, such as a quantum Otto cycle or an interaction-driven cycle.
A systematic approach to quantum friction suppression is provided by shortcuts to adiabaticity (STAs) \cite{GueryOdelin19}. 
STAs are techniques for the fast nonadiabatic driving of classical and quantum systems that yield the same final state as in an adiabatic protocol. Crucially, they achieve this without requiring slow driving. As such, they were recognized early on as a way out of the ``tragedy of finite-time thermodynamics'', i.e., the trade-off between efficiency and power of a quantum thermodynamic cycle run in finite time. In short, STAs enable the engineering of heat engines that operate at maximum efficiency and have a tunable output power \cite{delCampo2018}.

Several techniques fall under the umbrella of STAs. They include reverse-engineering scaling laws, Lewis-Riesenfeld invariants of motion, fast-forward techniques, etc. Among them, counterdiabatic driving (CD), also known as transitionless quantum driving, provides a systematic approach to engineering STAs in an arbitrary system \cite{delCampo2018,GueryOdelin19}. CD relies on auxiliary counterdiabatic fields that assist the time evolution, allowing one to run a ``fast motion video'' of a reference adiabatic trajectory \cite{delcampo13}. Identifying the required CD fields can be challenging and is notoriously complicated in many-body systems.  The original formulation of CD required knowledge of the spectral properties of the system, which is generally intractable in complex many-body systems. In addition, implementing such CD fields is generally difficult in the laboratory, as the required interactions are many-body and non-local. The recognition of this fact has motivated broad efforts for the efficient approximation of the CD terms \cite{Takahashi24}. While approaches based on approximate CD driving are being exhaustively explored in the context of quantum optimization and computation, their study in quantum thermodynamics remains to be elaborated.\\

\paragraph*{Current and future challenges.}
In isolated systems, using STAs is facilitated in self-similar evolutions displaying scale-invariance \cite{delcampo13}. The latter often arises in ultracold gases in time-dependent traps that can describe the working substance of a quantum machine. This symmetry holds approximately in strokes involving moderate compression and expansion ratios. 
In one dimension, it describes exactly the evolution in a harmonic trap of spin-polarized Fermi gases, the Tonks-Girardeau gas, and the Calogero-Sutherland gas (ideal gas of geons). It also applies to the interacting Bose gas in two dimensions (up to quantum anomalies) and, in the Thomas-Fermi regime, in any dimension. In three spatial dimensions, it also describes the unitary Fermi gas, used to demonstrate frictionless strokes in finite-time thermodynamics \cite{delCampo2018}. 

While the quantum Otto cycle is favored for its simplicity in theoretical studies of quantum machines, general thermodynamic cycles include non-isentropic strokes, leading to inherently open, non-unitary dynamics.  The same applies to continuously driven cycles or cycles driven by measurements.
Controlling a driven quantum system becomes more challenging when accounting for the contact with a surrounding environment, which makes it possible to exchange energy and heat. As discussed, fast control of isolated many-body systems is possible through STAs. The CD technique has been generalized to open systems: in addition to auxiliary CD Hamiltonian controls, fast-forwarding an open quantum trajectory generally requires implementing an auxiliary CD dissipator \cite{Vacanti2014,Dann2019,Alipour20,Kallush22}. 
Protocols for fast cooling and heating strokes have been designed in simple quantum mechanical systems, including the harmonic oscillator and a two-level system. Experimentally, an STA in an open system has been demonstrated in circuit quantum electrodynamics, using two coupled bosonic oscillators coupled to a transmon qubit \cite{Yin2022}. Progress at the many-particle level is currently limited.
A natural goal is thus the engineering of frictionless many-body quantum machines, such as quantum critical machines. 
 
Control protocols of thermodynamic devices and their cost have strong ties to information theory. Bounds illuminating the ultimate limits on the performance of quantum machines (e.g., governing the output power of an engine or the charging power of a battery) have been formulated using speed limits and information geometry, both in the classical and quantum domains. When applied to controlled quantum machines using CD and a single-particle working medium, such bounds are often saturated, and the performance admits a simple geometric understanding regarding the trajectory traced out in quantum state space.
However, these bounds need to be refined in a many-body setting. In particular, a direct application of quantum speed limits generally yields estimates for the minimum time scale for a process to unfold that are too conservative due to the orthogonality catastrophe. 

The cost of STAs and other control protocols has been assessed in terms of energy and work fluctuations, and operator norms involving the generator of evolution with the required CD terms \cite{delCampo2018}. Such efforts face the pitfall of ``definition-based physics'': ad hoc figures of merit without a solid motivation, which would ideally be based on physical grounds (rather than mathematical or computational), can be misleading and yield circumstantial and questionable conclusions. An example arises with alternative definitions of cycle efficiency, concluding the vanishing performance of STA-assisted quantum machines, refuted by experimental findings \cite{Hou2024}. From a complementary viewpoint, 
quantum speed limits allow us to distinguish classical and quantum contributions to the evolution speed \cite{GarciaPintos22}, making it possible to explore quantum advantage in quantum machines.

STAs in scale-invariant systems have been related to delta-kick cooling, and, more generally, it is possible to implement STAs by CD terms as an impulse. This potentially allows for quantum friction reduction in an arbitrarily fast process. This limit is amenable to a clear-cut study of the benefit-cost analysis of CD impulses. The generalization of such an analysis to include quantum superimpulses, as proposed by Jarzynski, provides an intriguing prospect without relying on self-similar dynamics \cite{Jarzynski24}.
To date, the extension of this approach to many-body and open systems remains unexplored. 

Engineering efficient quantum machines should consider alternative control schemes \cite{Koch2022}, including numerical methods such as the GRAPE and CRAB algorithms, Lyapunov control, and machine learning. Likewise, it can benefit from advantages and speedups achieved by other means, including physical phenomena such as synchronization, collective spontaneous emission, and the Mpemba effect.\\

\paragraph*{Broader perspective and relevance to other fields.}
Quantum thermodynamic processes are ubiquitous in Nature and technology, leading to a broad arena for their control. Let us illustrate some examples.

In the spirit of the pioneering work by Scovil and Schulz-DuBois, light harvesting in natural systems, as well as in artificial solar cells, has been described as a quantum heat engine \cite{Dorfman13}. It is yet to be seen whether the know-how in quantum control for the efficient engineering of quantum machines can be carried out to optimize light harvesting in such scenarios.

Control, as described in this contribution, is external and often guided by detailed knowledge of the performance of the uncontrolled machine.
An exciting open prospect involves the development of autonomous quantum machines with low or zero friction. Such a goal may be elusive, but progress may be guided by minimizing entropy production in their operation or, more generally, the minimization of an action principle in the spirit of the quantum brachistochrone problem and its generalizations. Beyond the engineering of a single efficient and autonomous quantum machine, one can envision the description of ensembles of such devices that would not only provide a quantum analog of classical active matter but may exhibit intrinsically quantum phenomenology. Developing the field of active quantum matter and optimizing the performance of swarms of autonomous quantum machines provides a new frontier for quantum thermodynamics. 

An exciting arena for engineering quantum machines, and indeed for quantum thermodynamics as a field, involves moving beyond an analog approach relying on specific platforms for the realization of quantum machines. As an alternative, one can embrace digital techniques in quantum simulation and computation and hybrid digital-analog approaches. Furthermore, specific quantum information tasks formulated in the circuit gate model can be described in the language of thermodynamic cycles, as in the case of quantum error correction with no syndrome measurements. The same holds for classical forms of computing, such as thermodynamic computing \cite{Lipka-Bartosik24}.
As in the classical case, the study of quantum thermodynamics in relation to the gate model opens the door to harnessing (classical and quantum) optimization algorithms for the engineering of quantum machines and to analyze the thermodynamic and energetic costs of quantum circuits, e.g., for digital quantum simulation, optimization, and computation. \\

\paragraph*{Concluding Remarks.}
It is widely recognized that the study of heat engines and thermodynamic devices played historically a key role in the development of thermodynamics, helping to identify its basic tenets. That the development of quantum thermodynamics follows a similar path may appear as a naive expectation, one that has nonetheless materialized to date. Exploring the ultimate performance of quantum machines and their engineering is likely to guide further developments that embrace the complexity of dissipative many-body quantum systems. In doing so, quantum control techniques may not only be required for achieving their optimal operation but may also facilitate implementations in the laboratory and their use in applications and quantum technologies.  It may further deepen our understanding of the fundamental interplay of information theory and quantum thermodynamics in complex systems.\\

\paragraph*{Acknowledgements.}
This project was supported by the Luxembourg National Research Fund (FNR Grant Nos.\ 17132054 and 16434093). It has also received funding from the QuantERA II Joint Programme and co-funding from the European Union’s Horizon 2020 research and innovation programme.

\clearpage

\section{Quantum batteries}
\label{sec:batteries}
\noindent
{\it Alan C. Santos}

\noindent
{Instituto de Física Fundamental, Consejo Superior de Investigaciones 
Científicas, Calle Serrano 113b, 28006 Madrid, Spain}\\

\noindent
{\it Chang-Kang Hu }

\noindent
{International Quantum Academy, Futian District, Shenzhen, Guangdong 
518048, China}\\
{Shenzhen Institute for Quantum Science and Engineering, Southern 
University of Science and Technology, Shenzhen, Guangdong 518055, 
China}\\
{Guangdong Provincial Key Laboratory of Quantum Science and 
Engineering, Southern University of Science and Technology, Shenzhen, 
Guangdong 518055, China}\\

\noindent
{\it Dario Rosa}

\noindent
{ICTP South American Institute for Fundamental Research, Instituto de 
F\'{i}sica Te\'{o}rica, UNESP - Univ. Estadual Paulista, Rua Dr. Bento 
Teobaldo Ferraz 271, 01140-070, S\~{a}o Paulo, SP, Brazil}\\

\noindent
{\it Dian Tan}

\noindent
{International Quantum Academy, Futian District, Shenzhen, Guangdong 
518048, China}\\
{Shenzhen Institute for Quantum Science and Engineering, Southern 
University of Science and Technology, Shenzhen, Guangdong 518055, 
China}\\
{Guangdong Provincial Key Laboratory of Quantum Science and 
Engineering, Southern University of Science and Technology, Shenzhen, 
Guangdong 518055, China}\\
{Shenzhen Branch, Hefei National Laboratory, Shenzhen 518048, China}\\

\paragraph*{State-of-the-art.} 
Quantum batteries (QBs) refers to quantum mechanical systems used as energy storage devices. In principle, every quantum system can serve as a QB. However, when dealing with quantum technologies, the challenge is to find quantum effects that enhance performance compared to a classical approach.
	
\begin{figure}[b]
\includegraphics[width=\linewidth]{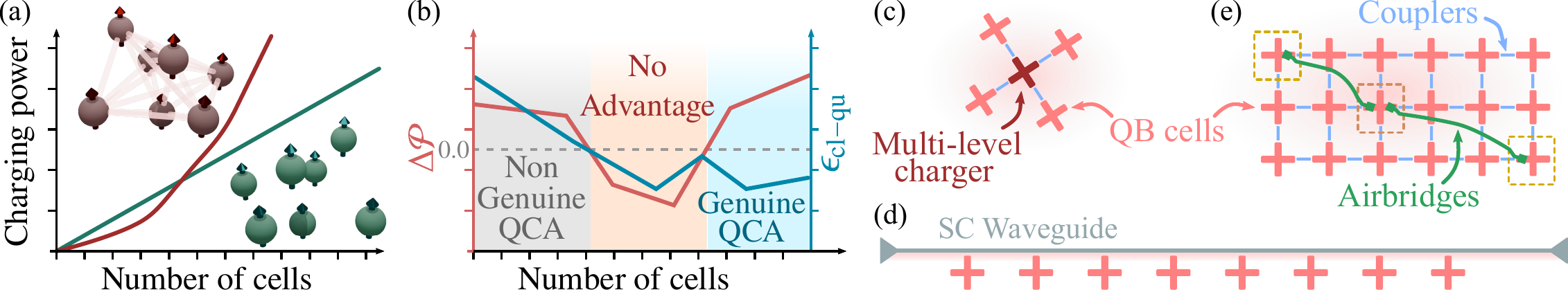}
\caption{(a) Example of the expected behavior of charging advantage between interacting and independent quantum cells. While the charging power for independent cells grows linearly, the collective behavior of interacting systems may provide enhanced charging scaling. (b) Diagram showing how to identify genuine quantum advantage. (c,d,e) Schematic representation of promising SC devices that can be used to experimentally verify genuine QCA.}
\label{Figure:QB}
\end{figure}

Among several figures of merits, the possibility of speeding up the charging process of a QB, resulting in a faster-than-linear scaling (in the number of cells) of its charging power, has received considerable attention, introducing the notion of \textit{quantum charging advantage} (QCA) \cite{campaioli2017enhancing}. It is of immediate verification, see Fig.~\ref{Figure:QB}(a), that a \textit{non-interacting} charging protocol does not lead to QCA. Therefore, it becomes imperative to design \textit{interacting} QBs and address the question: "What features are necessary to have a QCA?"
	
Theoretical evidence of QCA in solid-state quantum systems has been provided. In particular, the Sachdev-Ye-Kitaev QB~\cite{Rossini:20}, exploiting \textit{all-to-all} and \textit{global} interactions, has shown a robust QCA. Moreover, Ref.~\cite{Kim:22} has provided a mathematical proof of the \textit{necessary} condition for QCA: the charging protocol must couple distant (in energy) levels of the battery Hamiltonian, a requirement that in a many-body setup requires \textit{global} operations.
	
In addition, Ref.~\cite{Kim:22} has introduced two relevant quantities, which fully characterize the presence of a genuine QCA. These are the \textit{classical-quantum driving potential ratio}, $\epsilon_{\mathrm{cl-qu}}$, and the \textit{power enhancement}, $\Delta \mathcal{P}$, defined as
\begin{equation}
\epsilon_{\mathrm{cl-qu}} = 1 - \frac{v_{\mathrm{cl}}}{v_{\mathrm{qu}}} , \quad  \Delta \mathcal{P} = 1 - \frac{P^{\ast}_{\mathrm{cl}}}{P^{\ast}_{\mathrm{qu}}} ,
\end{equation}
with $v_{\mathrm{cl}}$ and $v_{\mathrm{qu}}$ the classical and quantum driving potential, respectively, while $P^{\ast}_{\mathrm{cl}}$ and $P^{\ast}_{\mathrm{qu}}$ are the classical and quantum maximum power, respectively, computed within the time interval of a single cycle of the classical charging. The role of the ratio $\epsilon_{\mathrm{cl-qu}}$ is related to the energy cost we spend to charge QBs, which is related to the thermodynamics cost of Hamiltonians, and it has to satisfy $\epsilon_{\mathrm{cl-qu}} \leq 0$ because the quantum charging cannot spend more energy than its classical counterpart in the context of genuine quantum advantage. On the other hand, $\Delta \mathcal{P}$ quantifies the gain in power of the quantum charging with respect to the classical one, and therefore $\Delta \mathcal{P} > 0$. As illustrated in Fig.~\ref{Figure:QB}(b), these two quantities together allow us to distinguish between \textit{valid} and \textit{non-genuine QCA}, avoiding then potential mischaracterized QCAs.

Complementary to QCA, the extraction of stored energy as useful work is a pivotal aspect of QBs. In this context, significant progress has been made both theoretically and experimentally. For example, single-qubit work extraction has been successfully implemented in circuit quantum electrodynamics, empirically demonstrating the existence of ergotropy-based bounds for work extraction. Notably, the findings in Ref.~\cite{Monsel:20} were followed by theoretical proposals for state-independent work extraction protocols~\cite{Dominik:23}, addressing a key limitation in earlier approaches that required prior knowledge of the quantum 
state of the battery. More recently, experiments with nitrogen-vacancy centers have further advanced the field by demonstrating the feasibility of extracting \emph{coherent ergotropy}~\cite{Zhibo:24}, i.e., the contribution of quantum state coherences to the total ergotropy~\cite{Francica:20}.

\paragraph*{Current and future challenges.}
The necessity of global interactions creates a critical theoretical hurdle in the advancement of QBs, since their practical implementation presents significant challenges. To address and potentially mitigate these technical issues, we are then led to other important theoretical questions: ``Can we identify alternatives that yield meaningful quantum advantages without relying on global interactions?''
	
In this regard, the use of localized pairwise interactions or networked systems might serve as pathways to achieve practical quantum charging advantages.  Other potential avenues to overcome the obstacle posed by global operations are under investigation. In Refs.~\cite{mondal2022periodically,Moraes_2024} (see also \cite{puri2024floquet} for recent developments) it has been proposed the use of \textit{Floquet charging protocols}, since Floquet systems, when observed at stroboscopic times, are described by time-independent Hamiltonian which can, in principle, be of very high $k$-locality. Another attempt, promoted in \cite{andolina2024genuine}, is to shift the ground where the QCA can be obtained to a \textit{single-particle} setup, where distant single-particle energy levels can be coupled more easily. Further exploration will be critical to bridging the gap between theoretical potential and practical applicability, making QBs more accessible for real-world applications.
	
On the experimental front, the demonstration of quantum advantage in charging and its scalability with the number of battery cells is a pivotal and open challenge to be addressed. While this issue is inherently linked to the theoretical question of global interactions, it requires practical solutions in quantum hardware. Notably, the feasibility of global interactions is not physically prohibited but remains a technical barrier. Recent advancements, such as the engineered five-body interaction in superconducting qubits~\cite{ZhangKe:22} offer a glimpse into the potential for overcoming this limitation. As depicted in Fig.~\ref{Figure:QB}(c), for example, this particular system can be used to reach a genuine QCA in a four-cell QB connected to a single multi-level quantum charger. While the practical realization may initially be limited to systems with four battery cells, such a demonstration would mark a major step forward in experimental QBs research.
	
Superconducting (SC) technologies provide other possibilities for achieving genuine QCA in QBs. For instance, in the context of cavity quantum electrodynamics, we can connect several quantum cells to the same SC waveguide, such that the all-to-all interactions required in Refs.~\cite{mondal2022periodically} are mediated by the waveguide modes, as shown in Fig.~\ref{Figure:QB}(d). In the same context, this regime of interactions also could be achieved through other techniques developed. The advances reported in~\cite{bu2025tantalum} allows us to physically connect two distant SC qubits, placed on the same chip, through SC airbridges, as it can be seen from Fig.~\ref{Figure:QB}(e). Overall, we believe the engineered interaction reported in Refs.~\cite{,mondal2022periodically,ZhangKe:22,bu2025tantalum} could be the most promising routes to observe maximum scaling in quantum charging advantage. 
	
By addressing these theoretical and experimental open problems, the field of QBs can make substantial progress. The development of feasible alternatives to global interactions and the experimental demonstration of scalable quantum advantage are key milestones. Together, these efforts will not only deepen our understanding of quantum systems but also pave the way for their practical implementation in next-generation energy storage technologies. \\

\paragraph*{Broader perspective and relevance to other fields.}
To support the broad application and relevance of QBs, at least four complementary discussions are crucial. These discussions aim to address foundational aspects of QB development while paving the way for their practical implementation and integration with other quantum technologies.
	
The first point arises from the assertion that ``quantum technologies need a quantum energy initiative", as highlighted by Auffèves~\cite{Auf23}. This perspective underscores the critical importance of addressing the energy demands inherent to quantum technologies, including the costs associated with key energy management processes such as extraction, injection, and transport. A robust quantum energy initiative would extend beyond advancing QB technology, encompassing innovative strategies for energy storage, optimized distribution, and the seamless, sustainable integration of quantum energy systems within existing quantum networks.
	
The second discussion revolves around open problems in QB research, as these can provide valuable insights and directions for future development. For instance, the requirement of global interactions in QB charging protocols represents a significant challenge. Addressing this requires advanced investigations into Hamiltonian engineering and quantum control. Progress in these areas could enable precise manipulation of energy transfer processes, thereby enhancing charging efficiencies.
	
Another critical challenge is the protection of energy storage against decoherence, the universal barrier for quantum technologies. In this sense, interdisciplinary approaches to address this problem could prove invaluable in mitigating its effects in QBs, with further application to other kinds of quantum devices. Researchers focused on extending qubit coherence times could contribute methodologies adaptable to QBs. Additionally, other fields, such as non-Hermitian physics and topological systems, offer innovative perspectives such as \textit{topological QBs}~\cite{Lu:24}. The use of topological effects, for example, may provide robust mechanisms to enhance the performance of QBs in terms of resilience, charging time, and energy storage~\cite{Catalano:24}.
	
Energy transmission is a critical issue following the charging and protection of energy in QBs. Two significant challenges arise: first, environment-induced decoherence leads to energy loss and QB aging; second, the coupling strength between the charger and QB decreases with distance, reducing transfer efficiency. Addressing these problems is essential to enable practical applications of QBs. One potential solution is highlighted in Ref.~\cite{Niu:23}, where low-loss interconnects using pure aluminum coaxial cables facilitate energy transfer between SC processors 0.25 meters apart. This architecture demonstrates the feasibility of transferring entangled states—and thereby some energy—over a distance, offering a pathway to overcoming energy transmission limitations in QBs and advancing energy-efficient quantum technologies. \\

\paragraph*{Concluding Remarks.}
QBs have garnered significant attention as potential energy storage devices leveraging quantum phenomena. Although experimental advancements have been made with SC qubits~\cite{Hu:22a} and spin systems~\cite{Joshi:22}, achieving a tangible quantum advantage remains an open challenge. Demonstrating this advantage in a well-characterized real experiment is crucial before declaring QBs as potentially useful technologies. To this end, devising methods to charge the battery and extract energy efficiently without relying on global interactions is mandatory, to avoid complexities in system design, scalability, and control for the practical implementation of QBs. Simplified approaches that circumvent these interactions are essential to ensure robustness and accessibility in real-world scenarios. Achieving such advancements would mark a significant step toward realizing functional QBs capable of delivering a true quantum advantage. \\

\paragraph*{Acknowledgements.}
ACS acknowledges the support by the Comunidad de Madrid through the research funding program Talento 2024 ``César` Nombela", under the grant No 2024-T1/COM-31530. C.-K.H. is supported by the National Natural Science Foundation of China (Grant No. 12205137). DR thanks FAPESP, for the ICTP-SAIFR grant 2021/14335-0 and the Young Investigator grant 2023/11832-9, and the Simons Foundation for the Targeted Grant to ICTP-SAIFR. DT acknowledges the support by the National Natural Science Foundation of China (Grants No. 12004167, 11934010).

\clearpage

\section{Quantum thermometry}
\label{sec:thermometry}
\noindent
{\it Mohammad Mehboudi}

\noindent
{Technische Universität Wien, 1020 Vienna, Austria}\\

\noindent
{\it Dmitrii Lvov and Sergei A. Lemziakov}

\noindent
{PICO group, QTF Centre of Excellence, Department of Applied Physics, Aalto University, P.O. Box 15100, FI-00076 Aalto, Finland}\\

\noindent
{\it Luis A. Correa}

\noindent
{Instituto Universitario de Estudios Avanzados (IUdEA), Universidad de La Laguna, La Laguna 38203, Spain,}\\
{Departamento de Física, Universidad de La Laguna, La Laguna 38203, Spain}\\

\paragraph*{State-of-the-art.}
Over the past decade, there has been a growing interest in establishing the fundamental precision limits of low-temperature thermometry, giving rise to the new theoretical field of `quantum thermometry' \cite{Mehboudi_2019}. Particular attention was paid to the scaling laws that govern thermometric precision close to the absolute zero \cite{Hovhannisyan2018measuring,potts2019fundamental}. Theoretical quantum thermometry is to be distinguished from the already mature field of thermometry with mesoscopic devices, such as quantum dots, Coulomb blockade thermometers or NIS-junctions, among others \cite{Giazotto2006}. The advent of upcoming quantum technologies has motivated new temperature-estimation experiments ever deeper in the quantum regime on platforms such as superconducting qubits, ultracold atomic gases or trapped ions \cite{lvov2024thermometry,olf2015thermometry,PhysRevX.10.011018,PRXQuantum.3.040330,PhysRevApplied.22.064022}. These have revealed a pressing need for better precision and accuracy at ultracold temperatures, which puts the recent theoretical advances on quantum thermometry in a whole new light. Quantum thermometry's utility extends beyond ultimate theoretical precision; it can inform the adaptive optimisation of experimental setups, and assist with the post-processing of the raw measured data into the most informative temperature estimates possible. In fact, the first steps in this direction have already been taken \cite{PRXQuantum.3.040330,PhysRevX.10.011018,lvov2024thermometry}. We believe that, in the coming years, quantum thermometry will unleash its full practical potential, becoming instrumental to beat current thermometric precision standards.

\begin{figure}
    \centering
    \includegraphics[width=0.5\linewidth]{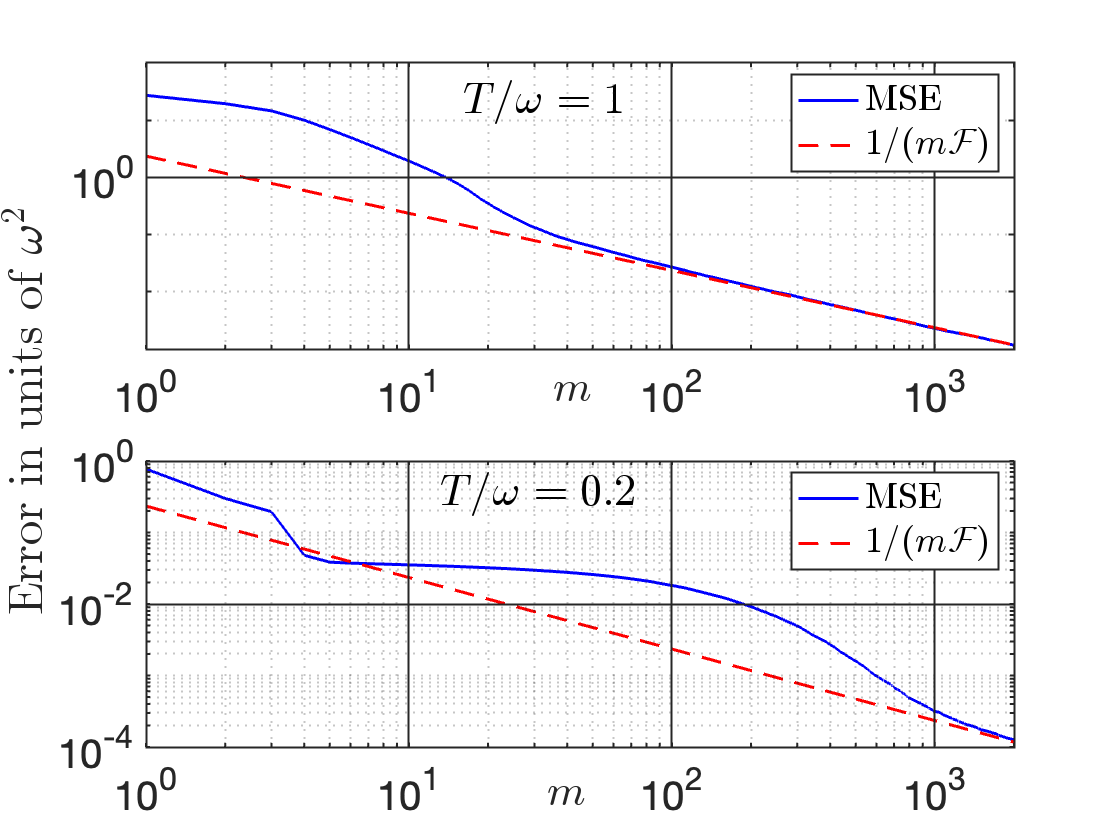}
    \caption{\textbf{A cautionary tale about QFI.} Comparison between the MSE of maximum-likelihood temperature estimates and the CRB from simulated population measurements of an equilibrium three-level system with energies $\{ 0, \omega, 2\omega\}$ ($\hbar=1$). While the  CRB must match the MSE asymptotically, in order to use the QFI as a reliable figure of merit, one needs to ensure that enough measurements have been performed. In this case, for instance, the minimum number of measurements needed for the QFI to be informative varies significantly with temperature, from just $m\approx 50$ at $T/\omega = 1$ to $m\approx 2000$ at $T/\omega = 0.2$.}
    \label{fig:MSE_vs_CRB}
\end{figure}

The central problem of quantum thermometry is to infer an unknown (cold) temperature either by post-processing outcomes of direct measurements on an equilibrium sample, or through the mediation of a probe that interacts with it. In order to assess precision and inform the optimal measurement strategies, one resorts to the framework of quantum parameter estimation. Most commonly, the Fisher information (FI) has been used as a figure of merit, as it lower-bounds the mean squared error (MSE) of the estimates, by virtue of the Cram\'er-Rao bound (CRB) \cite{Mehboudi_2019} 
\begin{equation*}
    \Delta {\tilde \vartheta}(x) \geq 1/\sqrt{\mathcal{F}(\pmb{\varrho}_T;M)}.
\end{equation*}
Here, $\pmb{\varrho}_T$ is the state of the probe (or that of the sample if this is measured directly) prior to performing the measurement $M$, with POVM elements $\{\pmb{M}_{x}\}$. ${\tilde \vartheta}(x)$ denotes the estimator which processes the measurement outcome $ x $ into an estimate of the unknown temperature $ T $. In turn, the FI may be computed from the likelihood $ p(x\,|T) = \tr(\pmb{\varrho}_T\,\pmb{M}_{x})$ as
\begin{equation*}
    \mathcal{F}(\pmb{\varrho}_T;M) = \int d x\,[\partial_T \log p(x\,|T)]^2. 
\end{equation*}
A further maximisation of the FI over all possible measurements $ M $ yields the \textit{quantum} Fisher information (QFI), i.e., $F(\pmb{\varrho}_T) = \max_{M} {\mathcal{F}(\pmb{\varrho}_T;M)}$. The QFI thus sets the ultimate scaling behaviour of the statistical uncertainty of temperature estimates, so long as the CRB is tight \cite{Mehboudi_2019}. This typically only happens asymptotically, as the number of performed measurements grows (cf. Fig.~\ref{fig:MSE_vs_CRB}). In this sense, QFI maximisation can be used as a guideline for improving the design of thermometric protocols. Indeed, much effort has been put into probe \cite{PhysRevLett.114.220405} and measurement optimisation \cite{PRXQuantum.2.020322}, as well as into the design of optimal temperature measurement protocols in finite time \cite{Mitchison2020}.

The Bayesian framework offers an alternative to the QFI-centered approach, that is particularly advantageous when estimating from finite data. Namely, one introduces a loss function $ \mathcal{L}[\tilde{\theta}(\mathsf{x}),\theta] $, penalising deviations between the temperature and the estimates $ \tilde{\theta}(\mathsf{x}) $ drawn from a vector $ \mathsf{x} $ of measurement outcomes \cite{PRXQuantum.3.040330}. Averaging gives
\begin{equation*}
    \langle\bar{\mathcal{L}}\rangle = \iint d\mathsf{x}\, d\theta\,p(\theta)\,p(\mathsf{x}\vert\theta)\,\mathcal{L}[\tilde{\theta}(\mathsf{x}),\theta], 
\end{equation*}
where, any \textit{a priori} information about the temperature has been encoded in the the prior $p(\theta)$. Here, $ p(\mathsf{x}\vert\theta) = \prod_i p(x_i\vert\theta)$. Explicit minimisation of $ \langle\bar{\mathcal{L}}\rangle $ yields an optimal estimator together with a Bayesian error bar, i.e., $ \tilde{\vartheta}(\mathsf{x}) \pm \Delta\tilde{\vartheta}(\mathsf{x})$, thus fully solving the problem. The quality of the resulting estimate will strongly depend on the particular loss function \cite{PhysRevA.104.052214}, as well as on the choice of prior $ p(\theta) $. Luckily, suitable choices and the rationale behind them have been extensively discussed in recent literature \cite{PhysRevA.104.052214,PhysRevA.105.042601,rubio2024first}. Also, explicit formulae for $ \tilde{\vartheta}(\mathsf{x}) $ and $ \Delta\tilde{\vartheta}(\mathsf{x}) $ are available. These can be applied to a broad class of \textit{location-isomorphic} estimation problems \cite{rubio2024first}.\\

\paragraph*{Current and future challenges.}
There are several obstacles preventing accurate thermometry deep in the quantum regime. First, it is known that the noise-to-signal ratio of temperature estimates diverges as $T\rightarrow 0$. In particular low-temperature thermometry on gapped systems suffers from exponential inefficiency, which may improve only to a power-law-like scaling in the gapless case \cite{Hovhannisyan2018measuring,potts2019fundamental}. Precision loss at low temperatures is also reminiscent of the Third Law of Thermodynamics, since thermal sensitivity is closely related to the heat capacity \cite{potts2019fundamental}. Hence, low temperature thermometry is inherently inaccurate. This shows, for instance, in the marked loss of precision of time-of-flight absorption imaging for thermometry on atomic clouds below the critical temperature. Specifically, the formation of a condensate masks the velocity distribution of the thermally excited fraction \cite{olf2015thermometry}. One may bypass this problem by adding a dilute minority gas of impurity probes to the sample \cite{olf2015thermometry,PhysRevX.10.011018}. After thermalisation, the (non-condensed) probe gas can be imaged separately, showcasing a clearer thermal profile down to nanokelvin temperatures \cite{olf2015thermometry}. The inherent inaccuracy of low-temperature thermometry calls for exploiting every tool at our disposal to optimise estimation protocols. For instance, in the case of impurity cold-atom thermometry, (Q)FI maximisation can inform how to harness the impurity--sample coupling for increased sensitivity, or reveal which measurements yield the best precision \cite{Mehboudi_2019}. 

Another issue may arise whenever a protocol relies on a large number of destructive measurements. This is usually the case in release--recapture thermometry, which offers a viable---albeit measurement-intensive---alternative to the default method of absorption imaging, whenever the latter becomes impractical; for instance, when the sample is made up of only one or few atoms. Luckily, the Bayesian framework can help in cases like this. Since the Bayesian formalism allows to quantify the average information gain per measurement \textit{a priori} \cite{rubio2024first}, it can be exploited to adaptively adjust the recapture time so that every shot provides the most informative outcomes (on average). This has been shown to substantially speed up convergence and enable enhanced precision in the final estimate \cite{PRXQuantum.3.040330}. Importantly, such an adaptive optimisation can be carried out on any experimental platform featuring tuneable control parameters. However, finding the relevant symmetries in order to apply the analytically tractable formalism of Ref.~\cite{rubio2024first} may prove challenging. 

\begin{figure}
    \centering
    \includegraphics[width=0.5\linewidth]{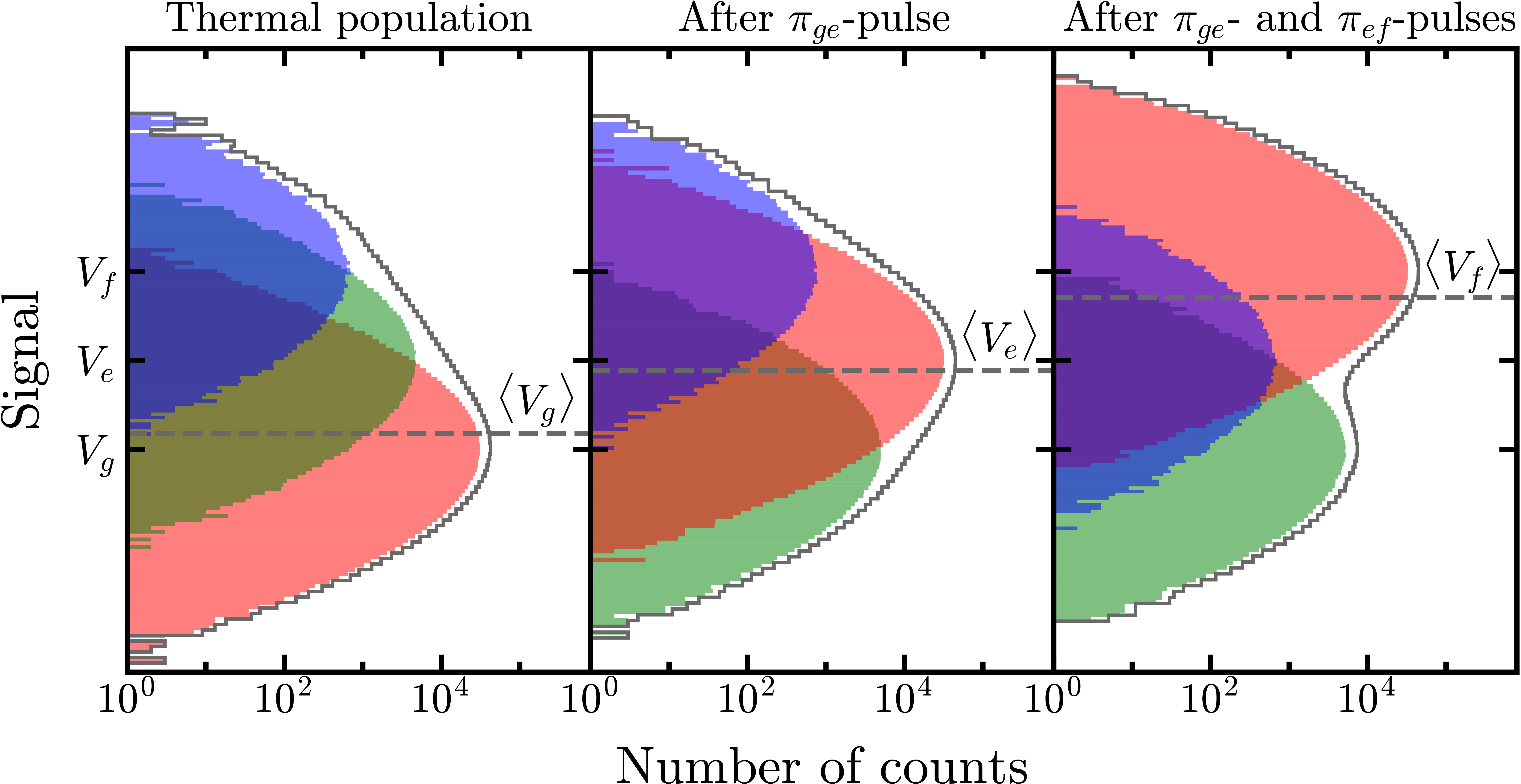}
    \caption{\textbf{Quantum thermometry on a transmon qubit.} Population distribution (numerical simulation for $10^6$ samples) for three lowest energy states of a transmon qubit with ground--excited transition frequency $\omega_{ge}/2\pi=4$ GHz and anharmonicity $\alpha/2\pi=200$ MHz. The qubit is assumed to reach a Gibbs state at the cryostat temperature of $100$ mK $(T/\omega_{ge} \approx 0.52$). Each qubit state $\ket{i}, i \in \{g,e,f\}$ corresponds to an unknown voltage $V_i$. Sequentially swapping the populations of these states by means of suitable $\pi$ pulses turns the average signal $\langle V_g\rangle = p_g V_g + p_e V_e + p_f V_f$ (left panel) into, e.g., $\langle V_e\rangle = p_e V_g + p_g V_e + p_f V_f$ (center panel) or $\langle V_f\rangle = p_e V_g + p_f V_e + p_g V_f$ (right panel), thus allowing to extract the equilibrium populations $ p_i $.}
    \label{fig:q_population}
\end{figure}

Recent progress in quantum thermometry on superconducting qubits clearly illustrates another critical issue in quantum thermometry \cite{lvov2024thermometry}. It is possible to estimate the effective temperature of such qubits, which are typically operated at dilution-refrigerator temperatures of 5--100 mK, by directly measuring their equilibrium populations (see Fig.~\ref{fig:q_population}). These temperatures can be measured very accurately without the need for qubit interrogation and hence, `quantum thermometry' is more of an environment-characterisation tool than a primary measurement goal. Interestingly, however, the measured effective qubit temperature often significantly exceeds the base cryostat temperature \cite{lvov2024thermometry}, which is likely due to uncontrolled dissipation. This observation underscores the importance of comprehensively modelling the probe and its coupling to the sample, as well as to any parasitic environments, possibly at various temperatures. This is inherently challenging, as details about the environment are rarely known; yet, deviations from the probe being in a local Gibbs state at the sample temperature can lead to an inaccurate likelihood $ p(x\,\vert T) $ and, eventually, to imprecise estimates $\tilde{\vartheta}(x)$. Even if only the desired sample couples to the probe, the finite strength of the coupling pushes the probe to a mean-force Gibbs state (cf. Sec.~\ref{sec:strongcoupling} on strong coupling), which differs from the canonical Gibbs state. Neglecting this difference can lead to systematic error in the temperature estimates. \\

\paragraph*{Broader perspective and relevance to other fields.}
Quantum thermometry borrows tools from quantum metrology, open quantum systems and quantum thermodynamics, but it has motivated new directions in these fields too. For instance, the framework of symmetry-informed parameter estimation was strongly influenced by progress on quantum thermometry \cite{rubio2024first}, as were novel adaptive approaches to quantum sensing on critical systems. Similarly, fundamental nonequilibrium energy--temperature uncertainty relations were established with the problem of temperature estimation in mind \cite{miller2018energy} (see also Sec.~\ref{sec:TURs}). In particular, these relations reflect the fact that the severe precision limitations hindering temperature estimation close to the absolute zero are thermodynamic in origin \cite{potts2019fundamental}. Furthermore, thinking operationally about how temperature is measured, can provide insight into the notion of local temperature in strongly-correlated many-body systems or non-equilibrium situations. Also, information geometry provides an alternative construction of the Bayesian thermometry framework, highlighting its close relationship with thermodynamic geometry (cf. Sec.~\ref{sec:thermogeometry}  on quantum thermodynamic geometry). 

As already mentioned, a mismatch between the expected temperature and the temperature read out from an equilibrated probe may be a smoking gun pointing to an incomplete modelling of dissipation. This can be used as a tool learn when it is necessary to refine the characterisation of the system--environment(s) interactions. Finally, we note that accurate characterisation of the initial temperature of an ultracold lattice gas is crucial for any quantum-simulation application, as is thermometry of the super-conducting qubits used for quantum-thermodynamic applications (cf. Sec.~\ref{sec:superconductors}), in which the qubits are coupled to mesoscopic heat baths. \\

\paragraph*{Concluding Remarks.}
The theory of quantum thermometry is a very active emergent field. While its primary aims have been establishing the ultimate bounds limiting the scaling of thermometric precision, it holds promise to substantially improve both accuracy and efficiency of practical temperature-measurement protocols in the lab. While, so far, only a handful of experiments have been directly informed by the information-theoretic toolbox of quantum thermometry, we expect it to become increasingly relevant in practice in the near future, helping to improve thermometric precision standards and enabling upcoming quantum-technological applications.\\

\paragraph*{Acknowledgements.}
We are grateful to Jukka P. Pekola for his support and feedback on the paper.
This research was funded in part by the Austrian Science Fund (FWF) [10.55776/PAT1969224] and by the European Research Council (Consolidator grant `Cocoquest' 101043705). MM and LAC acknowledge the University of La Laguna (ULL) the Consejería de Economía, Conocimiento y Empleo from the Canary Islands Government and the Spanish Ministry of Universities for supporting the \textit{``QSB Quantum Thermodynamics Visiting Programme''}. LAC acknowledges support from the Ministerio de Ciencia e Innovaci\'{o}n and European Union (FEDER) (PID2022-138269NB-I00) and from the Ram\'{o}n y Cajal fellowship (RYC2021-325804-I), funded by MCIN/AEI/10.13039/501100011033 and the EU “NextGenerationEU”/PRTR. DSL and SAL acknowledge the provision of facilities and technical support by Aalto University at OtaNano - Micronova Nanofabrication Centre and OtaNano - Low Temperature Laboratory. 

\clearpage

\section{Thermodynamics and quantum computing}
\label{sec:computing}
\noindent
{\it Lindsay Bassman Oftelie and Michele Campisi}

\noindent
{NEST, Istituto Nanoscienze-CNR and Scuola Normale Superiore, 56127 Pisa, Italy}\\

\paragraph*{State-of-the-art.}
Progress in both quantum computation and quantum thermodynamics (QT) has unfolded rapidly over the last few decades.  Their apparent co-development is not mere coincidence as each contributes to the advancement of the other.  The performance of a quantum computer (QC), as a quantum information processing device, is fundamentally bound by the laws of Thermodynamics as elucidated by Landauer.  Thus, a better understanding of QT, (i.e., the thermodynamics of systems and devices operating in the quantum regime) can inform best practices for the implementation and performance optimization of QCs.  At the same time, the QC, with its precise control over individual quantum constituents, offers a game-changing new platform for exploring QT. An elegant synergy therefore exists whereby results from QT may be used to improve the operation of QCs, and QCs can be used to improve our fundamental understanding of QT.

One of the core connections between QCs and QT is dissipation \cite{harrington2022engineered}. Dissipation is a central object of investigation in (quantum) thermodynamics.  Meanwhile, dissipation is both essential and toxic to quantum computation.  On the one hand, dissipation is required to cool/reset qubits to an initial fiducial state (one of DiVincenzo's criteria for QCs) \cite{buffoni2024collective, bassman2024dynamic, Gasparinetti2025}.  Furthermore, it has been shown that the more accuracy is demanded of an information processing task, the more dissipation is required \cite{chiribella2022nonequilibrium}.  On the other hand, dissipation tends to corrupt computational results.  We note that this may not universally be the case, as some work has explored how dissipation can be used for good, driving the qubits into some desired state through (possibly carefully engineered) interactions between system and environment \cite{cech2023thermodynamics}.  Generally, however, results with the highest fidelity are accomplished when dissipation during computation is minimized.  Here, QT can be exploited for minimizing, detecting, mitigating, and even correcting dissipation-induced errors by providing a characterization of the dissipation channels \cite{oftelie2024measurement, cattaneo2023quantum, Pijn2022}. The ability to characterize dissipative noise would enable the generation of better noise models, which in turn would lead to better error mitigation and correction techniques.

As the performance of QCs continues to improve, they become fruitful playgrounds for implementing QT experiments.  Initial investigations include verification of fluctuation relations \cite{Solfanelli21PRXQ2}; simulations of  energy storage \cite{razzoli2025cyclic}, work extraction \cite{elyasi2024experimental, melo2022implementation}, and thermalization \cite{yang2023simulating} in quantum systems quantum heat engines~\cite{flywheel2019} and refrigerators~\cite{Gasparinetti2025}; as well as the computation of free energy differences in quantum systems \cite{bassman2022computing}, all of which reveal the great potential and versatility of QCs to serve as experimental platforms for QT. \\

\paragraph*{Current and future challenges.}
Just as thermodynamics in the early 19th century helped drive major progress in the efficiency of machines, we believe that QT can have an analogously valuable impact on quantum machines.  QCs are currently one of the most prominent and promising quantum machines, and stand to benefit appreciably from  QT knowledge.  We see two main challenges in improving the performance of QCs. The first challenge is modeling the noise on quantum hardware. Accurate noise characterization is essential for developing noise reduction and error mitigation techniques in the near term, as well as for full-scale quantum error correction (QEC) in the future.  Central results in QT,  specifically, fluctuation relations, can be used to build accurate noise  profiles \cite{oftelie2024measurement}. 

The second challenge is the initialization of the qubits into a pure fiducial state. Algorithms presume the qubits to be initialized in their pure ground state, and QEC techniques generally rely on the injection of pure ancillary qubits throughout the computation.  In particular, the continuous supply of pure ancilla qubits demanded by QEC requires information erasure, which by the Landauer principle, necessitates heat generation. This heating will increase the error rate of the qubits, which in turn will require more rounds of error correction, generating even more heat.  Specific system parameters determine whether (1) this cycle snowballs out of control, resulting in runaway temperatures that prohibit quantum computation, or (2) heat generated by QEC is balanced by refrigeration, stabilizing qubit temperatures and bounding their error rates below the fault-tolerance threshold~\cite{bilokur2024thermodynamic}.  QT therefore sets a fundamental boundary between QCs that are and are not able to implement QEC based on their specific hardware parameters.  Here, optimal techniques for generating/resetting pure qubits must be developed that minimize heat dissipation. Meanwhile techniques based on statistical/thermodynamical tools have proven useful in achieving collective preparation of large registers of qubits with exceptional high global fidelity using dissipative quantum annealing \cite{buffoni2024collective}.

As QCs continue to improve, they can provide a test-ground for experiments seeking to better understand the fundamentals of QT. Open questions in QT that QCs could potentially be used to answer include: Can we experimentally validate the Jazrynski equality in open quantum systems? Is there an advantage to quantum batteries? Are quantum machines useful? What is the role of information in thermodynamic processes?  A major outstanding challenge in using QCs for QT experiments is the difficulty in preparing thermal states on QCs.  Thermal states are mixed states, which are inherently difficult to prepare on QCs as typically the user only has the ability to perform unitary operations.  Several techniques have been developed for approximating thermal states and require varying amounts of resources (e.g., ancilla qubits, variational optimization with a classical computer). This difficulty in preparing thermal states at a desired temperature may perhaps seem counter-intuitive since thermodynamics generally assumes thermal states to be available for free.  Perhaps some hardware-level operations could be developed that can automatically generate desired thermal states, in the spirit of how FPGA's perform specialized logic encoded in the hardware. \\

\paragraph*{Broader perspective and relevance to other fields.}
QCs offer the promise of revolutionizing a wide range of applications including optimization in financial and logistics settings, machine learning, and high-throughput simulations for drug development and new materials design.  However, in order to achieve meaningful breakthroughs, larger QCs with significantly better performance are required. The optimal design and implementation of QCs are therefore under active investigation within both academia and industry. The initial stages of development have been focused on optimizing the performance of QCs with respect to the fidelity of results.  Progress has therefore been concentrated on optimizing qubit decoherence times and gate fidelities, without much regard for the energetic resource costs.  In order for QCs to provide a meaningful advantage however, they will not only need to provide high quality results, but obtain such results within a reasonable energetic cost.  It is therefore timely to begin to ask a more nuanced question: can a quantum computational advantage be achieved at a reasonable energetic price?  QT provides tools to connect information processing tasks with physical energetic exchanges, and is therefore precisely the right framework with which to answer such a question. Some initial work has examined how to optimize the efficiency of quantum computation with, for example, the D-Wave quantum annealer \cite{smierzchalski2024efficiency}.  Maximizing the impact of QCs across all possible applications will crucially depend on optimizing their energy consumption, which in turn will depend heavily on the peculiar physics of QT.   

In general, energy consumption is minimized when dissipation is minimized.  However, as discussed above, dissipation is essential for quantum computation. QT, which characterizes dissipation in quantum systems, therefore must be employed to strike an optimal balance between performance and energy consumption of QCs.

In fact, new insights from QT could have an even broader impact on the performance of quantum technologies in general, used, for example, in sensing, metrology, and communication.  All such devices rely on components that obey the laws of quantum mechanics, and as such, their energetic footprint must be calculated based on quantum thermodynamics.  As electronic devices continue their march down to the quantum realm, it will become essential to have a solid understanding of thermodynamics at the quantum scale to evaluate and optimize their efficiency.  QCs provide an extremely useful platform for simulating the thermodynamics of quantum systems, thereby elucidating the fundamentals of QT. \\

\paragraph*{Concluding Remarks.}
While quantum computation and quantum thermodynamics each stand to benefit greatly from one another, currently, the two fields are relatively siloed, with experts in one field rarely well-versed in the other. Deliberate efforts should therefore be made to foster collaborations between researchers in both fields.  Skilled quantum computer programmers can develop specialized algorithms for performing simulations of quantum thermodynamic processes, providing much needed experimental results with which QT experts can build and extend theoretical models.  Likewise, scientists with a deep understanding of QT can aid in developing better noise models for QCs and better protocols for qubit reset, thereby improving the overall performance of QCs for their users. Perhaps the greatest impact, however, of bringing together researchers in these two fields will be determining if and how QCs can provide an advantage over their classical counterparts with reasonable energetic consumption, which we believe is one of the major unresolved questions in the field.  A small but slowly growing community of scientists at the nexus of quantum computation and quantum thermodynamics is actively capitalizing on the synergy between the two fields to push boundaries in knowledge and computational performance.  However, it is crucial that we continue to bring awareness to this specialized, interdisciplinary area to accelerate progress. \\

\paragraph*{Acknowledgements.}
LBO gratefully acknowledges funding from the European Union’s Horizon 2020 research and innovation program under the Marie Skłodowska-Curie grant agreement No 101063316.

\clearpage

\bibliography{master_references_roadmap}

@article{Saryal.2019.PRE
    , author = {Saryal, Sushant and Friedman, Hava Meira and Segal, Dvira and Agarwalla, Bijay Kumar}
    , doi = {10.1103/PhysRevE.100.042101}
    , journal = {Phys. Rev. E}
    , pages = {042101}
    , title = {{Thermodynamic uncertainty relation in thermal transport}}
    , volume = {100}
    , year = {2019}
}

@article{Brandner.2025.PRL
    , author = {Brandner, Kay and Saito, Keiji}
    , doi = {10.1103/6nww-8wcp}
    , journal = {Phys. Rev. Lett.}
    , pages = {046302}
    , title = {{Thermodynamic uncertainty relations for coherent transport}}
    , volume = {135}
    , year = {2025}
}

@article{Ptaszynski.2018.PRB
    , author = {Ptaszy\ifmmode \acute{n}\else \'{n}\fi{}ski, Krzysztof}
    , doi = {10.1103/PhysRevB.98.085425}
    , journal = {Phys. Rev. B}
    , pages = {085425}
    , title = {{Coherence-enhanced constancy of a quantum thermoelectric generator}}
    , volume = {98}
    , year = {2018}
}

@article{Kwon.2024.arxiv
    , author = {Euijoon Kwon and Jae Sung Lee}
    , journal = {arXiv:2412.04988}
    , title = {{A unified framework for classical and quantum uncertainty relations using stochastic representations}}
    , url = {https://arxiv.org/abs/2412.04988}
    , doi = {10.48550/arXiv.2412.04988}
    , year = {2024}
}

@article{Vu.2025.arxiv    , author = {Van Vu, Tan and Honma, Ryotaro and Saito, Keiji}
    , journal = {arXiv:2508.21567}
    , title = {{Universal precision limits in general open quantum systems}}
    , url = {https://arxiv.org/abs/2508.21567}
    , doi = {10.48550/arXiv.2508.21567}
    , year = {2025}
}

@article{Horowitz.2020.NP
    , author = {Horowitz, Jordan M. and Gingrich, Todd R.}
    , doi = {10.1038/s41567-019-0702-6}
    , url = {https://doi.org/10.1038/s41567-019-0702-6}
    , journal = {Nat. Phys.}
    , pages = {15--20}
    , title = {{Thermodynamic uncertainty relations constrain non-equilibrium fluctuations}}
    , volume = {16}
    , year = {2020}
}

@article{Gingrich.2016.PRL
    , author = {Gingrich, Todd R. and Horowitz, Jordan M. and Perunov, Nikolay and England, Jeremy L.}
    , doi = {10.1103/PhysRevLett.116.120601}
    , url = {https://doi.org/10.1103/PhysRevLett.116.120601}
    , journal = {Phys. Rev. Lett.}
    , pages = {120601}
    , title = {{Dissipation bounds all steady-state current fluctuations}}
    , volume = {116}
    , year = {2016}
}

@article{Terlizzi.2019.JPA
    , author = {Di Terlizzi, Ivan and Marco Baiesi}
    , doi = {10.1088/1751-8121/aaee34}
    , url = {https://doi.org/10.1088/1751-8121/aaee34}
    , journal = {J. Phys. A}
    , pages = {02LT03}
    , title = {{Kinetic uncertainty relation}}
    , volume = {52}
    , year = {2019}
}

@article{Vo.2022.JPA
    , author = {Vo, Van Tuan and Van Vu, Tan and Hasegawa, Yoshihiko}
    , doi = {10.1088/1751-8121/ac9099}
    , url = {http://doi.org/10.1088/1751-8121/ac9099}
    , journal = {J. Phys. A}
    , pages = {405004}
    , title = {{Unified thermodynamic-kinetic uncertainty relation}}
    , volume = {55}
    , year = {2022}
}

@article{Hasegawa.2019.PRL
    , author = {Hasegawa, Yoshihiko and Van Vu, Tan}
    , doi = {10.1103/PhysRevLett.123.110602}
    , url = {http://doi.org/10.1103/PhysRevLett.123.110602}
    , journal = {Phys. Rev. Lett.}
    , pages = {110602}
    , title = {{Fluctuation theorem uncertainty relation}}
    , volume = {123}
    , year = {2019}
}

@article{Timpanaro.2019.PRL
    , author = {Timpanaro, Andr\'e M. and Guarnieri, Giacomo and Goold, John and Landi, Gabriel T.}
    , doi = {10.1103/PhysRevLett.123.090604}
    , url = {http://doi.org/10.1103/PhysRevLett.123.090604}
    , journal = {Phys. Rev. Lett.}
    , pages = {090604}
    , title = {{Thermodynamic uncertainty relations from exchange fluctuation theorems}}
    , volume = {123}
    , year = {2019}
}

@article{Guarnieri.2019.PRR
    , author = {Guarnieri, Giacomo and Landi, Gabriel T. and Clark, Stephen R. and Goold, John}
    , doi = {10.1103/PhysRevResearch.1.033021}
    , url = {http://doi.org/10.1103/PhysRevResearch.1.033021}
    , journal = {Phys. Rev. Res.}
    , pages = {033021}
    , title = {{Thermodynamics of precision in quantum nonequilibrium steady states}}
    , volume = {1}
    , year = {2019}
}

@article{Hasegawa.2021.PRL
    , author = {Hasegawa, Yoshihiko}
    , doi = {10.1103/PhysRevLett.126.010602}
    , url = {http://doi.org/10.1103/PhysRevLett.126.010602}
    , journal = {Phys. Rev. Lett.}
    , pages = {010602}
    , title = {{Thermodynamic uncertainty relation for general open quantum systems}}
    , volume = {126}
    , year = {2021}
}

@article{Miller.2021.PRL.TUR
    , author = {Miller, Harry J. D. and Mohammady, M. Hamed and Perarnau-Llobet, Mart\'{\i} and Guarnieri, Giacomo}
    , doi = {10.1103/PhysRevLett.126.210603}
    , url = {http://doi.org/10.1103/PhysRevLett.126.210603}
    , journal = {Phys. Rev. Lett.}
    , pages = {210603}
    , title = {{Thermodynamic uncertainty relation in slowly driven quantum heat engines}}
    , volume = {126}
    , year = {2021}
}

@article{Hasegawa.2020.PRL
    , author = {Hasegawa, Yoshihiko}
    , doi = {10.1103/PhysRevLett.125.050601}
    , url = {http://doi.org/10.1103/PhysRevLett.125.050601}
    , journal = {Phys. Rev. Lett.}
    , pages = {050601}
    , title = {{Quantum thermodynamic uncertainty relation for continuous measurement}}
    , volume = {125}
    , year = {2020}
}

@article{Vu.2022.PRL.TUR
    , author = {Van Vu, Tan and Saito, Keiji}
    , doi = {10.1103/PhysRevLett.128.140602}
    , url = {http://doi.org/10.1103/PhysRevLett.128.140602}
    , journal = {Phys. Rev. Lett.}
    , pages = {140602}
    , title = {{Thermodynamics of precision in Markovian open quantum dynamics}}
    , volume = {128}
    , year = {2022}
}

@article{Prech.2025.PRL
    , author = {Prech, Kacper and Potts, Patrick P. and Landi, Gabriel T.}
    , doi = {10.1103/PhysRevLett.134.020401}
    , url = {http://doi.org/10.1103/PhysRevLett.134.020401}
    , journal = {Phys. Rev. Lett.}
    , pages = {020401}
    , title = {{Role of quantum coherence in kinetic uncertainty relations}}
    , volume = {134}
    , year = {2025}
}

@article{Vu.2025.PRXQ,
  title = {{Fundamental bounds on precision and response for quantum trajectory observables}},
  author = {Van Vu, Tan},
  journal = {PRX Quantum},
  volume = {6},
  issue = {1},
  pages = {010343},
  numpages = {22},
  year = {2025},
  month = {Mar},
  publisher = {American Physical Society},
  doi = {10.1103/PRXQuantum.6.010343},
  url = {https://link.aps.org/doi/10.1103/PRXQuantum.6.010343}
}

@article{Ptaszynski.2024.PRL
    , author = {Ptaszy\ifmmode \acute{n}\else \'{n}\fi{}ski, Krzysztof and Aslyamov, Timur and Esposito, Massimiliano}
    , doi = {10.1103/PhysRevLett.133.227101}
    , url = {http://doi.org/10.1103/PhysRevLett.133.227101}
    , journal = {Phys. Rev. Lett.}
    , pages = {227101}
    , title = {{Dissipation bounds precision of current response to kinetic perturbations}}
    , volume = {133}
    , year = {2024}
}

@article{Gingrich.2017.PRL
    , author = {Gingrich, Todd R. and Horowitz, Jordan M.}
    , doi = {10.1103/PhysRevLett.119.170601}
    , url = {http://doi.org/10.1103/PhysRevLett.119.170601}
    , journal = {Phys. Rev. Lett.}
    , pages = {170601}
    , title = {{Fundamental bounds on first passage time fluctuations for currents}}
    , volume = {119}
    , year = {2017}
}

@article{Majidy_NATSReview_2023,
author={Majidy, Shayan
and Braasch, William F.
and Lasek, Aleksander
and Upadhyaya, Twesh
and Kalev, Amir
and Yunger Halpern, Nicole},
title={Noncommuting conserved charges in quantum thermodynamics and beyond},
journal={Nature Reviews Physics},
year={2023},
month={Nov},
day={01},
volume={5},
number={11},
pages={689-698},
issn={2522-5820},
doi={10.1038/s42254-023-00641-9},
url={https://doi.org/10.1038/s42254-023-00641-9}
}

@article{Majidy_Entropy_2023,
  title = {Non-Abelian symmetry can increase entanglement entropy},
  author = {Majidy, Shayan and Lasek, Aleksander and Huse, David A. and Yunger Halpern, Nicole},
  journal = {Phys. Rev. B},
  volume = {107},
  issue = {4},
  pages = {045102},
  numpages = {13},
  year = {2023},
  month = {Jan},
  publisher = {American Physical Society},
  doi = {10.1103/PhysRevB.107.045102},
  url = {https://link.aps.org/doi/10.1103/PhysRevB.107.045102}
}

@article{Kranzl_NATS_2023,
  title = {Experimental Observation of Thermalization with Noncommuting Charges},
  author = {Kranzl, Florian and Lasek, Aleksander and Joshi, Manoj K. and Kalev, Amir and Blatt, Rainer and Roos, Christian F. and Yunger Halpern, Nicole},
  journal = {PRX Quantum},
  volume = {4},
  issue = {2},
  pages = {020318},
  numpages = {19},
  year = {2023},
  month = {Apr},
  publisher = {American Physical Society},
  doi = {10.1103/PRXQuantum.4.020318},
  url = {https://link.aps.org/doi/10.1103/PRXQuantum.4.020318}
}

@article{Murthy_NAETH_2023,
  title = {Non-Abelian Eigenstate Thermalization Hypothesis},
  author = {Murthy, Chaitanya and Babakhani, Arman and Iniguez, Fernando and Srednicki, Mark and Yunger Halpern, Nicole},
  journal = {Phys. Rev. Lett.},
  volume = {130},
  issue = {14},
  pages = {140402},
  numpages = {8},
  year = {2023},
  month = {Apr},
  publisher = {American Physical Society},
  doi = {10.1103/PhysRevLett.130.140402},
  url = {https://link.aps.org/doi/10.1103/PhysRevLett.130.140402}
}

@article{Znidaric_2011_Integrable,
  title = {Spin Transport in a One-Dimensional Anisotropic Heisenberg Model},
  author = {\ifmmode \check{Z}\else \v{Z}\fi{}nidari\ifmmode \check{c}\else \v{c}\fi{}, Marko},
  journal = {Phys. Rev. Lett.},
  volume = {106},
  issue = {22},
  pages = {220601},
  numpages = {4},
  year = {2011},
  month = {May},
  publisher = {American Physical Society},
  doi = {10.1103/PhysRevLett.106.220601},
  url = {https://link.aps.org/doi/10.1103/PhysRevLett.106.220601}
}

@article{Vasseur_2016_MBL,
  title = {Symmetry constraints on many-body localization},
  author = {Potter, Andrew C. and Vasseur, Romain},
  journal = {Phys. Rev. B},
  volume = {94},
  issue = {22},
  pages = {224206},
  numpages = {7},
  year = {2016},
  month = {Dec},
  publisher = {American Physical Society},
  doi = {10.1103/PhysRevB.94.224206},
  url = {https://link.aps.org/doi/10.1103/PhysRevB.94.224206}
}

@Article{Lucas_2021_Hydro,
	title={{Hydrodynamics in lattice models with continuous non-Abelian symmetries}},
	author={Paolo Glorioso and Luca V. Delacrétaz and Xiao Chen and Rahul M. Nandkishore and Andrew Lucas},
	journal={SciPost Phys.},
	volume={10},
	pages={015},
	year={2021},
	publisher={SciPost},
	doi={10.21468/SciPostPhys.10.1.015},
	url={https://scipost.org/10.21468/SciPostPhys.10.1.015}
}

@Article{NYH_2016_NATS,
author={Yunger Halpern, Nicole
and Faist, Philippe
and Oppenheim, Jonathan
and Winter, Andreas},
title={Microcanonical and resource-theoretic derivations of the thermal state of a quantum system with noncommuting charges},
journal={Nature Communications},
year={2016},
month={Jul},
day={07},
volume={7},
number={1},
pages={12051},
issn={2041-1723},
doi={10.1038/ncomms12051},
url={https://doi.org/10.1038/ncomms12051}
}

@article{Landi_2022_Entropy,
  title = {Non-Abelian Quantum Transport and Thermosqueezing Effects},
  author = {Manzano, Gonzalo and Parrondo, Juan M.R. and Landi, Gabriel T.},
  journal = {PRX Quantum},
  volume = {3},
  issue = {1},
  pages = {010304},
  numpages = {14},
  year = {2022},
  month = {Jan},
  publisher = {American Physical Society},
  doi = {10.1103/PRXQuantum.3.010304},
  url = {https://link.aps.org/doi/10.1103/PRXQuantum.3.010304}
}

@article{NYH_2024_Entropy,
  title = {Non-Abelian Transport Distinguishes Three Usually Equivalent Notions of Entropy Production},
  author = {Upadhyaya, Twesh and Braasch, William F. and Landi, Gabriel T. and Yunger Halpern, Nicole},
  journal = {PRX Quantum},
  volume = {5},
  issue = {3},
  pages = {030355},
  numpages = {25},
  year = {2024},
  month = {Sep},
  publisher = {American Physical Society},
  doi = {10.1103/PRXQuantum.5.030355},
  url = {https://link.aps.org/doi/10.1103/PRXQuantum.5.030355}
}

@article{Ott_2022_Gauge,
  title = {Thermalization of Gauge Theories from their Entanglement Spectrum},
  author = {Mueller, Niklas and Zache, Torsten V. and Ott, Robert},
  journal = {Phys. Rev. Lett.},
  volume = {129},
  issue = {1},
  pages = {011601},
  numpages = {7},
  year = {2022},
  month = {Jun},
  publisher = {American Physical Society},
  doi = {10.1103/PhysRevLett.129.011601},
  url = {https://link.aps.org/doi/10.1103/PhysRevLett.129.011601}
}

@Article{Guryanova_2016_thermo,
author={Guryanova, Yelena
and Popescu, Sandu
and Short, Anthony J.
and Silva, Ralph
and Skrzypczyk, Paul},
title={Thermodynamics of quantum systems with multiple conserved quantities},
journal={Nature Communications},
year={2016},
month={Jul},
day={07},
volume={7},
number={1},
pages={12049},
issn={2041-1723},
doi={10.1038/ncomms12049},
url={https://doi.org/10.1038/ncomms12049}
}

@article{Lostaglio_2017,
doi = {10.1088/1367-2630/aa617f},
url = {https://dx.doi.org/10.1088/1367-2630/aa617f},
year = {2017},
month = {apr},
publisher = {IOP Publishing},
volume = {19},
number = {4},
pages = {043008},
author = {Lostaglio, Matteo and Jennings, David and Rudolph, Terry},
title = {Thermodynamic resource theories, non-commutativity and maximum entropy principles},
journal = {New Journal of Physics}
}

@article{NYH_2020_BeverlandPRE,
  title = {Noncommuting conserved charges in quantum many-body thermalization},
  author = {Yunger Halpern, Nicole and Beverland, Michael E. and Kalev, Amir},
  journal = {Phys. Rev. E},
  volume = {101},
  issue = {4},
  pages = {042117},
  numpages = {13},
  year = {2020},
  month = {Apr},
  publisher = {American Physical Society},
  doi = {10.1103/PhysRevE.101.042117},
  url = {https://link.aps.org/doi/10.1103/PhysRevE.101.042117}
}

@Article{Majidy_2022_Hamiltonian,
author={Yunger Halpern, Nicole
and Majidy, Shayan},
title={How to build Hamiltonians that transport noncommuting charges in quantum thermodynamics},
journal={npj Quantum Information},
year={2022},
month={Jan},
day={27},
volume={8},
number={1},
pages={10},
issn={2056-6387},
doi={10.1038/s41534-022-00516-4},
url={https://doi.org/10.1038/s41534-022-00516-4}
}

@article{Mitchison2019, title={Quantum thermal absorption machines: refrigerators, engines and clocks}, volume={60}, ISSN={1366-5812}, url={http://dx.doi.org/10.1080/00107514.2019.1631555}, DOI={10.1080/00107514.2019.1631555}, number={2}, journal={Contemporary Physics}, publisher={Informa UK Limited}, author={Mitchison, Mark T.}, year={2019}, month=apr, pages={164–187} }

@article{Landi2024, title={Current Fluctuations in Open Quantum Systems: Bridging the Gap Between Quantum Continuous Measurements and Full Counting Statistics}, volume={5}, ISSN={2691-3399}, url={http://dx.doi.org/10.1103/PRXQuantum.5.020201}, pages={020201}, DOI={10.1103/prxquantum.5.020201}, number={2}, journal={PRX Quantum}, publisher={American Physical Society (APS)}, author={Landi, Gabriel T. and Kewming, Michael J. and Mitchison, Mark T. and Potts, Patrick P.}, year={2024}, month=apr }

@article{Hekking2013, title={Quantum Jump Approach for Work and Dissipation in a Two-Level System}, volume={111}, ISSN={1079-7114}, url={http://dx.doi.org/10.1103/PhysRevLett.111.093602}, pages={093602}, DOI={10.1103/physrevlett.111.093602}, number={9}, journal={Phys. Rev. Lett.}, publisher={American Physical Society (APS)}, author={Hekking, F. W. J. and Pekola, J. P.}, year={2013}, month=aug }

@article{Alonso2016, title={Thermodynamics of Weakly Measured Quantum Systems}, volume={116}, ISSN={1079-7114}, url={http://dx.doi.org/10.1103/PhysRevLett.116.080403}, pages={080403}, DOI={10.1103/physrevlett.116.080403}, number={8}, journal={Physical Review Letters}, publisher={American Physical Society (APS)}, author={Alonso, Jose Joaquin and Lutz, Eric and Romito, Alessandro}, year={2016}, month=feb }

@article{Rossi2020, title={Experimental Assessment of Entropy Production in a Continuously Measured Mechanical Resonator}, volume={125}, ISSN={1079-7114}, url={http://dx.doi.org/10.1103/PhysRevLett.125.080601}, DOI={10.1103/physrevlett.125.080601}, pages={080601}, number={8}, journal={Physical Review Letters}, publisher={American Physical Society (APS)}, author={Rossi, Massimiliano and Mancino, Luca and Landi, Gabriel T. and Paternostro, Mauro and Schliesser, Albert and Belenchia, Alessio}, year={2020}, month=aug }

@article{Murch2013, title={Observing single quantum trajectories of a superconducting quantum bit}, volume={502}, ISSN={1476-4687}, url={http://dx.doi.org/10.1038/nature12539}, DOI={10.1038/nature12539}, number={7470}, journal={Nature}, publisher={Springer Science and Business Media LLC}, author={Murch, K. W. and Weber, S. J. and Macklin, C. and Siddiqi, I.}, year={2013}, month=oct, pages={211–214} }

@article{Ro_nagel2016, title={A single-atom heat engine}, volume={352}, ISSN={1095-9203}, url={http://dx.doi.org/10.1126/science.aad6320}, DOI={10.1126/science.aad6320}, number={6283}, journal={Science}, publisher={American Association for the Advancement of Science (AAAS)}, author={Roßnagel, Johannes and Dawkins, Samuel T. and Tolazzi, Karl N. and Abah, Obinna and Lutz, Eric and Schmidt-Kaler, Ferdinand and Singer, Kilian}, year={2016}, month=apr, pages={325–329} }

@article{Pekola2013, title={Calorimetric measurement of work in a quantum system}, volume={15}, ISSN={1367-2630}, url={http://dx.doi.org/10.1088/1367-2630/15/11/115006}, DOI={10.1088/1367-2630/15/11/115006}, number={11}, journal={New Journal of Physics}, publisher={IOP Publishing}, author={Pekola, J P and Solinas, P and Shnirman, A and Averin, D V}, year={2013}, month=nov, pages={115006} }

@article{Manzano2022, title={Quantum thermodynamics under continuous monitoring: A general framework}, volume={4}, ISSN={2639-0213}, pages={025302}, url={http://dx.doi.org/10.1116/5.0079886}, DOI={10.1116/5.0079886}, number={2}, journal={AVS Quantum Science}, publisher={American Vacuum Society}, author={Manzano, Gonzalo and Zambrini, Roberta}, year={2022}, month=may }

@article{Landi2021,
  title = {Irreversible entropy production: From classical to quantum},
  author = {Landi, Gabriel T. and Paternostro, Mauro},
  journal = {Rev. Mod. Phys.},
  volume = {93},
  issue = {3},
  pages = {035008},
  numpages = {58},
  year = {2021},
  month = {Sep},
  publisher = {American Physical Society},
  doi = {10.1103/RevModPhys.93.035008},
  url = {https://link.aps.org/doi/10.1103/RevModPhys.93.035008}
}

@article{Esposito2009, title={Nonequilibrium fluctuations, fluctuation theorems, and counting statistics in quantum systems}, volume={81}, ISSN={1539-0756}, url={http://dx.doi.org/10.1103/RevModPhys.81.1665}, DOI={10.1103/revmodphys.81.1665}, number={4}, journal={Reviews of Modern Physics}, publisher={American Physical Society (APS)}, author={Esposito, Massimiliano and Harbola, Upendra and Mukamel, Shaul}, year={2009}, month=dec, pages={1665–1702} }

@article{Hegde2024,
  title = {Time-Resolved Stochastic Dynamics of Quantum Thermal Machines},
  author = {Hegde, Abhaya S. and Potts, Patrick P. and Landi, Gabriel T.},
  journal = {Phys. Rev. Lett.},
  volume = {134},
  issue = {15},
  pages = {150402},
  numpages = {7},
  year = {2025},
  month = {Apr},
  publisher = {American Physical Society},
  doi = {10.1103/PhysRevLett.134.150402},
  url = {https://link.aps.org/doi/10.1103/PhysRevLett.134.150402}
}

@article{Monsel2025, title={Autonomous demon exploiting heat and information at the trajectory level}, volume={111}, ISSN={2469-9969}, url={http://dx.doi.org/10.1103/PhysRevB.111.045419}, pages={045419}, DOI={10.1103/physrevb.111.045419}, number={4}, journal={Physical Review B}, publisher={American Physical Society (APS)}, author={Monsel, Juliette and Acciai, Matteo and Sánchez, Rafael and Splettstoesser, Janine}, year={2025}, month=jan }

@article{Esposito2012, title={Stochastic thermodynamics under coarse graining}, volume={85}, ISSN={1550-2376}, url={http://dx.doi.org/10.1103/PhysRevE.85.041125}, pages={041125}, DOI={10.1103/physreve.85.041125}, number={4}, journal={Physical Review E}, publisher={American Physical Society (APS)}, author={Esposito, Massimiliano}, year={2012}, month=apr }

@ARTICLE{Gyongyosi2018IEEE,
  author={Gyongyosi, Laszlo and Imre, Sandor and Nguyen, Hung Viet},
  journal={IEEE Communications Surveys \& Tutorials}, 
  title={A Survey on Quantum Channel Capacities}, 
  year={2018},
  volume={20},
  number={2},
  pages={1149-1205},
  doi={10.1109/COMST.2017.2786748}}

@article{Yadav2024arXiv,
  title={Minimal thermodynamic cost of communication},
  author={Yadav, Abhishek and Wolpert, David},
  journal={arXiv preprint arXiv:2410.14920},
url={https://arxiv.org/abs/2410.14920},
  year={2024}
}

@article{cavalcante2025,
  title = {Emergence of $X$ states in a quantum impurity model},
  author = {Cavalcante, Moallison F. and Bonan\ifmmode \mbox{\c{c}}\else \c{c}\fi{}a, Marcus V. S. and Miranda, Eduardo and Deffner, Sebastian},
  journal = {Phys. Rev. Res.},
  volume = {7},
  issue = {2},
  pages = {L022027},
  numpages = {7},
  year = {2025},
  month = {May},
  publisher = {American Physical Society},
  doi = {10.1103/PhysRevResearch.7.L022027},
  url = {https://link.aps.org/doi/10.1103/PhysRevResearch.7.L022027}
}

@Article{Lieb_1972,
author={Lieb, Elliott H.
and Robinson, Derek W.},
title={The finite group velocity of quantum spin systems},
journal={Communications in Mathematical Physics},
year={1972},
month={Sep},
day={01},
volume={28},
number={3},
pages={251-257},
issn={1432-0916},
doi={10.1007/BF01645779},
url={https://doi.org/10.1007/BF01645779}
}

@article{Rau_2009,
doi = {10.1088/1751-8113/42/41/412002},
url = {https://dx.doi.org/10.1088/1751-8113/42/41/412002},
year = {2009},
month = {sep},
publisher = {},
volume = {42},
number = {41},
pages = {412002},
author = {A R P Rau},
title = {Algebraic characterization of X-states in quantum information},
journal = {Journal of Physics A: Mathematical and Theoretical}
}

@article{Doucet24,
  title = {Classifying Two-Body Hamiltonians for Quantum Darwinism},
  author = {Doucet, Emery and Deffner, Sebastian},
  journal = {Phys. Rev. X},
  volume = {14},
  issue = {4},
  pages = {041064},
  numpages = {28},
  year = {2024},
  month = {Dec},
  publisher = {American Physical Society},
  doi = {10.1103/PhysRevX.14.041064},
  url = {https://link.aps.org/doi/10.1103/PhysRevX.14.041064}
}

@article{Zurek22,
  title = {Quantum Theory of the Classical: Einselection,  Envariance,  Quantum Darwinism and Extantons},
  volume = {24},
  ISSN = {1099-4300},
  url = {http://dx.doi.org/10.3390/e24111520},
  DOI = {10.3390/e24111520},
  number = {11},
  journal = {Entropy},
  publisher = {MDPI AG},
  author = {Zurek,  Wojciech Hubert},
  year = {2022},
  month = oct,
  pages = {1520}
}

@article{Hayden2007,
  title = {Black holes as mirrors: quantum information in random subsystems},
  volume = {2007},
  ISSN = {1029-8479},
  url = {http://dx.doi.org/10.1088/1126-6708/2007/09/120},
  DOI = {10.1088/1126-6708/2007/09/120},
  number = {09},
  journal = {Journal of High Energy Physics},
  publisher = {Springer Science and Business Media LLC},
  author = {Hayden,  Patrick and Preskill,  John},
  year = {2007},
  month = sep,
  pages = {120–120}
}

@book{leff1990maxwell,
  title={Maxwell's demon: entropy, information, computing},
  author={Leff, Harvey S and Rex, Andrew F},
  year={1990},
  publisher={Princeton University Press}
}

@article{Boyd2016PRL,
  title = {Maxwell Demon Dynamics: Deterministic Chaos, the Szilard Map, and the Intelligence of Thermodynamic Systems},
  author = {Boyd, Alexander B. and Crutchfield, James P.},
  journal = {Phys. Rev. Lett.},
  volume = {116},
  issue = {19},
  pages = {190601},
  numpages = {5},
  year = {2016},
  month = {May},
  publisher = {American Physical Society},
  doi = {10.1103/PhysRevLett.116.190601},
  url = {https://link.aps.org/doi/10.1103/PhysRevLett.116.190601}
}

@book{cover1999elements,
  title={Elements of information theory},
  author={Cover, Thomas M},
  year={1999},
  publisher={John Wiley \& Sons}
}

@article{Landauer1987APL,
    author = {Landauer, Rolf},
    title = {Energy requirements in communication},
    journal = {Applied Physics Letters},
    volume = {51},
    number = {24},
    pages = {2056-2058},
    year = {1987},
    month = {12},
    issn = {0003-6951},
    doi = {10.1063/1.98291},
    url = {https://doi.org/10.1063/1.98291}
}

@article{Touil2024EPL,
doi = {10.1209/0295-5075/ad4413},
url = {https://dx.doi.org/10.1209/0295-5075/ad4413},
year = {2024},
month = {jun},
publisher = {EDP Sciences, IOP Publishing and Società Italiana di Fisica},
volume = {146},
number = {4},
pages = {48001},
author = {Touil, Akram and Deffner, Sebastian},
title = {Information scrambling —A quantum thermodynamic perspective},
journal = {Europhysics Letters}
}

@article{Roberts2016PRL,
  title = {Lieb-Robinson Bound and the Butterfly Effect in Quantum Field Theories},
  author = {Roberts, Daniel A. and Swingle, Brian},
  journal = {Phys. Rev. Lett.},
  volume = {117},
  issue = {9},
  pages = {091602},
  numpages = {6},
  year = {2016},
  month = {Aug},
  publisher = {American Physical Society},
  doi = {10.1103/PhysRevLett.117.091602},
  url = {https://link.aps.org/doi/10.1103/PhysRevLett.117.091602}
}

@article{Verhulst2001,
    author = {Verhulst, Anne S. and Liivak, Oskar and Sherwood, Mark H. and Vieth, Hans-Martin and Chuang, Isaac L.},
    title = {Non-thermal nuclear magnetic resonance quantum computing using hyperpolarized xenon},
    journal = {Applied Physics Letters},
    volume = {79},
    number = {15},
    pages = {2480-2482},
    year = {2001},
    month = {10},
    issn = {0003-6951},
    doi = {10.1063/1.1409279},
    url = {https://doi.org/10.1063/1.1409279}
}

@article{Boykin2002,
author = {P. Oscar Boykin  and Tal Mor  and Vwani Roychowdhury  and Farrokh Vatan  and Rutger Vrijen },
title = {Algorithmic cooling and scalable NMR quantum computers},
journal = {Proceedings of the National Academy of Sciences},
volume = {99},
number = {6},
pages = {3388-3393},
year = {2002},
doi = {10.1073/pnas.241641898},
URL = {https://www.pnas.org/doi/abs/10.1073/pnas.241641898}
}

@article{Lin2024,
  title = {Thermodynamic analysis of algorithmic cooling protocols: Efficiency metrics and improved designs},
  author = {Lin, Junan and Rodr\'{\i}guez-Briones, Nayeli A. and Mart\'{\i}n-Mart\'{\i}nez, Eduardo and Laflamme, Raymond},
  journal = {Phys. Rev. A},
  volume = {110},
  issue = {2},
  pages = {022215},
  numpages = {25},
  year = {2024},
  month = {Aug},
  publisher = {American Physical Society},
  doi = {10.1103/PhysRevA.110.022215},
  url = {https://link.aps.org/doi/10.1103/PhysRevA.110.022215}
}

@article{Molpeceres2025,
  title = {Quantum algorithms for cooling: A simple case study},
  author = {Molpeceres, Daniel and Lu, Sirui and Cirac, J. Ignacio and Kraus, Barbara},
  journal = {Phys. Rev. Res.},
  volume = {7},
  issue = {3},
  pages = {033162},
  numpages = {36},
  year = {2025},
  month = {Aug},
  publisher = {American Physical Society},
  doi = {10.1103/4hx7-xnhw},
  url = {https://link.aps.org/doi/10.1103/4hx7-xnhw}
}

@article{Knill1998,
   title={Effective pure states for bulk quantum computation},
   volume={57},
   ISSN={1094-1622},
   url={http://dx.doi.org/10.1103/PhysRevA.57.3348},
   DOI={10.1103/physreva.57.3348},
   number={5},
   journal={Physical Review A},
   publisher={American Physical Society (APS)},
   author={Knill, E. and Chuang, I. and Laflamme, R.},
   year={1998}, pages={3348–3363} }

@book{SlichterLivro1990, 
    author = {Slichter, C. P.}, 
    title = {Principles of magnetic resonance}, 
    volume = {3}, 
    publisher = {Springer}, 
    address = {Berlin}, 
    year =  {1990}
}

@article{Mendonca2020,
  title = {Reservoir engineering for maximally efficient quantum engines},
  author = {Mendon\c{c}a, Taysa M. and Souza, Alexandre M. and de Assis, Rog\'erio J. and de Almeida, Norton G. and Sarthour, Roberto S. and Oliveira, Ivan S. and Villas-Boas, Celso J.},
  journal = {Phys. Rev. Research},
  volume = {2},
  issue = {4},
  pages = {043419},
  numpages = {10},
  year = {2020},
  month = {Dec},
  publisher = {American Physical Society},
  doi = {10.1103/PhysRevResearch.2.043419},
  url = {https://link.aps.org/doi/10.1103/PhysRevResearch.2.043419}
}

@article{Assis2019,
  title = {Efficiency of a Quantum Otto Heat Engine Operating under a Reservoir at Effective Negative Temperatures},
  author = {de Assis, R. J. and de Mendon\c{c}a, T. M. and Villas-Boas, C. J. and de Souza, A. M. and Sarthour, R. S. and Oliveira, I. S. and de Almeida, N. G.},
  journal = {Phys. Rev. Lett.},
  volume = {122},
  issue = {24},
  pages = {240602},
  numpages = {5},
  year = {2019},
  month = {Jun},
  publisher = {American Physical Society},
  doi = {10.1103/PhysRevLett.122.240602},
  url = {https://link.aps.org/doi/10.1103/PhysRevLett.122.240602}
}

@book{IvanLivroNMR, 
    author = {Oliveira, I. S. and Bonagamba, T. J. and Sarthour, R. S. and Freitas, J. C. C. and deAzevedo, E. R.}, 
    title = {NMR quantum information processing}, 
    publisher = {Elsevier}, 
    address = {Amsterdam}, 
    year =  {2007}
}

@article{Batalhao2014,
  title = {Experimental Reconstruction of Work Distribution and Study of Fluctuation Relations in a Closed Quantum System},
  author = {Batalh\~ao, Tiago B. and Souza, Alexandre M. and Mazzola, Laura and Auccaise, Ruben and Sarthour, Roberto S. and Oliveira, Ivan S. and Goold, John and De Chiara, Gabriele and Paternostro, Mauro and Serra, Roberto M.},
  journal = {Phys. Rev. Lett.},
  volume = {113},
  issue = {14},
  pages = {140601},
  numpages = {5},
  year = {2014},
  month = {Oct},
  publisher = {American Physical Society},
  doi = {10.1103/PhysRevLett.113.140601},
  url = {https://link.aps.org/doi/10.1103/PhysRevLett.113.140601}
}

@article{Peterson2019,
  title = {Experimental Characterization of a Spin Quantum Heat Engine},
  author = {Peterson, John P. S. and Batalh\~ao, Tiago B. and Herrera, Marcela and Souza, Alexandre M. and Sarthour, Roberto S. and Oliveira, Ivan S. and Serra, Roberto M.},
  journal = {Phys. Rev. Lett.},
  volume = {123},
  issue = {24},
  pages = {240601},
  numpages = {7},
  year = {2019},
  month = {Dec},
  publisher = {American Physical Society},
  doi = {10.1103/PhysRevLett.123.240601},
  url = {https://link.aps.org/doi/10.1103/PhysRevLett.123.240601}
}

@article{Serra2019,
author = {Kaonan Micadei and John P. S. Peterson and Alexandre M. Souza and Roberto S. Sarthour and Ivan S. Oliveira and Gabriel T. Landi and Batalh\~ao, Tiago B. and Roberto M. Serra and Eric Lutz},
title = {Reversing the direction of heat flow using quantum correlations},
journal = {Nat. Commun.},
volume = {10},
pages = {2456},
year = {2019},
doi = {https://doi.org/10.1038/s41467-019-10333-7}
}

@article{Laflamme2008,
  title = {Liquid-state nuclear magnetic resonance as a testbed for developing quantum control methods},
  author = {Ryan, C. A. and Negrevergne, C. and Laforest, M. and Knill, E. and Laflamme, R.},
  journal = {Phys. Rev. A},
  volume = {78},
  issue = {1},
  pages = {012328},
  numpages = {14},
  year = {2008},
  month = {Jul},
  publisher = {American Physical Society},
  doi = {10.1103/PhysRevA.78.012328},
  url = {https://link.aps.org/doi/10.1103/PhysRevA.78.012328}
}

@article{Suter2010,
  title = {NMR Quantum Simulation of Localization Effects Induced by Decoherence},
  author = {\'Alvarez, Gonzalo A. and Suter, Dieter},
  journal = {Phys. Rev. Lett.},
  volume = {104},
  issue = {23},
  pages = {230403},
  numpages = {4},
  year = {2010},
  month = {Jun},
  publisher = {American Physical Society},
  doi = {10.1103/PhysRevLett.104.230403},
  url = {https://link.aps.org/doi/10.1103/PhysRevLett.104.230403}
}

@article{Barret2018,
  title = {Observation of Discrete-Time-Crystal Signatures in an Ordered Dipolar Many-Body System},
  author = {Rovny, Jared and Blum, Robert L. and Barrett, Sean E.},
  journal = {Phys. Rev. Lett.},
  volume = {120},
  issue = {18},
  pages = {180603},
  numpages = {5},
  year = {2018},
  month = {May},
  publisher = {American Physical Society},
  doi = {10.1103/PhysRevLett.120.180603},
  url = {https://link.aps.org/doi/10.1103/PhysRevLett.120.180603}
}

@article{Suter2021,
  title = {Floquet Prethermalization with Lifetime Exceeding 90 s in a Bulk Hyperpolarized Solid},
  author = {Beatrez, William and Janes, Otto and Akkiraju, Amala and Pillai, Arjun and Oddo, Alexander and Reshetikhin, Paul and Druga, Emanuel and McAllister, Maxwell and Elo, Mark and Gilbert, Benjamin and Suter, Dieter and Ajoy, Ashok},
  journal = {Phys. Rev. Lett.},
  volume = {127},
  issue = {17},
  pages = {170603},
  numpages = {8},
  year = {2021},
  month = {Oct},
  publisher = {American Physical Society},
  doi = {10.1103/PhysRevLett.127.170603},
  url = {https://link.aps.org/doi/10.1103/PhysRevLett.127.170603}
}

@article{Beatrez2023,
   title={Critical prethermal discrete time crystal created by two-frequency driving},
   volume={19},
   ISSN={1745-2481},
   url={http://dx.doi.org/10.1038/s41567-022-01891-7},
   DOI={10.1038/s41567-022-01891-7},
   number={3},
   journal={Nature Physics},
   publisher={Springer Science and Business Media LLC},
   author={Beatrez, William and Fleckenstein, Christoph and Pillai, Arjun and de Leon Sanchez, Erica and Akkiraju, Amala and Diaz Alcala, Jesus and Conti, Sophie and Reshetikhin, Paul and Druga, Emanuel and Bukov, Marin and Ajoy, Ashok},
   year={2023},
   month={jan},
   pages={407–413}
}

@article{Hernandez2022,
   title={Autonomous Dissipative Maxwell’s Demon in a Diamond Spin Qutrit},
   volume={3},
   ISSN={2691-3399},
   url={http://dx.doi.org/10.1103/PRXQuantum.3.020329},
   DOI={10.1103/prxquantum.3.020329},
   number={2},
pages={020329},
   journal={PRX Quantum},
   publisher={American Physical Society (APS)},
   author={Hernández-Gómez, S. and Gherardini, S. and Staudenmaier, N. and Poggiali, F. and Campisi, M. and Trombettoni, A. and Cataliotti, F.S. and Cappellaro, P. and Fabbri, N.},
   year={2022},
   month=may }

@book{livro_DeffnerSteve,
  title={Quantum Thermodynamics: An introduction to the thermodynamics of quantum information},
  author={Sebastian Deffner and Steve Campbell},
  year={2019},
  address = {Bristol},
  publisher={IOP Publishing}
}

@article{Jones2011,
title = {Quantum computing with NMR},
journal = {Progress in Nuclear Magnetic Resonance Spectroscopy},
volume = {59},
number = {2},
pages = {91-120},
year = {2011},
issn = {0079-6565},
doi = {https://doi.org/10.1016/j.pnmrs.2010.11.001},
url = {https://www.sciencedirect.com/science/article/pii/S0079656510001111},
author = {Jonathan A. Jones}
}

@article{Monsel:20,
  title = {The Energetic Cost of Work Extraction},
  author = {Monsel, Juliette and Fellous-Asiani, Marco and Huard, Benjamin and Auff\`eves, Alexia},
  journal = {Phys. Rev. Lett.},
  volume = {124},
  issue = {13},
  pages = {130601},
  numpages = {6},
  year = {2020},
  month = {Mar},
  publisher = {American Physical Society},
  doi = {10.1103/PhysRevLett.124.130601},
  url = {https://link.aps.org/doi/10.1103/PhysRevLett.124.130601}
}

@article{Dominik:23,
  title = {Work Extraction from Unknown Quantum Sources},
  author = {\v{S}afr\'anek, Dominik and Rosa, Dario and Binder, Felix C.},
  journal = {Phys. Rev. Lett.},
  volume = {130},
  issue = {21},
  pages = {210401},
  numpages = {9},
  year = {2023},
  month = {May},
  publisher = {American Physical Society},
  doi = {10.1103/PhysRevLett.130.210401},
  url = {https://link.aps.org/doi/10.1103/PhysRevLett.130.210401}
}

@article{Zhibo:24,
  title = {Experimental Investigation of Coherent Ergotropy in a Single Spin System},
  author = {Niu, Zhibo and Wu, Yang and Wang, Yunhan and Rong, Xing and Du, Jiangfeng},
  journal = {Phys. Rev. Lett.},
  volume = {133},
  issue = {18},
  pages = {180401},
  numpages = {6},
  year = {2024},
  month = {Oct},
  publisher = {American Physical Society},
  doi = {10.1103/PhysRevLett.133.180401},
  url = {https://link.aps.org/doi/10.1103/PhysRevLett.133.180401}
}

@article{Francica:20,
  title = {Quantum Coherence and Ergotropy},
  author = {Francica, G. and Binder, F. C. and Guarnieri, G. and Mitchison, M. T. and Goold, J. and Plastina, F.},
  journal = {Phys. Rev. Lett.},
  volume = {125},
  issue = {18},
  pages = {180603},
  numpages = {8},
  year = {2020},
  month = {Oct},
  publisher = {American Physical Society},
  doi = {10.1103/PhysRevLett.125.180603},
  url = {https://link.aps.org/doi/10.1103/PhysRevLett.125.180603}
}

@ARTICLE{Kim:22,
	title = {Quantum Charging Advantage Cannot Be Extensive without Global Operations},
	author = {Gyhm, Ju-Yeon and \ifmmode \check{S}\else \v{S}\fi{}afr\'anek, Dominik and Rosa, Dario},
	journal = {Phys. Rev. Lett.},
	volume = {128},
	issue = {14},
	pages = {140501},
	numpages = {6},
	year = {2022},
	month = {Apr},
	publisher = {American Physical Society},
	doi = {10.1103/PhysRevLett.128.140501},
	url = {https://link.aps.org/doi/10.1103/PhysRevLett.128.140501}
}

@article{Moraes_2024,
	doi = {10.1088/2058-9565/ad71ed},
	url = {https://dx.doi.org/10.1088/2058-9565/ad71ed},
	year = {2024},
	month = {aug},
	publisher = {IOP Publishing},
	volume = {9},
	number = {4},
	pages = {045033},
	author = {de Moraes, L F C and Duriez, Alan C and Saguia, A and Santos, Alan C and Sarandy, M S},
	title = {Quantum battery supercharging via counter-diabatic dynamics},
	journal = {Quantum Science and Technology}
}

@ARTICLE{ZhangKe:22,
	title = {Synthesizing Five-Body Interaction in a Superconducting Quantum Circuit},
	author = {Zhang, Ke and Li, Hekang and Zhang, Pengfei and Yuan, Jiale and Chen, Jinyan and Ren, Wenhui and Wang, Zhen and Song, Chao and Wang, Da-Wei and Wang, H. and Zhu, Shiyao and Agarwal, Girish S. and Scully, Marlan O.},
	journal = {Phys. Rev. Lett.},
	volume = {128},
	issue = {19},
	pages = {190502},
	numpages = {6},
	year = {2022},
	month = {May},
	publisher = {American Physical Society},
	doi = {10.1103/PhysRevLett.128.190502},
	url = {https://link.aps.org/doi/10.1103/PhysRevLett.128.190502}
}

@article{Joshi:22,
	title = {Experimental investigation of a quantum battery using star-topology NMR spin systems},
	author = {Joshi, Jitendra and Mahesh, T. S.},
	journal = {Phys. Rev. A},
	volume = {106},
	issue = {4},
	pages = {042601},
	numpages = {8},
	year = {2022},
	month = {Oct},
	publisher = {American Physical Society},
	doi = {10.1103/PhysRevA.106.042601},
	url = {https://link.aps.org/doi/10.1103/PhysRevA.106.042601}
}

@ARTICLE{Hu:22a,
	doi = {10.1088/2058-9565/ac8444},
	url = {https://doi.org/10.1088/2058-9565/ac8444},
	year = 2022,
	month = {aug},
	publisher = {{IOP} Publishing},
	volume = {7},
	number = {4},
	pages = {045018},
	author = {Chang-Kang Hu and Jiawei Qiu and Paulo J P Souza and Jiahao Yuan and Yuxuan Zhou and Libo Zhang and Ji Chu and Xianchuang Pan and Ling Hu and Jian Li and Yuan Xu and Youpeng Zhong and Song Liu and Fei Yan and Dian Tan and R Bachelard and C J Villas-Boas and Alan C Santos and Dapeng Yu},
	title = {Optimal charging of a superconducting quantum battery},
	journal = {Quantum Science and Technology}
}

@article{Niu:23,
	title={Low-loss interconnects for modular superconducting quantum processors},
	author = {{Niu}, Jingjing and {Zhang}, Libo and {Liu}, Yang and {Qiu}, Jiawei and {Huang}, Wenhui and {Huang}, Jiaxiang and {Jia}, Hao and {Liu}, Jiawei and {Tao}, Ziyu and {Wei}, Weiwei and {Zhou}, Yuxuan and {Zou}, Wanjing and {Chen}, Yuanzhen and {Deng}, Xiaowei and {Deng}, Xiuhao and {Hu}, Changkang and {Hu}, Ling and {Li}, Jian and {Tan}, Dian and {Xu}, Yuan and {Yan}, Fei and {Yan}, Tongxing and {Liu}, Song and {Zhong}, Youpeng and {Cleland}, Andrew N. and {Yu}, Dapeng},
	journal={Nature Electronics},
	pages={235–241},
	url = {https://doi.org/10.1038/s41928-023-00925-z},
	doi = {10.1038/s41928-023-00925-z},
	year={2023},
	publisher={Nature Publishing Group UK London}
}

@article{Rossini:20,
	title = {Quantum Advantage in the Charging Process of Sachdev-Ye-Kitaev Batteries},
	author = {Rossini, Davide and Andolina, Gian Marcello and Rosa, Dario and Carrega, Matteo and Polini, Marco},
	journal = {Phys. Rev. Lett.},
	volume = {125},
	issue = {23},
	pages = {236402},
	numpages = {6},
	year = {2020},
	month = {Dec},
	publisher = {American Physical Society},
	doi = {10.1103/PhysRevLett.125.236402},
	url = {https://link.aps.org/doi/10.1103/PhysRevLett.125.236402}
}

@article{bu2025tantalum,
	title={Tantalum airbridges for scalable superconducting quantum processors},
 author = {{Bu}, Kunliang and {Huai}, Sainan and {Zhang}, Zhenxing and {Li}, Dengfeng and {Li}, Yuan and {Hu}, Jingjing and {Yang}, Xiaopei and {Dai}, Maochun and {Cai}, Tianqi and {Zheng}, Yi-Cong and {Zhang}, Shengyu},
	journal={npj Quantum Information},
	volume={11},
	number={1},
	pages={17},
	year={2025},
	doi = {https://doi.org/10.1038/s41534-025-00972-8},
	publisher={Nature Publishing Group UK London}
}

@article{Catalano:24,
	title = {Frustrating Quantum Batteries},
	author = {Catalano, A.G. and Giampaolo, S.M. and Morsch, O. and Giovannetti, V. and Franchini, F.},
	journal = {PRX Quantum},
	volume = {5},
	issue = {3},
	pages = {030319},
	numpages = {15},
	year = {2024},
	month = {Jul},
	publisher = {American Physical Society},
	doi = {10.1103/PRXQuantum.5.030319},
	url = {https://link.aps.org/doi/10.1103/PRXQuantum.5.030319}
}

@article{Lu:24,
         title = {Topological Quantum Batteries},
         author = {Zhi-Guang Lu and Guoqing Tian and Xin-You Lü and Cheng Shang },
         year = {2024},
         eprint = {2405.03675v4},
         archivePrefix = {arXiv},
         primaryClass ={quant-ph}
        }

@article{campaioli2017enhancing,
	title = {Enhancing the Charging Power of Quantum Batteries},
	author = {Campaioli, Francesco and Pollock, Felix A. and Binder, Felix C. and C\'eleri, Lucas and Goold, John and Vinjanampathy, Sai and Modi, Kavan},
	journal = {Phys. Rev. Lett.},
	volume = {118},
	issue = {15},
	pages = {150601},
	numpages = {6},
	year = {2017},
	month = {Apr},
	publisher = {American Physical Society},
	doi = {10.1103/PhysRevLett.118.150601},
	url = {https://link.aps.org/doi/10.1103/PhysRevLett.118.150601}
}

@article{mondal2022periodically,
	title = {Periodically driven many-body quantum battery},
	author = {Mondal, Saikat and Bhattacharjee, Sourav},
	journal = {Phys. Rev. E},
	volume = {105},
	issue = {4},
	pages = {044125},
	numpages = {8},
	year = {2022},
	month = {Apr},
	publisher = {American Physical Society},
	doi = {10.1103/PhysRevE.105.044125},
	url = {https://link.aps.org/doi/10.1103/PhysRevE.105.044125}
}

@article{puri2024floquet,
         title = {Floquet driven long-range interactions induce super-extensive scaling in quantum battery},
         author = {Stavya Puri and Tanoy Kanti Konar and Leela Ganesh Chandra Lakkaraju and Aditi Sen De },
         year = {2024},
         eprint = {2412.00921v1},
         archivePrefix = {arXiv},
         primaryClass ={quant-ph}
        }

@article{andolina2024genuine,
         title = {Genuine quantum advantage in non-linear bosonic quantum batteries},
         author = {Gian Marcello Andolina and Vittoria Stanzione and Vittorio Giovannetti and Marco Polini },
         year = {2024},
         eprint = {2409.08627v1},
         archivePrefix = {arXiv},
         primaryClass ={quant-ph}
        }

@article{Bhandari2020,
  title = {Geometric properties of adiabatic quantum thermal machines},
  volume = {102},
  ISSN = {2469-9969},
  url = {http://dx.doi.org/10.1103/PhysRevB.102.155407},
  DOI = {10.1103/physrevb.102.155407},
  number = {15},
  pages = {155407},
  journal = {Physical Review B},
  publisher = {American Physical Society (APS)},
  author = {Bhandari,  Bibek and Alonso,  Pablo Terrén and Taddei,  Fabio and von Oppen,  Felix and Fazio,  Rosario and Arrachea,  Liliana},
  year = {2020},
  month = oct 
}

@article{PRXVanVu2023,
  title = {Thermodynamic Unification of Optimal Transport: Thermodynamic Uncertainty Relation, Minimum Dissipation, and Thermodynamic Speed Limits},
  author = {Van Vu, Tan and Saito, Keiji},
  journal = {Phys. Rev. X},
  volume = {13},
  issue = {1},
  pages = {011013},
  numpages = {45},
  year = {2023},
  month = {Feb},
  publisher = {American Physical Society},
  doi = {10.1103/PhysRevX.13.011013},
  url = {https://link.aps.org/doi/10.1103/PhysRevX.13.011013}
}

@article{Bettmann2025,
  title = {Information geometry approach to quantum stochastic thermodynamics},
  volume = {111},
  ISSN = {2470-0053},
  pages = {014133},
  url = {http://dx.doi.org/10.1103/PhysRevE.111.014133},
  DOI = {10.1103/physreve.111.014133},
  number = {1},
  journal = {Physical Review E},
  publisher = {American Physical Society (APS)},
  author = {Bettmann,  Laetitia P. and Goold,  John},
  year = {2025},
  month = jan 
}

@article{Scandi2025,
  title = {Quantum Fisher information and its dynamical nature},
  volume = {88},
  ISSN = {1361-6633},
  url = {http://dx.doi.org/10.1088/1361-6633/ade453},
  DOI = {10.1088/1361-6633/ade453},
  number = {7},
  journal = {Reports on Progress in Physics},
  publisher = {IOP Publishing},
  author = {Scandi,  Matteo and Abiuso,  Paolo and Surace,  Jacopo and De Santis,  Dario},
  year = {2025},
  month = jul,
  pages = {076001}
}

@article{rolandi2023collective,
  title = {Collective Advantages in Finite-Time Thermodynamics},
  volume = {131},
  ISSN = {1079-7114},
  pages = {210401},
  url = {http://dx.doi.org/10.1103/PhysRevLett.131.210401},
  DOI = {10.1103/physrevlett.131.210401},
  number = {21},
  journal = {Physical Review Letters},
  publisher = {American Physical Society (APS)},
  author = {Rolandi,  Alberto and Abiuso,  Paolo and Perarnau-Llobet,  Martí},
  year = {2023},
  month = nov 
}

@article{salamon1983thermodynamic,
  title = {Thermodynamic Length and Dissipated Availability},
  author = {Salamon, Peter and Berry, R. Stephen},
  journal = {Phys. Rev. Lett.},
  volume = {51},
  issue = {13},
  pages = {1127--1130},
  numpages = {0},
  year = {1983},
  month = {Sep},
  publisher = {American Physical Society},
  doi = {10.1103/PhysRevLett.51.1127},
  url = {https://link.aps.org/doi/10.1103/PhysRevLett.51.1127}
}

@article{carollo_2020,
title     = "Geometry of quantum phase transitions",
author    = "Carollo, Angelo and Valenti, Davide and Spagnolo, Bernardo",
journal   = "Phys. Rep.",
publisher = "Elsevier BV",
volume    =  838,
pages     = "1--72",
month     =  jan,
doi= {10.1016/j.physrep.2019.11.002},
url={https://doi.org/10.1016/j.physrep.2019.11.002},
year      =  2020
}

@article{sivak2012thermodynamic,
  title = {Thermodynamic Metrics and Optimal Paths},
  author = {Sivak, David A. and Crooks, Gavin E.},
  journal = {Phys. Rev. Lett.},
  volume = {108},
  issue = {19},
  pages = {190602},
  numpages = {5},
  year = {2012},
  month = {May},
  publisher = {American Physical Society},
  doi = {10.1103/PhysRevLett.108.190602},
  url = {https://link.aps.org/doi/10.1103/PhysRevLett.108.190602}
}

@article{eglinton2023thermodynamic,
doi = {10.1088/1367-2630/acc966},
url = {https://dx.doi.org/10.1088/1367-2630/acc966},
year = {2023},
month = {apr},
publisher = {IOP Publishing},
volume = {25},
number = {4},
pages = {043014},
author = {Eglinton, Joshua and Pyhäranta, Tuomas and Saito, Keiji and Brandner, Kay},
title = {Thermodynamic geometry of ideal quantum gases: a general framework and a geometric picture of BEC-enhanced heat engines},
journal = {New Journal of Physics}
}

@article{rolandi2023finite,
  doi = {10.22331/q-2023-11-03-1161},
  url = {https://doi.org/10.22331/q-2023-11-03-1161},
  title = {Finite-time {L}andauer principle beyond weak coupling},
  author = {Rolandi, Alberto and Perarnau-Llobet, Mart{\'{i}}},
  journal = {{Quantum}},
  issn = {2521-327X},
  publisher = {{Verein zur F{\"{o}}rderung des Open Access Publizierens in den Quantenwissenschaften}},
  volume = {7},
  pages = {1161},
  month = nov,
  year = {2023}
}

@article{mayer2020nonequilibrium,
  title = {Nonequilibrium thermodynamics and optimal cooling of a dilute atomic gas},
  author = {Mayer, Daniel and Schmidt, Felix and Haupt, Steve and Bouton, Quentin and Adam, Daniel and Lausch, Tobias and Lutz, Eric and Widera, Artur},
  journal = {Phys. Rev. Res.},
  volume = {2},
  issue = {2},
  pages = {023245},
  numpages = {13},
  year = {2020},
  month = {May},
  publisher = {American Physical Society},
  doi = {10.1103/PhysRevResearch.2.023245},
  url = {https://link.aps.org/doi/10.1103/PhysRevResearch.2.023245}
}

@article{miller2020geometry,
  title = {Geometry of Work Fluctuations versus Efficiency in Microscopic Thermal Machines},
  author = {Miller, Harry J. D. and Mehboudi, Mohammad},
  journal = {Phys. Rev. Lett.},
  volume = {125},
  issue = {26},
  pages = {260602},
  numpages = {6},
  year = {2020},
  month = {Dec},
  publisher = {American Physical Society},
  doi = {10.1103/PhysRevLett.125.260602},
  url = {https://link.aps.org/doi/10.1103/PhysRevLett.125.260602}
}

@article{abiuso2020geometric,
AUTHOR = {Abiuso, Paolo and Miller, Harry J. D. and Perarnau-Llobet, Martí and Scandi, Matteo},
TITLE = {Geometric Optimisation of Quantum Thermodynamic Processes},
JOURNAL = {Entropy},
VOLUME = {22},
YEAR = {2020},
NUMBER = {10},
pages = {1076},
URL = {https://www.mdpi.com/1099-4300/22/10/1076},
PubMedID = {33286845},
ISSN = {1099-4300},
DOI = {10.3390/e22101076}
}

@article{brandner2020thermodynamic,
  title = {Thermodynamic Geometry of Microscopic Heat Engines},
  author = {Brandner, Kay and Saito, Keiji},
  journal = {Phys. Rev. Lett.},
  volume = {124},
  issue = {4},
  pages = {040602},
  numpages = {7},
  year = {2020},
  month = {Jan},
  publisher = {American Physical Society},
  doi = {10.1103/PhysRevLett.124.040602},
  url = {https://link.aps.org/doi/10.1103/PhysRevLett.124.040602}
}

@article{Aifer2024,
  title = {Thermodynamic linear algebra},
  volume = {1},
  ISSN = {3004-8672},
  url = {http://dx.doi.org/10.1038/s44335-024-00014-0},
  DOI = {10.1038/s44335-024-00014-0},
  pages = {13},
  journal = {npj Unconventional Computing},
  publisher = {Springer Science and Business Media LLC},
  author = {Aifer,  Maxwell and Donatella,  Kaelan and Gordon,  Max Hunter and Duffield,  Samuel and Ahle,  Thomas and Simpson,  Daniel and Crooks,  Gavin and Coles,  Patrick J.},
  year = {2024},
  month = nov 
}

@article{blaber2023optimal,
doi = {10.1088/2399-6528/acbf04},
url = {https://dx.doi.org/10.1088/2399-6528/acbf04},
year = {2023},
month = {mar},
publisher = {IOP Publishing},
volume = {7},
number = {3},
pages = {033001},
author = {Blaber, Steven and Sivak, David A},
title = {Optimal control in stochastic thermodynamics},
journal = {Journal of Physics Communications}
}

@article{lacerda2025information,
  title = {Information geometry of transitions between quantum nonequilibrium steady states},
  volume = {112},
  ISSN = {2470-0053},
  url = {http://dx.doi.org/10.1103/9f6l-d766},
  DOI = {10.1103/9f6l-d766},
  number = {2},
  pages = {L022101},
  journal = {Physical Review E},
  publisher = {American Physical Society (APS)},
  author = {Lacerda,  Artur M. and Bettmann,  Laetitia P. and Goold,  John},
  year = {2025},
  month = aug 
}

@article{erdman2022identifying,
  title = {Identifying optimal cycles in quantum thermal machines with reinforcement-learning},
  volume = {8},
  ISSN = {2056-6387},
  url = {http://dx.doi.org/10.1038/s41534-021-00512-0},
  DOI = {10.1038/s41534-021-00512-0},
  pages = {1},
  journal = {npj Quantum Information},
  publisher = {Springer Science and Business Media LLC},
  author = {Erdman,  Paolo A. and Noé,  Frank},
  year = {2022},
  month = jan 
}

@article{van2021geometrical,
  title = {Geometrical Bounds of the Irreversibility in Markovian Systems},
  author = {Van Vu, Tan and Hasegawa, Yoshihiko},
  journal = {Phys. Rev. Lett.},
  volume = {126},
  issue = {1},
  pages = {010601},
  numpages = {7},
  year = {2021},
  month = {Jan},
  publisher = {American Physical Society},
  doi = {10.1103/PhysRevLett.126.010601},
  url = {https://link.aps.org/doi/10.1103/PhysRevLett.126.010601}
}

@article{Huang2020shadows,
  title={Predicting many properties of a quantum system from very few measurements},
  author={Huang, Hsin-Yuan and Kueng, Richard and Preskill, John},
  journal={Nature Physics},
  volume={16},
  number={10},
  pages={1050--1057},
  year={2020},
  publisher={Nature Publishing Group UK London},
doi={https://doi.org/10.1038/s41567-020-0932-7}
}

@article{Zawadzki2022,
  title={Approximating quantum thermodynamic properties using DFT},
  author={Zawadzki, Krissia and Skelt, AH and D’Amico, Irene},
  journal={Journal of Physics: Condensed Matter},
  volume={34},
  number={27},
  pages={274002},
  year={2022},
  publisher={IOP Publishing},
  url={https://doi.org/10.1088/1361-648X/ac6648},
  doi={10.1088/1361-648X/ac6648}
}

@article{Ludmila2025,
      title={Quantum Time Crystal Clock and its Performance}, 
      author={Ludmila Viotti and Marcus Huber and Rosario Fazio and Gonzalo Manzano},
      year={2025},
      eprint={2505.08276},
      archivePrefix={arXiv},
      primaryClass={quant-ph},
      url={https://arxiv.org/abs/2505.08276}, 
}

@article{campisi16the,
author={Campisi, Michele
and Fazio, Rosario},
title={The power of a critical heat engine},
journal={Nature Communications},
year={2016},
month={Jun},
day={20},
volume={7},
number={1},
pages={11895},
issn={2041-1723},
doi={10.1038/ncomms11895},
url={https://doi.org/10.1038/ncomms11895}
}

@article{niedenzu18cooperative,
doi = {10.1088/1367-2630/aaed55},
url = {https://dx.doi.org/10.1088/1367-2630/aaed55},
year = {2018},
month = {nov},
publisher = {IOP Publishing},
volume = {20},
number = {11},
pages = {113038},
author = {Niedenzu, Wolfgang and Kurizki, Gershon},
title = {Cooperative many-body enhancement of quantum thermal machine power},
journal = {New Journal of Physics}
}

@article{altman21quantum,
  title = {Quantum Simulators: Architectures and Opportunities},
  author = {Altman, Ehud and Brown, Kenneth R. and Carleo, Giuseppe and Carr, Lincoln D. and Demler, Eugene and Chin, Cheng and DeMarco, Brian and Economou, Sophia E. and Eriksson, Mark A. and Fu, Kai-Mei C. and Greiner, Markus and Hazzard, Kaden R.A. and Hulet, Randall G. and Koll\'ar, Alicia J. and Lev, Benjamin L. and Lukin, Mikhail D. and Ma, Ruichao and Mi, Xiao and Misra, Shashank and Monroe, Christopher and Murch, Kater and Nazario, Zaira and Ni, Kang-Kuen and Potter, Andrew C. and Roushan, Pedram and Saffman, Mark and Schleier-Smith, Monika and Siddiqi, Irfan and Simmonds, Raymond and Singh, Meenakshi and Spielman, I.B. and Temme, Kristan and Weiss, David S. and Vu\ifmmode \check{c}\else \v{c}\fi{}kovi\ifmmode \acute{c}\else \'{c}\fi{}, Jelena and Vuleti\ifmmode \acute{c}\else \'{c}\fi{}, Vladan and Ye, Jun and Zwierlein, Martin},
  journal = {PRX Quantum},
  volume = {2},
  issue = {1},
  pages = {017003},
  numpages = {19},
  year = {2021},
  month = {Feb},
  publisher = {American Physical Society},
  doi = {10.1103/PRXQuantum.2.017003},
  url = {https://link.aps.org/doi/10.1103/PRXQuantum.2.017003}
}

@article{beau16scaling,
AUTHOR = {Beau, Mathieu and Jaramillo, Juan and Del Campo, Adolfo},
TITLE = {Scaling-Up Quantum Heat Engines Efficiently via Shortcuts to Adiabaticity},
JOURNAL = {Entropy},
VOLUME = {18},
YEAR = {2016},
NUMBER = {5},
pages = {168},
URL = {https://www.mdpi.com/1099-4300/18/5/168},
ISSN = {1099-4300},
DOI = {10.3390/e18050168}
}

@Inbook{goold18the,
author="Goold, John
and Plastina, Francesco
and Gambassi, Andrea
and Silva, Alessandro",
editor="Binder, Felix
and Correa, Luis A.
and Gogolin, Christian
and Anders, Janet
and Adesso, Gerardo",
title="The Role of Quantum Work Statistics in Many-Body Physics",
bookTitle="Thermodynamics in the Quantum Regime: Fundamental Aspects and New Directions",
year="2018",
publisher="Springer International Publishing",
address="Cham",
pages="317--336",
isbn="978-3-319-99046-0",
doi="10.1007/978-3-319-99046-0_13",
url="https://doi.org/10.1007/978-3-319-99046-0_13"
}

@article{watanabe20quantum,
  title = {Quantum Statistical Enhancement of the Collective Performance of Multiple Bosonic Engines},
  author = {Watanabe, Gentaro and Venkatesh, B. Prasanna and Talkner, Peter and Hwang, Myung-Joong and del Campo, Adolfo},
  journal = {Phys. Rev. Lett.},
  volume = {124},
  issue = {21},
  pages = {210603},
  numpages = {6},
  year = {2020},
  month = {May},
  publisher = {American Physical Society},
  doi = {10.1103/PhysRevLett.124.210603},
  url = {https://link.aps.org/doi/10.1103/PhysRevLett.124.210603}
}

@article{quach22superabsorption,
author = {James Q. Quach  and Kirsty E. McGhee  and Lucia Ganzer  and Dominic M. Rouse  and Brendon W. Lovett  and Erik M. Gauger  and Jonathan Keeling  and Giulio Cerullo  and David G. Lidzey  and Tersilla Virgili },
title = {Superabsorption in an organic microcavity: Toward a quantum battery},
journal = {Science Advances},
volume = {8},
number = {2},
pages = {eabk3160},
year = {2022},
doi = {10.1126/sciadv.abk3160},
URL = {https://www.science.org/doi/abs/10.1126/sciadv.abk3160}
}

@misc{gribben2024boundarytimecrystalsac,
      title={Boundary Time Crystals as AC sensors: enhancements and constraints}, 
      author={Dominic Gribben and Anna Sanpera and Rosario Fazio and Jamir Marino and Fernando Iemini},
      year={2024},
      eprint={2406.06273},
      archivePrefix={arXiv},
      primaryClass={quant-ph},
      url={https://arxiv.org/abs/2406.06273}, 
}

@article{deffner2019qtddevices,
  title = {Quantum thermodynamic devices: From theoretical proposals to experimental reality},
  author={Myers, Nathan M. and Abah, Obinna and Deffner, Sebastian},
  journal = {AVS Quantum Science},
  volume = {4},
  number = {2},
  pages = {027101},
  year = {2022},
  doi = {10.1116/5.0083192},
  url = {https://doi.org/10.1116/5.0083192}
}

@book{binder2019bookqtd,
  title={Thermodynamics in the Quantum Regime: Fundamental Aspects and New Directions},
  author={Binder, F. and Correa, L.A. and Gogolin, C. and Anders, J. and Adesso, G.},
  isbn={9783319990460},
  series={Fundamental Theories of Physics},
  url={https://books.google.com.br/books?id=5uWPDwAAQBAJ},
  year={2019},
  publisher={Springer International Publishing},
    doi={10.1007/978-3-319-99046-0}
}

@article{mukherjee2024promises,
  title={The promises and challenges of many-body quantum technologies: A focus on quantum engines},
  author={Mukherjee, Victor and Divakaran, Uma},
  journal={Nature Communications},
  volume={15},
  number={1},
  pages={3170},
  year={2024},
  url={https://doi.org/10.1038/s41467-024-47638-1},
  doi={10.1038/s41467-024-47638-1},
  publisher={Nature Publishing Group UK London}
}

@article{solfanelli2025universal,
  title = {Universal Work Statistics in Long-Range Interacting Quantum Systems},
  author = {Solfanelli, Andrea and Defenu, Nicol\`o},
  journal = {Phys. Rev. Lett.},
  volume = {134},
  issue = {3},
  pages = {030402},
  numpages = {7},
  year = {2025},
  month = {Jan},
  publisher = {American Physical Society},
  doi = {10.1103/PhysRevLett.134.030402},
  url = {https://link.aps.org/doi/10.1103/PhysRevLett.134.030402}
}

@article{popescu2006entanglement,
  title={Entanglement and the foundations of statistical mechanics},
  author={Popescu, Sandu and Short, Anthony J and Winter, Andreas},
  journal={Nature Physics},
  volume={2},
  number={11},
  pages={754--758},
  year={2006},
  publisher={Nature Publishing Group UK London},
  url={https://doi.org/10.1038/nphys444},
  doi={10.1038/nphys444}
}

@article{GoldsteinPRL2006,
  title = {Canonical Typicality},
  author = {Goldstein, Sheldon and Lebowitz, Joel L. and Tumulka, Roderich and Zangh\`{\i}, Nino},
  journal = {Phys. Rev. Lett.},
  volume = {96},
  issue = {5},
  pages = {050403},
  numpages = {3},
  year = {2006},
  month = {Feb},
  publisher = {American Physical Society},
  doi = {10.1103/PhysRevLett.96.050403},
  url = {https://link.aps.org/doi/10.1103/PhysRevLett.96.050403}
}

@article{d2016quantum,
  title={From quantum chaos and eigenstate thermalization to statistical mechanics and thermodynamics},
  author={D'Alessio, Luca and Kafri, Yariv and Polkovnikov, Anatoli and Rigol, Marcos},
  journal={Advances in Physics},
  volume={65},
  number={3},
  pages={239--362},
  year={2016},
  url={https://doi.org/10.1080/00018732.2016.1198134},
  doi={10.1080/00018732.2016.1198134},
  publisher={Taylor \& Francis}
}

@article{mi2024stable,
  title={Stable quantum-correlated many-body states through engineered dissipation},
  author={Mi, Xiao and Michailidis, AA and Shabani, Sara and Miao, KC and Klimov, PV and Lloyd, J and Rosenberg, E and Acharya, R and Aleiner, I and Andersen, TI and others},
  journal={Science},
  volume={383},
  number={6689},
  pages={1332--1337},
  year={2024},
  url={https://doi.org/10.1126/science.adh9932},
doi={10.1126/science.adh9932},
  publisher={American Association for the Advancement of Science}
}

@article{scully2003extracting,
  title={Extracting work from a single heat bath via vanishing quantum coherence},
  author={Scully, Marlan O and Zubairy, M Suhail and Agarwal, Girish S and Walther, Herbert},
  journal={Science},
  volume={299},
  number={5608},
  pages={862--864},
  year={2003},
  url={https://doi.org/10.1126/science.1078955},
  doi={10.1126/science.1078955},
  publisher={American Association for the Advancement of Science}
}

@article{PhysRevX.4.041048,
  title = {Long-time Behavior of Isolated Periodically Driven Interacting Lattice Systems},
  author = {D'Alessio, Luca and Rigol, Marcos},
  journal = {Phys. Rev. X},
  volume = {4},
  issue = {4},
  pages = {041048},
  numpages = {12},
  year = {2014},
  month = {Dec},
  publisher = {American Physical Society},
  doi = {10.1103/PhysRevX.4.041048},
  url = {https://link.aps.org/doi/10.1103/PhysRevX.4.041048}
}

@article{PhysRevLett.126.103401,
  title = {Preparation of Low Entropy Correlated Many-Body States via Conformal Cooling Quenches},
  author = {Zaletel, Michael P. and Kaufman, Adam and Stamper-Kurn, Dan M. and Yao, Norman Y.},
  journal = {Phys. Rev. Lett.},
  volume = {126},
  issue = {10},
  pages = {103401},
  numpages = {7},
  year = {2021},
  month = {Mar},
  publisher = {American Physical Society},
  doi = {10.1103/PhysRevLett.126.103401},
  url = {https://link.aps.org/doi/10.1103/PhysRevLett.126.103401}
}

@article{PhysRevApplied.21.044050,
  title = {Floquet analysis of a superradiant many-qutrit refrigerator},
  author = {Kolisnyk, Dmytro and Quei\ss{}er, Friedemann and Schaller, Gernot and Sch\"utzhold, Ralf},
  journal = {Phys. Rev. Appl.},
  volume = {21},
  issue = {4},
  pages = {044050},
  numpages = {24},
  year = {2024},
  month = {Apr},
  publisher = {American Physical Society},
  doi = {10.1103/PhysRevApplied.21.044050},
  url = {https://link.aps.org/doi/10.1103/PhysRevApplied.21.044050}
}

@article{PhysRevE.90.012110,
  title = {Equilibrium states of generic quantum systems subject to periodic driving},
  author = {Lazarides, Achilleas and Das, Arnab and Moessner, Roderich},
  journal = {Phys. Rev. E},
  volume = {90},
  issue = {1},
  pages = {012110},
  numpages = {6},
  year = {2014},
  month = {Jul},
  publisher = {American Physical Society},
  doi = {10.1103/PhysRevE.90.012110},
  url = {https://link.aps.org/doi/10.1103/PhysRevE.90.012110}
}

@article{PRXQuantum.5.010201,
  title = {Scrambling Dynamics and Out-of-Time-Ordered Correlators in Quantum Many-Body Systems},
  author = {Xu, Shenglong and Swingle, Brian},
  journal = {PRX Quantum},
  volume = {5},
  issue = {1},
  pages = {010201},
  numpages = {46},
  year = {2024},
  month = {Jan},
  publisher = {American Physical Society},
  doi = {10.1103/PRXQuantum.5.010201},
  url = {https://link.aps.org/doi/10.1103/PRXQuantum.5.010201}
}

@article{Serbyn_2021,
   title={Quantum many-body scars and weak breaking of ergodicity},
   volume={17},
   ISSN={1745-2481},
   url={http://dx.doi.org/10.1038/s41567-021-01230-2},
   DOI={10.1038/s41567-021-01230-2},
   number={6},
   journal={Nature Physics},
   publisher={Springer Science and Business Media LLC},
   author={Serbyn, Maksym and Abanin, Dmitry A. and Papić, Zlatko},
   year={2021},
   month=may, pages={675–685} }

@article{PhysRevLett.119.030601,
  title = {Systematic Construction of Counterexamples to the Eigenstate Thermalization Hypothesis},
  author = {Shiraishi, Naoto and Mori, Takashi},
  journal = {Phys. Rev. Lett.},
  volume = {119},
  issue = {3},
  pages = {030601},
  numpages = {6},
  year = {2017},
  month = {Jul},
  publisher = {American Physical Society},
  doi = {10.1103/PhysRevLett.119.030601},
  url = {https://link.aps.org/doi/10.1103/PhysRevLett.119.030601}
}

@article{PhysRevE.90.052105,
  title = {Testing whether all eigenstates obey the eigenstate thermalization hypothesis},
  author = {Kim, Hyungwon and Ikeda, Tatsuhiko N. and Huse, David A.},
  journal = {Phys. Rev. E},
  volume = {90},
  issue = {5},
  pages = {052105},
  numpages = {8},
  year = {2014},
  month = {Nov},
  publisher = {American Physical Society},
  doi = {10.1103/PhysRevE.90.052105},
  url = {https://link.aps.org/doi/10.1103/PhysRevE.90.052105}
}

@article{odonovan2024quantummasterequationeigenstate,
      title={Quantum master equation from the eigenstate thermalization hypothesis}, 
      author={Peter O'Donovan and Philipp Strasberg and Kavan Modi and John Goold and Mark T. Mitchison},
      year={2024},
      eprint={2411.07706},
      archivePrefix={arXiv},
      primaryClass={quant-ph},
      url={https://arxiv.org/abs/2411.07706}, 
}

@article{PhysRevLett.119.150602,
  title = {How a Small Quantum Bath Can Thermalize Long Localized Chains},
  author = {Luitz, David J. and Huveneers, Fran\ifmmode \mbox{\c{c}}\else \c{c}\fi{}ois and De Roeck, Wojciech},
  journal = {Phys. Rev. Lett.},
  volume = {119},
  issue = {15},
  pages = {150602},
  numpages = {6},
  year = {2017},
  month = {Oct},
  publisher = {American Physical Society},
  doi = {10.1103/PhysRevLett.119.150602},
  url = {https://link.aps.org/doi/10.1103/PhysRevLett.119.150602}
}

@article{PhysRevLett.131.220405,
  title = {Stochastic Thermodynamics of a Quantum Dot Coupled to a Finite-Size Reservoir},
  author = {Moreira, Saulo V. and Samuelsson, Peter and Potts, Patrick P.},
  journal = {Phys. Rev. Lett.},
  volume = {131},
  issue = {22},
  pages = {220405},
  numpages = {7},
  year = {2023},
  month = {Dec},
  publisher = {American Physical Society},
  doi = {10.1103/PhysRevLett.131.220405},
  url = {https://link.aps.org/doi/10.1103/PhysRevLett.131.220405}
}

@article{PhysRevE.92.022104,
  title = {Quantum heat baths satisfying the eigenstate thermalization hypothesis},
  author = {Fialko, O.},
  journal = {Phys. Rev. E},
  volume = {92},
  issue = {2},
  pages = {022104},
  numpages = {5},
  year = {2015},
  month = {Aug},
  publisher = {American Physical Society},
  doi = {10.1103/PhysRevE.92.022104},
  url = {https://link.aps.org/doi/10.1103/PhysRevE.92.022104}
}

@article{berry1977regular,
  title={Regular and irregular semiclassical wavefunctions},
  author={Berry, Michael V},
  journal={Journal of Physics A: Mathematical and General},
  volume={10},
  number={12},
  pages={2083},
  year={1977},
  doi={10.1088/0305-4470/10/12/016},
  url={https://doi.org/10.1088/0305-4470/10/12/016},
  publisher={IOP Publishing}
}

@article{Gogolin_2016,
   title={Equilibration, thermalisation, and the emergence of statistical mechanics in closed quantum systems},
   volume={79},
   ISSN={1361-6633},
   url={http://dx.doi.org/10.1088/0034-4885/79/5/056001},
   DOI={10.1088/0034-4885/79/5/056001},
   number={5},
   journal={Reports on Progress in Physics},
   publisher={IOP Publishing},
   author={Gogolin, Christian and Eisert, Jens},
   year={2016},
   month=apr, pages={056001} }

@article{Srednicki_1996,
   title={Thermal fluctuations in quantized chaotic systems},
   volume={29},
   ISSN={1361-6447},
   url={http://dx.doi.org/10.1088/0305-4470/29/4/003},
   DOI={10.1088/0305-4470/29/4/003},
   number={4},
   journal={Journal of Physics A: Mathematical and General},
   publisher={IOP Publishing},
   author={Srednicki, Mark},
   year={1996},
   month=feb, pages={L75–L79} }

@article{rigol2008thermalization,
  title={Thermalization and its mechanism for generic isolated quantum systems},
  author={Rigol, Marcos and Dunjko, Vanja and Olshanii, Maxim},
  journal={Nature},
  volume={452},
  number={7189},
  pages={854--858},
  year={2008},
  url={https://doi.org/10.1038/nature06838},
  doi={10.1038/nature06838},
  publisher={Nature Publishing Group UK London}
}

@article{Kumasaki2025,
title = {Thermodynamic approach to quantum cooling limit of continuous Gaussian feedback},
author = {Kumasaki, Kousuke and Yada, Toshihiro and Funo, Ken and Sagawa, Takahiro},
journal = {Phys. Rev. Res.},
volume = {7},
issue = {4},
pages = {043147},
numpages = {15},
year = {2025},
month = {Nov},
publisher = {American Physical Society},
doi = {10.1103/5cz5-n6jt},
url = {https://link.aps.org/doi/10.1103/5cz5-n6jt}
}

@book{munnig_schmidt_design_2014,
	title = {The Design of High Performance Mechatronics: High-Tech Functionality by Multidisciplinary System Integration},
	publisher = {Delft University Press},
	author = {R. M. Schmidt and G. Schitter and A. Rankers and J. van Eijk},
	year = {2014},
}

@book{bechhoefer,
	title = {Control Theory for Physicists},
	publisher = {Cambridge University Press},
	author = {J. Bechhoefer},
	year = {2021}
}

@book{wiseman_quantum_2009,
	title = {Quantum {Measurement} and {Control}},
	publisher = {Cambridge University Press},
	author = {H. M. Wiseman and G. J. Milburn},
	year = {2009},
}

@article{magrini_real-time_2021,
	title = {Real-time optimal quantum control of mechanical motion at room temperature},
	volume = {595},
	doi = {10.1038/s41586-021-03602-3},
	number = {7867},
	journal = {Nature},
	author = {L. Magrini and P. Rosenzweig and C. Bach and A. Deutschmann-Olek and S. Hofer and S. Hong and N. Kiesel and A. Kugi and M. Aspelmeyer},
	year = {2021},
	pages = {373--377},
}

@article{
Review_Levitation,
author = {C. Gonzalez-Ballestero  and M. Aspelmeyer  and L. Novotny  and R. Quidant  and O. Romero-Isart },
title = {Levitodynamics: Levitation and control of microscopic objects in vacuum},
journal = {Science},
volume = {374},
number = {6564},
pages = {eabg3027},
year = {2021},
doi = {10.1126/science.abg3027},
}

@article{ferri-cortes_conditional_2025,
  title = {Conditional fluctuation theorems and entropy production for monitored quantum systems under imperfect detection},
  volume = {7},
  doi = {10.1103/PhysRevResearch.7.013077},
  number = {1},
  journal = {Phys. Rev. Research},
  author = {Mar Ferri-Cortés and José A. Almanza-Marrero and Rosa López and Roberta Zambrini and Gonzalo Manzano},
  year = {2025},
  pages = {013077}
}

@article{RevModPhys,
  title = {Cavity optomechanics},
  author = {M. Aspelmeyer and T. J. Kippenberg and F. Marquardt},
  journal = {Rev. Mod. Phys.},
  volume = {86},
  issue = {4},
  pages = {1391--1452},
  numpages = {62},
  year = {2014},
  month = {Dec},
  publisher = {American Physical Society},
  doi = {10.1103/RevModPhys.86.1391},
  url = {https://link.aps.org/doi/10.1103/RevModPhys.86.1391}
}

@article{jordan_measurement_2020,
  title = {Quantum measurement engines and their relevance for quantum interpretations},
  volume = {7},
  doi = {10.1007/s40509-019-00217-2},
  journal = {Quantum Stud.: Math. Found.},
  author = {Andrew N. Jordan and Cyril Elouard and Alexia Auffèves},
  year = {2020},
  pages = {203–215}
}

@article{SagawaUeda,
  title = {Minimal Energy Cost for Thermodynamic Information Processing: Measurement and Information Erasure},
  author = {T. Sagawa and M. Ueda},
  journal = {Phys. Rev. Lett.},
  volume = {102},
  issue = {25},
  pages = {250602},
  numpages = {4},
  year = {2009},
  month = {Jun},
  publisher = {American Physical Society},
  doi = {10.1103/PhysRevLett.102.250602},
  url = {https://link.aps.org/doi/10.1103/PhysRevLett.102.250602}
}

@article{engtanglement,
      title={Steady-state entanglement of interacting masses in free space through optimal feedback control}, 
      author={K. Winkler and A. V. Zasedatelev and B. A. Stickler and U. Delić and A. Deutschmann-Olek and M. Aspelmeyer},
      year={2024},
      eprint={2408.07492},
      archivePrefix={arXiv},
      primaryClass={quant-ph},
      url={https://arxiv.org/abs/2408.07492}, 
}

@article{RossiSchliesser,
      title={Measurement-based quantum control of mechanical motion}, 
  author = {M. Rossi and D. Mason and J. Chen and A. Tsaturyan and A. Schliesser},
  journal = {Nature},
  volume = {563},
  pages = {53},
  numpages = {4},
  year = {2018},
  month = {Jun},
  doi = {doi.org/10.1038/s41586-018-0643-8},
  url = {https://doi.org/10.1038/s41586-018-0643-8}
}

@article{sayrin_real-time_2011,
	title = {Real-time quantum feedback prepares and stabilizes photon number states},
	volume = {477},
	number = {7362},
	journal = {Nature},
	author = {Sayrin, C. and Dotsenko, I. and Zhou, X. and Peaudecerf, B. and Rybarczyk, T. and Gleyzes, S. and Rouchon, P. and Mirrahimi, M. and Amini, H. and Brune, M. and Raimond, J.-M. and Haroche, S.},
	year = {2011},
	pages = {73--77},
  url={https://doi.org/10.1038/nature10376},
doi={10.1038/nature10376}
}

@article{Debiossac,
      title={Non-Markovian Feedback Control and Acausality: An Experimental Study}, 
  author = {M. Debiossac and M. L. Rosinberg and E. Lutz and N. Kiesel},
  journal = {Phys. Rev. Lett.},
  volume = {128},
  pages = {200601},
  year = {2022},
 doi = {10.1103/PhysRevLett.128.200601},
 url={https://doi.org/10.1103/PhysRevLett.128.200601}
}

@article{gieseler_millen_levitated_2018,
    title = {Levitated nanoparticles for microscopic thermodynamics—A review},
    volume = {20},
    doi = {10.3390/e20050326},
    number = {5},
    journal = {Entropy},
    author = {J. Gieseler and J. Millen},
    year = {2018},
    pages = {326},
}

@article{Belavkin,
      title={Optimal filtering of markov signals with quantum white noise}, 
      author={V. P. Belavkin},
      year={1980},
  journal = {Radio Eng. Electron. Phys. (USSR)},
  volume = {25},
  pages = {1445},
  url={https://doi.org/10.1007/978-1-4899-1391-3_37},
  doi={10.1007/978-1-4899-1391-3_37}
}

@article{omran_generation_2019,
	title = {Generation and manipulation of {Schrödinger} cat states in {Rydberg} atom arrays},
	volume = {365},
	doi = {10.1126/science.aax9743},
	number = {6453},
	journal = {Science},
	author = {A. Omran and H. Levine and A. Keesling and G. Semeghini and T. T. Wang and S. Ebadi and H. Bernien and A. S. Zibrov and H. Pichler and S. Choi and J. Cui and M. Rossignolo and P. Rembold and S. Montangero and T. Calarco and M. Endres and M. Greiner and V. Vuletić and M. D. Lukin},
	year = {2019},
	pages = {570--574},
}

@article{tebbenjohanns_quantum_2021,
    title = {Quantum control of a nanoparticle optically levitated in cryogenic free space},
    volume = {595},
    doi = {10.1038/s41586-021-03617-w},
    number = {7867},
    journal = {Nature},
    author = {F. Tebbenjohanns and M. L. Mattana and M. Rossi and M. Frimmer and L. Novotny},
    year = {2021},
    pages = {378--382},
}

@article{tourchette,
	title = {Information-Theoretic Limits of Control},
	volume = {84},
        pages = {1156},
	doi = {doi.org/10.1103/PhysRevLett.84.1156},
	journal = {Phys. Rev. Lett.},
	author = {H. Touchette and S. Lloyd},
	year = {2009},
}

@article{Yamazaki2025Sep,
	author = {Yamazaki, Hikaru and Uemura, Masashi and Tanaka, Haruhi and Hata, Tokuro and Lin, Chaojing and Akiho, Takafumi and Muraki, Koji and Fujisawa, Toshimasa},
	title = {{Efficient heat-energy conversion from a non-thermal Tomonaga-Luttinger liquid}},
	journal = {Commun. Phys.},
	volume = {8},
	number = {387},
	pages = {1--10},
	year = {2025},
	month = sep,
	issn = {2399-3650},
	publisher = {Nature Publishing Group},
	doi = {10.1038/s42005-025-02297-6}
}

@article{Monsel2025Jan,
	author = {Monsel, Juliette and Acciai, Matteo and S{\ifmmode\acute{a}\else\'{a}\fi}nchez, Rafael and Splettstoesser, Janine},
	title = {{Autonomous demon exploiting heat and information at the trajectory level}},
	journal = {Phys. Rev. B},
	volume = {111},
	number = {4},
	pages = {045419},
	year = {2025},
	month = jan,
	publisher = {American Physical Society},
	doi = {10.1103/PhysRevB.111.045419}
}

@article{Benenti2017Jun,
	author = {Benenti, Giuliano and Casati, Giulio and Saito, Keiji and Whitney, Robert S.},
	title = {{Fundamental aspects of steady-state conversion of heat to work at the nanoscale}},
	journal = {Phys. Rep.},
	volume = {694},
	pages = {1--124},
	year = {2017},
	month = jun,
	issn = {0370-1573},
	publisher = {North-Holland},
	doi = {10.1016/j.physrep.2017.05.008}
}

@article{Parrondo2015Feb,
	author = {Parrondo, Juan M. R. and Horowitz, Jordan M. and Sagawa, Takahiro},
	title = {{Thermodynamics of information}},
	journal = {Nat. Phys.},
	volume = {11},
	pages = {131--139},
	year = {2015},
	month = feb,
	issn = {1745-2481},
	publisher = {Nature Publishing Group},
	doi = {10.1038/nphys3230}
}

@article{Sanchez2019Nov,
	author = {S{\ifmmode\acute{a}\else\'{a}\fi}nchez, Rafael and Splettstoesser, Janine and Whitney, Robert S.},
	title = {{Nonequilibrium System as a Demon}},
	journal = {Phys. Rev. Lett.},
	volume = {123},
	number = {21},
	pages = {216801},
	year = {2019},
	month = nov,
	publisher = {American Physical Society},
	doi = {10.1103/PhysRevLett.123.216801}
}

@article{Heremans2013Jul,
	author = {Heremans, Joseph P. and Dresselhaus, Mildred S. and Bell, Lon E. and Morelli, Donald T.},
	title = {{When thermoelectrics reached the nanoscale}},
	journal = {Nat. Nanotechnol.},
	volume = {8},
	number = {7},
	pages = {471--473},
	year = {2013},
	month = jul,
	issn = {1748-3395},
	publisher = {Nature Publishing Group},
	doi = {10.1038/nnano.2013.129}
}

@article{Josefsson2018Oct,
	author = {Josefsson, Martin and Svilans, Artis and Burke, Adam M. and Hoffmann, Eric A. and Fahlvik, Sofia and Thelander, Claes and Leijnse, Martin and Linke, Heiner},
	title = {{A quantum-dot heat engine operating close to the thermodynamic efficiency limits}},
	journal = {Nat. Nanotechnol.},
	volume = {13},
	number = {10},
	pages = {920--924},
	year = {2018},
	month = oct,
	issn = {1748-3395},
	publisher = {Nature Publishing Group},
	doi = {10.1038/s41565-018-0200-5}
}

@article{Thierschmann2015Oct,
	author = {Thierschmann, Holger and S{\ifmmode\acute{a}\else\'{a}\fi}nchez, Rafael and Sothmann, Bj{\ifmmode\ddot{o}\else\"{o}\fi}rn and Arnold, Fabian and Heyn, Christian and Hansen, Wolfgang and Buhmann, Hartmut and Molenkamp, Laurens W.},
	title = {{Three-terminal energy harvester with coupled quantum dots}},
	journal = {Nat. Nanotechnol.},
	volume = {10},
	number = {10},
	pages = {854--858},
	year = {2015},
	month = oct,
	issn = {1748-3395},
	publisher = {Nature Publishing Group},
	doi = {10.1038/nnano.2015.176}
}

@Article{Sivre2018,
author={Sivre, E.
and Anthore, A.
and Parmentier, F. D.
and Cavanna, A.
and Gennser, U.
and Ouerghi, A.
and Jin, Y.
and Pierre, F.},
title={{Heat Coulomb blockade of one ballistic channel}},
journal={Nature Physics},
year={2018},
month={Feb},
day={01},
volume={14},
number={2},
pages={145-148},
issn={1745-2481},
doi={10.1038/nphys4280},
url={https://doi.org/10.1038/nphys4280}
}

@article{Giazotto2006,
  title = {Opportunities for mesoscopics in thermometry and refrigeration: Physics and applications},
  author = {Giazotto, Francesco and Heikkil\"a, Tero T. and Luukanen, Arttu and Savin, Alexander M. and Pekola, Jukka P.},
  journal = {Rev. Mod. Phys.},
  volume = {78},
  issue = {1},
  pages = {217--274},
  numpages = {0},
  year = {2006},
  month = {Mar},
  publisher = {American Physical Society},
  doi = {10.1103/RevModPhys.78.217},
  url = {https://link.aps.org/doi/10.1103/RevModPhys.78.217}
}

@article{Chiaracane2020Jan,
	author = {Chiaracane, Cecilia and Mitchison, Mark T. and Purkayastha, Archak and Haack, G{\ifmmode\acute{e}\else\'{e}\fi}raldine and Goold, John},
	title = {{Quasiperiodic quantum heat engines with a mobility edge}},
	journal = {Phys. Rev. Res.},
	volume = {2},
	number = {1},
	pages = {013093},
	year = {2020},
	month = jan,
	publisher = {American Physical Society},
	doi = {10.1103/PhysRevResearch.2.013093}
}

@article{Carrega2024Jun,
	author = {Carrega, Matteo and Razzoli, Luca and Erdman, Paolo Andrea and Cavaliere, Fabio and Benenti, Giuliano and Sassetti, Maura},
	title = {{Dissipation-induced collective advantage of a quantum thermal machine}},
	journal = {AVS Quantum Sci.},
	volume = {6},
	number = {2},
	year = {2024},
        pages={025001},
	month = jun,
	publisher = {AIP Publishing},
	doi = {10.1116/5.0190340},
        url={https://doi.org/10.1116/5.0190340}
}

@article{Tesser2024May,
	author = {Tesser, Ludovico and Splettstoesser, Janine},
	title = {{Out-of-Equilibrium Fluctuation-Dissipation Bounds}},
	journal = {Phys. Rev. Lett.},
	volume = {132},
	number = {18},
	pages = {186304},
	year = {2024},
	month = may,
	publisher = {American Physical Society},
	doi = {10.1103/PhysRevLett.132.186304}
}

@article{Dutta2019Jan,
	author = {Dutta, Bivas and Majidi, Danial and Garc{\ifmmode\acute{\imath}\else\'{\i}\fi}a Corral, Alvaro and Erdman, Paolo A. and Florens, Serge and Costi, Theo A. and Courtois, Herv{\ifmmode\acute{e}\else\'{e}\fi} and Winkelmann, Clemens B.},
	title = {{Direct Probe of the Seebeck Coefficient in a Kondo-Correlated Single-Quantum-Dot Transistor}},
	journal = {Nano Lett.},
	volume = {19},
	number = {1},
	pages = {506--511},
	year = {2019},
	month = jan,
	issn = {1530-6984},
	publisher = {American Chemical Society},
	doi = {10.1021/acs.nanolett.8b04398}
}

@article{Pietzonka2018May,
	author = {Pietzonka, Patrick and Seifert, Udo},
	title = {{Universal Trade-Off between Power, Efficiency, and Constancy in Steady-State Heat Engines}},
	journal = {Phys. Rev. Lett.},
	volume = {120},
	number = {19},
	pages = {190602},
	year = {2018},
	month = may,
	publisher = {American Physical Society},
	doi = {10.1103/PhysRevLett.120.190602}
}

@Article{Karimi2020NatComms,
author={Karimi, Bayan
and Brange, Fredrik
and Samuelsson, Peter
and Pekola, Jukka P.},
title={Reaching the ultimate energy resolution of a quantum detector},
journal={Nature Communications},
year={2020},
month={Jan},
day={17},
volume={11},
number={1},
pages={367},
issn={2041-1723},
doi={10.1038/s41467-019-14247-2},
url={https://doi.org/10.1038/s41467-019-14247-2}
}

@article{MailletPRL2019,
  title = {Optimal Probabilistic Work Extraction beyond the Free Energy Difference with a Single-Electron Device},
  author = {Maillet, Olivier and Erdman, Paolo A. and Cavina, Vasco and Bhandari, Bibek and Mannila, Elsa T. and Peltonen, Joonas T. and Mari, Andrea and Taddei, Fabio and Jarzynski, Christopher and Giovannetti, Vittorio and Pekola, Jukka P.},
  journal = {Phys. Rev. Lett.},
  volume = {122},
  issue = {15},
  pages = {150604},
  numpages = {6},
  year = {2019},
  month = {Apr},
  publisher = {American Physical Society},
  doi = {10.1103/PhysRevLett.122.150604},
  url = {https://link.aps.org/doi/10.1103/PhysRevLett.122.150604}
}

@article{kleeorin2019measure,
  title={How to measure the entropy of a mesoscopic system via thermoelectric transport},
  author={Kleeorin, Yaakov and Thierschmann, Holger and Buhmann, Hartmut and Georges, Antoine and Molenkamp, Laurens W and Meir, Yigal},
  journal={Nature {C}ommunications},
  volume={10},
  number={1},
  pages={5801},
  year={2019},
  publisher={Nature Publishing Group UK London},
  url={https://doi.org/10.1038/s41467-019-13630-3}
}

@Article{Pekola-Hekking,
  author    = {Pekola, J. P. and Hekking, F. W. J.},
  journal   = {Phys. Rev. Lett.},
  title     = {Normal-Metal-Superconductor Tunnel Junction as a Brownian Refrigerator},
  year      = {2007},
  month     = {May},
  pages     = {210604},
  volume    = {98},
  doi       = {10.1103/PhysRevLett.98.210604},
  issue     = {21},
  numpages  = {4},
  publisher = {American Physical Society},
  url       = {https://link.aps.org/doi/10.1103/PhysRevLett.98.210604},
}

@Article{PK-heatbath,
  author         = {Pekola, Jukka P. and Karimi, Bayan},
  journal        = {Entropy},
  title          = {Heat Bath in a Quantum Circuit},
  year           = {2024},
  issn           = {1099-4300},
  number         = {5},
  volume         = {26},
  pages = {429},
  doi            = {10.3390/e26050429},
  url            = {https://www.mdpi.com/1099-4300/26/5/429},
}

@Article{PK-cQTD,
  author    = {Pekola, Jukka P. and Karimi, Bayan},
  journal   = {Rev. Mod. Phys.},
  title     = {Colloquium: Quantum heat transport in condensed matter systems},
  year      = {2021},
  month     = {Oct},
  pages     = {041001},
  volume    = {93},
  doi       = {10.1103/RevModPhys.93.041001},
  issue     = {4},
  numpages  = {25},
  publisher = {American Physical Society},
  url       = {https://link.aps.org/doi/10.1103/RevModPhys.93.041001},
}

@Article{Pekola-Averin2010,
  author    = {Averin, Dmitri V. and Pekola, Jukka P.},
  journal   = {Phys. Rev. Lett.},
  title     = {Violation of the Fluctuation-Dissipation Theorem in Time-Dependent Mesoscopic Heat Transport},
  year      = {2010},
  month     = {Jun},
  pages     = {220601},
  volume    = {104},
  doi       = {10.1103/PhysRevLett.104.220601},
  issue     = {22},
  numpages  = {4},
  publisher = {American Physical Society},
  url       = {https://link.aps.org/doi/10.1103/PhysRevLett.104.220601},
}

@Article{SHEVCHENKO20101,
  author   = {S.N. Shevchenko and S. Ashhab and Franco Nori},
  journal  = {Physics Reports},
  title    = {Landau–Zener–Stückelberg interferometry},
  year     = {2010},
  issn     = {0370-1573},
  number   = {1},
  pages    = {1-30},
  volume   = {492},
  doi      = {https://doi.org/10.1016/j.physrep.2010.03.002},
  url      = {https://www.sciencedirect.com/science/article/pii/S0370157310000815},
}

@Article{KoskiPekola2014,
  author   = {Jonne V. Koski and Ville F. Maisi and Jukka P. Pekola and Dmitri V. Averin},
  journal  = {Proc. Natl. Acad. Sci. U.S.A},
  title    = {Experimental realization of a Szilard engine with a single electron},
  year     = {2014},
  number   = {38},
  pages    = {13786-13789},
  volume   = {111},
  doi      = {10.1073/pnas.1406966111},
  url      = {https://www.pnas.org/doi/abs/10.1073/pnas.1406966111}
}

@Article{Nakamura1999,
  author  = {Y. Nakamura and Yu. A. Pashkin and J. S. Tsai},
  journal = {Nature},
  title   = {Coherent control of macroscopic quantum states in a single-Cooper-pair box},
  year    = {1999},
  pages   = {786–788},
  volume  = {398},
  doi     = {https://doi.org/10.1038/19718},
  url     = {https://doi.org/10.1038/19718},
}

@Article{Landauer1961,
  author  = {Landauer, R.},
  journal = {IBM Journal of Research and Development},
  title   = {Irreversibility and Heat Generation in the Computing Process},
  year    = {1961},
  number  = {3},
  pages   = {183-191},
  volume  = {5},
  doi     = {10.1147/rd.53.0183},
}

@Article{Pendry1983,
  author  = {J. Pendry},
  journal = {Journal of Physics A},
  title   = {Quantum limits to the flow of information and entropy},
  year    = {1983},
  pages   = {2161-2171},
  volume  = {16},
  doi     = {10.1088/0305-4470/16/10/012},
}

@Article{Luca2025,
  author    = {Magazzù, L and Paladino, E and Grifoni, M},
  journal   = {Quantum Science and Technology},
  title     = {Heat transport in the quantum Rabi model: universality and ultrastrong coupling effects},
  year      = {2025},
  month     = {feb},
  number    = {2},
  pages     = {025013},
  volume    = {10},
  doi       = {10.1088/2058-9565/adae2c},
  publisher = {IOP Publishing},
  url       = {https://dx.doi.org/10.1088/2058-9565/adae2c},
}

@Article{Saito2013,
  author    = {Saito, Keiji and Kato, Takeo},
  journal   = {Phys. Rev. Lett.},
  title     = {Kondo Signature in Heat Transfer via a Local Two-State System},
  year      = {2013},
  month     = {Nov},
  pages     = {214301},
  volume    = {111},
  doi       = {10.1103/PhysRevLett.111.214301},
  issue     = {21},
  numpages  = {5},
  publisher = {American Physical Society},
  url       = {https://link.aps.org/doi/10.1103/PhysRevLett.111.214301},
}

@article{Mikko2025,
         title = {Experimental realization of a quantum heat engine based on dissipation-engineered superconducting circuits},
         author = {Tuomas Uusnäkki and Timm Mörstedt and Wallace Teixeira and Miika Rasola and Mikko Möttönen },
         year = {2025},
         eprint = {2502.20143v1},
         archivePrefix = {arXiv},
         primaryClass ={quant-ph}
        }

@Article{Mori2018,
  author        = {Mori, Takashi and Ikeda, Tatsuhiko N. and Kaminishi, Eriko and Ueda, Masahito},
  journal       = {J. Phys. B},
  title         = {{Thermalization and prethermalization in isolated quantum systems: a theoretical overview}},
  year          = {2018},
  number        = {11},
  pages         = {112001},
  volume        = {51},
  doi           = {10.1088/1361-6455/aabcdf}
}

@Article{Giazotto2012,
  author  = {Francesco Giazotto and María José Martínez-Pérez},
  journal = {Nature},
  title   = {The Josephson heat interferometer},
  year    = {2012},
  pages   = {401–405},
  volume  = {492},
  doi     = {https://doi.org/10.1038/nature11702},
  url     = {https://doi.org/10.1038/nature11702},
}

@Article{Gasparinetti2025,
  author  = {M. A. Aamir and P. J. Suria and J. A. M. Guzmán and C. Castillo-Moreno and J. M. Epstein and N. Yunger Halpern and S. Gasparinetti},
  journal = {Nat. Phys.},
  title   = {Thermally driven quantum refrigerator autonomously resets a superconducting qubit},
  year    = {2025},
  pages   = {318–323},
  volume  = {21},
  doi     = {https://doi.org/10.1038/s41567-024-02708-5},
  url     = {https://doi.org/10.1038/s41567-024-02708-5},
}

@Article{Dicke1954,
  author    = {Dicke, R. H.},
  journal   = {Phys. Rev.},
  title     = {Coherence in Spontaneous Radiation Processes},
  year      = {1954},
  month     = {Jan},
  pages     = {99--110},
  volume    = {93},
  doi       = {10.1103/PhysRev.93.99},
  issue     = {1},
  numpages  = {0},
  publisher = {American Physical Society},
  url       = {https://link.aps.org/doi/10.1103/PhysRev.93.99},
}

@Article{Koch2022,
author={Koch, Christiane P.
and Boscain, Ugo
and Calarco, Tommaso
and Dirr, Gunther
and Filipp, Stefan
and Glaser, Steffen J.
and Kosloff, Ronnie
and Montangero, Simone
and Schulte-Herbr{\"u}ggen, Thomas
and Sugny, Dominique
and Wilhelm, Frank K.},
title={Quantum optimal control in quantum technologies. Strategic report on current status, visions and goals for research in Europe},
journal={EPJ Quantum Technology},
year={2022},
month={Jul},
day={20},
volume={9},
number={1},
pages={19},
issn={2196-0763},
doi={10.1140/epjqt/s40507-022-00138-x},
url={https://doi.org/10.1140/epjqt/s40507-022-00138-x}
}

@Article{Chen2019,
author={Chen, Yang-Yang
and Watanabe, Gentaro
and Yu, Yi-Cong
and Guan, Xi-Wen
and del Campo, Adolfo},
title={An interaction-driven many-particle quantum heat engine and its universal behavior},
journal={npj Quantum Information},
year={2019},
month={Oct},
day={17},
volume={5},
number={1},
pages={88},
issn={2056-6387},
doi={10.1038/s41534-019-0204-5},
url={https://doi.org/10.1038/s41534-019-0204-5}
}

@article{Jarzynski24,
  title = {Theory of Quantum Super Impulses},
  author = {Jarzynski, Christopher},
  journal = {PRX Quantum},
  volume = {5},
  issue = {1},
  pages = {010322},
  numpages = {18},
  year = {2024},
  month = {Feb},
  publisher = {American Physical Society},
  doi = {10.1103/PRXQuantum.5.010322},
  url = {https://link.aps.org/doi/10.1103/PRXQuantum.5.010322}
}

@article{Dorfman13,
author = {Konstantin E. Dorfman  and Dmitri V. Voronine  and Shaul Mukamel  and Marlan O. Scully },
title = {Photosynthetic reaction center as a quantum heat engine},
journal = {Proceedings of the National Academy of Sciences},
volume = {110},
number = {8},
pages = {2746-2751},
year = {2013},
doi = {10.1073/pnas.1212666110},
URL = {https://www.pnas.org/doi/abs/10.1073/pnas.1212666110}}

@article{Lipka-Bartosik24,
author = {Patryk Lipka-Bartosik  and Martí Perarnau-Llobet  and Nicolas Brunner},
title = {Thermodynamic computing via autonomous quantum thermal machines},
journal = {Science Advances},
volume = {10},
number = {36},
pages = {eadm8792},
year = {2024},
doi = {10.1126/sciadv.adm8792}}

@article{Takahashi24,
  title = {Shortcuts to Adiabaticity in Krylov Space},
  author = {Takahashi, Kazutaka and del Campo, Adolfo},
  journal = {Phys. Rev. X},
  volume = {14},
  issue = {1},
  pages = {011032},
  numpages = {23},
  year = {2024},
  month = {Feb},
  publisher = {American Physical Society},
  doi = {10.1103/PhysRevX.14.011032},
  url = {https://link.aps.org/doi/10.1103/PhysRevX.14.011032}
}

@Article{Yin2022,
author={Yin, Zelong
and Li, Chunzhen
and Allcock, Jonathan
and Zheng, Yicong
and Gu, Xiu
and Dai, Maochun
and Zhang, Shengyu
and An, Shuoming},
title={Shortcuts to adiabaticity for open systems in circuit quantum electrodynamics},
journal={Nature Communications},
year={2022},
month={Jan},
day={10},
volume={13},
number={1},
pages={188},
issn={2041-1723},
doi={10.1038/s41467-021-27900-6},
url={https://doi.org/10.1038/s41467-021-27900-6}
}

@article{Kallush22,
author = {Shimshon Kallush  and Roie Dann  and Ronnie Kosloff },
title = {Controlling the uncontrollable: Quantum control of open-system dynamics},
journal = {Science Advances},
volume = {8},
number = {44},
pages = {eadd0828},
year = {2022},
doi = {10.1126/sciadv.add0828},
URL = {https://www.science.org/doi/abs/10.1126/sciadv.add0828},
}

@article{Hou2024,
         title = {An energy efficient quantum-enhanced machine},
         author = {Waner Hou and Xingyu Zhao and Kamran Rehan and Yi Li and Yue Li and Eric Lutz and Yiheng Lin and Jiangfeng Du },
         year = {2024},
         eprint = {2404.15075v1},
         archivePrefix = {arXiv},
         primaryClass ={quant-ph}
        }

@article{Vacanti2014,
doi = {10.1088/1367-2630/16/5/053017},
url = {https://dx.doi.org/10.1088/1367-2630/16/5/053017},
year = {2014},
month = {may},
publisher = {IOP Publishing},
volume = {16},
number = {5},
pages = {053017},
author = {G. Vacanti and R. Fazio and S. Montangero and G. M. Palma and M. Paternostro and V. Vedral},
title = {Transitionless quantum driving in open quantum systems},
journal = {New Journal of Physics}
}

@article{Alipour20,
  doi = {10.22331/q-2020-09-28-336},
  url = {https://doi.org/10.22331/q-2020-09-28-336},
  title = {Shortcuts to {A}diabaticity in {D}riven {O}pen {Q}uantum {S}ystems: {B}alanced {G}ain and {L}oss and {N}on-{M}arkovian {E}volution},
  author = {Alipour, Sahar and Chenu, Aurelia and Rezakhani, Ali T. and del Campo, Adolfo},
  journal = {{Quantum}},
  issn = {2521-327X},
  publisher = {{Verein zur F{\"{o}}rderung des Open Access Publizierens in den Quantenwissenschaften}},
  volume = {4},
  pages = {336},
  month = sep,
  year = {2020}
}

@article{delcampo13,
  title = {Shortcuts to Adiabaticity by Counterdiabatic Driving},
  author = {del Campo, Adolfo},
  journal = {Phys. Rev. Lett.},
  volume = {111},
  issue = {10},
  pages = {100502},
  numpages = {5},
  year = {2013},
  month = {Sep},
  publisher = {American Physical Society},
  doi = {10.1103/PhysRevLett.111.100502},
  url = {https://link.aps.org/doi/10.1103/PhysRevLett.111.100502}
}

@article{GueryOdelin19,
  title = {Shortcuts to adiabaticity: Concepts, methods, and applications},
  author = {Gu\'ery-Odelin, D. and Ruschhaupt, A. and Kiely, A. and Torrontegui, E. and Mart\'{\i}nez-Garaot, S. and Muga, J. G.},
  journal = {Rev. Mod. Phys.},
  volume = {91},
  issue = {4},
  pages = {045001},
  numpages = {54},
  year = {2019},
  month = {Oct},
  publisher = {American Physical Society},
  doi = {10.1103/RevModPhys.91.045001},
  url = {https://link.aps.org/doi/10.1103/RevModPhys.91.045001}
}

@Inbook{delCampo2018,
author="del Campo, Adolfo
and Chenu, Aur{\'e}lia
and Deng, Shujin
and Wu, Haibin",
editor="Binder, Felix
and Correa, Luis A.
and Gogolin, Christian
and Anders, Janet
and Adesso, Gerardo",
title="Friction-Free Quantum Machines",
bookTitle="Thermodynamics in the Quantum Regime: Fundamental Aspects and New Directions",
year="2018",
publisher="Springer International Publishing",
address="Cham",
pages="127--148",
isbn="978-3-319-99046-0",
doi="10.1007/978-3-319-99046-0_5",
url="https://doi.org/10.1007/978-3-319-99046-0_5"
}

@Article{Kosloff17,
AUTHOR = {Kosloff, Ronnie and Rezek, Yair},
TITLE = {The Quantum Harmonic Otto Cycle},
JOURNAL = {Entropy},
VOLUME = {19},
YEAR = {2017},
NUMBER = {4},
pages = {136},
URL = {https://www.mdpi.com/1099-4300/19/4/136},
ISSN = {1099-4300},
DOI = {10.3390/e19040136}
}

@article{GarciaPintos22,
  title = {Unifying Quantum and Classical Speed Limits on Observables},
  author = {Garc\'{\i}a-Pintos, Luis Pedro and Nicholson, Schuyler B. and Green, Jason R. and del Campo, Adolfo and Gorshkov, Alexey V.},
  journal = {Phys. Rev. X},
  volume = {12},
  issue = {1},
  pages = {011038},
  numpages = {22},
  year = {2022},
  month = {Feb},
  publisher = {American Physical Society},
  doi = {10.1103/PhysRevX.12.011038},
  url = {https://link.aps.org/doi/10.1103/PhysRevX.12.011038}
}

@article{olf2015thermometry,
  title={Thermometry and cooling of a Bose gas to 0.02 times the condensation temperature},
  author={Olf, Ryan and Fang, Fang and Marti, G Edward and MacRae, Andrew and Stamper-Kurn, Dan M},
  journal={Nat. Phys.},
  volume={11},
  number={9},
  pages={720--723},
  year={2015},
  publisher={Nature Publishing Group UK London},
  doi = {10.1038/nphys3408}
}

@article{rubio2024first,
  title={First-principles construction of symmetry-informed quantum metrologies},
  author={Rubio, Jes{\'u}s},
  journal={Phys. Rev. A},
  volume={110},
  number={3},
  pages={L030401},
  year={2024},
  publisher={APS},
  doi = {10.1103/PhysRevA.110.L030401}
}

@article{hovhannisyan2018measuring,
  title={Measuring the temperature of cold many-body quantum systems},
  author={Hovhannisyan, Karen V and Correa, Luis A},
  journal={Physical Review B},
  volume={98},
  number={4},
  pages={045101},
  year={2018},
  publisher={APS}
}

@article{potts2019fundamental,
  title={Fundamental limits on low-temperature quantum thermometry with finite resolution},
  author={Potts, Patrick P and Brask, Jonatan Bohr and Brunner, Nicolas},
  journal={Quantum},
  volume={3},
  pages={161},
  year={2019},
  publisher={Verein zur F{\"o}rderung des Open Access Publizierens in den Quantenwissenschaften},
  doi = {10.22331/q-2019-07-09-161}
}

@article{PhysRevA.105.042601,
  title = {Bayesian quantum thermometry based on thermodynamic length},
  author = {J\o{}rgensen, Mathias R. and Ko\l{}ody\ifmmode \acute{n}\else \'{n}\fi{}ski, Jan and Mehboudi, Mohammad and Perarnau-Llobet, Mart\'{\i} and Brask, Jonatan B.},
  journal = {Phys. Rev. A},
  volume = {105},
  issue = {4},
  pages = {042601},
  numpages = {11},
  year = {2022},
  month = {Apr},
  publisher = {American Physical Society},
  doi = {10.1103/PhysRevA.105.042601},
  url = {https://link.aps.org/doi/10.1103/PhysRevA.105.042601}
}

@article{miller2018energy,
  title={Energy-temperature uncertainty relation in quantum thermodynamics},
  author={Miller, Harry JD and Anders, Janet},
  journal={Nat. Commun.},
  volume={9},
  number={1},
  pages={2203},
  year={2018},
  publisher={Nature Publishing Group UK London},
doi={10.1038/s41467-018-04536-7}
}

@article{PhysRevA.104.052214,
  title = {Uninformed Bayesian quantum thermometry},
  author = {Boeyens, Julia and Seah, Stella and Nimmrichter, Stefan},
  journal = {Phys. Rev. A},
  volume = {104},
  issue = {5},
  pages = {052214},
  numpages = {11},
  year = {2021},
  month = {Nov},
  publisher = {American Physical Society},
  doi = {10.1103/PhysRevA.104.052214},
  url = {https://link.aps.org/doi/10.1103/PhysRevA.104.052214}
}

@article{PhysRevApplied.22.064022,
  title = {Thermometry of trapped ions based on bichromatic driving},
  author = {Li, Xie-Qian and Tao, Yi and Chen, Ting and Wu, Wei and Xie, Yi and Wu, Chun-Wang and Chen, Ping-Xing},
  journal = {Phys. Rev. Appl.},
  volume = {22},
  issue = {6},
  pages = {064022},
  numpages = {10},
  year = {2024},
  month = {Dec},
  publisher = {American Physical Society},
  doi = {10.1103/PhysRevApplied.22.064022},
  url = {https://link.aps.org/doi/10.1103/PhysRevApplied.22.064022}
}

@article{lvov2024thermometry,
         title = {Thermometry Based on a Superconducting Qubit},
         author = {Dmitrii S. Lvov and Sergei A. Lemziakov and Elias Ankerhold and Joonas T. Peltonen and Jukka P. Pekola },
         year = {2024},
         eprint = {2409.02784v3},
         archivePrefix = {arXiv},
         primaryClass ={quant-ph}
        }

@article{PRXQuantum.3.040330,
  title = {Optimal Cold Atom Thermometry Using Adaptive Bayesian Strategies},
  author = {Glatthard, Jonas and Rubio, Jes\'us and Sawant, Rahul and Hewitt, Thomas and Barontini, Giovanni and Correa, Luis A.},
  journal = {PRX Quantum},
  volume = {3},
  issue = {4},
  pages = {040330},
  numpages = {15},
  year = {2022},
  month = {Dec},
  publisher = {American Physical Society},
  doi = {10.1103/PRXQuantum.3.040330},
  url = {https://link.aps.org/doi/10.1103/PRXQuantum.3.040330}
}

@article{PhysRevX.10.011018,
  title = {Single-Atom Quantum Probes for Ultracold Gases Boosted by Nonequilibrium Spin Dynamics},
  author = {Bouton, Quentin and Nettersheim, Jens and Adam, Daniel and Schmidt, Felix and Mayer, Daniel and Lausch, Tobias and Tiemann, Eberhard and Widera, Artur},
  journal = {Phys. Rev. X},
  volume = {10},
  issue = {1},
  pages = {011018},
  numpages = {13},
  year = {2020},
  month = {Jan},
  publisher = {American Physical Society},
  doi = {10.1103/PhysRevX.10.011018},
  url = {https://link.aps.org/doi/10.1103/PhysRevX.10.011018}
}

@article{Mehboudi_2019,
doi = {10.1088/1751-8121/ab2828},
url = {https://dx.doi.org/10.1088/1751-8121/ab2828},
year = {2019},
month = {jul},
publisher = {IOP Publishing},
volume = {52},
number = {30},
pages = {303001},
author = {Mohammad Mehboudi and Anna Sanpera and Luis A Correa},
title = {Thermometry in the quantum regime: recent theoretical progress},
journal = {J. Phys. A}
}

@article{PhysRevLett.114.220405,
  title = {Individual Quantum Probes for Optimal Thermometry},
  author = {Correa, Luis A. and Mehboudi, Mohammad and Adesso, Gerardo and Sanpera, Anna},
  journal = {Phys. Rev. Lett.},
  volume = {114},
  issue = {22},
  pages = {220405},
  numpages = {5},
  year = {2015},
  month = {Jun},
  publisher = {American Physical Society},
  doi = {10.1103/PhysRevLett.114.220405},
  url = {https://link.aps.org/doi/10.1103/PhysRevLett.114.220405}
}

@article{PRXQuantum.2.020322,
  title = {Optimal Quantum Thermometry with Coarse-Grained Measurements},
  author = {Hovhannisyan, Karen V. and J\o{}rgensen, Mathias R. and Landi, Gabriel T. and Alhambra, \'Alvaro M. and Brask, Jonatan B. and Perarnau-Llobet, Mart\'{\i}},
  journal = {PRX Quantum},
  volume = {2},
  issue = {2},
  pages = {020322},
  numpages = {28},
  year = {2021},
  month = {May},
  publisher = {American Physical Society},
  doi = {10.1103/PRXQuantum.2.020322},
  url = {https://link.aps.org/doi/10.1103/PRXQuantum.2.020322}
}

@article{bilokur2024thermodynamic,
         title = {Thermodynamic limitations on fault-tolerant quantum computing},
         author = {Mykhailo Bilokur and Sarang Gopalakrishnan and Shayan Majidy },
         year = {2024},
         eprint = {2411.12805v2},
         archivePrefix = {arXiv},
         primaryClass ={quant-ph}
}

@article{bassman2022computing,
  title={Computing free energies with fluctuation relations on quantum computers},
  author={Bassman Oftelie, Lindsay and Klymko, Katherine and Liu, Diyi and Tubman, Norm M and de Jong, Wibe A},
  journal={Physical review letters},
  volume={129},
  number={13},
  pages={130603},
  year={2022},
  url = {https://doi.org/10.1103/PhysRevLett.129.130603},
  doi = {10.1103/PhysRevLett.129.130603},
  publisher={APS}
}

@article{bassman2024dynamic,
  title={Dynamic Cooling on Contemporary Quantum Computers},
  author={Bassman Oftelie, Lindsay and De Pasquale, Antonella and Campisi, Michele},
  journal={PRX Quantum},
  volume={5},
  number={3},
  pages={030309},
  year={2024},
  url={https://doi.org/10.1103/PRXQuantum.5.030309},
  doi={10.1103/PRXQuantum.5.030309},
  publisher={APS}
}

@article{oftelie2024measurement, 
title={Measurement of the work statistics of an open quantum system using a quantum computer}, volume={10}, 
url={http://dx.doi.org/10.1088/2058-9565/adbd6c}, 
DOI={10.1088/2058-9565/adbd6c}, 
number={2}, journal={Quantum Science and Technology}, 
publisher={IOP Publishing}, 
author={Bassman Oftelie, Lindsay and Campisi, Michele}, 
year={2025}, 
pages={025045} 
}

@article{Solfanelli21PRXQ2,
	author = {Solfanelli, Andrea and Santini, Alessandro and Campisi, Michele},
	doi = {10.1103/PRXQuantum.2.030353},
	journal = {PRX Quantum},
	pages = {030353},
	title = {Experimental Verification of Fluctuation Relations with a Quantum Computer},
	volume = {2},
	year = {2021}}

@article{buffoni2024collective,
  title={Collective preparation of large quantum registers with high fidelity},
  author={Buffoni, Lorenzo and Campisi, Michele},
  journal={Quantum Science and Technology},
  volume={10},
  number={2},
  pages={025053},
  year={2025},
  publisher={IOP Publishing},
  url={https://doi.org/10.1088/2058-9565/adc3bb},
  doi={10.1088/2058-9565/adc3bb}
}

@article{cattaneo2023quantum,
  title={Quantum simulation of dissipative collective effects on noisy quantum computers},
  author={Cattaneo, Marco and Rossi, Matteo AC and Garc{\'\i}a-P{\'e}rez, Guillermo and Zambrini, Roberta and Maniscalco, Sabrina},
  journal={PRX quantum},
  volume={4},
  number={1},
  pages={010324},
  year={2023},
  url={https://doi.org/10.1103/PRXQuantum.4.010324},
  doi={10.1103/PRXQuantum.4.010324},
  publisher={APS}
}

@article{cech2023thermodynamics,
  title={Thermodynamics of quantum trajectories on a quantum computer},
  author={Cech, Marcel and Lesanovsky, Igor and Carollo, Federico},
  journal={Physical Review Letters},
  volume={131},
  number={12},
  pages={120401},
  year={2023},
  url={https://doi.org/10.1103/PhysRevLett.131.120401},
  doi={10.1103/PhysRevLett.131.120401},
  publisher={APS}
}

@article{chiribella2022nonequilibrium,
  title={The nonequilibrium cost of accurate information processing},
  author={Chiribella, Giulio and Meng, Fei and Renner, Renato and Yung, Man-Hong},
  journal={Nature Communications},
  volume={13},
  number={1},
  pages={7155},
  year={2022},
  url={https://doi.org/10.1038/s41467-022-34541-w},
  doi={10.1038/s41467-022-34541-w},
  publisher={Nature Publishing Group UK London}
}

@article{elyasi2024experimental,
  title={Experimental simulation of daemonic work extraction in open quantum batteries on a digital quantum computer},
  author={Elyasi, Seyed Navid and Rossi, Matteo and Genoni, Marco G},
  journal={Quantum Science and Technology},
  volume={10},
  pages = {025017},
  url= {https://doi.org/10.1088/2058-9565/adae2d},
  doi= {10.1088/2058-9565/adae2d},
  year={2025}
}

@article{harrington2022engineered,
  title={Engineered dissipation for quantum information science},
  author={Harrington, Patrick M and Mueller, Erich J and Murch, Kater W},
  journal={Nature Reviews Physics},
  volume={4},
  number={10},
  pages={660--671},
  year={2022},
  url={https://doi.org/10.1038/s42254-022-00494-8},
  doi={10.1038/s42254-022-00494-8},
  publisher={Nature Publishing Group UK London}
}

@article{melo2022implementation,
  title={Implementation of a two-stroke quantum heat engine with a collisional model},
  author={Melo, Filipe V and S{\'a}, Nahum and Roditi, Itzhak and Souza, Alexandre M and Oliveira, Ivan S and Sarthour, Roberto S and Landi, Gabriel T},
  journal={Physical Review A},
  volume={106},
  number={3},
  pages={032410},
  year={2022},
  url = {https://doi.org/10.1103/PhysRevA.106.032410},
  doi={10.1103/PhysRevA.106.032410},
  publisher={APS}
}

@article{razzoli2025cyclic,
  title={Cyclic solid-state quantum battery: Thermodynamic characterization and quantum hardware simulation},
  author={Razzoli, Luca and Gemme, Giulia and Khomchenko, Ilia and Sassetti, Maura and Ouerdane, Henni and Ferraro, Dario and Benenti, Giuliano},
  journal={Quantum Sci. Technol.},
  volume={10},
  number={1},
  pages={015064},
  year={2025},
  url={https://doi.org/10.1088/2058-9565/ad9ed4},
  doi={10.1088/2058-9565/ad9ed4},
  publisher={IOP Publishing}
}

@article{smierzchalski2024efficiency,
  title={Efficiency optimization in quantum computing: balancing thermodynamics and computational performance},
  author={{\'S}mierzchalski, Tomasz and Mzaouali, Zakaria and Deffner, Sebastian and Gardas, Bart{\l}omiej},
  journal={Scientific Reports},
  volume={14},
  number={1},
  pages={4555},
  year={2024},
  url={https://doi.org/10.1038/s41598-024-55314-z},
  doi={10.1038/s41598-024-55314-z},
  publisher={Nature Publishing Group UK London}
}

@article{yang2023simulating,
  title={Simulating prethermalization using near-term quantum computers},
  author={Yang, Yilun and Christianen, Arthur and Coll-Vinent, Sandra and Smelyanskiy, Vadim and Ba{\~n}uls, Mari Carmen and O’Brien, Thomas E and Wild, Dominik S and Cirac, J Ignacio},
  journal={PRX Quantum},
  volume={4},
  number={3},
  pages={030320},
  year={2023},
  url = {https://doi.org/10.1103/PRXQuantum.4.030320},
  doi={10.1103/PRXQuantum.4.030320},
  publisher={APS}
}

@article{field1993measurements,
  title={Measurements of {C}oulomb blockade with a noninvasive voltage probe},
  author={Field, M and Smith, CG and Pepper, Michael and Ritchie, DA and Frost, JEF and Jones, GAC and Hasko, DG},
  journal={Physical {R}eview {L}etters},
  volume={70},
  number={9},
  pages={1311},
  year={1993},
  publisher={APS},
  doi = {10.1103/PhysRevLett.70.1311},
  url = {https://doi.org/10.1103/PhysRevLett.70.1311}
}

@article{adam2024entropy,
  title={Entropy spectroscopy of a bilayer graphene quantum dot},
  author={Adam, Christoph and Duprez, Hadrien and Lehmann, Natalie and Yglesias, Antoni and Cances, Solenn and Ruckriegel, Max Josef and Masseroni, Michele and Tong, Chuyao and Denisov, Artem Olegovich and Huang, Wei Wister and others},
  journal={ar{X}iv preprint ar{X}iv:2412.18000},
  year={2024},
      url={https://arxiv.org/abs/2412.18000}
}

@article{gustavsson2006counting,
  title={Counting statistics of single electron transport in a quantum dot},
  author={Gustavsson, S and Leturcq, R and Simovi{\v{c}}, B and Schleser, R and Ihn, T and Studerus, P and Ensslin, K and Driscoll, DC and Gossard, AC},
  journal={Physical {R}eview {L}etters},
  volume={96},
  number={7},
  pages={076605},
  year={2006},
  publisher={APS},
  doi = {10.1103/PhysRevLett.96.076605},
  url = {https://doi.org/10.1103/PhysRevLett.96.076605}
}

@article{piquard2023observing,
  title={Observing the universal screening of a {K}ondo impurity},
  author={Piquard, C and Glidic, P and Han, C and Aassime, A and Cavanna, A and Gennser, U and Meir, Y and Sela, E and Anthore, A and Pierre, F},
  journal={Nature {C}ommunications},
  volume={14},
  number={1},
  pages={7263},
  year={2023},
  publisher={Nature {P}ublishing {G}roup {UK} {L}ondon},
  doi = {10.1038/s41467-023-42857-4},
  url = {https://doi.org/10.1038/s41467-023-42857-4}
}

@article{bayer2025real,
  title={Real-time detection and control of correlated charge tunneling in a quantum dot},
  author={Bayer, Johannes C and Brange, Fredrik and Schmidt, Adrian and Wagner, Timo and Rugeramigabo, Eddy P and Flindt, Christian and Haug, Rolf J},
  journal={Physical {R}eview {L}etters},
  volume={134},
  number={4},
  pages={046303},
  year={2025},
  publisher={APS},
  doi = {10.1103/PhysRevLett.134.046303},
  url = {https://doi.org/10.1103/PhysRevLett.134.046303}
}

@article{hofmann2016equilibrium,
  title={Equilibrium free energy measurement of a confined electron driven out of equilibrium},
  author={Hofmann, Andrea and Maisi, Ville F and R{\"o}ssler, Clemens and Basset, Julien and Kr{\"a}henmann, Tobias and M{\"a}rki, Peter and Ihn, Thomas and Ensslin, Klaus and Reichl, Christian and Wegscheider, Werner},
  journal={Physical {R}eview {B}},
  volume={93},
  number={3},
  pages={035425},
  year={2016},
  publisher={APS},
  doi = {10.1103/PhysRevB.93.035425},
  url = {https://doi.org/10.1103/PhysRevB.93.035425}
}

@article{hofmann2017heat,
  title={Heat dissipation and fluctuations in a driven quantum dot},
  author={Hofmann, Andrea and Maisi, Ville F and Basset, Julien and Reichl, Christian and Wegscheider, Werner and Ihn, Thomas and Ensslin, Klaus and Jarzynski, Christopher},
  journal={physica status solidi (b)},
  volume={254},
  number={3},
  pages={1600546},
  year={2017},
  doi = {10.1002/pssb.201600546},
  url = {https://doi.org/10.1002/pssb.201600546}
}

@article{kung2012irreversibility,
  title={Irreversibility on the level of single-electron tunneling},
  author={K{\"u}ng, Bruno and R{\"o}ssler, Clemens and Beck, Mattias and Marthaler, M and Golubev, DS and Utsumi, Yasuhiro and Ihn, T and Ensslin, Klaus},
  journal={Physical {R}eview {X}},
  volume={2},
  number={1},
  pages={011001},
  year={2012},
  publisher={APS},
  doi = {10.1103/PhysRevX.2.011001},
  url = {https://doi.org/10.1103/PhysRevX.2.011001}
}

@article{koski2013distribution,
  title={Distribution of entropy production in a single-electron box},
  author={Koski, JV and Sagawa, T and Saira, OP and Yoon, Y and Kutvonen, A and Solinas, P and M{\"o}tt{\"o}nen, M and Ala-Nissila, T and Pekola, JP},
  journal={Nature {P}hysics},
  volume={9},
  number={10},
  pages={644},
  year={2013},
  publisher={Nature {P}ublishing {G}roup {UK} {L}ondon},
  doi = {10.1038/nphys2711},
  url = {https://doi.org/10.1038/nphys2711}
}

@article{saira2012test,
  title={Test of the {J}arzynski and {C}rooks fluctuation relations in an electronic system},
  author={Saira, O-P and Yoon, Y and Tanttu, T and M{\"o}tt{\"o}nen, Mikko and Averin, DV and Pekola, Jukka P},
  journal={Physical {R}eview {L}etters},
  volume={109},
  number={18},
  pages={180601},
  year={2012},
  publisher={APS},
  doi = {10.1103/PhysRevLett.109.180601},
  url = {https://doi.org/10.1103/PhysRevLett.109.180601}
}

@article{ma2025quantum,
  title={Quantum work statistics across a critical point: full crossover from sudden quench to the adiabatic limit},
  author={Ma, Zhanyu and Mitchell, Andrew K and Sela, Eran},
  journal={Physical {R}eview {L}etters},
  volume={135},
  number={13},
  pages={130402},
  year={2025},
  publisher={APS},
  url={https://doi.org/10.1103/vn83-mt2v}
}

@article{han2024quantum,
  title={Quantum limitation on experimental testing of nonequilibrium fluctuation theorems},
  author={Han, Cheolhee and Cohen, Doron and Sela, Eran},
  journal={Physical {R}eview {B}},
  volume={110},
  number={11},
  pages={115153},
  year={2024},
  publisher={APS},
  doi = {10.1103/PhysRevB.110.115153},
  url = {https://doi.org/10.1103/PhysRevB.110.115153}
}

@article{annby2020maxwell,
  title={Maxwell's demon in a double quantum dot with continuous charge detection},
  author={Annby-Andersson, Bj{\"o}rn and Samuelsson, Peter and Maisi, Ville F and Potts, Patrick P},
  journal={Physical {R}eview {B}},
  volume={101},
  number={16},
  pages={165404},
  year={2020},
  publisher={APS},
  doi = {10.1103/PhysRevB.101.165404},
  url = {https://doi.org/10.1103/PhysRevB.101.165404}
}

@article{barker2022experimental,
  title={Experimental verification of the work fluctuation-dissipation relation for information-to-work conversion},
  author={Barker, David and Scandi, Matteo and Lehmann, Sebastian and Thelander, Claes and Dick, Kimberly A and Perarnau-Llobet, Mart{\'\i} and Maisi, Ville F},
  journal={Physical {R}eview {L}etters},
  volume={128},
  number={4},
  pages={040602},
  year={2022},
  publisher={APS},
  doi = {10.1103/PhysRevLett.128.040602},
  url = {https://doi.org/10.1103/PhysRevLett.128.040602}
}

@article{scandi2022minimally,
  title={Minimally dissipative information erasure in a quantum dot via thermodynamic length},
  author={Scandi, Matteo and Barker, David and Lehmann, Sebastian and Dick, Kimberly A and Maisi, Ville F and Perarnau-Llobet, Mart{\'\i}},
  journal={Physical {R}eview {L}etters},
  volume={129},
  number={27},
  pages={270601},
  year={2022},
  publisher={APS},
  doi = {10.1103/PhysRevLett.129.270601},
  url = {https://doi.org/10.1103/PhysRevLett.129.270601}
}

@article{hartman2018direct,
  title={Direct entropy measurement in a mesoscopic quantum system},
  author={Hartman, Nikolaus and Olsen, Christian and L{\"u}scher, Silvia and Samani, Mohammad and Fallahi, Saeed and Gardner, Geoffrey C and Manfra, Michael and Folk, Joshua},
  journal={Nature {P}hysics},
  volume={14},
  number={11},
  pages={1083},
  year={2018},
  publisher={Nature {P}ublishing {G}roup {UK} {L}ondon},
  doi = {10.1038/s41567-018-0250-5},
  url = {https://doi.org/10.1038/s41567-018-0250-5}
}

@article{child2022entropy,
  title={Entropy measurement of a strongly coupled quantum dot},
  author={Child, Timothy and Sheekey, Owen and L{\"u}scher, Silvia and Fallahi, Saeed and Gardner, Geoffrey C and Manfra, Michael and Mitchell, Andrew and Sela, Eran and Kleeorin, Yaakov and Meir, Yigal and others},
  journal={Physical {R}eview {L}etters},
  volume={129},
  number={22},
  pages={227702},
  year={2022},
  publisher={APS},
  doi = {10.1103/PhysRevLett.129.227702},
  url = {https://doi.org/10.1103/PhysRevLett.129.227702}
}

@article{iftikhar2018tunable,
  title={Tunable quantum criticality and super-ballistic transport in a charge {K}ondo circuit},
  author={Iftikhar, Z and Anthore, A and Mitchell, AK and Parmentier, FD and Gennser, U and Ouerghi, A and Cavanna, A and Mora, C and Simon, P and Pierre, F},
  journal={Science},
  volume={360},
  number={6395},
  pages={1315},
  year={2018},
  publisher={American Association for the Advancement of Science},
  doi = {10.1126/science.aan5592},
  url = {https://doi.org/10.1126/science.aan5592}
}

@article{karki2023z,
  title={Z3 parafermion in the double charge {K}ondo model},
  author={Karki, DB and Boulat, Edouard and Pouse, Winston and Goldhaber-Gordon, David and Mitchell, Andrew K and Mora, Christophe},
  journal={Physical {R}eview {L}etters},
  volume={130},
  number={14},
  pages={146201},
  year={2023},
  publisher={APS},
  doi = {10.1103/PhysRevLett.130.146201},
  url = {https://doi.org/10.1103/PhysRevLett.130.146201}
}

@Article{Masanes2017,
author={Masanes, Llu{\'i}s
and Oppenheim, Jonathan},
title={A general derivation and quantification of the third law of thermodynamics},
journal={Nature Communications},
year={2017},
month={Mar},
day={14},
volume={8},
number={1},
pages={14538},
issn={2041-1723},
doi={10.1038/ncomms14538},
url={https://doi.org/10.1038/ncomms14538}
}

@article{Whitty_2020,
title = {Quantum control via enhanced shortcuts to adiabaticity},
author = {Whitty, C. and Kiely, A. and Ruschhaupt, A.},
journal = {Phys. Rev. Res.},
volume = {2},
issue = {2},
pages = {023360},
numpages = {9},
year = {2020},
month = {Jun},
publisher = {American Physical Society},
doi = {10.1103/PhysRevResearch.2.023360},
url = {https://link.aps.org/doi/10.1103/PhysRevResearch.2.023360}
}

@Article{Li2024,
author={Li, Xi
and Wang, Shuai
and Luo, Xiang
and Zhou, Yu-Yang
and Xie, Ke
and Shen, Hong-Chi
and Nie, Yu-Zhao
and Chen, Qijin
and Hu, Hui
and Chen, Yu-Ao
and Yao, Xing-Can
and Pan, Jian-Wei},
title={Observation and quantification of the pseudogap in unitary Fermi gases},
journal={Nature},
year={2024},
month={Feb},
day={01},
volume={626},
number={7998},
pages={288-293},
issn={1476-4687},
doi={10.1038/s41586-023-06964-y},
url={https://doi.org/10.1038/s41586-023-06964-y}
}

@article{Mitchison2020,
  title = "{In Situ Thermometry of a Cold Fermi Gas via Dephasing Impurities}",
  author = {Mitchison, M.T. and Fogarty, T. and Guarnieri, G. and Campbell, S. and Busch, Th. and Goold, J.},
  journal = {Phys. Rev. Lett.},
  volume = {125},
  issue = {8},
  pages = {080402},
  numpages = {7},
  year = {2020},
  month = {Aug},
  publisher = {American Physical Society},
  doi = {10.1103/PhysRevLett.125.080402},
  url = {https://link.aps.org/doi/10.1103/PhysRevLett.125.080402}
}

@article{Khan2022,
  title = "{Subnanokelvin thermometry of an interacting $d$-dimensional homogeneous Bose gas}",
  author = {Khan, M.M. and Mehboudi, M. and Ter\ifmmode \mbox{\c{c}}\else \c{c}\fi{}as, H. and Lewenstein, M. and Garcia-March, M.A.},
  journal = {Phys. Rev. Res.},
  volume = {4},
  issue = {2},
  pages = {023191},
  numpages = {12},
  year = {2022},
  month = {Jun},
  publisher = {American Physical Society},
  doi = {10.1103/PhysRevResearch.4.023191},
  url = {https://link.aps.org/doi/10.1103/PhysRevResearch.4.023191}
}

@article{flywheel2019,
  title = "{Spin Heat Engine Coupled to a Harmonic-Oscillator Flywheel}",
  author = {von Lindenfels, D. and Gr\"ab, O. and Schmiegelow, C. T. and Kaushal, V. and Schulz, J. and Mitchison, M.T. and Goold, J. and Schmidt-Kaler, F. and Poschinger, U. G.},
  journal = {Phys. Rev. Lett.},
  volume = {123},
  issue = {8},
  pages = {080602},
  numpages = {6},
  year = {2019},
  month = {Aug},
  publisher = {American Physical Society},
  doi = {10.1103/PhysRevLett.123.080602},
  url = {https://link.aps.org/doi/10.1103/PhysRevLett.123.080602}
}

@article{Brown_2016,
doi = {10.1088/1367-2630/18/11/113028},
url = {https://dx.doi.org/10.1088/1367-2630/18/11/113028},
year = {2016},
month = {nov},
publisher = {IOP Publishing},
volume = {18},
number = {11},
pages = {113028},
author = {Brown, E.G. and Friis, N. and Huber, M.},
title = "{Passivity and practical work extraction using Gaussian operations}",
journal = {New Journal of Physics}
}

@article{Dann2019,
  title = "{Shortcut to Equilibration of an Open Quantum System}",
  author = {Dann, R. and Tobalina, A. and Kosloff, R.},
  journal = {Phys. Rev. Lett.},
  volume = {122},
  issue = {25},
  pages = {250402},
  numpages = {6},
  year = {2019},
  month = {Jun},
  publisher = {American Physical Society},
  doi = {10.1103/PhysRevLett.122.250402},
  url = {https://link.aps.org/doi/10.1103/PhysRevLett.122.250402}
}

@article{Boubakour_2025,
doi = {10.1088/2058-9565/adbcce},
url = {https://dx.doi.org/10.1088/2058-9565/adbcce},
year = {2025},
month = {mar},
publisher = {IOP Publishing},
volume = {10},
number = {2},
pages = {025036},
author = {Boubakour, M. and Endo, S. and Fogarty, T. and Busch, Th.},
title = "{Dynamical invariant based shortcut to equilibration in open quantum systems}",
journal = {Quantum Science and Technology}
}

@article{COLD2023,
  title = "{Counterdiabatic Optimized Local Driving}",
  author = {\ifmmode \check{C}\else \v{C}\fi{}epait\ifmmode \dot{e}\else \.{e}\fi{}, I. and Polkovnikov, A. and Daley, A.J. and Duncan, C.W.},
  journal = {PRX Quantum},
  volume = {4},
  issue = {1},
  pages = {010312},
  numpages = {21},
  year = {2023},
  month = {Jan},
  publisher = {American Physical Society},
  doi = {10.1103/PRXQuantum.4.010312},
  url = {https://link.aps.org/doi/10.1103/PRXQuantum.4.010312}
}

@article{Simmons2023,
  title = "{Thermodynamic engine with a quantum degenerate working fluid}",
  author = {Simmons, E.Q. and Sajjad, R. and Keithley, K. and Mas, H. and Tanlimco, J.L. and Nolasco-Martinez, E. and Bai, Y. and Fredrickson, G.H. and Weld, D.M.},
  journal = {Phys. Rev. Res.},
  volume = {5},
  issue = {4},
  pages = {L042009},
  numpages = {8},
  year = {2023},
  month = {Oct},
  publisher = {American Physical Society},
  doi = {10.1103/PhysRevResearch.5.L042009},
  url = {https://link.aps.org/doi/10.1103/PhysRevResearch.5.L042009}
}

@article{Carollo2020,
  title = "{Nonequilibrium Quantum Many-Body Rydberg Atom Engine}",
  author = {Carollo, F. and Gambetta, F.M. and Brandner, K. and Garrahan, J.P. and Lesanovsky, I.},
  journal = {Phys. Rev. Lett.},
  volume = {124},
  issue = {17},
  pages = {170602},
  numpages = {7},
  year = {2020},
  month = {Apr},
  publisher = {American Physical Society},
  doi = {10.1103/PhysRevLett.124.170602},
  url = {https://link.aps.org/doi/10.1103/PhysRevLett.124.170602}
}

@article{Carollo_2020_2,
  title = {Nonequilibrium Many-Body Quantum Engine Driven by Time-Translation Symmetry Breaking},
  author = {Carollo, Federico and Brandner, Kay and Lesanovsky, Igor},
  journal = {Phys. Rev. Lett.},
  volume = {125},
  issue = {24},
  pages = {240602},
  numpages = {6},
  year = {2020},
  month = {Dec},
  publisher = {American Physical Society},
  doi = {10.1103/PhysRevLett.125.240602},
  url = {https://link.aps.org/doi/10.1103/PhysRevLett.125.240602}
}

@article{Adam2022,
  title = "{Coherent and Dephasing Spectroscopy for Single-Impurity Probing of an Ultracold Bath}",
  author = {Adam, D. and Bouton, Q. and Nettersheim, J. and Burgardt, S. and Widera, A.},
  journal = {Phys. Rev. Lett.},
  volume = {129},
  issue = {12},
  pages = {120404},
  numpages = {6},
  year = {2022},
  month = {Sep},
  publisher = {American Physical Society},
  doi = {10.1103/PhysRevLett.129.120404},
  url = {https://link.aps.org/doi/10.1103/PhysRevLett.129.120404}
}

@article{Cetina2016,
author = {M. Cetina  and M. Jag  and R.S. Lous  and I. Fritsche  and J.T.M. Walraven  and R. Grimm  and J. Levinsen  and M.M. Parish  and R. Schmidt  and M. Knap  and E. Demler },
title = "{Ultrafast many-body interferometry of impurities coupled to a Fermi sea}",
journal = {Science},
volume = {354},
number = {6308},
pages = {96-99},
year = {2016},
doi = {10.1126/science.aaf5134},
URL = {https://www.science.org/doi/abs/10.1126/science.aaf5134}}

@Article{Koch2023,
author={Koch, J.
and Menon, K.
and Cuestas, E.
and Barbosa, S.
and Lutz, E.
and Fogarty, T.
and Busch, Th.
and Widera, A.},
title="{A quantum engine in the BEC--BCS crossover}",
journal={Nature},
year={2023},
month={Sep},
day={01},
volume={621},
number={7980},
pages={723-727},
issn={1476-4687},
doi={10.1038/s41586-023-06469-8},
url={https://doi.org/10.1038/s41586-023-06469-8}
}

@Article{Bouton2021,
author={Bouton, Q.
and Nettersheim, J.
and Burgardt, S.
and Adam, D.
and Lutz, E.
and Widera, A.},
title="{A quantum heat engine driven by atomic collisions}",
journal={Nature Communications},
year={2021},
month={Apr},
day={06},
volume={12},
number={1},
pages={2063},
issn={2041-1723},
doi={10.1038/s41467-021-22222-z},
url={https://doi.org/10.1038/s41467-021-22222-z}
}

@article{Becher_2020,
  title = "{Measurement of Identical Particle Entanglement and the Influence of Antisymmetrization}",
  author = {Becher, J. H. and Sindici, E. and Klemt, R. and Jochim, S. and Daley, A. J. and Preiss, P. M.},
  journal = {Phys. Rev. Lett.},
  volume = {125},
  issue = {18},
  pages = {180402},
  numpages = {6},
  year = {2020},
  month = {Oct},
  publisher = {American Physical Society},
  doi = {10.1103/PhysRevLett.125.180402},
  url = {https://link.aps.org/doi/10.1103/PhysRevLett.125.180402}
}

@article{Nascimbene_2010,
doi = {10.1088/1367-2630/12/10/103026},
url = {https://dx.doi.org/10.1088/1367-2630/12/10/103026},
year = {2010},
month = {oct},
publisher = {},
volume = {12},
number = {10},
pages = {103026},
author = {Nascimb{\`e}ne, S. and Navon, N. and Chevy, F. and Salomon, C.},
title = "{The equation of state of ultracold Bose and Fermi gases: a few examples}",
journal = {New Journal of Physics}
}

@article{Auf23,
  title = {Quantum Technologies Need a Quantum Energy Initiative},
  author = {Auff\`eves, Alexia},
  journal = {PRX Quantum},
  volume = {3},
  issue = {2},
  pages = {020101},
  numpages = {12},
  year = {2022},
  month = {Jun},
  publisher = {American Physical Society},
  doi = {10.1103/PRXQuantum.3.020101},
  url = {https://link.aps.org/doi/10.1103/PRXQuantum.3.020101}
}

@article{ADA+24, title={Introduction to theoretical and experimental aspects of quantum optimal control}, volume={57}, ISSN={1361-6455}, url={http://dx.doi.org/10.1088/1361-6455/ad46a5}, DOI={10.1088/1361-6455/ad46a5}, number={13}, journal={J. Phys. B: At. Mol. Opt. Phys.}, publisher={IOP Publishing}, author={Ansel, Q and Dionis, E and Arrouas, F and Peaudecerf, B and Guérin, S and Guéry-Odelin, D and Sugny, D}, year={2024}, month=jun, pages={133001} }

@article{BCS+21,
  title = {Two-Qubit Engine Fueled by Entanglement and Local Measurements},
  author = {Bresque, L\'ea and Camati, Patrice A. and Rogers, Spencer and Murch, Kater and Jordan, Andrew N. and Auff\`eves, Alexia},
  journal = {Phys. Rev. Lett.},
  volume = {126},
  issue = {12},
  pages = {120605},
  numpages = {6},
  year = {2021},
  month = {Mar},
  publisher = {American Physical Society},
  doi = {10.1103/PhysRevLett.126.120605},
  url = {https://link.aps.org/doi/10.1103/PhysRevLett.126.120605}
}

@article{EHC+17, title={The role of quantum measurement in stochastic thermodynamics}, volume={3}, ISSN={2056-6387}, url={http://dx.doi.org/10.1038/s41534-017-0008-4}, DOI={10.1038/s41534-017-0008-4}, number={1}, journal={npj Quantum Information}, pages={9}, publisher={Springer Science and Business Media LLC}, author={Elouard, Cyril and Herrera-Martí, David A. and Clusel, Maxime and Auffèves, Alexia}, year={2017}, month=mar }

@article{EL23,
  title = {Extending the Laws of Thermodynamics for Arbitrary Autonomous Quantum Systems},
  author = {Elouard, Cyril and Lombard Latune, Camille},
  journal = {PRX Quantum},
  volume = {4},
  issue = {2},
  pages = {020309},
  numpages = {19},
  year = {2023},
  month = {Apr},
  publisher = {American Physical Society},
  doi = {10.1103/PRXQuantum.4.020309},
  url = {https://link.aps.org/doi/10.1103/PRXQuantum.4.020309}
}

@article{ERW+24,
         title = {Equilibration of objective observables in a dynamical model of quantum measurements},
         author = {Sophie Engineer and Tom Rivlin and Sabine Wollmann and Mehul Malik and Maximilian P. E. Lock },
         year = {2024},
         eprint = {2403.18016v1},
         archivePrefix = {arXiv},
         primaryClass ={quant-ph}
        }

@article{FAC+23,
  title = {Optimizing Resource Efficiencies for Scalable Full-Stack Quantum Computers},
  author = {Fellous-Asiani, Marco and Chai, Jing Hao and Thonnart, Yvain and Ng, Hui Khoon and Whitney, Robert S. and Auff\`eves, Alexia},
  journal = {PRX Quantum},
  volume = {4},
  issue = {4},
  pages = {040319},
  numpages = {40},
  year = {2023},
  month = {Oct},
  publisher = {American Physical Society},
  doi = {10.1103/PRXQuantum.4.040319},
  url = {https://link.aps.org/doi/10.1103/PRXQuantum.4.040319}
}

@article{Gea02,
  title = {Minimum Energy Requirements for Quantum Computation},
  author = {Gea-Banacloche, Julio},
  journal = {Phys. Rev. Lett.},
  volume = {89},
  issue = {21},
  pages = {217901},
  numpages = {4},
  year = {2002},
  month = {Nov},
  publisher = {American Physical Society},
  doi = {10.1103/PhysRevLett.89.217901},
  url = {https://link.aps.org/doi/10.1103/PhysRevLett.89.217901}
}

\end{document}